\documentclass[twocolumn, trackchanges]{aastex61}

\usepackage{amsmath}
\usepackage{graphicx}   
\usepackage{bm}         
\usepackage{natbib}
\usepackage{etoolbox}
\usepackage{rotating}
\usepackage{array}
\usepackage{booktabs}
\usepackage{amssymb}
\usepackage{CJK}

\makeatletter 
\patchcmd{\NAT@citex}
  {\@citea\NAT@hyper@{%
     \NAT@nmfmt{\NAT@nm}%
\hyper@natlinkbreak{\NAT@aysep\NAT@spacechar}{\@citeb\@extra@b@citeb}%
     \NAT@date}}
  {\@citea\NAT@nmfmt{\NAT@nm}%
   \NAT@aysep\NAT@spacechar\NAT@hyper@{\NAT@date}}{}{}

\patchcmd{\NAT@citex}
  {\@citea\NAT@hyper@{%
     \NAT@nmfmt{\NAT@nm}%
\hyper@natlinkbreak{\NAT@spacechar\NAT@@open\if*#1*\else#1\NAT@spacechar\fi}%
       {\@citeb\@extra@b@citeb}%
     \NAT@date}}
  {\@citea\NAT@nmfmt{\NAT@nm}%
\NAT@spacechar\NAT@@open\if*#1*\else#1\NAT@spacechar\fi\NAT@hyper@{\NAT@date}}
  {}{}
\makeatother

\newcolumntype{X}[1]{>{\begin{turn}{90}\begin{minipage}{#1}\scriptsize}l%
<{\end{minipage}\end{turn}}%
}

\newcolumntype{C}[1]{>{\centering\arraybackslash}p{#1}}

\newcommand{\lsun}{\mbox{L$_\odot$}}

\newcommand{\kms}{\mbox{km~s$^{-1}$}}

\newcommand{\lsmm}{\mbox{$L_{\rm smm}$}} 
\newcommand{\lbol}{\mbox{$L_{\rm bol}$}} 
\newcommand{\tbol}{\mbox{$T_{\rm bol}$}} 
\newcommand{\trot}{\mbox{$T_{\rm rot}$}} 
\newcommand{\ee}[1]{\mbox{${} \times 10^{#1}$}}

\newcommand{\water}{\mbox{H$_{2}$O}}
\newcommand{\OI}{\mbox{[\ion{O}{1}]}}
\newcommand{\CI}{\mbox{[\ion{C}{1}]}}
\newcommand{\CII}{\mbox{[\ion{C}{2}]}}
\newcommand{\NII}{\mbox{[\ion{N}{2}]}}

\newcommand{\jj}[2]{\mbox{$J = #1\rightarrow#2$}}

\newcommand{\Jup}{\mbox{$J_{\rm up}$}}

\shorttitle{COPS-SPIRE}
\shortauthors{Yang et al.}
\begin{document}

\title{CO in Protostars (COPS): \textit{Herschel}-SPIRE Spectroscopy of Embedded Protostars
\footnote{\textit{Herschel} is an
ESA space observatory with science instruments provided by European-led
Principal Investigator consortia and with important participation from NASA.}}

\author{Yao-Lun Yang}
\affiliation{The University of Texas at Austin, Department of Astronomy, 2515 Speedway, Stop C1400, Austin, TX 78712, USA}

\author{Joel D. Green}
\affiliation{Space Telescope Science Institute, 3700 San Martin Dr., Baltimore, MD 02138, USA}
\affiliation{The University of Texas at Austin, Department of Astronomy, 2515 Speedway, Stop C1400, Austin, TX 78712, USA}

\author{Neal J. Evans II}
\affiliation{The University of Texas at Austin, Department of Astronomy, 2515 Speedway, Stop C1400, Austin, TX 78712, USA}
\affiliation{Korea Astronomy and Space Science Institute, 776 Daedeokdae-ro, Yuseong-gu, Daejeon 34055, Korea}
\affiliation{Humanitas College, Global Campus, Kyung Hee University, Yongin-shi 17104, Korea}

\author{Jeong-Eun Lee}
\affiliation{Department of Astronomy \&\ Space Science, Kyung Hee University, Gyeonggi 446-701, Korea  \\
School of Space Research, Kyung Hee University, Yongin-shi, Kyungki-do 449-701, Korea}

\author{Jes K. J{\o}rgensen}
\affiliation{Centre for Star and Planet Formation, Niels Bohr Institute and Natural History Museum of Denmark, University of Copenhagen, {\O}ster Voldgade 5-7, DK-1350 Copenhagen K., Denmark}

\author{Lars E. Kristensen}
\affiliation{Centre for Star and Planet Formation, Niels Bohr Institute and Natural History Museum of Denmark, University of Copenhagen, {\O}ster Voldgade 5-7, DK-1350 Copenhagen K., Denmark}

\author{Joseph C. Mottram}
\affiliation{Max Planck Institute for Astronomy, K\"{o}nigstuhl 17, 69117 Heidelberg, Germany}

\author{Gregory Herczeg}
\affiliation{Kavli Institute for Astronomy and Astrophysics, Peking University, Yi He Yuan Lu 5, Haidian Qu, 100871 Beijing, China}

\author{Agata Karska}
\affiliation{Centre for Astronomy, Faculty of Physics, Astronomy and Informatics, Nicolaus Copernicus University, Grudziadzka 5, 87-100 Torun, Poland}

\author{Odysseas Dionatos}
\affiliation{Department of Astrophysics, University of Vienna, Tuerkenschanzstrasse 17, 1180 Vienna, Austria}

\author{Edwin A. Bergin}
\affiliation{Department of Astronomy, University of Michigan, 1085 S. University Avenue, Ann Arbor, MI 48109, USA}

\author{Jeroen Bouwman}
\affiliation{Max Planck Institute for Astronomy, K\"{o}nigstuhl 17, 69117 Heidelberg, Germany}

\author{Ewine F. van Dishoeck}
\affiliation{Leiden Observatory, Leiden University, Netherlands}
\affiliation{Max Planck Institute for Extraterrestrial Physics, Garching, Germany}

\author{Tim A. van Kempen}
\affiliation{SRON Netherlands Institute for Space Research, Sorbonnelaan 2, 3584 CA Utrecht, Netherlands}
\affiliation{Leiden Observatory, Leiden University, Netherlands}

\author{Rebecca L. Larson}
\affiliation{The University of Texas at Austin, Department of Astronomy, 2515 Speedway, Stop C1400, Austin, TX 78712, USA}

\author{Umut A. Y{\i}ld{\i}z}
\affiliation{Jet Propulsion Laboratory, California Institute of Technology, 4800 Oak Grove Drive, Pasadena, CA 91109, USA}

\correspondingauthor{Yao-Lun Yang}
\email{yaolun@astro.as.utexas.edu}

\begin{abstract}
We present full spectral scans from 200--670~\micron\ of 26 Class 0+I protostellar sources, obtained with \textit{Herschel}-SPIRE, as part of the "COPS-SPIRE" Open Time program, complementary to the DIGIT and WISH Key programs.  Based on our nearly continuous, line-free spectra from 200--670~\micron, the calculated bolometric luminosities (\lbol) increase by 50\%\ on average, and the bolometric temperatures (\tbol) decrease by 10\%\ on average, in comparison with the measurements without \textit{Herschel}.  Fifteen protostars have the same Class using \tbol\ and \lbol/\lsmm.
We identify rotational transitions of CO lines from \jj{4}{3} to \jj{13}{12}, along with emission lines of $^{13}$CO, HCO$^+$, \water, and \CI. The ratios of $^{12}$CO to $^{13}$CO indicate that $^{12}$CO emission remains optically thick for \Jup\ $<$ 13.  We fit up to four components of temperature from the rotational diagram with flexible break points to separate the components.
The distribution of rotational temperatures shows a primary population around 100~K with a secondary population at $\sim$350~K.  We quantify the correlations of each line pair found in our dataset, and find the strength of correlation of CO lines decreases as the difference between $J$-level between two CO lines increases. The multiple origins of CO emission previously revealed by velocity-resolved profiles are consistent with this smooth distribution if each physical component contributes to a wide range of CO lines with significant overlap in the CO ladder.  We investigate the spatial extent of CO emission and find that the morphology is more centrally peaked and less bipolar at high-$J$ lines.  We find the CO emission observed with SPIRE related to outflows, which consists two components, the entrained gas and shocked gas, as revealed by our rotational diagram analysis as well as the studies with velocity-resolved CO emission.
\end{abstract}

\section{Introduction}
Large samples of protostars in relatively nearby ($d \leq 300$ pc) clouds have been developed through recent surveys with the \textit{Spitzer} Space Telescope \citep[e.g.][]{evans09,2015ApJS..220...11D} as well as the \textit{Herschel} Space Observatory \citep[e.g.][]{2010AA...518L.102A,vandishoeck11,2013MNRAS.432.1424K,green13b,2016ApJ...831...69M,2017AA...600A..99M}, along with ground-based surveys \citep[e.g.][]{jorgensen09}.  Over recent decades, infrared and submillimeter studies of such samples have allowed significant advances in our understanding of the properties, structure and evolution of such protostars.

ISO/LWS spanned the 20--200 \micron\ spectral region and was well-suited to study the warm (T $>$ 100 K) region of protostellar envelopes, distinguishable from the ambient cloud typically probed in ground-based millimeter studies. ISO-LWS detected gas phase H$_2$O, high-$J$ CO rotational transitions, and fine structure emission lines toward protostars and related structures \citep[e.g.,][]{lorenzetti99,giannini99,ceccarelli99,lorenzetti00,giannini01,nisini02}.
These lines were considered to originate from the outflows \citep{giannini99,giannini01,nisini02}, or the inner envelope \citep{ceccarelli99}; however, recent observations of \textit{Herschel} Space Observatory clearly show that these lines are dominated by outflow activity \citep{2010AA...521L..30K,kristensen12,2014AA...572A..21M}.

\textit{Herschel} was an European Space Agency (ESA) space-based far-infrared/submillimeter telescope with a 3.5-meter primary mirror \citep{pilbratt10}.  For the first time, \textit{Herschel}-SPIRE \citep[Spectral and Photometric Imaging REceiver, 194--670 \micron;][]{griffin10} enabled low resolution spectroscopy of the entire submillimeter domain.  These wavelengths are sensitive to dust continuum, and provide access to the full suite of mid-$J$ CO, HCO$^+$, $^{13}$CO and several H$_2$O emission lines.

In the previous analysis with data from the Dust, Gas, and Ice In Time \textit{Herschel} Key Program (DIGIT), we used the PACS spectrograph (50--200 \micron, \citealt{poglitsch10}) to characterize a sample of well-studied protostars, selected from the c2d sample, including both Class 0 and Class I objects \citep{green13b}.  Similar studies also used PACS data to characterize the properties of protostars in different regions with data from the Water in Star-forming Regions with Herschel (WISH) Key Program \citep{karska13} and the Herschel Orion Protostar Survey (HOPS) Key Program \citep{manoj12,2016ApJ...831...69M}.  The analysis of the combined PACS and SPIRE spectra was also presented for specific sources (e.g. Serpens~SMM1, \citealt{goi12}).

The full \textit{Herschel} bands contain numerous pure rotational transitions of CO, as well as lines of H$_2$O, OH, HCO$^+$, and atomic lines (\CI, \CII, \NII, and \OI), all potential tracers of gas content and properties.
\textit{Herschel} enabled access to the CO ladder toward higher energy levels (\Jup=4--48), providing an opportunity to constrain the origin of CO emission entirely.  At least two rotational temperatures are typically found with the PACS spectra toward embedded protostars \citep[e.g.][]{green13b,manoj12,karska13}.  The SPIRE spectra reveal the colder CO component, which has a different physical origin than the one for the higher-$J$ CO lines, suggested from velocity-resolved observations (see \citealt{2017arXiv170510269K}).
\citet{visser12} argued that C-type shocks dominate the high-$J$ CO lines at embedded sources with the the velocity-unresolved PACS data, while \citet{yildiz12} found the signature of shocked gas in the broad line profiles in the CO~\jj{6}{5} line, suggesting a different origin for the higher-$J$ CO lines compared to the CO lines that are typically accessed from the ground (\Jup\ $\leq$ 3).  \citet{2017arXiv170510269K}, who have a similar source list, showed that the CO~\jj{16}{15} line contains a broad component at the source velocity and a narrow component offsetting from the source velocity.

The emission of water has a similar line profile to the broad component found in CO~\jj{3}{2} and \jj{10}{9} lines  \citep{kristensen12,2014AA...572A..21M,2017AA...600A..99M}.  Both the line profiles and spatial extent of water emission suggest its close relation to outflows \citep{santangelo12,vasta12,kristensen12,2014AA...572A..21M,2017AA...600A..99M}.
Detailed modelings of water emission indicate a similar shock origin to the high-$J$ CO lines (\citealt{karska14,manoj12}, Karska et al. to be accepted)  Thus, these lines make excellent diagnostics of opacity, density, temperature, and shock velocities \citep[e.g.,][]{kaufman96,flower10,neufeld12} of the gas surrounding these systems \citep{2013AA...557A..23K,2014AA...572A..21M}.

In this work, we present \textit{Herschel}-SPIRE observations from the ``CO in Protostars-SPIRE'' (COPS-SPIRE) survey (PI: J. Green), of 27 protostars taken from the ``DIGIT'' and ``WISH'' samples.  In Section 2, we describe the sample, and provide an archive of 1--1000~\micron\ spectral energy distributions (SEDs), combining the \textit{Spitzer}-IRS, PACS, and SPIRE spectra, along with the description of the data processing pipeline.
In Section 3, we present the SEDs and the line fitting results, as well as the effect of emission lines on photometry.  We characterize the molecular and atomic lines with a customized line fitting pipeline previously optimized for \textit{Herschel} spectra, and provide the detection statistics and limits.  In Section 4, we derive the optical depth of CO and discuss the uncertainties.  We perform rotational diagram analysis of CO and HCO$^{+}$.
Furthermore, we characterize the correlations of each pair of lines detected in our sample, and discuss the origin of CO gas.
In Section 5, we consider the classification system in the context of the origin of the CO emission.  We also compare our analyses to those of FU Orionis objects, T Tauri stars, and Herbig Ae/Be stars \citep{green13c,2013AA...559A..77F}.  We summarize our conclusions in Section 6.

\section{Observations}
\label{sec:obs}

\subsection{The Sample}
Our sample of 27 ``COPS'' protostars contains 20 sources overlapping with the DIGIT (PI: N. Evans; \citealt{green13b,2013AA...559A..77F,2013AA...559A..84M,2014ApJS..214...21L}) and 17 overlapping with the WISH (PI: E. van Dischoek; \citealt{vandishoeck11}; see also \citealt{nisini10,kristensen12,wampfler13,karska13,2013AA...553A.125S,yildiz13}) \textit{Herschel} Key programs.
The sources were originally chosen to be well-studied \citep[e.g.][]{2002AA...389..908J,2004AA...416..603J,jorgensen07a}, nearby (within 450~pc), and spanning a range of luminosities.
The Class 0 protostars were originally chosen from the sample of \citet{2000prpl.conf...59A}.  The selected protostars have a wide range of bolometric temperatures (33.2~K--592.0~K) and bolometric luminosities (0.33~\lsun--70.4~\lsun).  Additionally, the sources were carefully chosen to have well-studied \textit{Spitzer} data to complement our observations \citep{evans09,2015ApJS..220...11D}, and drawn from both larger clouds and isolated environments (see Column 2 in Table~\ref{sourcelist}).  The data for all sources were reprocessed with identical techniques detailed in the next section.
The full list of sources appears by region in Table \ref{sourcelist}, and by observation date in Table \ref{obslog}.  We have updated the distances of several sources based on the recent studies (see the references in Table~\ref{sourcelist}).  In particular, we update the distance of L1551~IRS5 to 147~pc due to its proximity to T~Tau and XZ~Tau, for which the distance is measured by Galli et al. (in prep.).  The characteristics of the SEDs (e.g. bolometric luminosity and temperature) are updated with the \textit{Herschel} data presented in this study (Table~\ref{evolutionary}).
A nearly identical sample was observed in CO~\jj{16}{15} with HIFI (PI: L. Kristensen) summarized in \citet{2017arXiv170510269K}.

\begin{deluxetable*}{l r r r r r r r}
\tabletypesize{\scriptsize}
\tablecaption{Source List \label{sourcelist}}
\tablewidth{0pt}
\tablehead{
    \colhead{Source} & \colhead{Cloud} & \colhead{ Dist. (pc)} & \colhead{RA (J2000)} & \colhead{Dec (J2000)} & \colhead{DIGIT/WISH} & \colhead{Ref.} & \colhead{Dist. Ref.}
    }
\startdata
IRAS 03245+3002 & Per  & 235 & 03h27m39.1s &  +30d13m03.1s       & D & c2d & 1 \\
L1455~IRS3      & Per  & 250 & 03h28m00.4s &  +30d08m01.3s       & D & c2d & 2,3,4,5\\
IRAS 03301+3111 & Per  & 250 & 03h33m12.8s &  +31d21m24.2s       & D & c2d & 2,3,4,5\\
B1-a            & Per  & 250 & 03h33m16.7s &  +31d07m55.2s       & D & c2d & 2,3,4,5\\
B1-c            & Per  & 250 & 03h33m17.9s &  +31d09m31.9s       & D & c2d & 2,3,4,5\\
L1551~IRS5      & Tau  & 147 & 04h31m34.1s &  +18d08m04.9s       & D/W & 2MASS & 6 \\
TMR 1           & Tau  & 140 & 04h39m13.9s &  +25d53m20.6s       & D/W & H & 7\\
TMC 1A          & Tau  & 140 & 04h39m35.0s &  +25d41m45.5s       & D/W & H & 7\\
TMC 1           & Tau  & 140 & 04h41m12.7s &  +25d46m35.9s       & D/W & A & 7\\
HH~46           & Core & 450 & 08h25m43.9s & --51d00m36.0s       & W & vD & 8,9,10 \\
Ced110~IRS4     & ChaI  & 150 & 11h06m47.0s & --77d22m32.4s       & W & vD & 11,12,13 \\
BHR 71          & Core & 200 & 12h01m36.3s & --65d08m53.0s       & D/W & c2d & 14,15 \\
DK Cha          & ChaII  & 178 & 12h53m17.2s & --77d07m10.7s       & D/W & c2d & 11 \\
IRAS 15398-3359 & Core & 130 & 15h43m01.3s & --34d09m15.0s       & W & vD & 16,17 \\
GSS~30~IRS1     & Oph  & 137 & 16h26m21.4s & --24d23m04.3s & D/W & c2d & 18 \\
VLA 1623$-$243    & Oph  & 137 & 16h26m26.4s & --24d24m30.0s & D & c2d & 18 \\
WL~12           & Oph  & 137 & 16h26m44.2s & --24d34m48.4s & D & c2d & 18 \\
RNO~91          & Oph  & 130 & 16h34m29.3s & --15d47m01.4s       & W & vD & 19 \\
L483            & Aqu  & 200 & 18h17m29.9s & --04d39m39.5s       & W & vD & 20 \\
RCrA~IRS5A      & CrA  & 130 & 19h01m48.1s & --36d57m22.7s       & D/W & N & 21,22,23 \\
HH~100          & CrA  & 130 & 19h01m49.1s & --36d58m16.0s       & W & vD & 21,22,23 \\
RCrA~IRS7C      & CrA  & 130 & 19h01m55.3s & --36d57m17.0s       & D & L & 21,22,23 \\
RCrA~IRS7B      & CrA  & 130 & 19h01m56.4s & --36d57m28.3s       & D & L & 21,22,23 \\
L723~MM         & Core & 300 & 19h17m53.7s &  +19d12m20.0s       & W & vD & 24 \\
B335            & Core & 106 & 19h37m00.9s & +07d34m09.7s  & D/W & PROSAC & 25 \\
L1157           & Core & 325 & 20h39m06.3s &  +68d02m16.0s       & D/W & PROSAC & 26 \\
L1014           & Core & 200 & 21h24m07.5s &  +49d59m09.0s       & D & Y & 27,28 \\
\enddata
\tablecomments{
List of protostellar sources discussed in this
work by region, sorted by RA.  Coordinate reference code: D $=$ \citet{dunham06};
Y $=$ \citet{young04};
L $=$ \citet{lindberg11}; N $=$ \citet{nisini05}; H $=$ \citet{haisch04};
A $=$ \citet{apai05}; B $=$ \citet{brinch07}; c2d $=$ \citet{evans09};
PROSAC $=$ \citet{jorgensen09}; vD $=$ \citet{vandishoeck11}. \\
List of references for distance: 1 $=$ \citet{hirota11};
2 $=$ \citet{1993BaltA...2..214C};
3 $=$ \citet{2002AA...387..117B};
4 $=$ \citet{2003BaltA..12..301C}; 5 $=$ \citet{enoch06};
6 $=$ Galli et al. (in prep.); 7 $=$ \citet{1994AJ....108.1872K};
8 $=$ \citet{1971ApJ...163L..99B}; 9 $=$ \citet{1976ApJ...203..151R};
10 $=$ \citet{1980ApJ...238..919E};
11 $=$ \citet{1997AA...327.1194W};
12 $=$ \citet{1999AA...352..574B};
13 $=$ \citet{2008hsf2.book..169L};
14 $=$ \citet{1989AA...225..192S};
15 $=$ \citet{1994BaltA...3..199S};
16 $=$ \citet{1998AA...338..897K};
17 $=$ \citet{vandishoeck11};
18 $=$ \citet{ortiz-leon17};
19 $=$ \citet{1990AA...231..137D}
20 $=$ \citet{1985ApJ...297..751D}; 21 $=$ \citet{1998AJ....115.1617C};
22 $=$ \citet{1999AJ....117..354D}; 23 $=$ \citet{neuhauser08};
24 $=$ \citet{1984ApJ...286..599G}; 25 $=$ \citet{olofsson09};
26 $=$ \citet{1992BaltA...1..149S}; 27 $=$ \citet{young04};
28 $=$ \citet{2004MNRAS.355.1272M}.}
\end{deluxetable*}

\begin{deluxetable*}{l l l l r r r}
\tabletypesize{\scriptsize}
\tablecaption{Observing Log \label{obslog}}
\tablewidth{\textwidth}
\tablehead{
    \colhead{Source} & \colhead{Other Name} & \colhead{OBSID} & \colhead{Date Obs.} & \colhead{PACS 1D aperture$^{1}$} & \colhead{Notes}
    }
\startdata
IRAS 03245+3002  & L1455~IRS1      & 1342249053 & 04 Aug 2012 & 31.8 & \\
L1455~IRS3       &                 & 1342249474 & 13 Aug 2012 & A    & \\
IRAS 03301+3111  & Perseus Bolo76  & 1342249477 & 13 Aug 2012 & A    & \\
B1-a             &                 & 1342249475 & 13 Aug 2012 & A    & \\
B1-c             &                 & 1342249476 & 13 Aug 2012 & 36.8 & \\
L1551~IRS5       &                 & 1342249470 & 12 Aug 2012 & 24.8 & \\
TMR 1            & IRAS 04361+2547 & 1342250509 & 02 Sep 2012 & 51.8 & \\
TMC 1A           & IRAS 04362+2535 & 1342250510 & 02 Sep 2012 & 41.8 & \\
TMC 1            & IRAS 04381+2540 & 1342250512 & 02 Sep 2012 & A    & \\
HH~46             &                 & 1342245084 & 28 Apr 2012 & N/A  & linescan \\
Ced110~IRS4      &                 & 1342248246 & 17 Jul 2012 & N/A  & linescan \\
BHR 71           &                 & 1342248249 & 17 Jul 2012 & 29.8 & \\
DK Cha           & IRAS 12496--7650& 1342254037 & 28 Oct 2012 & 31.8 & linescan \\
IRAS 15398--3359 & B228            & 1342250515 & 02 Sep 2012 & N/A  & \\
GSS~30~IRS1       &                 & 1342251286 & 23 Sep 2012 & A    & \\
VLA~1623$-$243      &                 & 1342251287 & 23 Sep 2012 & 41.8 & \\
WL~12             &                 & 1342251290 & 23 Sep 2012 & A    & \\
RNO~91            &                 & 1342251285 & 23 Sep 2012 & N/A  & linescan \\
L483             & IRAS 18140--0440& 1342253649 & 19 Oct 2012 & N/A  & linescan \\
RCrA~IRS5A       &                 & 1342253646 & 19 Oct 2012 & A    & \\
HH~100            &                 & 1342252897 & 07 Oct 2012 & N/A  & mis-pointed \\
RCrA~IRS7C       &                 & 1342242621 & 11 Mar 2012 & 41.8 & mult. sources \\
RCrA~IRS7B       &                 & 1342242620 & 11 Mar 2012 & 41.8 & mult. sources \\
L723~MM          &                 & 1342245094 & 28 Apr 2012 & N/A  & linescan \\
B335             &                 & 1342253652 & 19 Oct 2012 & 24.8 & \\
L1157            &                 & 1342247625 & 02 Jul 2012 & 21.8 & \\
L1014            &                 & 1342245857 & 16 May 2012 & A    & \\
\enddata
\tablecomments{Observations log for protostellar sources discussed in this work.  The ``mode'' indicates the spatial coverage of the observation.\\
$^1$The unit is arcsec.  If all 25 PACS spaxels are used for extracting 1D spectrum, it will be denoted as ``A,'' and ``N/A'' indicates that no PACS 1D spectrum is extracted.}
\end{deluxetable*}

\subsection{Data Processing Pipeline}
The data processing pipeline is based on the method described in \citep[][hereafter the CDF archive]{green16a} with the modifications presented in \citet{yang17}.  The major differences between the data presented here and the CDF archive are the PACS 1D spectra and the version of Herschel Interactive Processing Environment (\textsc{HIPE}, \citealt{ott10}).  The PACS 1D spectra were extracted from the central 3$\times$3 spaxels for the CDF archive, whereas we sum over the emission within a circular aperture determined from the flux agreement with the SPIRE 1D spectra to extract the PACS 1D spectra.  We also adopt the same method shown in \citet{yang17} to choose the apertures for measuring photometry that is consistent with the spectroscopy.  The detailed procedures of the reduction are described in the following sections.

\label{sec:pipeline}
\subsubsection{SPIRE}
The SPIRE-FTS (Fourier Transform Spectrometer) data were taken in a single pointing with sparse image sampling in 1 hr of integration time per source.  The spectra were taken with the high resolution (HR) mode and are divided into two orders covering the spectral ranges of 194 -- 325 \micron\ (``SSW''; Spectrograph Short Wavelengths) and 320 -- 690 \micron\ (``SLW''; Spectrograph Long Wavelengths), with a spectral resolution element $\Delta\nu$ of 2.16 GHz after the apodization ($\lambda$/$\Delta\lambda$ $\sim$ 200--670, or $\Delta v \sim$ 400--1500~\kms, \citealt{griffin10}).  The SPIRE-FTS has a field-of-view of 180\arcsec$\times$180\arcsec, with spatial pixel (spaxel) separations of 33\arcsec\ and 51\arcsec\ for SSW and SLW, respectively.
The SPIRE beam size ranges from 17\arcsec--40\arcsec, equivalent to physical sizes of $\sim$3200--7600 AU at the mean distance of the COPS sources (189~pc), comparable to the size of a typical core \citep{ward07} but smaller than the typical length of an outflow \citep{2007prpl.conf..245A,2015AA...576A.109Y}.  The SPIRE beam size increases with wavelength. But the beam size also jumps at 300~\micron\ \citep{2013ApOpt..52.3864M} due to the complex modes of waveguide at the short wavelength end of the SLW module.
SPIRE used an onboard calibration source for flux calibration, resulting in $<$ 6\%\ calibration uncertainty updated in \citet{swinyard14}.

Each module was reduced separately within \textsc{HIPE} version 14.0.3446 with the SPIRE calibration dataset \texttt{spire\_cal\_14\_3}.  We applied an apodization of 1.5, which reduces the resolution by a factor of 1.5 to 2.16~GHz but suppresses baseline variation.
The SPIRE data were extracted using the ``extended source'' calibration pipeline, as this produced a smoother continuum between modules, better S/N, and fewer spectral artifacts than the ``point source'' pipeline.  The extracted SPIRE data contain the spectra of each spaxel from two modules, which is used for analyzing the spatial distribution of spectral lines in this study.  We further extracted the 1D spectrum of each source with the \textsc{SemiExtendedSourceCorrector} (SECT) script from \textsc{HIPE} version 14.0.3446, using SPIRE calibration dataset \texttt{spire\_cal\_14\_3}.  Most of the COPS sources are partially resolved, therefore neither a standard point source extraction nor an extended source extraction is suitable to extract a single spectrum of each source, which would result in a mismatch between the spectra of the SLW and SSW modules.
This script fits a ``source size'' that produces a smooth spectrum, and then normalizes the 1D spectrum with a given aperture \citep{2013AA...556A.116W}.  We adopted the prescription developed by \citet{makiwa16} to extract the 1D spectrum of the entire source.  The SECT script failed to produce the 1D spectrum for HH~100, which is mis-pointed and contaminated by nearby RCrA~IRS7C.  No background subtraction was performed (unlike, for example in \citealt{wiel14}, who examined point-like disk sources and subtracted non-central pixels from the center pixel), whereas, for PACS, a spectrum was taken at off-source position for the background subtraction.  Thus, possible contribution from extended background emission should be considered when interpreting the spectra.  We note cases where extended emission was seen in Section~\ref{sec:seds}.

The methodology of the SECT script provides the calibrated spectrum that best describes the emission from entire source.  When comparing the photometry with the SECT-corrected spectra, we find a good agreement when we use the convolution of the fitted source size and the beam sizes of SPIRE as the apertures.  The convolved aperture is larger at long wavelengths due to the beam profile of SPIRE.  Therefore, the SECT-corrected spectrum has more emission in the longer wavelengths compared to the spectrum extracted with a single aperture, resulting in shallower slopes at long wavelengths.

The observed source size is a strong function of wavelength for embedded protostars arises from cooler dust that is farther from the source.  To characterize the entire source, it is necessary to use larger apertures at longer wavelengths.  Therefore, we argue that using the SPIRE spectra with semi-extended source correction to calibrate the extraction of PACS 1D spectra best represents the emission from entire protostars, which yielding realistic estimates of \lbol\ and \tbol.

\subsubsection{PACS}
\label{sec:pipeline_pacs}
We also collect and reduce PACS spectroscopy and photometry data from the CDF archive for comparison.  Of the 27 COPS sources, 21 sources have PACS rangescan spectroscopy from the DIGIT program, 16 sources have PACS linescan spectroscopy from the WISH program, and all have photometry available from the Herschel Science Archive (HSA) except for HH~46.  The PACS reduction was updated with \textsc{HIPE} version 14.0.3446 (calibration version 72), using the Telescope Background Correction algorithm, including a PSF correction and a correction for telescope jitter (changes in pointing offset during the observation), when possible, to produce the best flux calibration.  The jitter correction was only applicable to rangescans (linescans are too narrow to properly determine the source centroid), and thus for the 6 sources where only linescan data is available, we use the reductions from \citet{karska13}.  We only consider the linescan data for calculating the properties of the SED.  Note that HH~100 was not observed with PACS spectroscopy.

The PACS 1D spectra are extracted using the method described in \citet{yang17}, which is different from the one used in the CDF archive.  This method calculates the total flux density within a circular aperture.  Using the same aperture size for all sources results in flux mismatches between PACS and SPIRE for several sources.  To provide useful PACS products for comparison, we modify the aperture size used for the PACS data of each source until the PACS and SPIRE spectra agree within 5\%\ of the flux density at the conjunction of PACS and SPIRE, where we take the median flux from 185--190~\micron, and 195--200~\micron, respectively.  However, five sources have PACS and SPIRE spectra that disagree by more than 5\%\ even if all PACS spaxels are included (see Section~\ref{sec:seds}).  A direct comparison in the overlapping wavelengths is prohibited due to the increase of noise toward the end of band.  The fitted aperture sizes are listed in Table~\ref{obslog}.

\section{Results}
\label{sec:results}

\subsection{Source Confusion in the RCrA region}
Our data processing pipeline successfully reduced all sources, except for HH~100, where the reduction of the SPIRE 1D spectrum failed due to the contamination from RCrA~IRS7C.  We found that HH~100 was mis-pointed by about 30\arcsec\ toward the contaminating source, confusing the emission from HH~100; therefore, we exclude HH~100 from our analysis.  Furthermore, our SPIRE observation does not completely resolve some sources in the RCrA region, such as RCrA~IRS7B and RCrA~IRS7C.  The separation between two sources is only 17\arcsec.  Thus, we adopt the bolometric temperature and luminosity from \citet{2014AA...565A..29L} for RCrA~IRS7B and RCrA~IRS7C, where the PACS data are carefully deconvolved.  We exclude these two sources from most of the analyses, and in few cases, we use the spectrum of RCrA~IRS7C as the combined spectrum of both sources, noted as RCrA~IRS7B/C.  In the end, we present the data of 25 protostars including RCrA~IRS7B/C.

\subsection{Comparing Spectra with Photometry}
Our data processing pipeline produces flux-calibrated spectra, which typically match photometric observations, as well as line-free continua for SEDs.  As a cross-check on the flux calibration of our method, we collected archival PACS and SPIRE imaging from the Herschel Science Archive.  We list the OBSIDs used for the photometry in Appendix~\ref{photref}.  The photometry was then extracted with HIPE version 14.0.3446 using \textsc{daophot} \citep{1987PASP...99..191S}, from a top-hat circular aperture.  For SPIRE, the apertures were chosen to be the convolved size of the source size fitted by SECT and the beam sizes at 250, 350, and 500~\micron\ (18.4\arcsec, 25.2\arcsec, and 36.7\arcsec).  For PACS, we adopted the aperture used for extracting the 1D spectra (see Section~\ref{sec:pipeline_pacs}).  We apply color corrections to SPIRE photometry with the power law index fitted from spectra.  The color correction at PACS 160~\micron\ is 3--4\%\ without any systematic increase or decrease.  Due to the increasing contribution of hot dust at shorter wavelengths and the gap of spectra at 100~\micron, we add an uncertainty of 3\%\ to PACS photometry instead of applying color correction directly.

The photometry, both PACS and SPIRE, was typically observed in two or more OBSIDs, which we averaged together and used the standard deviation of the observations as its uncertainty (typically 1--3\%).  We choose a sky annulus between 100\arcsec\ and 150\arcsec\ in radius to avoid any extended emission from sources but include the emission from the surrounding filamentary structure.  Using this technique to minimize contamination, the sky emission is less than 10\% of the source flux at any given wavelength, except for a few cases where the background is as high as 20--30\%.

In addition to PACS and SPIRE photometry and spectroscopy, we collected 2MASS JHK, WISE Bands 3.4, 4.6, 12, and 22~\micron, \textit{Spitzer}-IRAC 3.6, 4.5, 5.8, and 8~\micron, MIPS 24~\micron\ and 70~\micron\ photometry, and archival photometry for all sources where available.  It became apparent upon comparison to the PACS 70~\micron, that the MIPS 70~\micron\ photometry was saturated in most sources, and thus we discard the MIPS 70~\micron\ photometry in those cases.  Additionally, we collected millimeter data where available. The detailed reference for the photometry used in this study is shown in the Appendix (Table~\ref{phot_reference}).

\subsection{The SEDs before and after \textit{Herschel}}
\label{sec:pre_post_herschel}
The bolometric luminosities, which include both spectroscopic and photometric measurements, and the bolometric temperatures calculated with the prescription proposed by \citet{myers93} and \citet{chen95} are shown in Table~\ref{evolutionary}.
We compute \lbol\ following the trapezoidal summation method used in \citet{dunham10}. In summary, we integrate over the finitely sampled SED by treating each datapoint as a trapezoid covering half of the wavelength range to its nearest neighbor, at the given value. We then add the areas of trapezoids together. Similarly, we compute \tbol\ as the blackbody that peaks at the flux-weighted average wavelength of the SED (using the trapezoidal summation method), and translate into a temperature via the Planck blackbody law. The difference is then the values of \lbol\ and \tbol\ computed with this method, with and without the \textit{Herschel} data.  Photometry is treated as a single flux density at its representing wavelength, and inserted into the spectra.  A single photometry only contributes to the area within its trapezoid (i.e. its width is smaller than the wavelength channel of spectra)  Thus the values of \lbol\ and \tbol\ are dominated by spectra, rather than photometric points. Thus, the spectral data dominate the uncertainties in cases of overlapping wavelength coverage.

The excellent sampling from \textit{Herschel} spectroscopy has decreased the uncertainties of \lbol\ and \tbol\ substantially.  Models of star formation predict the bolometric luminosity of protostars.  However, the bolometric luminosity can be underestimated by 35--40\%\ if a significant portion of data between 70~\micron\ and 850~\micron\ is missing \citep{dunham13}.  \textit{Herschel} fills this gap perfectly.  To compare the change of bolometric luminosity resulted from the addition of \textit{Herschel} data, we collected the bolometric luminosities measured primarily with \textit{Spitzer} \citep{2008ApJS..176..184F,2015ApJS..220...11D} for 15 COPS sources where bolometric luminosities were calculated, and compare with the bolometric luminosities measured from our PACS and SPIRE data.
We found a mean increase of 50\%\ with the \textit{Herschel} data, where 12 of 15 protostars have their luminosities increased.

The addition of \textit{Herschel} observations affect the bolometric temperatures as well.  With the same sample \citep{2008ApJS..176..184F,2015ApJS..220...11D}, we found the bolometric temperature decreases by 10\%\ on average after including the \textit{Herschel} spectroscopy, suggesting that the protostars would be systematically less evolved with \textit{Herschel} under the classification of \tbol.

Some sources have photometry that deviate significantly from spectroscopy (see the discussion in Section~\ref{sec:seds}).  Although the spectroscopy dominates the values of \lbol\ and \tbol, the mismatch between photometry and spectroscopy may suggest potential calibration problems, introducing systematic uncertainties.  If we simply calibrate the spectroscopy to photometry, the \lbol\ and \tbol\ can vary by as much as +90\%/-60\%\ and +65\%/-15\%\, respectively, but no systematic offset is found as the average differences for \lbol\ and \tbol\ among the COPS sources are 2\%\ and 4\%\, respectively.  Among the 16 sources that we have photometry to scale the spectroscopy, only three (one) sources have \lbol\ (\tbol) varies more than $\pm$20\%.  However, the calibration between spectroscopy and photometry is not likely to be the problem since the systematic uncertainties of PACS and SPIRE are only 1\%\ and 3\%, and our extracted spectra agree with photometry at long wavelengths.
If we further calibrate the SPIRE spectra to the PACS spectra, assuming that the PACS spectra are less confused by extended emission, \lbol\ becomes smaller while \tbol\ becomes larger, since the SPIRE spectra are always greater than the PACS spectra when there is a mismatch.  On average, \lbol\ decreases by 5\%, while \tbol\ increases by 5\%.  L1455~IRS3 is the most extreme case, whose \lbol\ and \tbol\ vary by -30\%\ and +40\%.  \citet{2015AJ....150..175R} found that the mid-infrared fluxes of Class 0 sources are highly variable by 10--15\%\ across 6--7 years baseline.  Thus, the mismatches between photometry and spectroscopy may be due to the intrinsic variability of the sources.

For the sources whose SPIRE spectrum disagrees with their PACS spectrum (see the discussion in Section~\ref{sec:seds}), their \lbol\ and \tbol\ can differ up to -25\%\ and +25\%, respectively, if we manually scale the SPIRE spectra to match the PACS spectra.

\begin{deluxetable*}{l r r r r r r r r r r}
\tabletypesize{\scriptsize}
\tablecaption{Evolutionary Indicators \label{evolutionary}}
\tablewidth{0pt}
\tablehead{
    \colhead{Source} & \colhead{\lbol\ (\lsun)} & \colhead{\lsmm\ (\lsun)} & \colhead{$L_{\rm CO}$ (\lsun)} & \colhead{\tbol\ (K)} & \colhead{$\alpha_{\rm NIR}$} & \colhead{Class$_{\alpha_{\rm NIR}}$} &
    \colhead{Class$_{\lsmm}$} & \colhead{Class$_{\tbol}$} & \colhead{Ref.}
    }
\startdata
IRAS 03245+3002    & 6.06 & 0.05  & 0.012  & 48.2  & 2.54$\pm$0.79        & 0+I & 0 & 0 & D \\
L1455~IRS3         & 0.55 & 0.05  & 0.0012 & 128.0 & 1.18$\pm$0.29        & 0+I & 0 & I & D \\
IRAS 03301+3111    & 3.91 & 0.05  & 0.0083 & 354.0 & 0.37$\pm$0.06        & 0+I & 0 & I/flat$^{1}$ & D \\
B1-a               & 2.47 & 0.09  & 0.048  & 79.9  & 1.87$\pm$0.45        & 0+I & 0 & I & D \\
B1-c               & 4.41 & 0.11  & 0.021  & 55.9  & 3.01$\pm$1.47        & 0+I & 0 & 0 & D \\
L1551~IRS5         & 25.9 & 0.09  & 0.0087 & 110.0 & 1.43$\pm$0.17        & 0+I & 0 & I & G \\
TMR 1              & 2.0  & 0.015 & 0.0064 & 125.0 & 1.02$\pm$0.33        & 0+I & 0 & I & R \\
TMC 1A             & 2.62 & 0.015 & 0.0023 & 159.0 & -0.15$\pm$0.84       & flat & 0 & I & R \\
TMC 1              & 0.79 & 0.015 & 0.0042 & 149.0 & 0.55$\pm$0.04        & 0+I & 0 & I & R \\
HH46               & 23.2 & 0.18  & 0.079  & 111.0 & 0.71$\pm$0.03        & 0+I & 0 & I & G \\
Ced110~IRS4        & 1.28 & 0.03  & 0.0064 & 53.6  & 1.99$\pm$0.54        & 0+I & 0 & 0 & G \\
BHR 71             & 13.5 & 0.16  & 0.039  & 51.1  & 1.95$\pm$0.33        & 0+I & 0 & 0 & G \\
DK Cha             & 35.1 & 0.025 & 0.017  & 592.0 & -0.05$\pm$0.34       & flat & I & flat & D \\
IRAS 15398$-$3359  & 1.49 & 0.032 & 0.0097 & 43.2  & 1.32$\pm$0.15        & 0+I & 0 & 0 & G \\
GSS~30~IRS1        & 19.7 & 0.07  & 0.064  & 129.0 & 1.58$\pm$0.30        & 0+I & I & I & D \\
VLA 1623$-$243     & 5.36 & 0.15  & 0.034  & 33.2  & 2.34$\pm$0.07        & 0+I & 0 & 0 & c2d \\
WL~12              & 2.23 & 0.039 & 0.010  & 210.0 & 2.93$\pm$0.60        & 0+I & 0 & I & D \\
RNO~91             & 2.53 & 0.016 & 0.0030 & 349.0 & 0.48$\pm$0.16        & 0+I & 0 & I/flat$^{1}$ & c2d \\
L483               & 8.78 & 0.11  & 0.015  & 49.3  & 2.05$\pm$1.12        & 0+I & 0 & 0 & G \\
RCrA~IRS5A$^{2}$   & 1.7  & 0.013 & 0.039 & 209.0  & 0.40$\pm$0.30       & 0+I & 0 & 0 & P, L \\
RCrA~IRS7C$^{2}$   & 9.1  & 0.092 & 0.18 & 79.0   & 2.65$\pm$0.71        & 0+I & I & 0 & P, L \\
RCrA~IRS7B$^{2}$   & 4.6  & 0.096 & 0.13 & 89  & 2.68$\pm$1.19          & 0+I & 0 & 0 & P, L \\
L723~MM            & 3.3  & 0.065 & 0.016  & 66.8  & 1.50$\pm$0.32        & 0+I & 0 & 0 & G \\
B335               & 0.57 & 0.012 & 0.0032 & 45.5  & 0.74$\pm$0.18        & 0+I & 0 & 0 & A \\
L1157              & 5.26 & 0.11  & 0.031  & 40.1  & 0.87$\pm$0.38        & 0+I & 0 & 0 & D \\
L1014              & 0.33 & 0.024 & \nodata$^{3}$ & 63.4  & 0.75$\pm$0.24        & 0+I & 0 & 0 & Y \\
\enddata
\tablecomments{\lbol\ and \tbol\ measured from the SEDs presented herein.  \lsmm is defined as the bolometric luminosity of the spectrum $>$ 350 \micron.  The $\alpha_{\rm NIR}$ is re-calculated with photometric fluxes collected from literatures.  If the photometric fluxes are not found, we use the spectrophotometric fluxes extracted from the \textit{Spitzer}-IRS spectra. \\
Reference code: c2d = \citet{evans09}, D = \citet{2015ApJS..220...11D}, L = \citet{2014AA...565A..29L}, P = \citet{2011ApJS..194...43P}, R = \citet{2010ApJS..186..259R}, A = \citet{2008ApJ...687..389S},
Y = \citet{young04}, G = \citet{green13b}. \\
$^{1}$ The sources have \tbol\ close to the criteria dividing between Class I and flat spectrum. \\
$^{2}$ The \lbol, \tbol, and \lsmm\ are collected from \citet{2014AA...565A..29L}.  Other quantities are calculated from the SEDs with the COPS spectra and data from the CDF archive, which have obvious source contaminations. \\
$^{3}$ No CO line is detected toward L1014.
}
\end{deluxetable*}

\subsection{Detection Limits}
\label{sec:limits}

\begin{figure}[htbp!]
    \centering
    \includegraphics[width=0.45\textwidth]{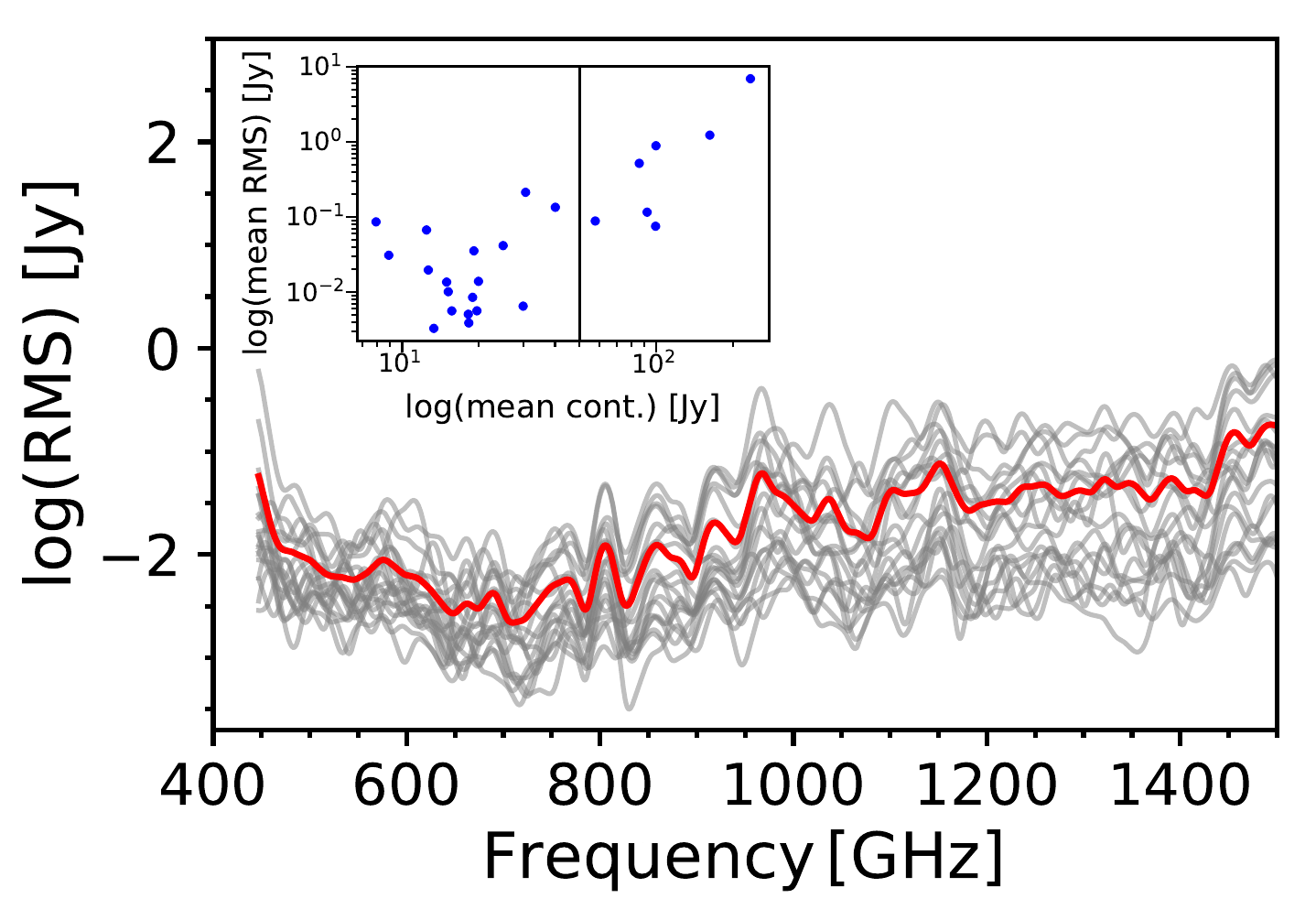}
    \caption{The RMS noises of the sources with their mean continuum fluxes lower than 50~Jy are shown in gray lines, while the mean RMS noise of those sources is shown in red line.  The RMS noise is convolved with a Gaussian that has a width of 20 wavelength channels for better visualization.  The inset figure shows the relation between the continuum flux and the RMS noise averaging over frequencies for the COPS sources.  The relation suggests a positive correlation when the mean continuum flux is greater than 50~Jy, and no specific trend is found when the mean continuum flux is lower than 50~Jy.}
    \label{fig:rmslimit}
\end{figure}

We measured the continuum RMS in the line-free spectrum, where the noise is dominated by the correlated baseline variation due to the nature of the FTS spectrograph and the apodization.  The continuum RMS scales roughly with the continuum flux when the continuum flux is greater than 50~Jy (Figure~\ref{fig:rmslimit}).  We found a mean RMS between 0.01~Jy to 0.1~Jy for the sources with their mean continuum fluxes lower than 50~Jy, whose RMS noise does not scale with the continuum flux.  The RMS noise significantly increases toward the edges of spectra.  We detect line fluxes down to 6.9\ee{-18}~W~m$^{-2}$ (equivalent to a peak flux of 0.28~Jy for unresolved lines) and 3.0\ee{-17}~W~m$^{-2}$ (equivalent to a peak flux of 1.6~Jy for unresolved lines), in PACS and SPIRE, respectively, for typical DIGIT/COPS integration times.

\subsection{Spectral Energy Distributions}
\label{sec:seds}

\begin{figure*}[htbp!]
    \centering
    \includegraphics[width=0.45\textwidth]{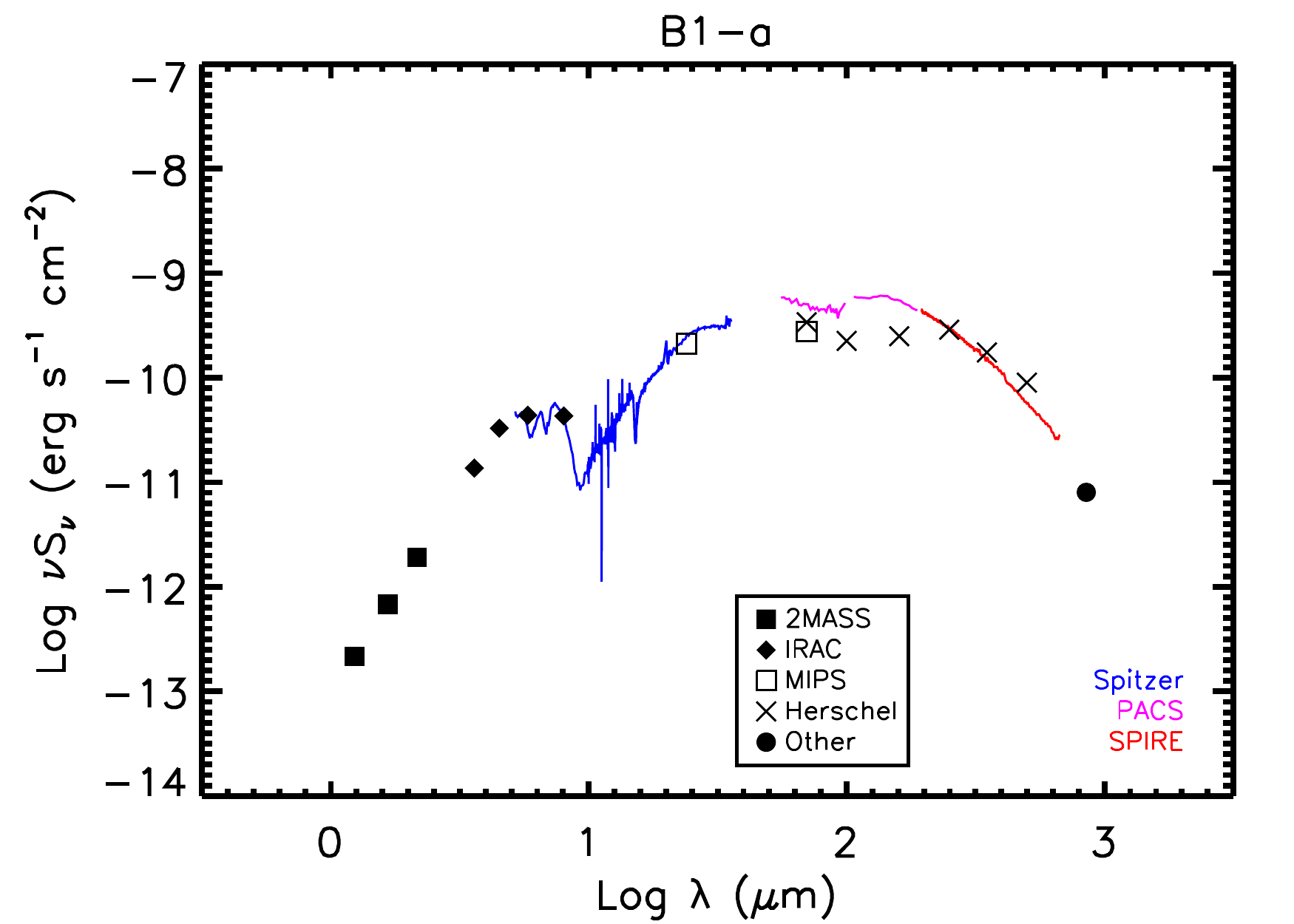}
    \includegraphics[width=0.45\textwidth]{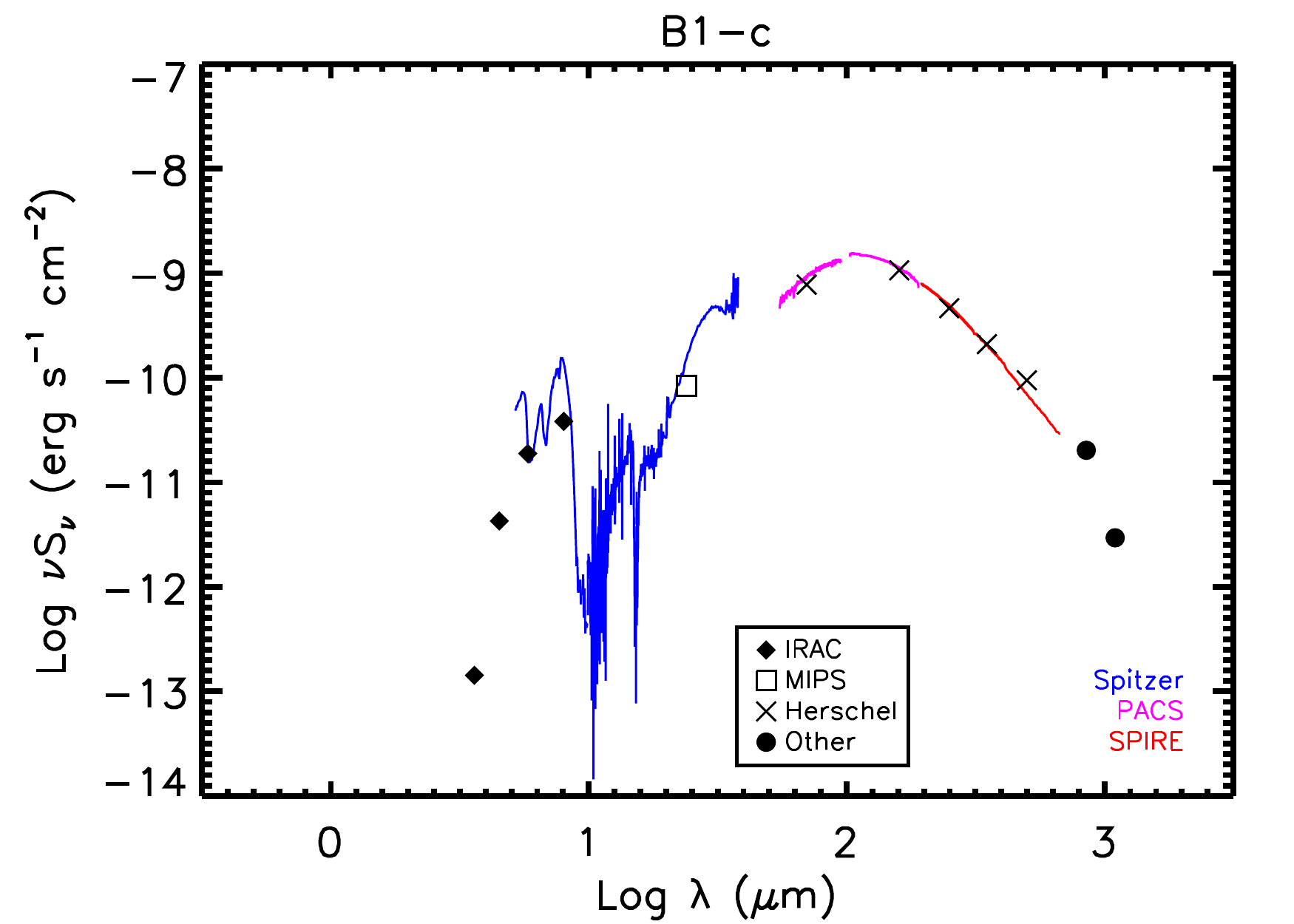}
    \includegraphics[width=0.45\textwidth]{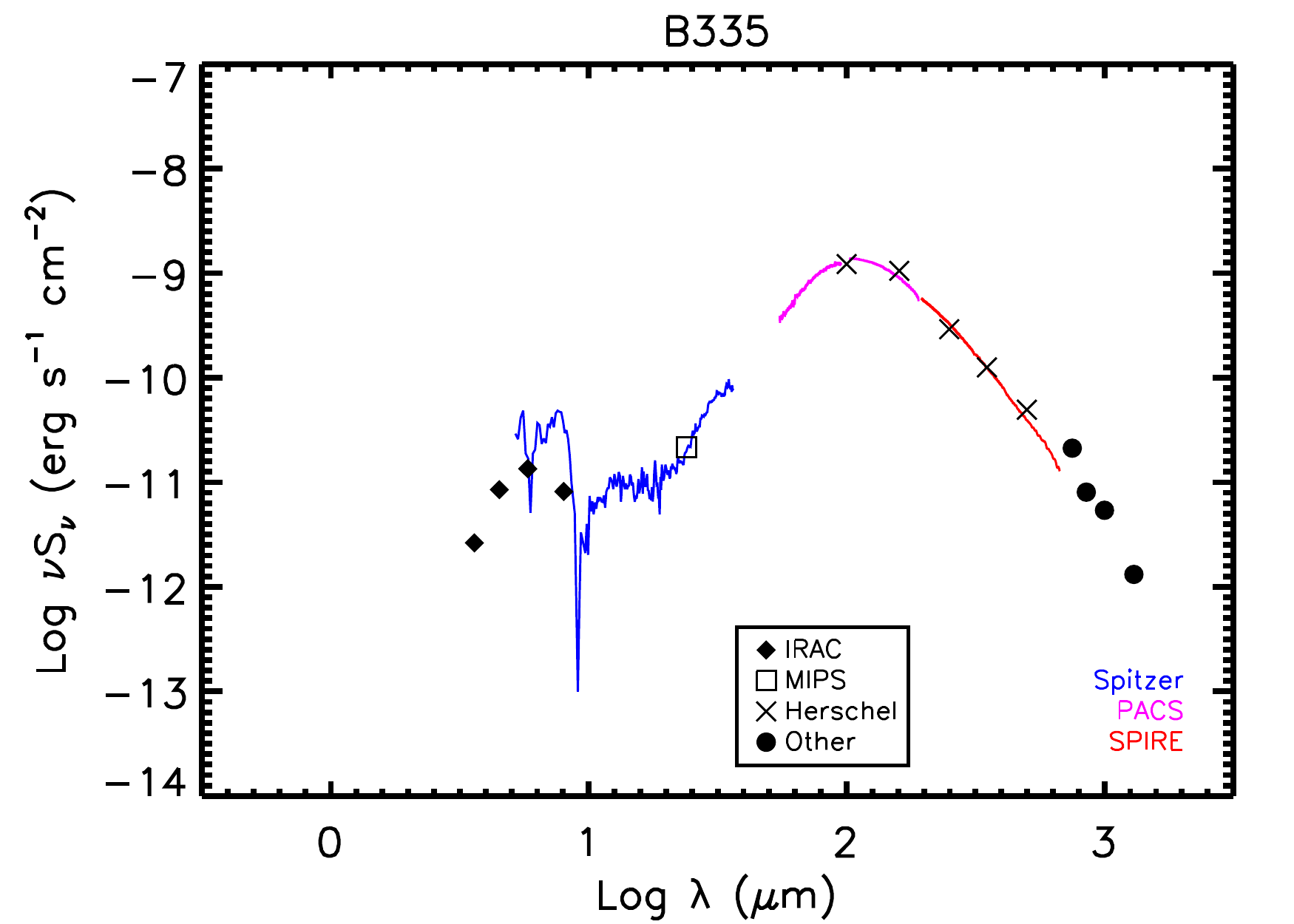}
    \includegraphics[width=0.45\textwidth]{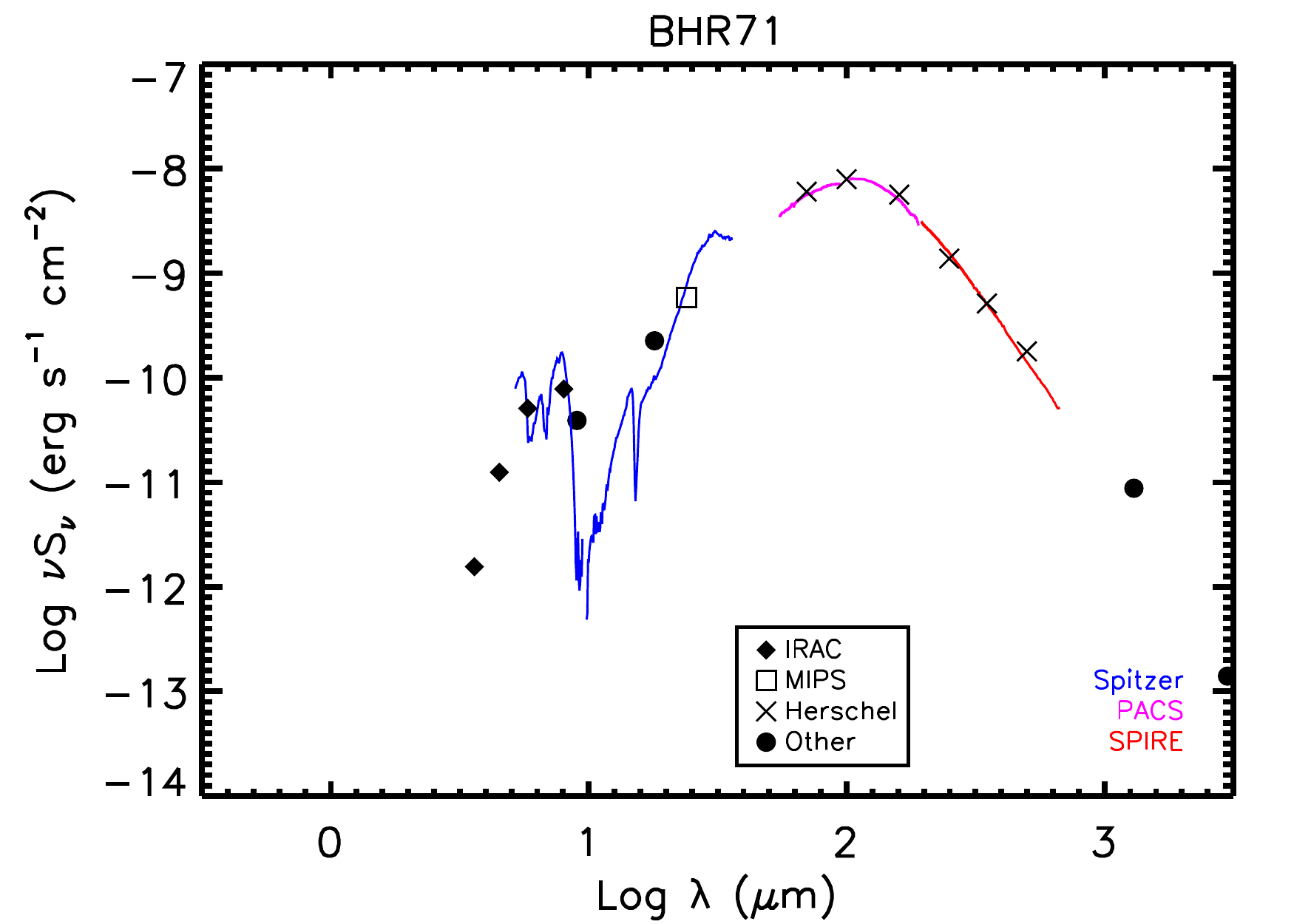}
    \includegraphics[width=0.45\textwidth]{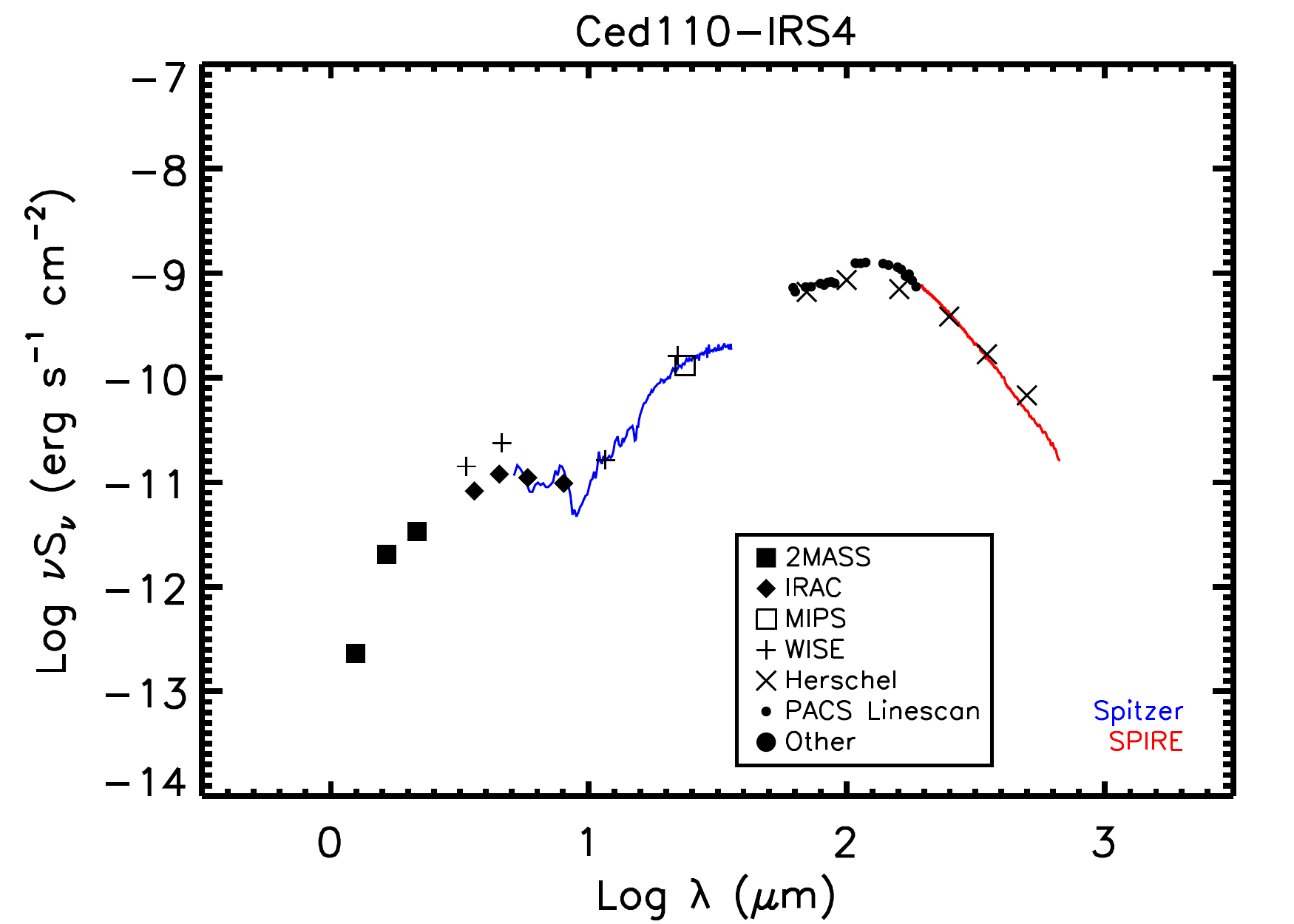}
    \includegraphics[width=0.45\textwidth]{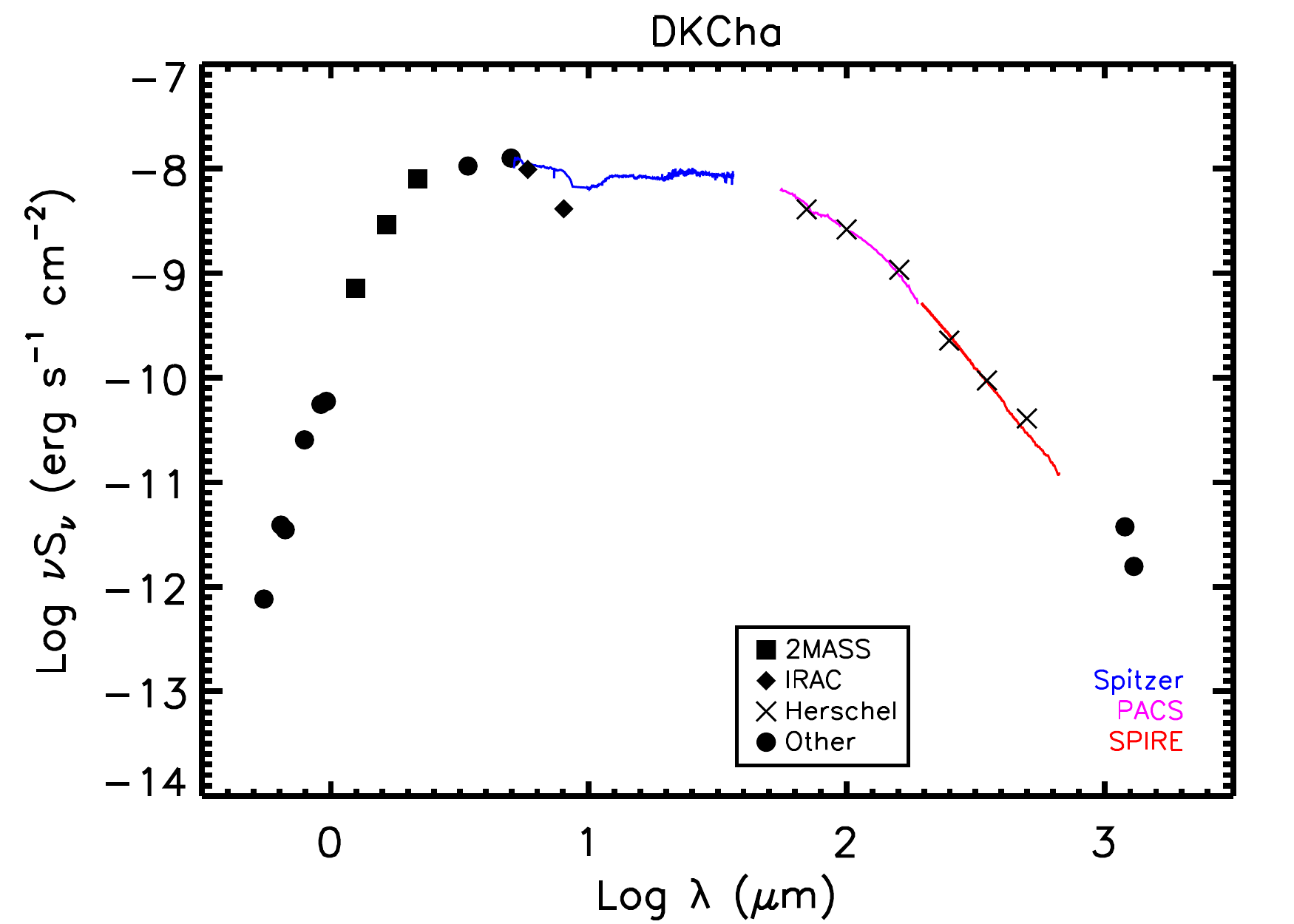}
    \caption{The SEDs of the COPS sources.  The spectra of \textit{Spitzer}-IRS, PACS, and SPIRE are shown in blue, magenta, and red, respectively.  We only show the line-free continuum for PACS and SPIRE, while the continuum-free SPIRE spectra are shown in the Appendix.  The black filled circles illustrate the linescan data where the rangescan PACS data is unavailable.  The sources of the photometry are shown in the legend.  For a better visualization, the PACS spectra of B1-a, DK~Cha, IRAS~03301+3002, and TMC1 are rebinned to R=100, while the PACS spectra of L1014 and L1455~IRS3 are rebinned to R=50.}
    \label{fig:sed1}
\end{figure*}

The full 1--1000~\micron\ SEDs of 26 protostars, excluding HH~100, observed in the COPS program are shown in Figure~\ref{fig:sed1}, along with the \textit{Herschel} PACS spectra observed in the DIGIT program \citep{green13b}, archival \textit{Spitzer}-IRS spectra, and archival photometry.  We show the SEDs of both RCrA~IRS7B and RCrA~IRS7C for their difference at shorter wavelength, although only RCrA~IRS7C (noted as RCrA~IRS7B/C) is considered in some of the analyses.  The agreement between the PACS and SPIRE spectra are improved with our extraction of the PACS 1D spectra.  In five sources (TMC~1, WL~12, L1014, L1455~IRS3, and IRAS~03301+3111), the total flux at 200~\micron\ derived from all 25 PACS spaxels is significantly less than the flux measured from the SPIRE data at the same wavelength. We found that the SPIRE emission in all five sources is extended, and therefore the observed SPIRE flux is derived from an even greater spatial extent than the entire PACS footprint.  However, we do not see such large-scale extended emission in either PACS spectroscopy or imaging, suggesting that the SPIRE flux may be contaminated by unrelated emission.
Note that the SPIRE-FTS calibrated the background emission with an on-board instrumental feed instead of chopping between source and background, which was used in PACS background calibration.  However, our data are insufficient to determine the origin of the extended emission; therefore, we do not correct for the extended emission.

The SPIRE images are available for 24 protostars, for which we performed aperture photometry to extract their photometric fluxes to compare with their 1D spectra.  The SPIRE spectroscopic and photometric flux, when extracted in the same size aperture for each band, match to within 15\%\ for 21 of 24 protostars at 250~\micron\ (TMC~1, BHR~71, and DK~Cha have 17\%, 16\%, and 17\%\ agreements, respectively) and 23 of 24 protostars at 350~\micron\ (IRAS~03301+3111 has a 19\%\ agreement), and they match to within 35\%\ for all 24 protostars at 500~\micron.  The median percentage differences between SPIRE photometry and SPIRE spectroscopy are 8\%, 3\%, and 24\%\ for 250~\micron, 350~\micron, and 500~\micron, respectively.  All sources show higher photometric fluxes at 500~\micron\ than the spectroscopic fluxes at the same wavelength, suggesting the presence of extended emission that is not considered by the SECT correction.  As a comparison, the PACS spectra presented in this study, which were originally published in \citet{green16a}, have median percentage differences between photometry and spectroscopy of 16\%, 8\%, and 10\%\ at 70~\micron, 100~\micron, and 160~\micron, respectively.

\renewcommand{\thefigure}{\arabic{figure} (Cont.)}
\addtocounter{figure}{-1}

\begin{figure*}[htbp!]
    \centering
    \includegraphics[width=0.45\textwidth]{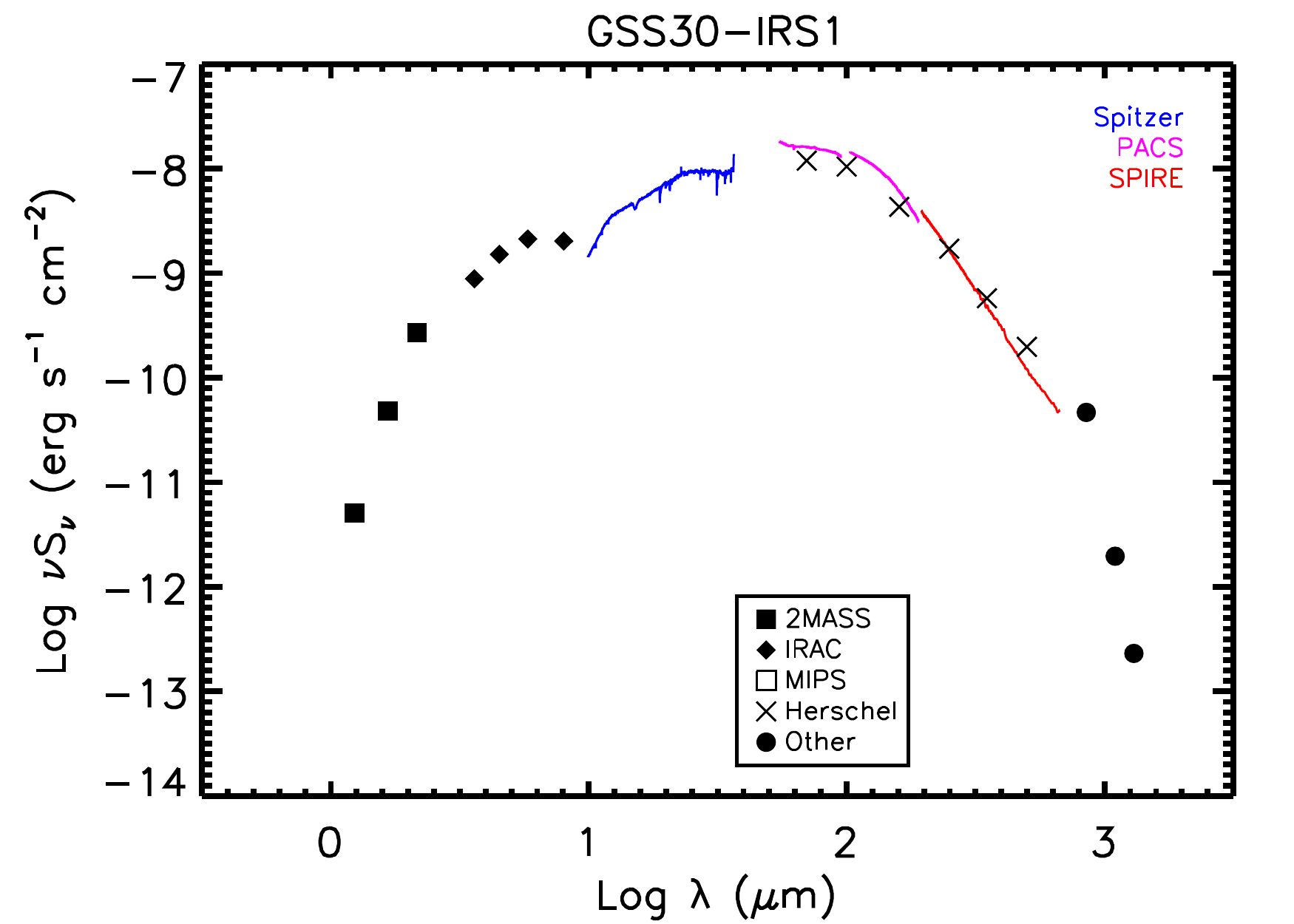}
    \includegraphics[width=0.45\textwidth]{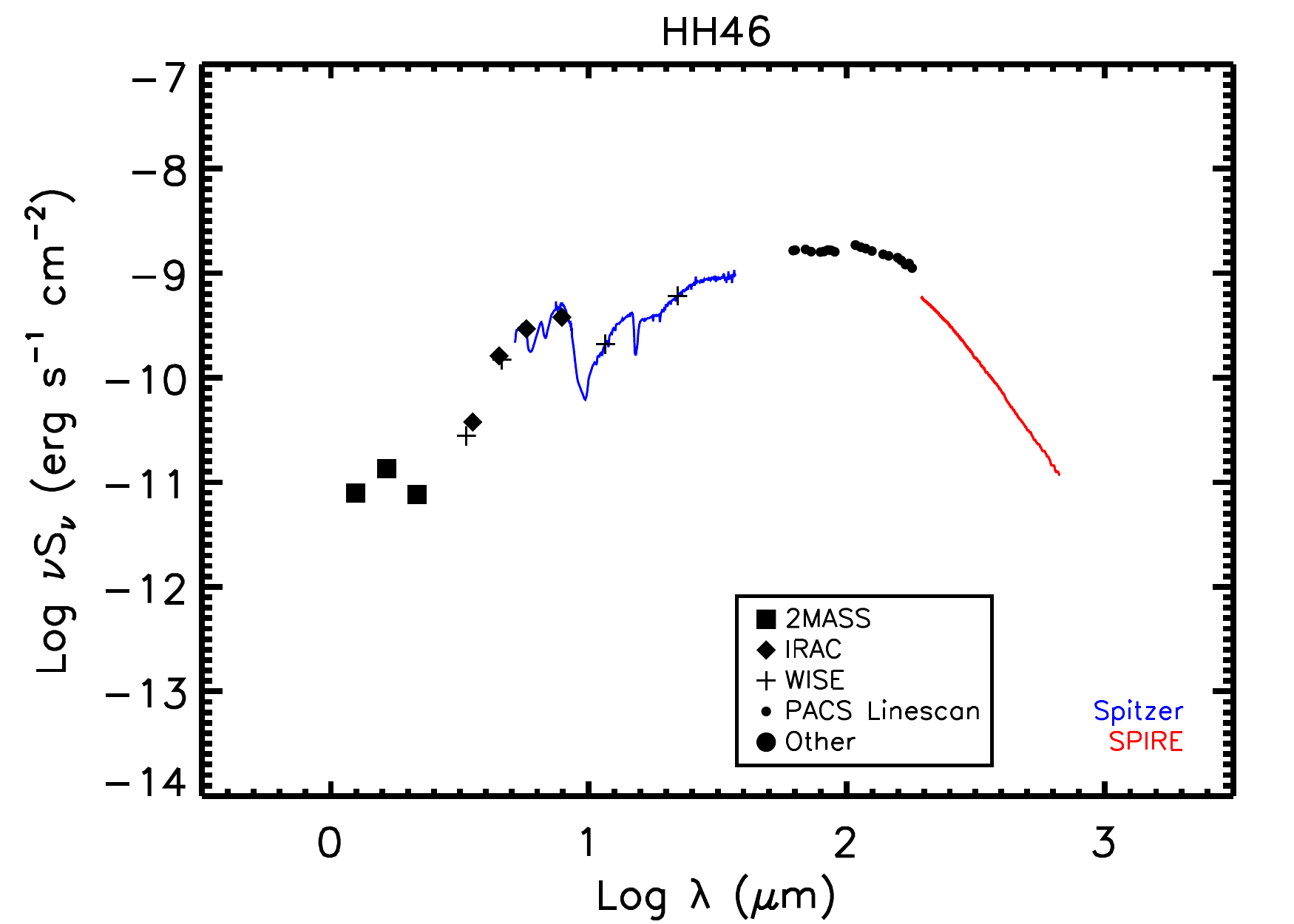}
    \includegraphics[width=0.45\textwidth]{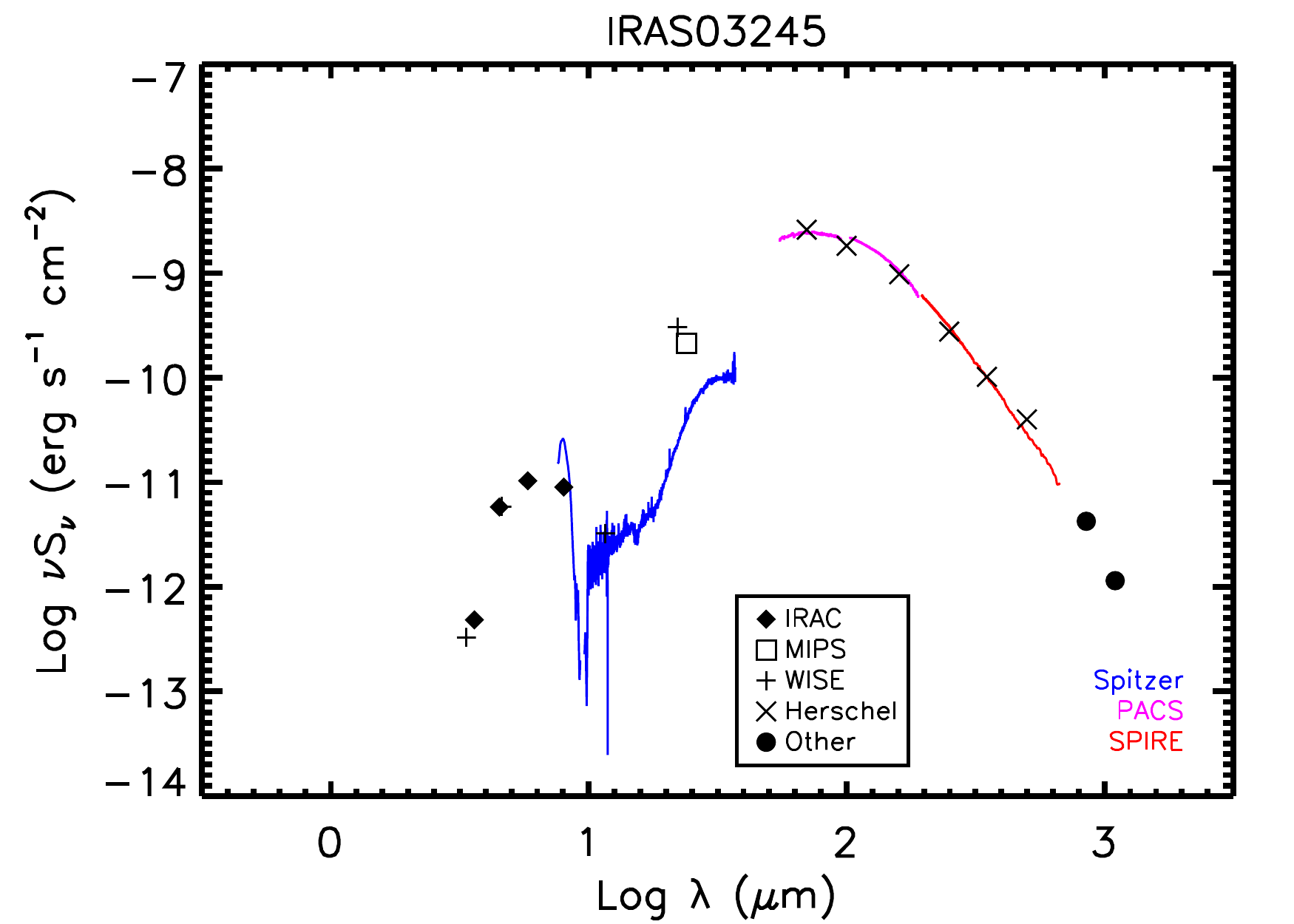}
    \includegraphics[width=0.45\textwidth]{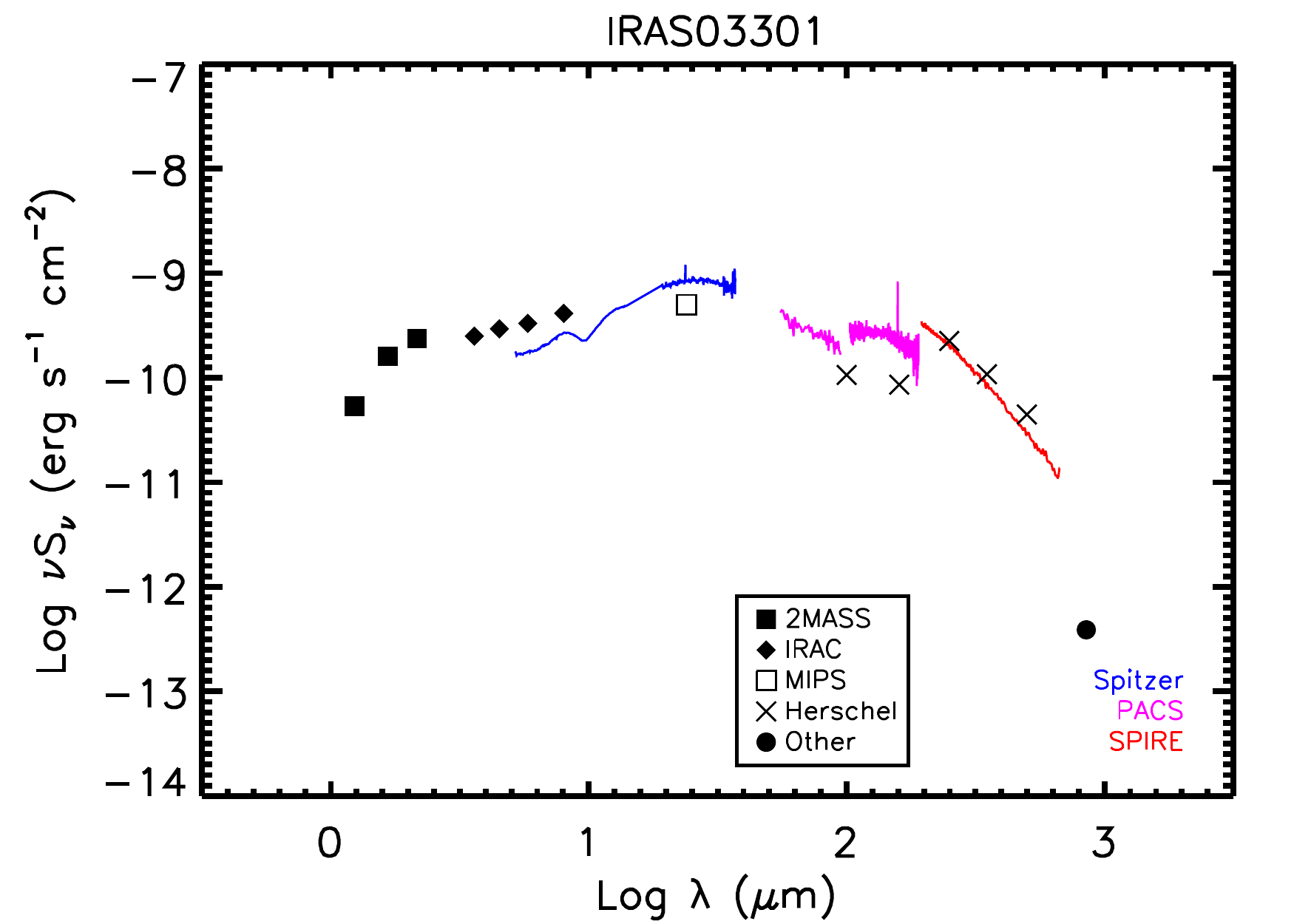}
    \includegraphics[width=0.45\textwidth]{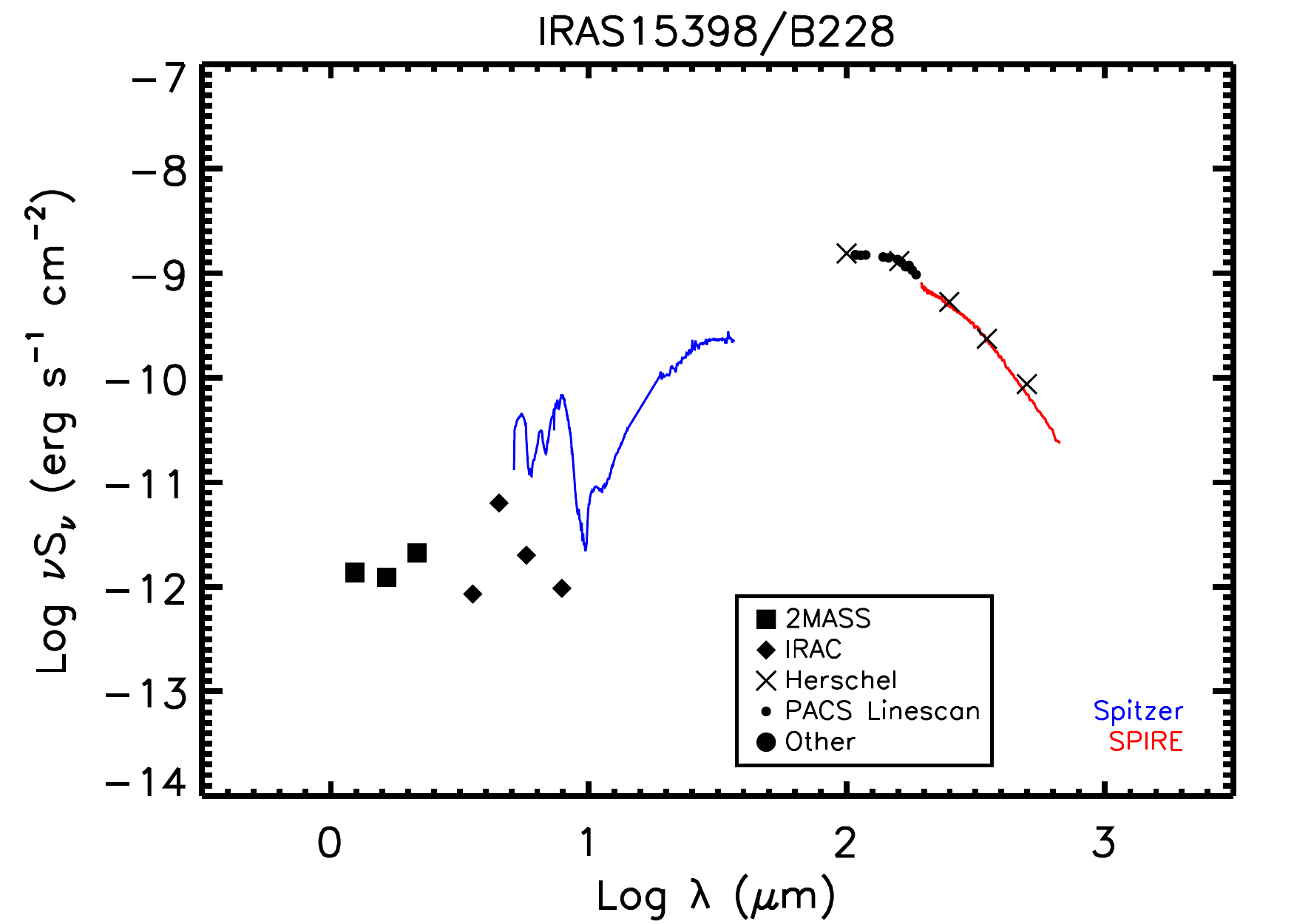}
    \includegraphics[width=0.45\textwidth]{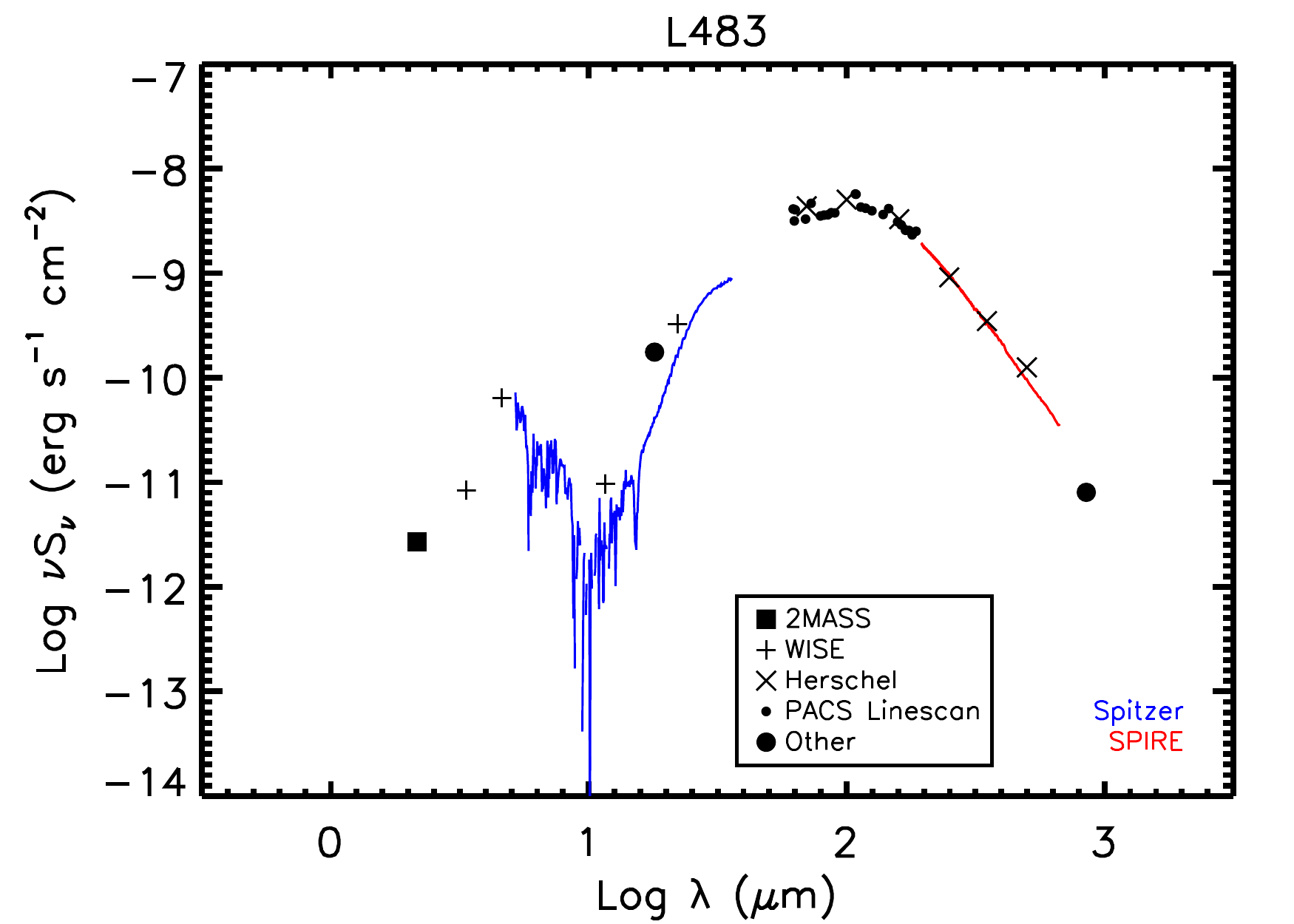}
    \caption{}
\end{figure*}
\addtocounter{figure}{-1}
\begin{figure*}[htbp!]
    \centering
    \includegraphics[width=0.45\textwidth]{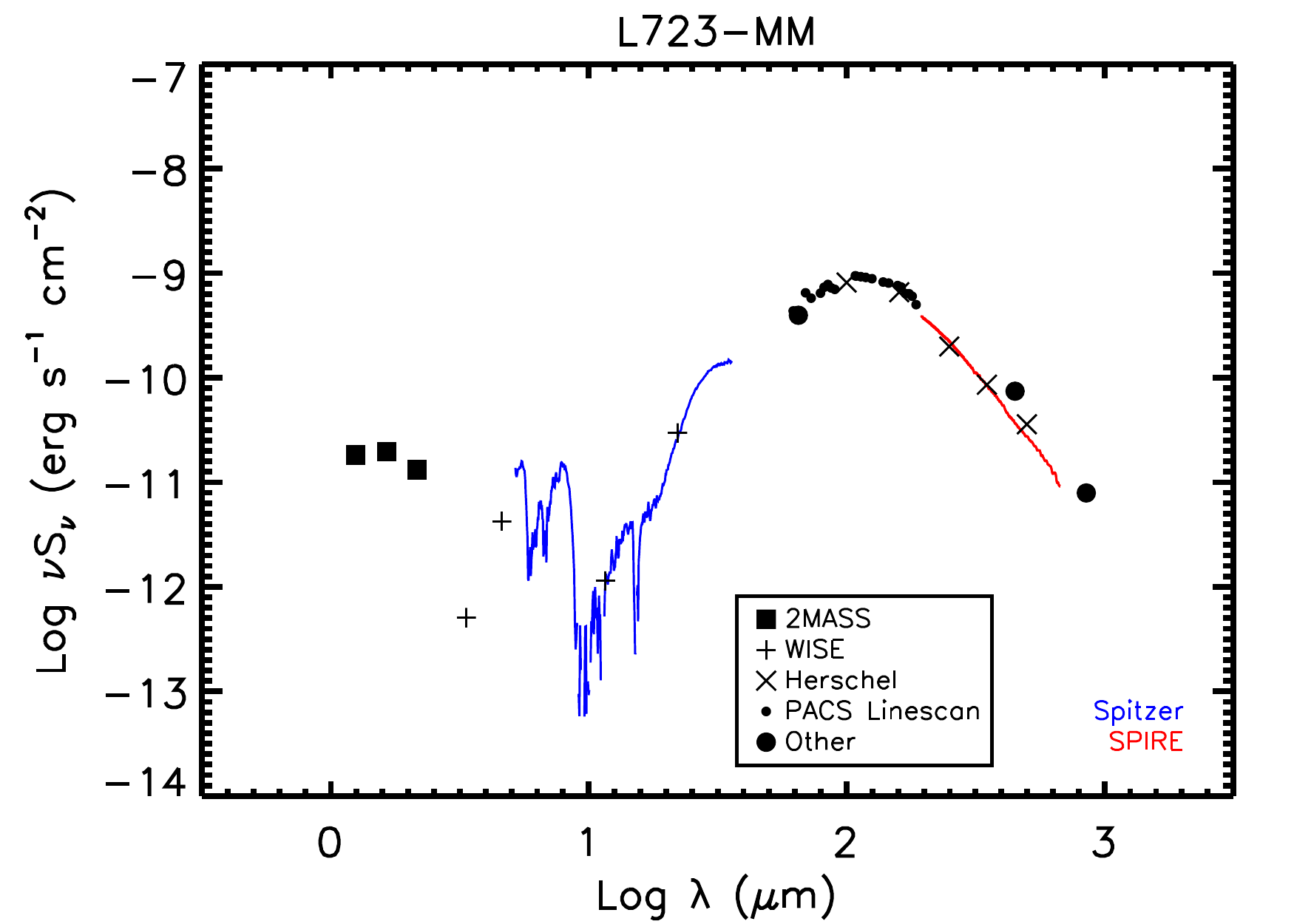}
    \includegraphics[width=0.45\textwidth]{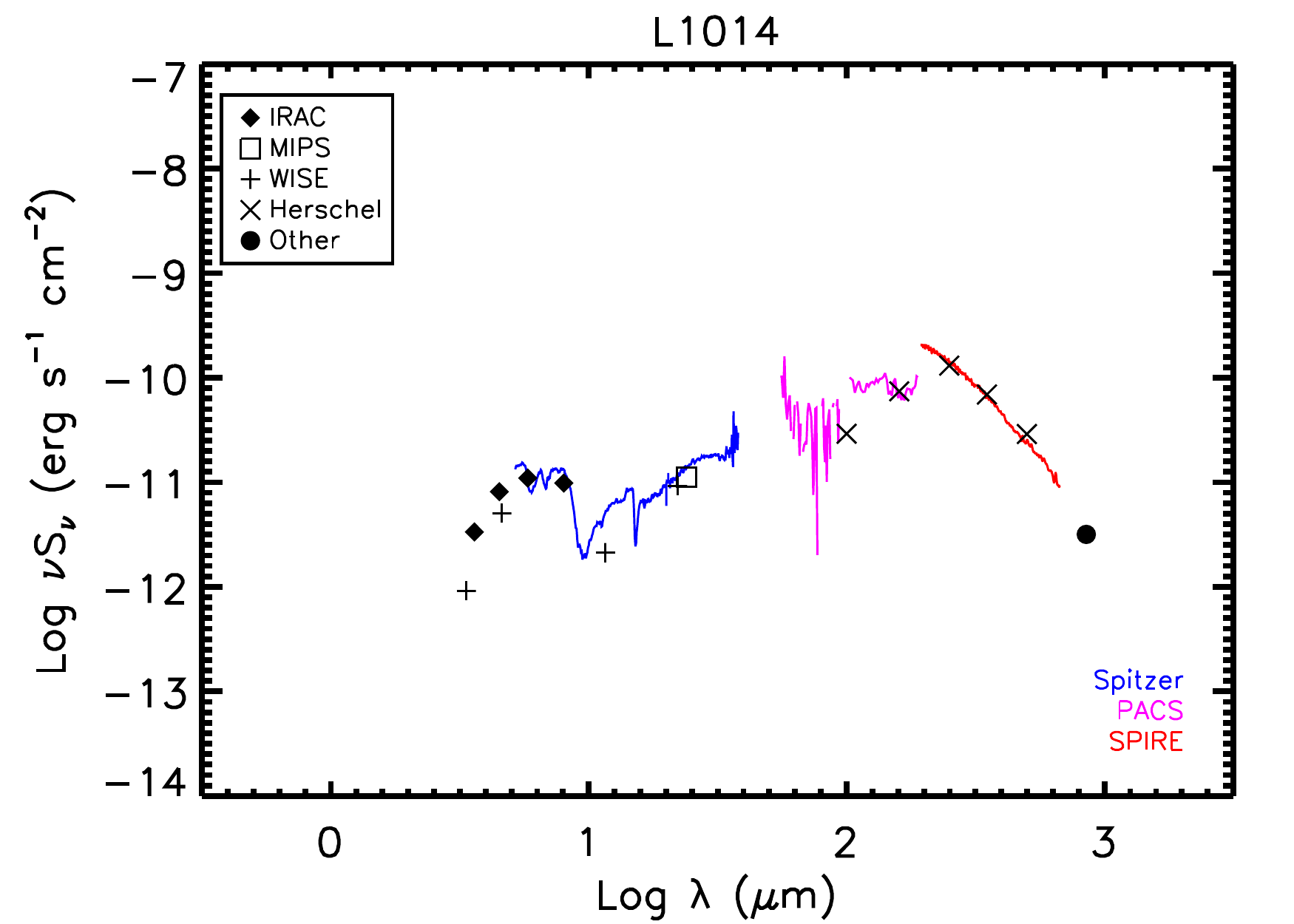}
    \includegraphics[width=0.45\textwidth]{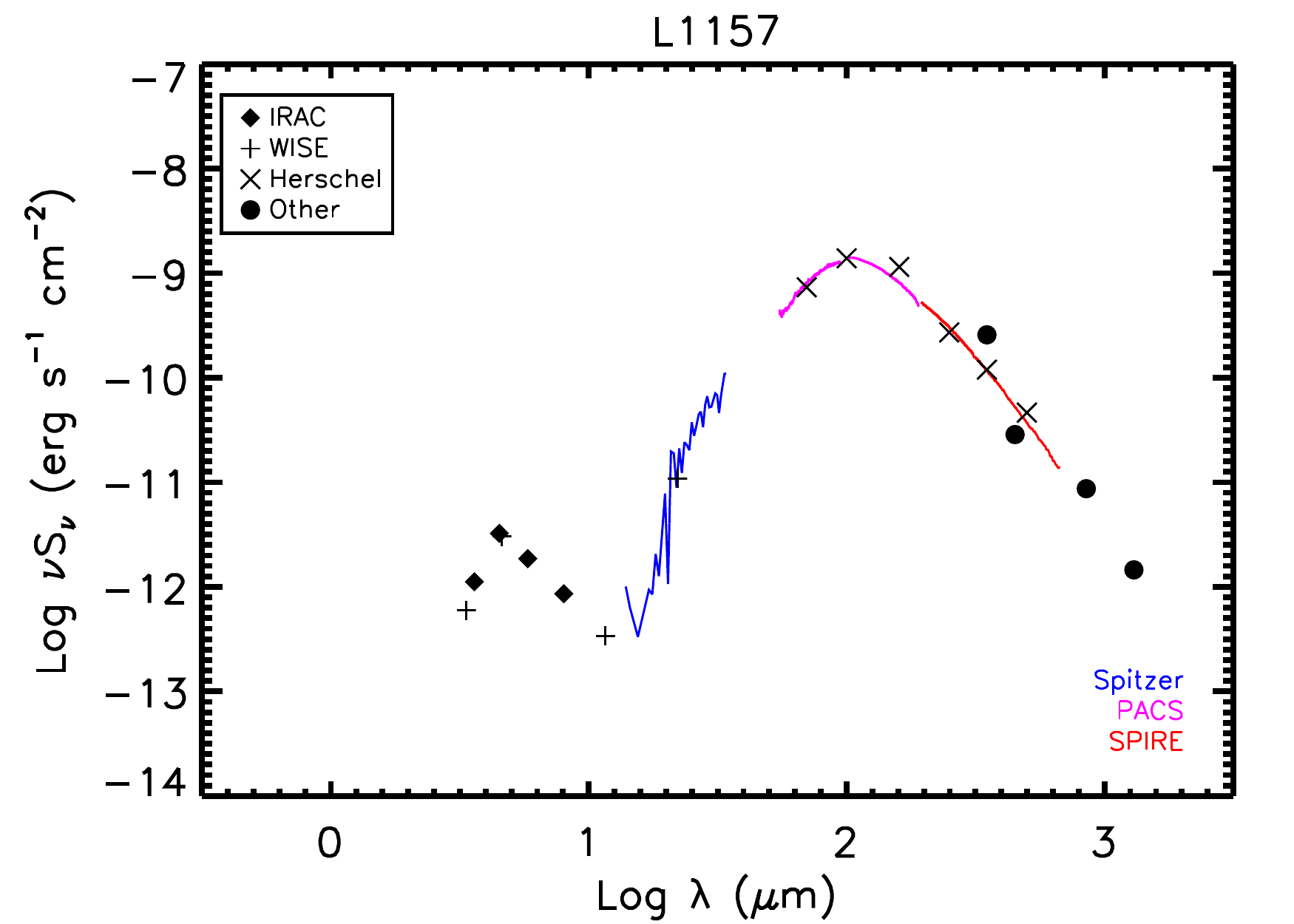}
    \includegraphics[width=0.45\textwidth]{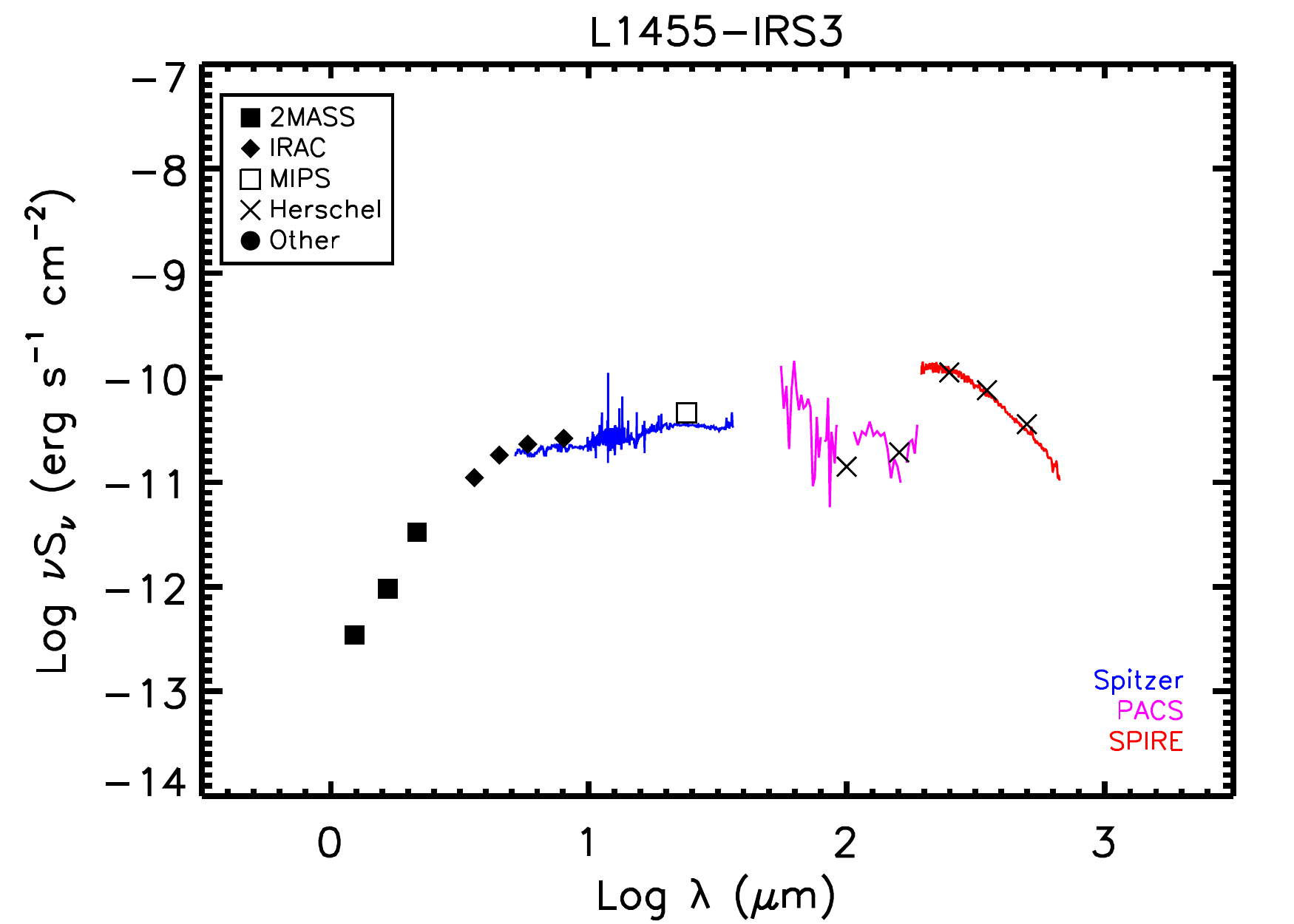}
    \includegraphics[width=0.45\textwidth]{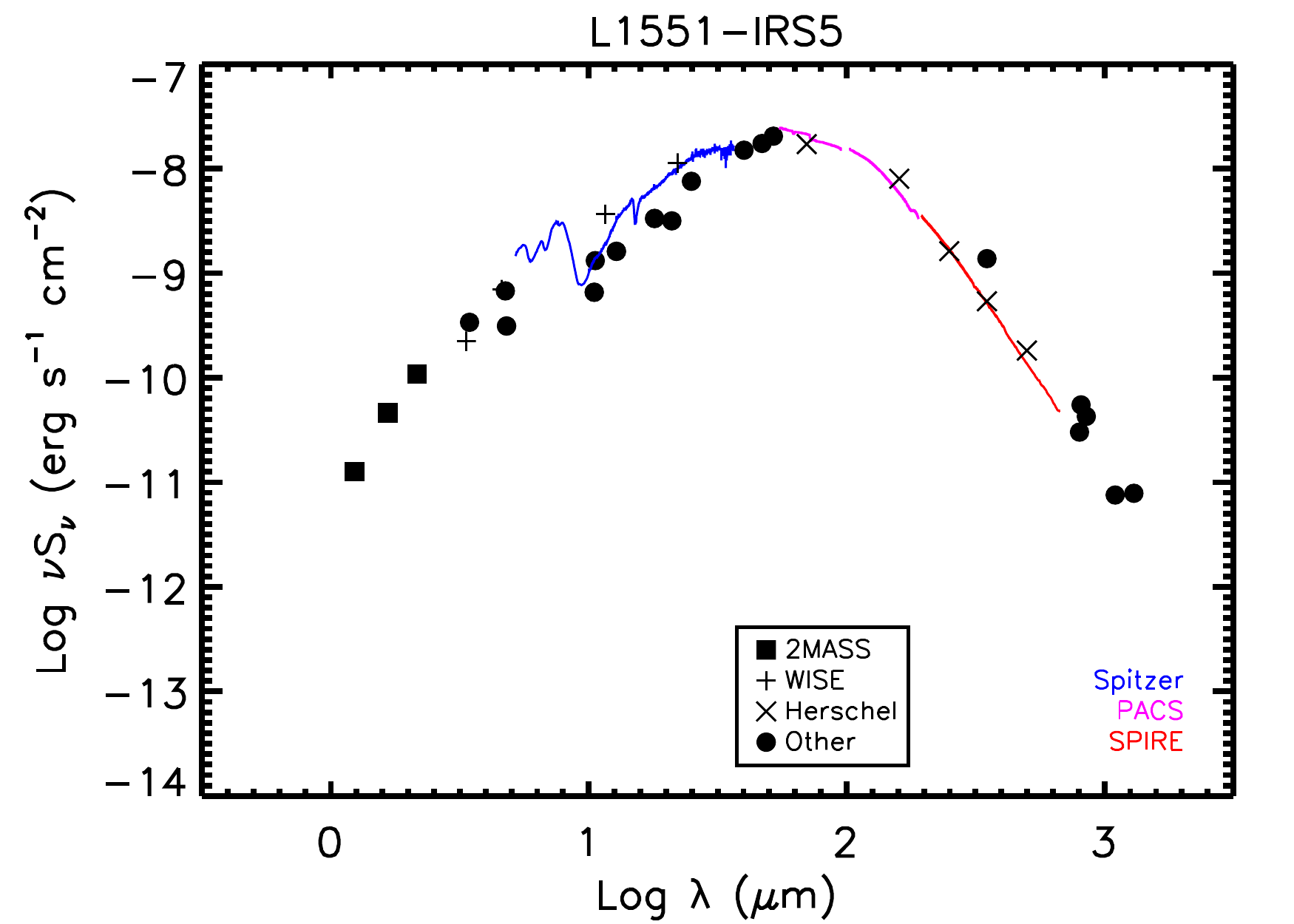}
    \includegraphics[width=0.45\textwidth]{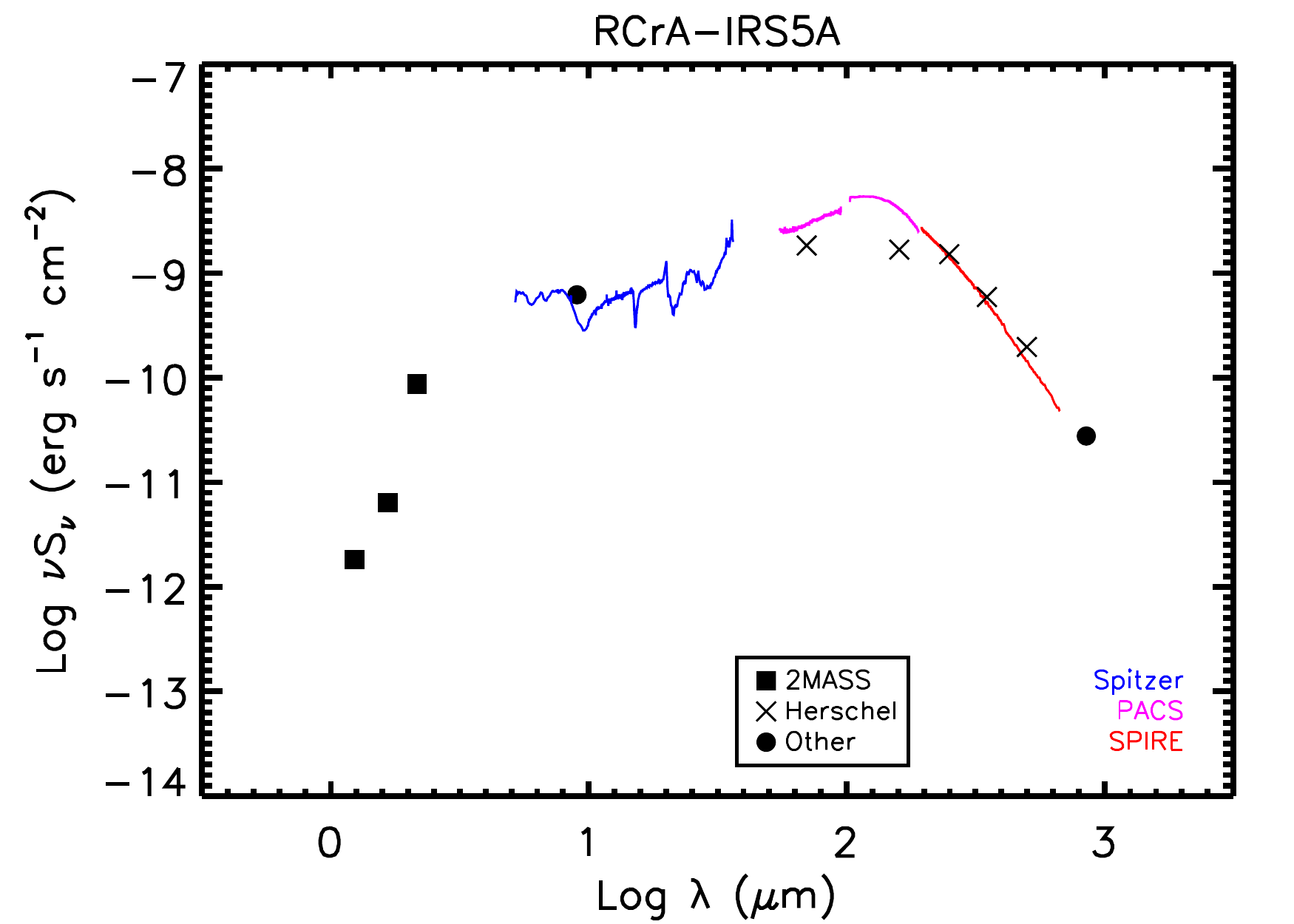}

    \caption{}
\end{figure*}
\addtocounter{figure}{-1}

\begin{figure*}[htbp!]
    \centering
    \includegraphics[width=0.45\textwidth]{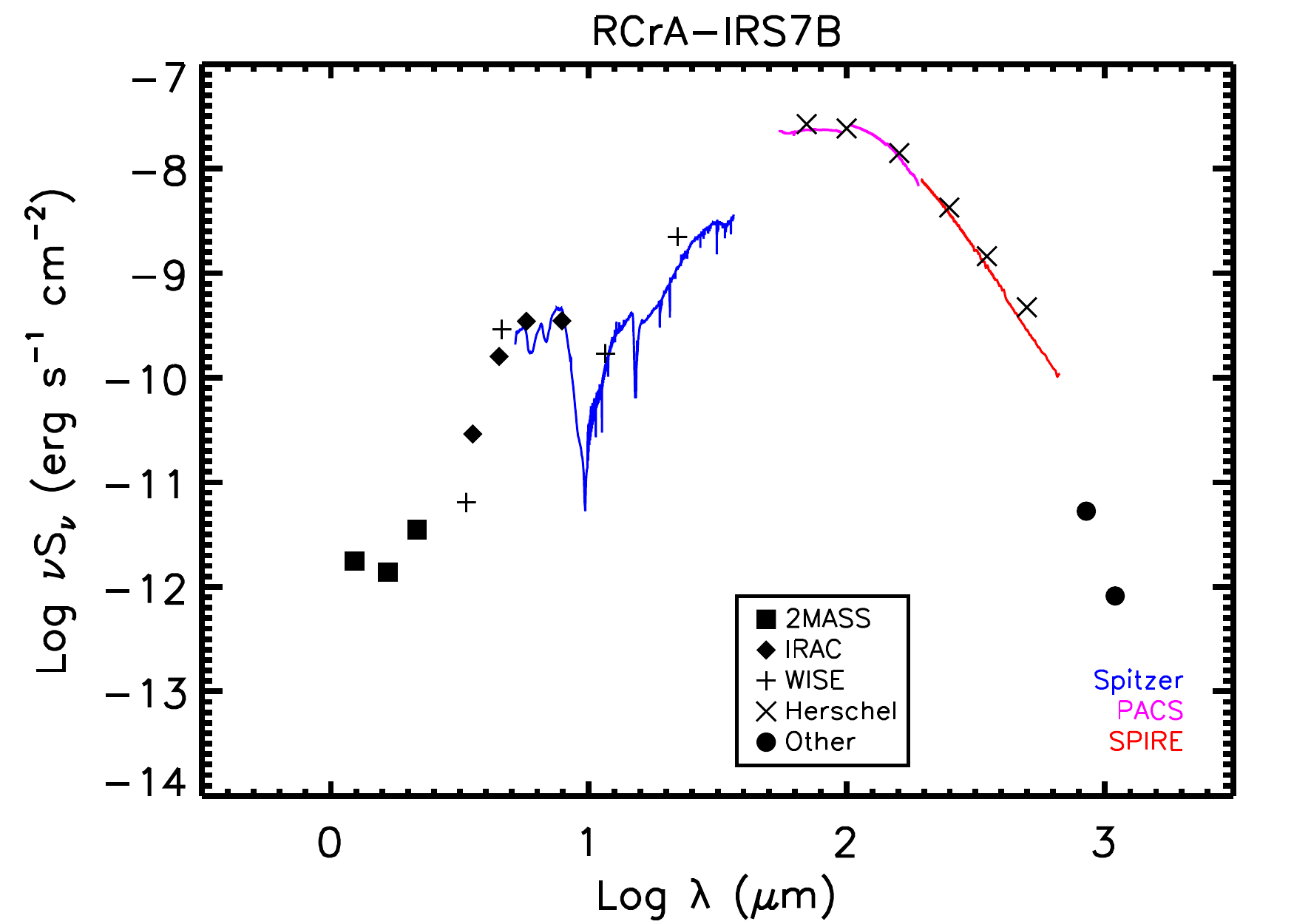}
    \includegraphics[width=0.45\textwidth]{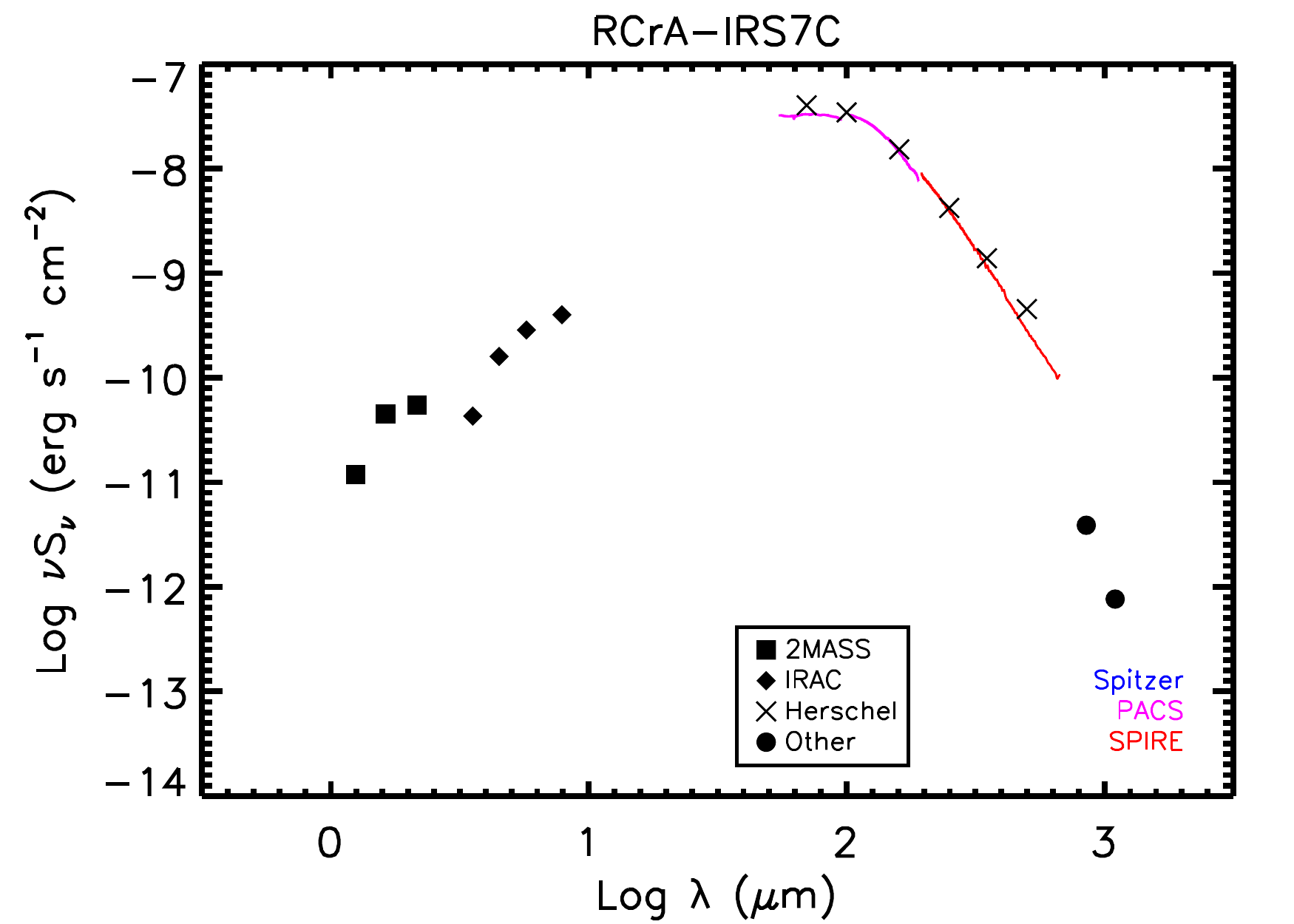}
    \includegraphics[width=0.45\textwidth]{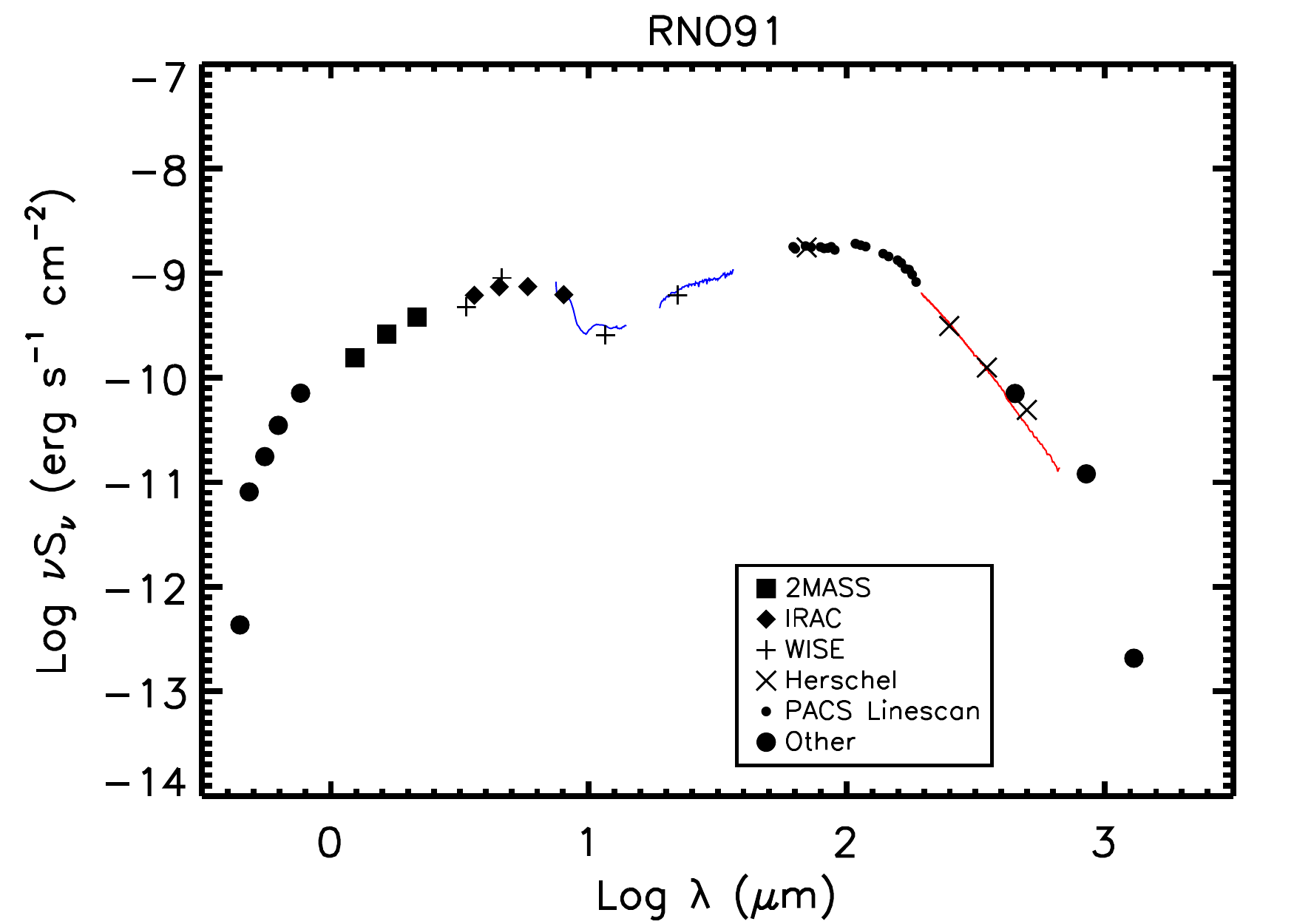}
    \includegraphics[width=0.45\textwidth]{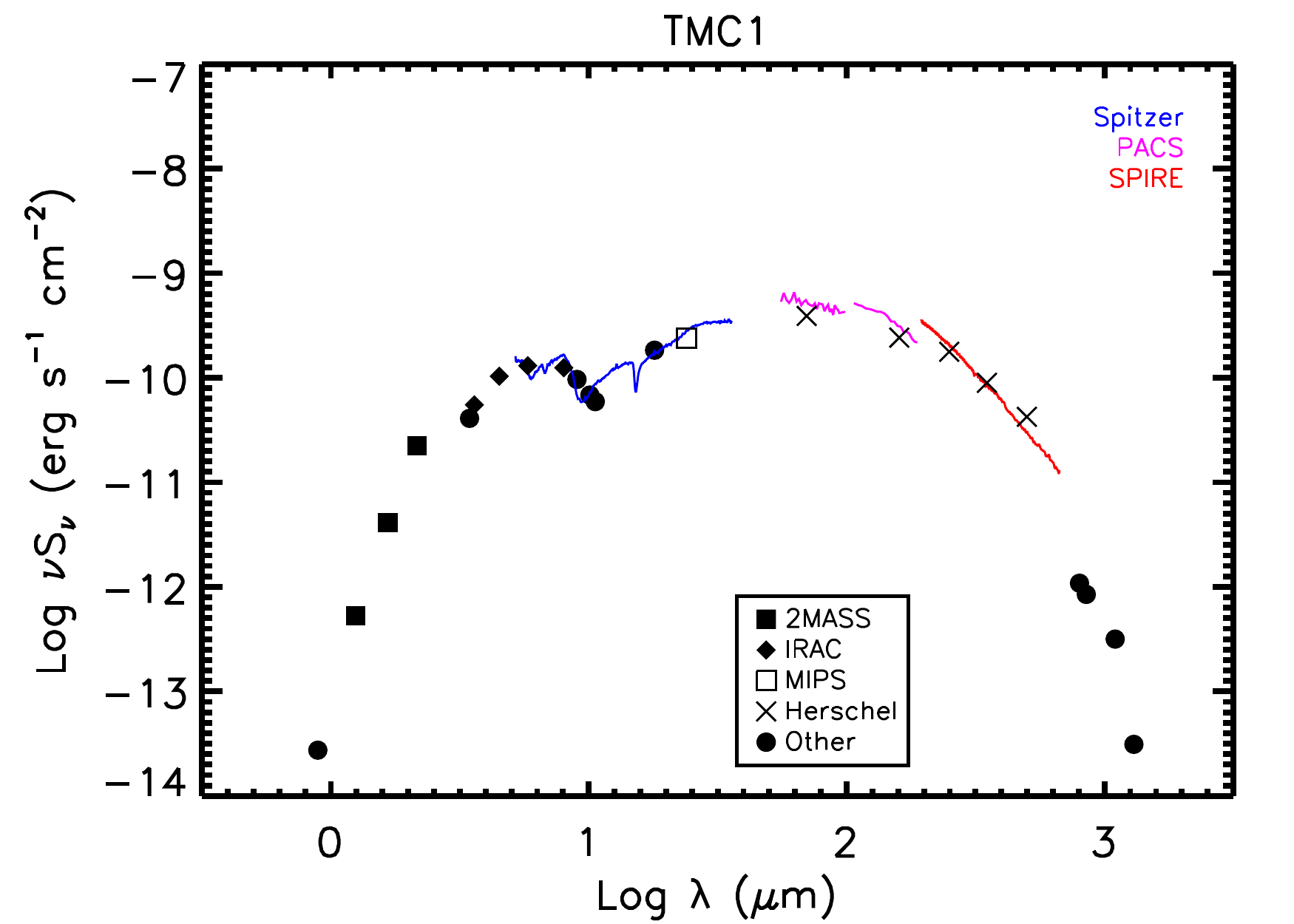}
    \caption{}
\end{figure*}
\addtocounter{figure}{-1}

\begin{figure*}[htbp!]
    \centering
    \includegraphics[width=0.45\textwidth]{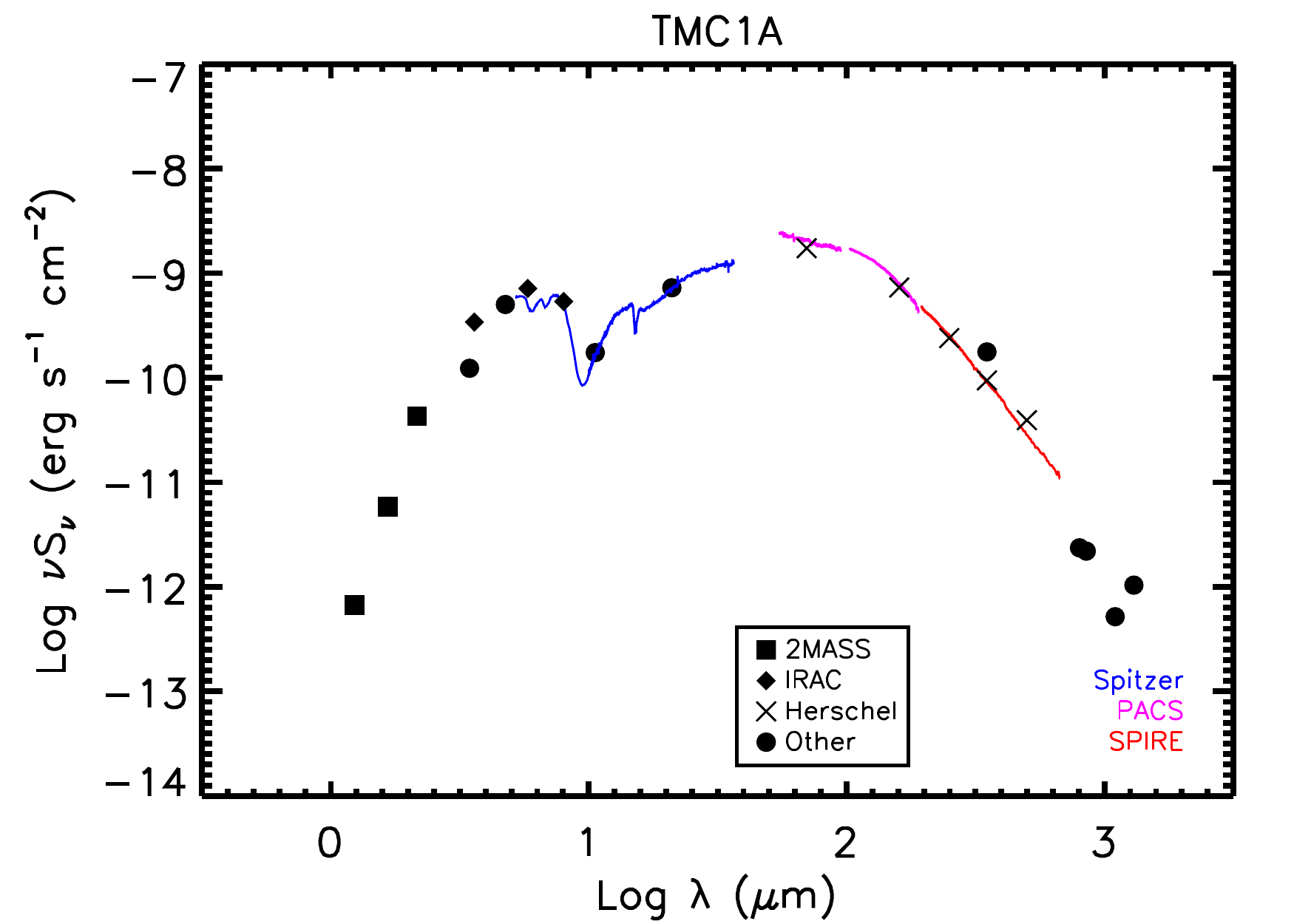}
    \includegraphics[width=0.45\textwidth]{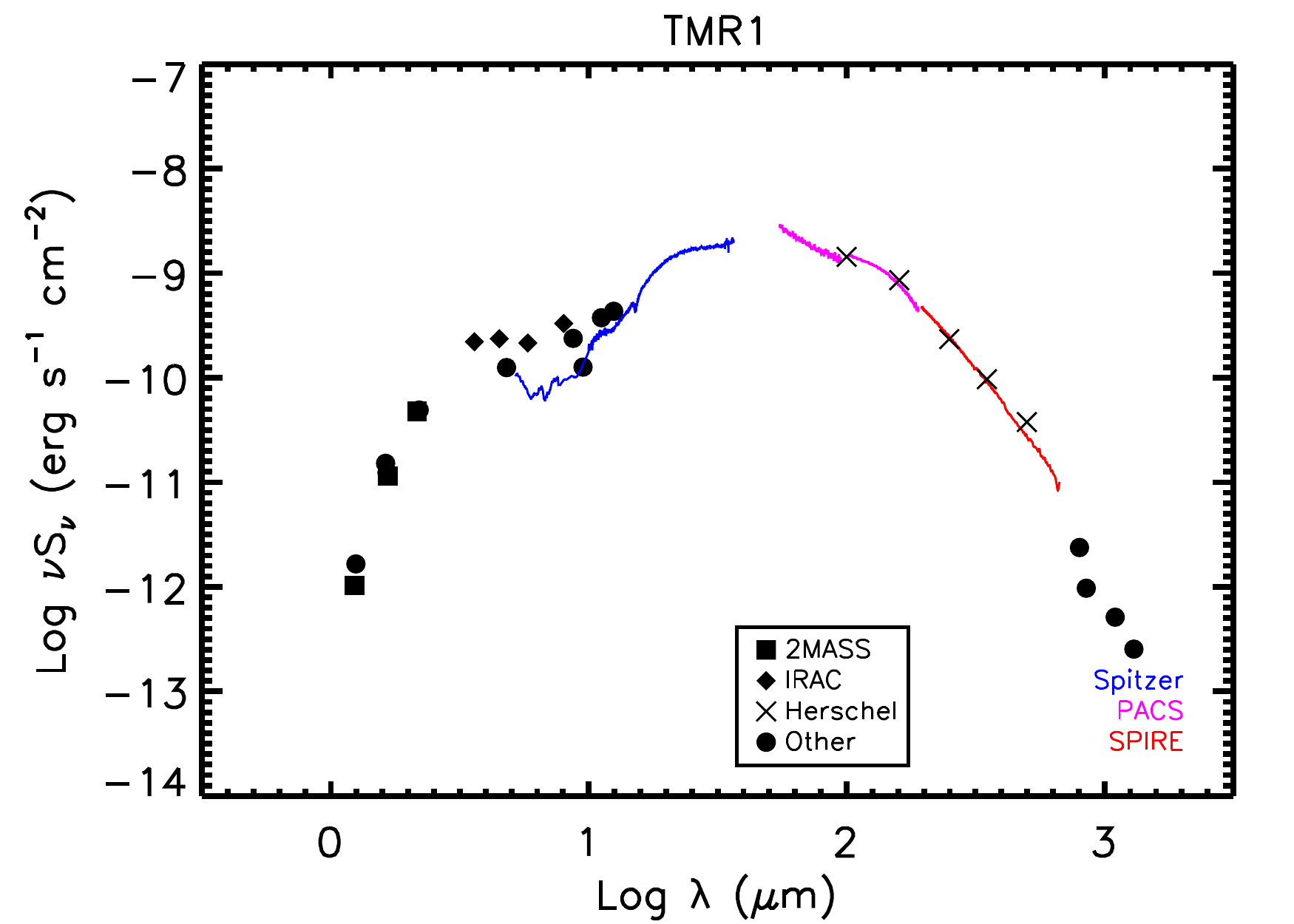}
    \includegraphics[width=0.45\textwidth]{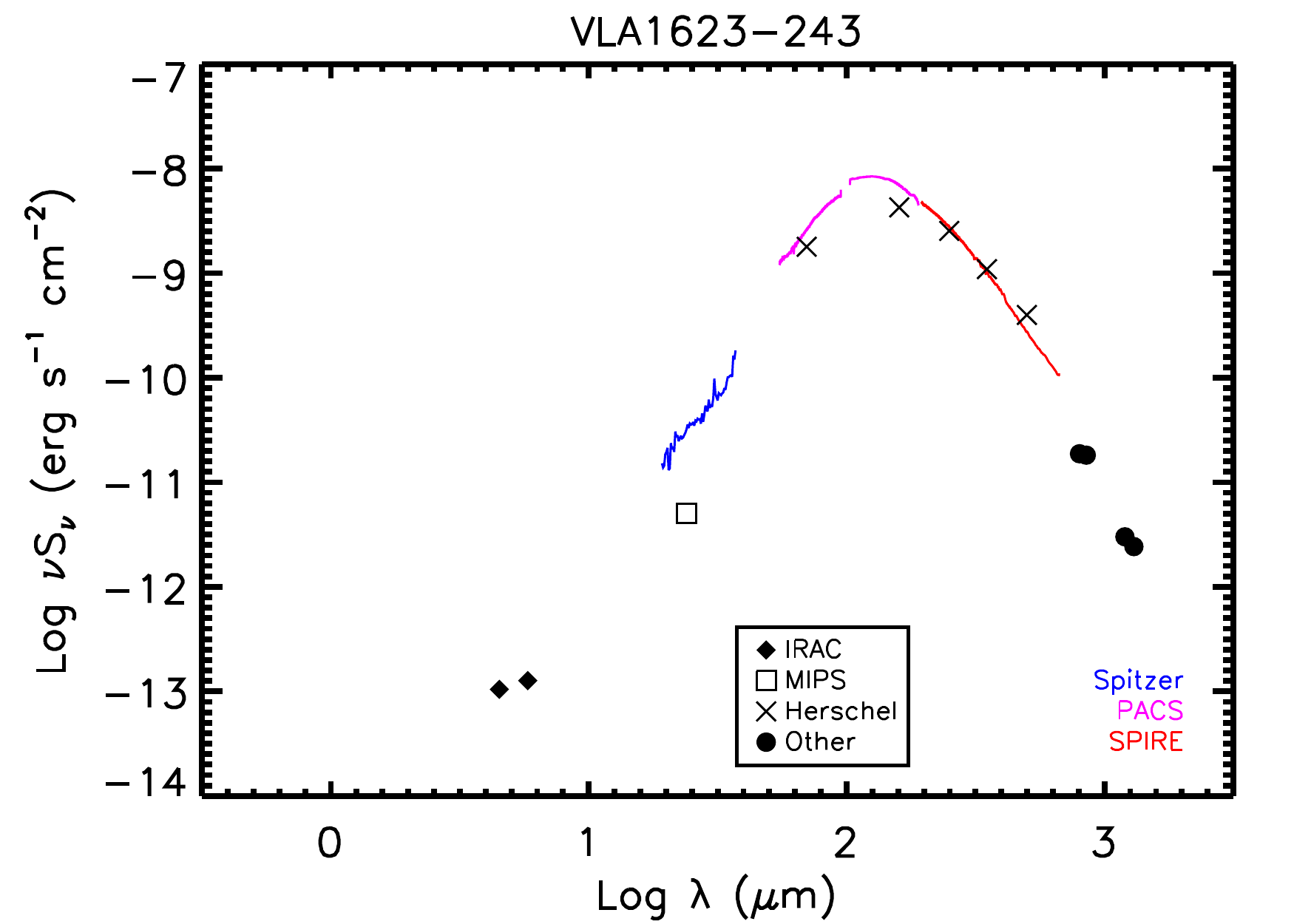}
    \includegraphics[width=0.45\textwidth]{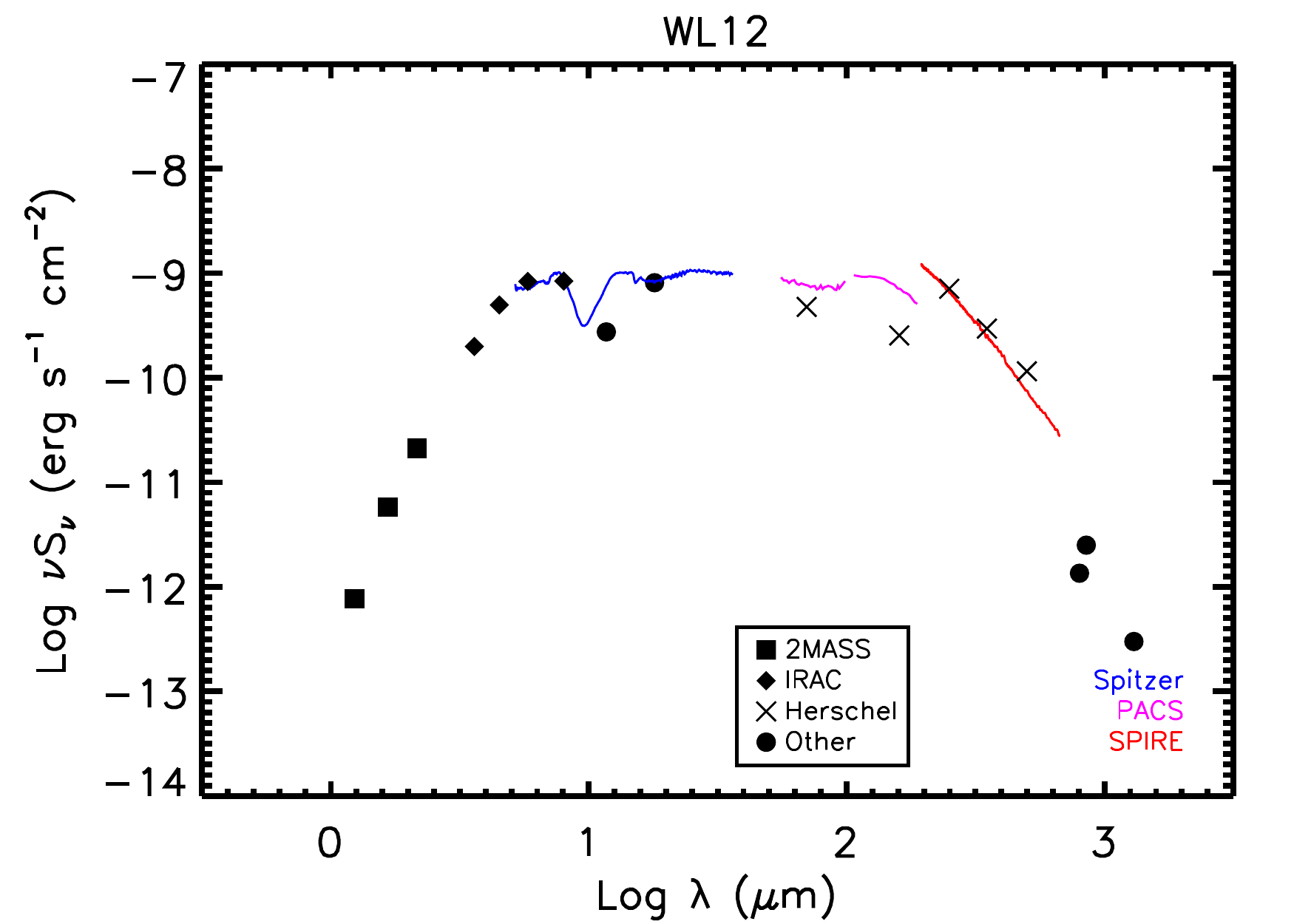}
    \caption{}
\end{figure*}

\renewcommand{\thefigure}{\arabic{figure}}

\subsection{Lines: Atomic and Molecular Emission}
\label{sec:lines}

With the improved reduction, we performed an automatic line fitting, which was developed for the CDF archive \citep{green16a}, on the spectra of each spaxel and the 1D spectra.  The line fluxes were updated from the CDF archive based on the updated reduction.
Here we briefly describe the concept of this automatic line fitting routine.

The routine utilizes a pre-defined line list \citep[see Table~2--6 in][]{green16a}, including the emission of molecules and atoms, such as CO, $^{13}$CO, \water, HCO$^{+}$, \CI, and \OI, to fit the local baselines around all potential lines.  After subtracting the baseline, we fit a Gaussian profile around the theoretical line centroid with limited flexibility on the exact wavelength of line centroids.  The full width at half maximum (FWHM) is fixed at the instrumental resolution for PACS spectra, while the FWHM is allowed to vary within $\pm$30\%\ of the instrumental resolution to better fit the apodized SPIRE spectra.  In some cases where two lines are blended together, we fit two Gaussian profiles simultaneously when the signal-to-noise ratio (SNR) is sufficiently large.  Then, we re-evaluate the uncertainties of the fitted parameters by re-fitting the baselines on the original spectra with the fitted line profiles subtracted.  Finally, the line fitting is performed for the third time to include the updated uncertainties, improving the fitting results.  Most of the lines are unresolved except for \OI~$^{3}P_{1}\rightarrow^{3}P_{2}$ at 63~$\mu$m, which can be as wide as 1.5$\times$FWHM, a limit set by the line fitting pipeline.  The line fitting pipeline allows us to separate the continuum from the spectrum.  We present the line-free continua in the SEDs (Figure~\ref{fig:sed1}), while showing the continuum-free spectra in Appendix~\ref{sec:a_spire_1d} (Figure~\ref{fig:spire_1d}).

Our fitting pipeline produces more than just the 1D spectra with the line taken out; it further smoothes the line-free continuum by 20 wavelength channels, resulting in continuum spectra with $\lambda/\Delta \lambda$=16--50.  The smoothing process has little effect on the mean photometry across the whole sample, only a 0.04\%\ decrease in SPIRE 500~\micron\ band.  However, the difference of the photometry ranges from $-0.6$\%\ to 0.4\%.

\subsubsection{Detection Statistics for the SPIRE Spectra}
\label{sec:linestats}

Quasi-periodic baseline variations at scales between 0.3--1~\micron\ are found in the SPIRE spectra because of the nature of the Fourier Transform Spectrograph (FTS) even if the side-lobes of the sinc function have been suppressed by the apodization, which convolves the spectra with a taper function.  We ran the same fitting pipeline with the line centroids shifted to supposedly line-free regions to quantify the false-positive rate of detections as a function of the SNR threshold.  We found that SNR thresholds of 3, 4, and 5 yield false-positive rates of 1.3\%, 0.98\%, and 0.84\%, respectively.  The false-positive rate with an SNR threshold of 3 is greater than the rate assuming Gaussian noise, which predicts a false-positive rate of 0.3\%, indicating that the apodization on the FTS spectra indeed introduces non-Gaussian noise.  To obtain a robust analysis, we choose a threshold of 1\%\ for the false-positive rate, corresponding to a SNR threshold of 4 for the SPIRE spectra, while we consider lines as detections with SNRs greater than 3 for the PACS spectra.  We summarize the detections of lines in Table~\ref{detection}.  We attempted to fit 28 lines of interest (see the full list in \citealt{green16a}) for each source.

The lowest excitation line of water, o-\water~1$_{10}\rightarrow1_{01}$ at 557 GHz (538 \micron; $E_{\rm{up}}/k$ = 61~K) was surveyed with HIFI for the DIGIT/WISH sources, which include all of the COPS sources \citep{kristensen12,green13b,2014AA...572A..21M}.
The line was detected by HIFI in 24 sources, while we detected the line toward 8 sources in the SPIRE spectra.  The main reason for this difference in detection rates is the difference of sensitivities of two instruments/programs.  The SPIRE-FTS spectral resolution ($\Delta\nu = $2.16~GHz after apodization, corresponding to $\Delta v$ = 1162~km~s$^{-1}$ at 557~GHz) is much greater than the linewidth measured with HIFI ($<$ 180 km s$^{-1}$); thus, the water lines remain unresolved and therefore spectrally diluted.  \citet{2014AA...572A..21M} found an RMS noise of $\sim$20~mK for the same line, while our data show an RMS noise of $\sim$0.7~K, which is about 35 times larger.
Overall, we detect up to six transitions of \water\ in the SPIRE bands, with at least one detection in 17 protostars, while \water\ in PACS bands is detected in almost all COPS protostars (24 sources) with PACS.

\subsubsection{Line Fitting Results}
\label{sec:fitting_results}
The line fitting results are written to an ASCII file for further analyses.  Table~\ref{table:report} shows an example of the line fitting results used in this study.  The full table can be accessed online as a machine readable table.  Each column follows the same terminology described in \citet{green16a}, which will be outlined briefly in the following.  Table~\ref{terminology} lists all of the column names along with their descriptions.  Special numbers, \texttt{-998} and \texttt{-999}, can be found under \texttt{Sig\_Cen(um)}, \texttt{Sig\_str(W/cm2)}, and \texttt{Sig\_FWHM(um)} columns. The \texttt{-998} indicates that the fitted parameter can be used, but the uncertainty must be extrapolated from other nearby fitted lines.  The \texttt{-999} indicates that the fitted parameter is not well-constrained.
The \texttt{Pixel\_No.} shows the spaxel name of the fitted spectrum.  In the case of our extracted 1D spectrum, it will list as \texttt{c}.  The blending flag highlights if there is any possible line in our line list within one resolution to the fitted line centroid.  Note that whether the nearby line is detected is not considered in reporting the blending flag.
We further selected several pairs of nearby lines to perform double Gaussian fittings, which are reported as \texttt{DoubleGaussian} in the blending flag.  Finally, the validity flag suggests whether the fitting result should be used.  If \texttt{-999} flag is found in any column of the fitting result, the validity will be flagged as \texttt{0}, indicating the line is not well-constrained; otherwise, the validity flag is \texttt{1}.

\begin{table*}[htbp!]
	\small
	\caption{A portion of the line fitting results.}
	\tt
	\begin{tabular}{r r r r r r r}
		\toprule
		  Object  &           Line   &      LabWL(um) &        ObsWL(um) &     Sig\_Cen(um) &       Str(W/cm2) &  Sig\_str(W/cm2) \\
          B1-a    & o-H2O8\_27-7\_16 &       55.13238 &         55.09925 &       -999.00000 &     5.136442e-21 &     1.588567e-21 \\
          B1-a    & o-H2O4\_41-4\_32 &       94.70758 &         94.68837 &          0.00613 &    -1.408871e-21 &     4.175396e-22 \\
          B1-a    & p-H2O6\_42-6\_33 &      103.91892 &        103.93085 &       -999.00000 &    -2.282466e-22 &     3.550550e-22 \\
          B1-a    & p-H2O6\_15-6\_06 &      103.94278 &        104.01953 &       -999.00000 &    -5.827662e-22 &     3.471036e-22 \\
          B1-a    & o-H2O6\_34-6\_25 &      104.09629 &        104.04658 &          0.03396 &    -6.743942e-22 &     3.439754e-22 \\
          L1157   & p-H2O6\_42-5\_51 &      636.66803 &        636.52551 &          0.64353 &    -5.939747e-25 &     1.666154e-25 \\
          L1157   &          CO4-3   &      650.26788 &        650.03816 &          0.49030 &     1.179826e-24 &     4.539972e-25 \\
          L1157   &      CI3P2-3P1   &      370.42438 &        370.42438 &       -998.00000 &     0.000000e+00 &    -9.980000e+02 \\
          B335    & o-H2O4\_23-4\_14 &      132.41173 &        132.36006 &          0.03324 &     7.017522e-22 &     3.188232e-22 \\
          L723-MM &        HCO+7-6   &      480.28812 &        480.43103 &          0.28217 &     2.894226e-22 &     1.482337e-22 \\
          \hline
          FWHM(um) &  Sig\_FWHM(um) &   Base(W/cm2/um) &  Noise(W/cm2/um) &              SNR &          E\_u(K) & A(s-1)      \\
          0.03899  &     -998.00000 &    -5.290667e-20 &     7.962727e-20 &         1.554990 &        1274.2000 & 1.89700e+00 \\
          0.03444  &     -998.00000 &     2.251379e-20 &     2.661830e-20 &         1.444368 &         702.3000 & 1.52800e-01 \\
          0.11076  &     -998.00000 &     4.141600e-21 &     6.019659e-21 &         0.321756 &        1090.3000 & 2.27200e-01 \\
          0.11077  &     -998.00000 &     3.302900e-21 &     5.905709e-21 &         0.837283 &         781.1000 & 1.36000e-01 \\
          0.11084  &     -998.00000 &     3.093281e-21 &     5.437168e-21 &         1.051746 &         933.7000 & 2.18100e-01 \\
          3.81872  &     -999.00000 &    -5.091552e-26 &     1.091546e-25 &         1.339266 &        1090.3000 & 3.18300e-05 \\
          3.96728  &        1.15402 &    -9.832410e-26 &     1.527495e-25 &         1.829799 &          55.3200 & 6.12600e-06 \\
          0.99437  &     -998.00000 &     8.266724e-25 &     1.427042e-25 &         0.000000 &          62.4620 & 2.65000e-07 \\
          0.12178  &     -998.00000 &     9.048764e-19 &     4.785423e-21 &         1.131773 &         432.2000 & 8.08400e-02 \\
          1.71562  &        0.66422 &     6.333804e-21 &     1.026068e-22 &         1.545234 &         119.8400 & 2.04020e-02 \\
          \hline
                    g &          RA(deg) &         Dec(deg) & Pixel\_No. &          Blend &   Validity & \\
                   51 &       53.3244040 &       31.1379872 &          1 &              x &          0 & \\
                   27 &       53.3244638 &       31.1379599 &          1 &              x &          1 & \\
                   13 &       53.3243090 &       31.1380844 &          1 &            Red &          0 & \\
                   13 &       53.3243077 &       31.1380844 &          1 &       Red/Blue &          0 & \\
                   39 &       53.3243048 &       31.1380791 &          1 &       Red/Blue &          0 & \\
                   13 &      309.7550000 &       68.0161000 &      SLWA2 &       Red/Blue &          0 & \\
                    9 &      309.7550000 &       68.0161000 &      SLWA2 &              x &          1 & \\
                    5 &      309.7550000 &       68.0161000 &      SLWA2 & DoubleGaussian &          1 & \\
                   27 &      294.2535754 &        7.5692374 &          c &              x &          1 & \\
                   15 &      289.4740000 &       19.2062000 &          c &              x &          1 & \\
   		\bottomrule
        \multicolumn{7}{p{\textwidth}}{\rm The table in the ASCII file have the same columns and style except that the rows are chopped into three parts here for better display.  Also this table has selected lines from different parts of the original results to demonstrate different flags. As mentioned in Section~\ref{sec:fitting_results} , any column with \texttt{-999} indicates a fitting result that is not well-constrained.  Therefore, the \texttt{Validity} flag is set to be 0.  The \texttt{ Pixel\_No.} column lists \texttt{c} for the 1-D spectrum measurements, and the specific pixel number/name for individual spaxels. This table (all line measurements for all sources) is published in its entirety online as a machine readable table.}
   		\label{table:report}
	\end{tabular}
\end{table*}

\begin{table}[htbp!]
    \centering
    \caption{The definitions of columns in the line fitting result}
    \label{terminology}
    \begin{tabular}{rp{2.1in}}
        \toprule
        Column name & Description \\
        \hline
        \texttt{Object}               & object name \\
        \texttt{Line}                 & line name \\
        \texttt{LabWL(um)}            & theoretical line centroid \\
        \texttt{ObsWL(um)}            & fitted line centroid \\
        \texttt{Sig\_Cen(um)}         & uncertainty on the fitted line centroid \\
        \texttt{Str(W/cm2)}           & fitted line strength \\
        \texttt{Sig\_str(W/cm2)}      & uncertainty on the fitted line strength \\
        \texttt{FWHM(um)}             & fitted full width at half-maximum \\
        \texttt{Sig\_FWHM(um)}        & uncertainty on the fitted FWHM \\
        \texttt{Base(W/cm2/um)}       & fitted baseline intensity \\
        \texttt{Noise(W/cm2/um)}      & residual intensity \\
        \texttt{SNR}                  & the signal-to-noise ratio \\
        \texttt{E\_u(K)}              & the upper energy level from LAMDA$^{1}$ \\
        \texttt{A(s-1)}               & the Einstein-A value from LAMDA$^{1}$ \\
        \texttt{g}                    & the multiplicity of the upper energy from LAMDA$^{1}$ \\
        \texttt{RA(deg)}              & right ascension \\
        \texttt{Dec(deg)}             & declination \\
        \texttt{Pixel\_No.}           & the pixel label \\
        \texttt{Blend}                & the blending flag which highlight any other line within one resolution element to the fitted line centroid \\
        \texttt{Validity}             & the validity flag which determines the overall certainty of the fitting for a given line \\
        \bottomrule
        \multicolumn{2}{p{3in}}{$^{1}$Leiden Atomic and Molecular Database \citep{schoier05}.}
    \end{tabular}
\end{table}

\subsubsection{The Effect of Lines on Photometry}
\label{sec:effect_photometry}
The line fitting pipeline provides us a chance to investigate the impact of the emission lines on the broad band photometry, which inevitably includes lines.  We calculated the spectrophotometry at SPIRE 250, 350, and 500~\micron\ bands with the corresponding filters for the spectra with and without lines.  After the removal of lines, the photometry decreases by 0.8\%, 1.0\%, and 0.7\% on average, respectively, while the maximum decrease of photometry is 2--3\%.  At longer wavelength, \citet{2012MNRAS.426...23D} found the emission of CO~\jj{3}{2} typically contributes to less then 20 \%\ of the 850~\micron\ continuum, but the contamination can be as high as 79\%\ in the regions dominated by outflows.  They also suggest the contribution of CO to the continuum would be small at shorter wavelengths (e.g. 450~\micron), because the continuum is brighter.

\section{Analysis}
\label{sec:analysis}

\subsection{CO Optical Depth}
\label{sec:co_opt_depth}

\begin{figure*}[htbp!]
    \centering
    \includegraphics[height=2.55in]{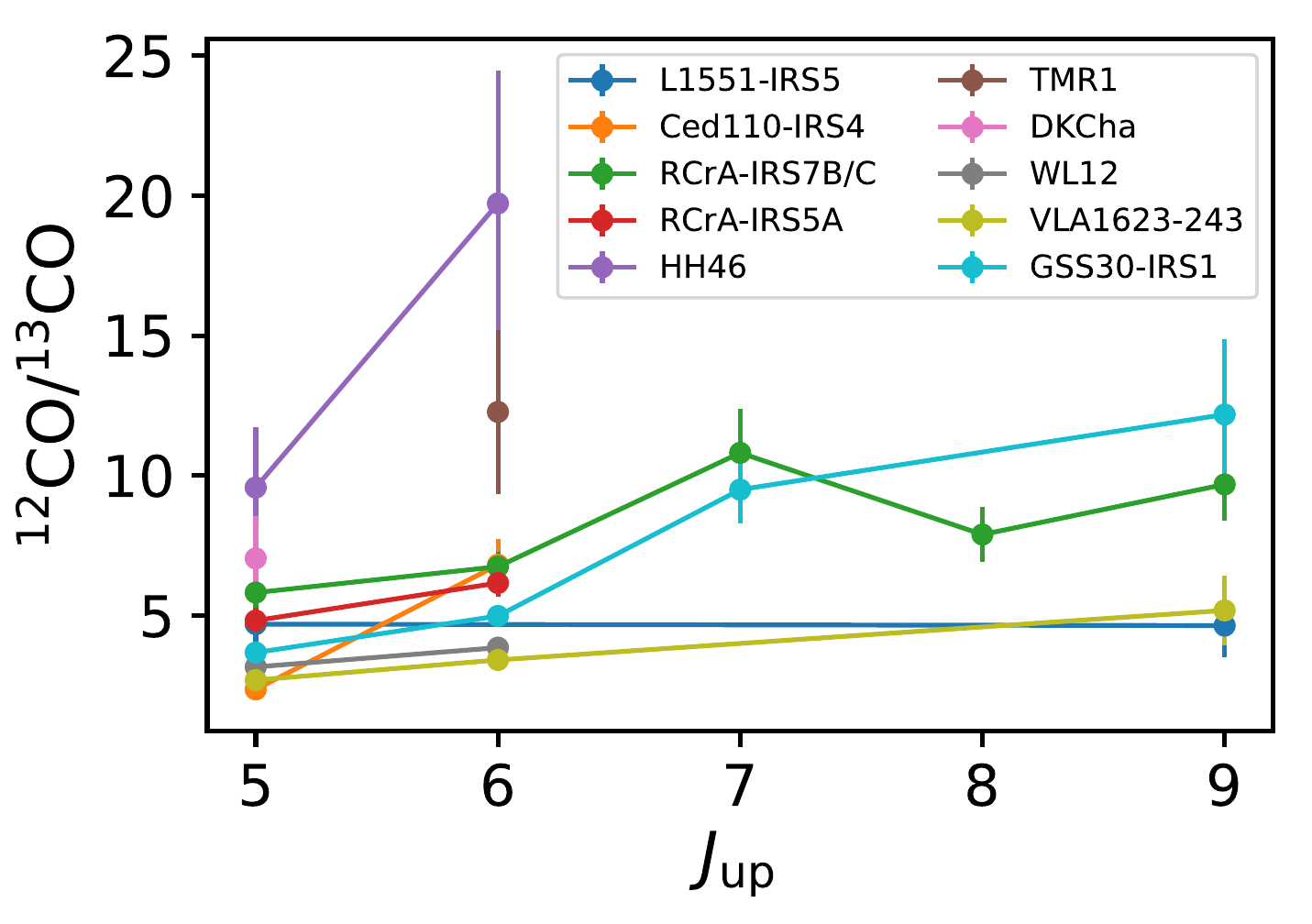}
    \includegraphics[height=2.7in]{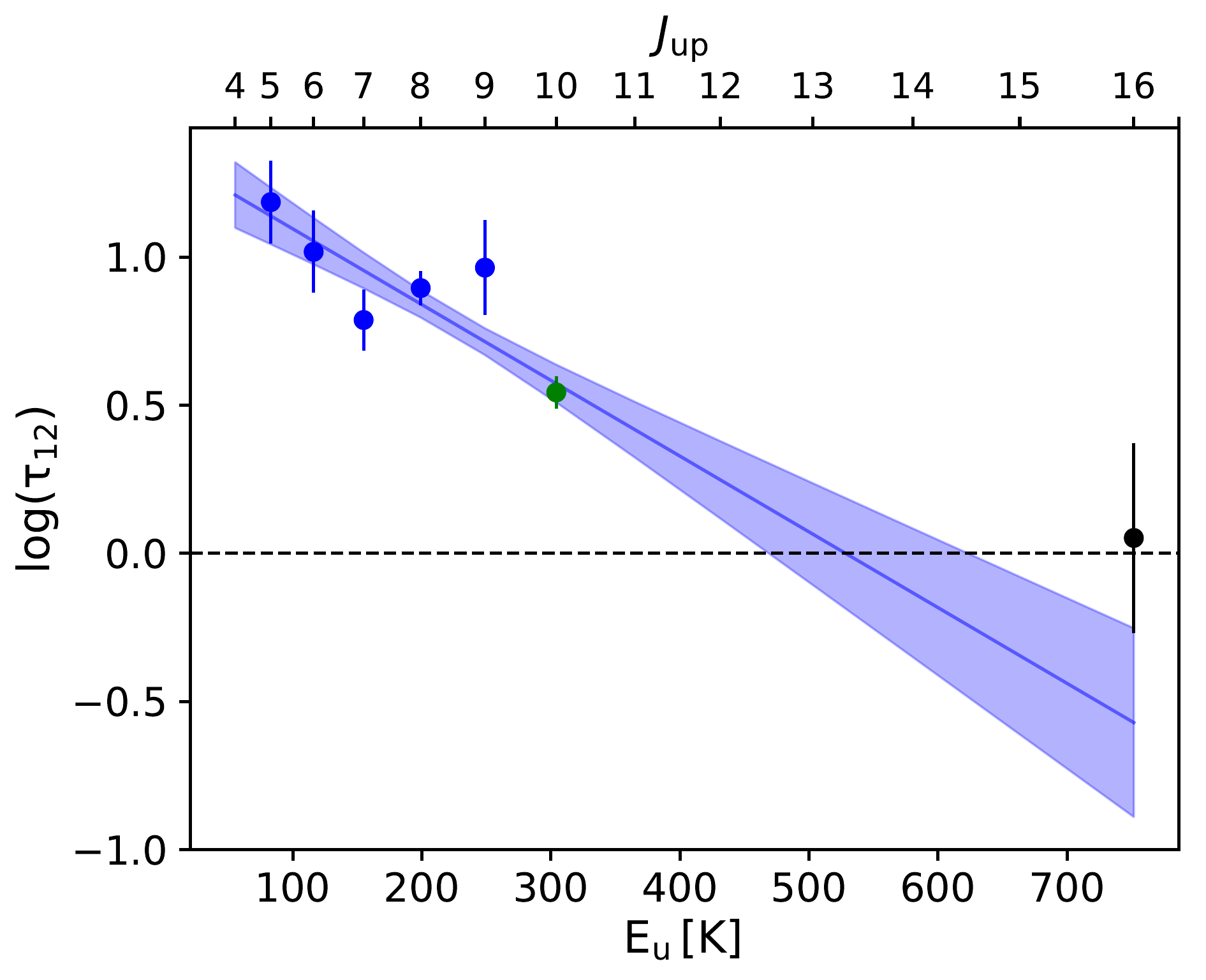}
    \caption{\textbf{Left: }The flux ratios of $^{12}$CO and $^{13}$CO originating from the same \Jup-level.  \textbf{Right: }The derived optical depth of $^{12}$CO with the corresponding upper energy.  The green point indicates the optical depth derived from the averaged $^{12}$CO/$^{13}$CO measured by \citet{2013AA...553A.125S}, while the black point shows the optical depth derived from the $^{12}$CO/$^{13}$CO of Serpens SMM1 measured by \citet{goi12}.  The blue line indicates the fitting result, while the shaded region indicates the 1-$\sigma$ uncertainty of the fit.}
    \label{fig:co_ratios}
\end{figure*}

The $^{13}$CO line is usually taken to be optically thin for the transitions detected with SPIRE, while $^{12}$CO is typically optically thick at low-$J$ before becoming optically thin at high-$J$ \citep[e.g.][]{1984ApJ...286..599G}.  For the COPS sources with $^{13}$CO detections, we tested this assumption.  Figure~\ref{fig:co_ratios} (left) shows the ratios of integrated fluxes of $^{12}$CO and $^{13}$CO as a function of $J$-level.
If both $^{12}$CO and $^{13}$CO are optically thin at a certain $J$-level, we should find the ratio of two lines approaching the intrinsic isotope ratio of 62 \citep{1993ApJ...408..539L}; however, the ratio reaches only 10--20 for the transitions where $^{13}$CO was detected.
We can further derive the optical depth from the following expression:
\begin{equation}
    \tau_{12} = R \frac{J_{\nu}(T_{\rm rot, ^{12}CO})}{J_{\nu}(T_{\rm rot, ^{13}CO})} \frac{F_{\rm 13}}{F_{\rm 12}},
    \label{eq:tau}
\end{equation}
where $R$ is the abundance ratio of $^{12}$CO to $^{13}$CO, $J_{\nu}(T)$ is the Planck function, $T_{\rm rot}$ is the rotational temperature, and $F_{12}$ and $F_{13}$ are the integrated line fluxes of $^{12}$CO and $^{13}$CO lines.  We also assume that both $^{12}$CO and $^{13}$CO have the same excitation temperature and the same line profile.
We average the measurements of $^{12}$CO/$^{13}$CO shown in Figure~\ref{fig:co_ratios} (left) to better constrain the relation.  We also collect other measurements of $^{12}$CO/$^{13}$CO from literatures to better constrain the optical depth of the entire CO ladder.  We calculated the averaged $^{12}$CO/$^{13}$CO at \jj{10}{9} measured by \citet{2013AA...553A.125S}, which includes 11 COPS sources.
Figure~\ref{fig:co_ratios} (right) shows the distribution of the averaged optical depth of $^{12}$CO versus the upper energy levels, derived from Equation~\ref{eq:tau}, assuming R=62.

From a simple two-level atom assumption, the relation between the optical depth and upper energy can be approximated as $\tau_{12} \propto E_{\rm u}^{-a}$, a relationship supported by the data (Figure~\ref{fig:co_ratios}, right).  The optical depth has a larger scatter between \Jup=7--9 due to the low detection rates of $^{13}$CO among the COPS sources.  Figure~\ref{fig:co_ratios} (right) also shows the optical depth of CO~\jj{16}{15} of Serpens SMM1 (black), a massive low-mass embedded protostar, derived from the $^{12}$CO/$^{13}$CO measured by \citet{goi12}.
We included all data other than the data of Serpens SMM1 for the fitting of the relation of optical depth as a function of the upper energy (Figure~\ref{fig:co_ratios}, right), and found the relation can be described as
\begin{equation}
    log(\tau_{12}) = -(2.6\pm0.6)\ee{-3}\times E_{\rm u}+(1.4\pm0.1).
    \label{eq:fitted_tau}
\end{equation}
We further extrapolated the fitted relation to find out that $\tau_{\rm 12}$ approaches to 1 as $E_{\rm u}$ reaches 522.3~K, which makes \Jup=13 the highest \Jup\ level requiring correction.  The uncertainty of the extrapolation for $\tau_{12}=1$ is $-55.6$~K/+86.8~K.  Before we adopt this result, some other uncertainties need to be considered.

The optical depth of resolved CO lines for low to mid-$J$ is known to be high at the peak of the line profile, and the lines become optically thin in the line wing \citep[e.g.][]{2001ApJ...554..132A,2014ApJ...783...29D}.
Both $^{12}$CO and $^{13}$CO lines are spectrally unresolved with SPIRE, which we have to consider when adopting our derived relation of optical depth.  \citet{yildiz12,yildiz13} presented the optical depth of CO as a function of velocity and the integrated fluxes of $^{12}$CO and $^{13}$CO toward NGC1333~IRAS4A and 4B.  Using our method to calculate the optical depth with the integrated fluxes of NGC1333~IRAS4A and 4B, we found the derived optical depths are equal to or greater than the highest optical depth found with the resolved line profiles by up to a factor of 4.  The overestimation of the optical depth is likely because the self-absorption of $^{12}$CO lines cannot be excluded from our analysis with the unresolved lines.
Self-absorption is rarely seen in $^{13}$CO lines due to the lower optical depth \citep{yildiz12,2013AA...553A.125S}; therefore, the derived optical depth is higher due to the self-absorption at the center of $^{12}$CO lines.
On the other hand, the broad components are rarely seen in the observed line profiles of $^{13}$CO \citep{2013AA...553A.125S}, possibly due to their low sensitivity, and thus signal-to-noise ratio.  However, the optical depth derived from unresolved lines is less affected by the undetected line wings compared to the studies using resolved lines, because all components of line profiles contributing to the measured line flux.

Despite the uncertainties of the optical depth, our fitted relation is consistent with the results of \citet{goi12}, which suggests an optical depth of 0.88 for the CO~\jj{16}{15} line.  Our relation suggests an optical depth of 0.25$^{+1.0}_{-0.1}$ for the CO~\jj{16}{15} line.  In the following analysis, we adopt the fitted relation of the optical depth as a function the upper energy, suggesting that CO lines with \Jup\ lower or equal to 13 need to be corrected for optical depth.  The correction is applied to all COPS sources regardless of whether $^{13}$CO was detected.

\subsection{CO Rotational Diagrams}
\label{sec:co_rot}
One useful tool in analyzing molecular emission is the rotational diagram; a detailed review can be found in \citet{goldsmith99}; brief reviews in the context of \textit{Herschel} spectroscopy are in \citet{green13b,manoj12,karska13}.  Here we focus only on the rotational diagram analysis for CO.
Other molecular species, such as OH, and \water, also probe the molecular budget of embedded protostars but require detailed analyses with extensive radiative transfer modeling \citep[e.g.][]{goi12,herczeg12,karska13,wampfler13,2013AA...557A..23K,2014AA...572A..21M,karska14}, which is beyond the scope of this study.

As noted in many previous studies \citep[e.g.][]{vankempen10,manoj12,goi12,dionatos13,karska13,green13b,yildiz13, karska14,2015AA...578A..20M}, the CO transitions from \jj{14}{13} to \jj{40}{39} can be approximated with two components:
a ``warm'' component (\trot\ $\sim$ 300 K) and a ``hot'' component (\trot\ $\sim$ 600--800 K); when detected, the hot component is approximately an order of magnitude less abundant, independent of the source luminosity \citep{green13b}.  Karska et al. (to be accepted) further constrained the median temperatures of the two components as  324~K and 719~K, using the largest sample of sources studies to date, with a broader range of temperature, 600--1100~K, for the ``hot'' component.  The higher-$J$ CO lines, up to the highest detected in these sources (\jj{48}{47}), have been fitted with an even hotter component \citep{goi12,manoj12,karska14}.

If we use only the two components inferred from the PACS data, the ``warm'' component alone under-predicts the intensity of the CO lines from \Jup$\leq$ 10.  The addition of the ``low-$J$'' CO (\Jup=4--13) from SPIRE requires additional one (``cool'') or two (``cold'' and ``cool'') component(s), noted in a few previous studies derived from \textit{Herschel}-HIFI, SPIRE, and ground-based observations \citep[e.g.][]{vankempen09,goi12,yildiz13,yang17}.

The rotational diagrams of the COPS sources all show positive curvature, which is always seen in the CO lines observed with \textit{Herschel} toward protostars (e.g. \citealt{vankempen10short,goi12,herczeg12,green13b}; Karska et al. to be accepted).  The positive curvature suggests that the CO gas has multiple rotational temperatures increasing with the energy levels, or that the transitions were sub-thermally excited in certain conditions.  \citet{neufeld12} demonstrated that a low density ($n<$~10$^{4.8}$~cm$^{-3}$) and high temperature ($\sim$2000~K) isothermal medium or a power-law distribution of the kinetic temperature produces positive curvature in the CO rotational diagram.
The isothermal medium is unlikely to fully describe the CO gas as many studies have shown heterogeneous environments toward protostars \citep[e.g.][]{2014ApJ...783...29D}, and the velocity-resolved line profiles show multiple velocity components \citep[e.g.][]{2013AA...553A.125S,2017arXiv170510269K}.
While the rotational temperatures need not equal the kinetic temperatures, we follow most other work in assuming that they do (LTE).  In this case, the changing $T_{\rm rot}$ correspond to changes in gas temperature.  While the power-law distribution of temperature can also explain the rotational diagrams \citep{manoj12}, we focus here on multiple discrete temperature components.

As we have learned from Section~\ref{sec:co_opt_depth}, CO lines remain optically thick up to \Jup=13.
The observed fluxes increase in proportion to the optical depth after the correction of optical depth; therefore, the temperatures decrease with increases of the number of molecules.  Thus, we need to correct for the effect of optical depth before fitting the rotational diagrams.  We use the fitted optical depth as a function of the upper energy to determine the optical depth at each $J$-level (see Equation~\ref{eq:fitted_tau}).  Finally, we add the systematic uncertainties derived from the differences between the spectroscopy and photometry (3\%--16\%, see Section~\ref{sec:pipeline}).

The previous analyses of rotational temperatures adopted fixed break points to break the CO ladder into pre-determined regions for fitting the temperatures \citep{green13b,karska13,manoj12}.  We adopt the fixed break points from \citet{green13b} with additional break points at the gap of PACS and SPIRE and the boundary between the two SPIRE modules.  The fixed break points separate the CO ladder into \Jup=4--8, \Jup=9--13, \Jup=14--25, and \Jup=28--48.
We adopt the choice of breakpoints that has been widely used in many \textit{Herschel} studies \citep[e.g.]{manoj12,green13b,karska13} so that the fitted rotational temperatures can be consistently compared with other studies.  There are only three detections of CO lines at wavelengths shorter than 70~\micron\ (B2A module); therefore, we combine the B2A module with the B2B module (70--100~\micron).
We recover the same division of the rotational temperatures that have been found previously (Figure~\ref{fig:co_rot_distribution}, left).  Here we label the temperature components by their \Jup-levels, corresponding to the four regions separated by the fixed break points.

However, fixing the break points may restrict the range of the temperatures over which each component can vary.  We take this opportunity to investigate the change if the break points remain flexible with a CO ladder with a maximum of 45 lines when combining the CO lines detected in both SPIRE and PACS spectra.  We attempt to fit up to four temperature components as well as the break points only if the $\chi^{2}$ is further minimized by increasing the number of temperature components.
The number of free parameters scales with the number of components we fit.  We start with a single component fitting, which has two free parameters; by adding one more component, we add three free parameters, one for the break point and two for the line.  Thus, the fitting can have a maximum of 11 free parameters.  We also require at least three data points for each component.  Each temperature component is fitted separately and simultaneously after determining the break points that separate those components.  Finally, the goodness-of-fit is estimated by the reduced $\chi^{2}$ by considering all fits and the total degrees of freedom from all components.

We include the systematic uncertainties based on the differences between spectroscopy and photometry at given photometry bands when we fit for the rotational temperatures.  Systematic uncertainties are applied independently.  There is no additional calibration performed for the fitting of rotational temperatures.

With flexible break points, we found that distributions of the four components overlap more with each other (Figure~\ref{fig:co_rot_distribution}, right), indicating that the fixed break points indeed restricts the range of fitted temperatures.
Two populations are seen from the distribution of $T_{\rm rot}$, a primary population at $\sim$100~K, and a secondary population at $\sim$350~K.  Here we do not label the temperature components because there is no restriction on the range of $T_{\rm rot}$ at certain \Jup-level.  Many sources have a component around 100~K contributed by the entrained gas in outflows.  \citet{yildiz13} found similar components labelled as ``cold'' and ``cool'' (41--68~K and 109--229~K) with CO data from \jj{2}{1} to \jj{10}{9}, and our primary population has a similar range of temperature.  For the case of flexible break points, the ``warm'' and ``hot'' components, which were seen in the $T_{\rm rot}$ distribution with fixed break points, merge together to form the secondary population at $\sim$350~K (centered on 356~K with a 1-$\sigma$ width of 126~K when fitted with a Gaussian distribution between 200~K and 700~K), where the number of the ``very hot'' ($\sim$1000~K) components reduces from 3 to 1 compared to the distribution with fixed break points.
The secondary population is likely to represent the ``warm'' component that is ubiquitously found toward embedded protostars (e.g. \citealt{green13b,karska13}, to be accepted; \citealt{2017arXiv170510269K});
\citet{2017arXiv170510269K} suggested that the universal ``warm'' component results from CO being the dominant coolant when H$_{2}$ cooling becomes insufficient around 300~K because of its widely-spaced energy levels ($E_{\rm u}$=510~K for the \jj{2}{0} transition).

Although fitting more temperature components to rotational diagrams may further reduce the $\chi^{2}$ value, we do not expect to find more components in the distribution of rotational temperatures.
The exercise of flexible breakpoints was intended to test whether a specific number of components was consistently indicated across the source sample.  Figure~\ref{fig:co_rot_distribution} (right) shows the distributions of rotational temperatures from fitting of a maximum of three and four temperature components.  We find that the distribution is well-characterized not by an increasing number of fixed breakpoint components, but more simply by two primary components with flexible breakpoints. This is a simpler solution than the larger number of fixed breakpoint components.

\begin{figure*}[htbp!]
    \centering
    \includegraphics[width=0.45\textwidth]{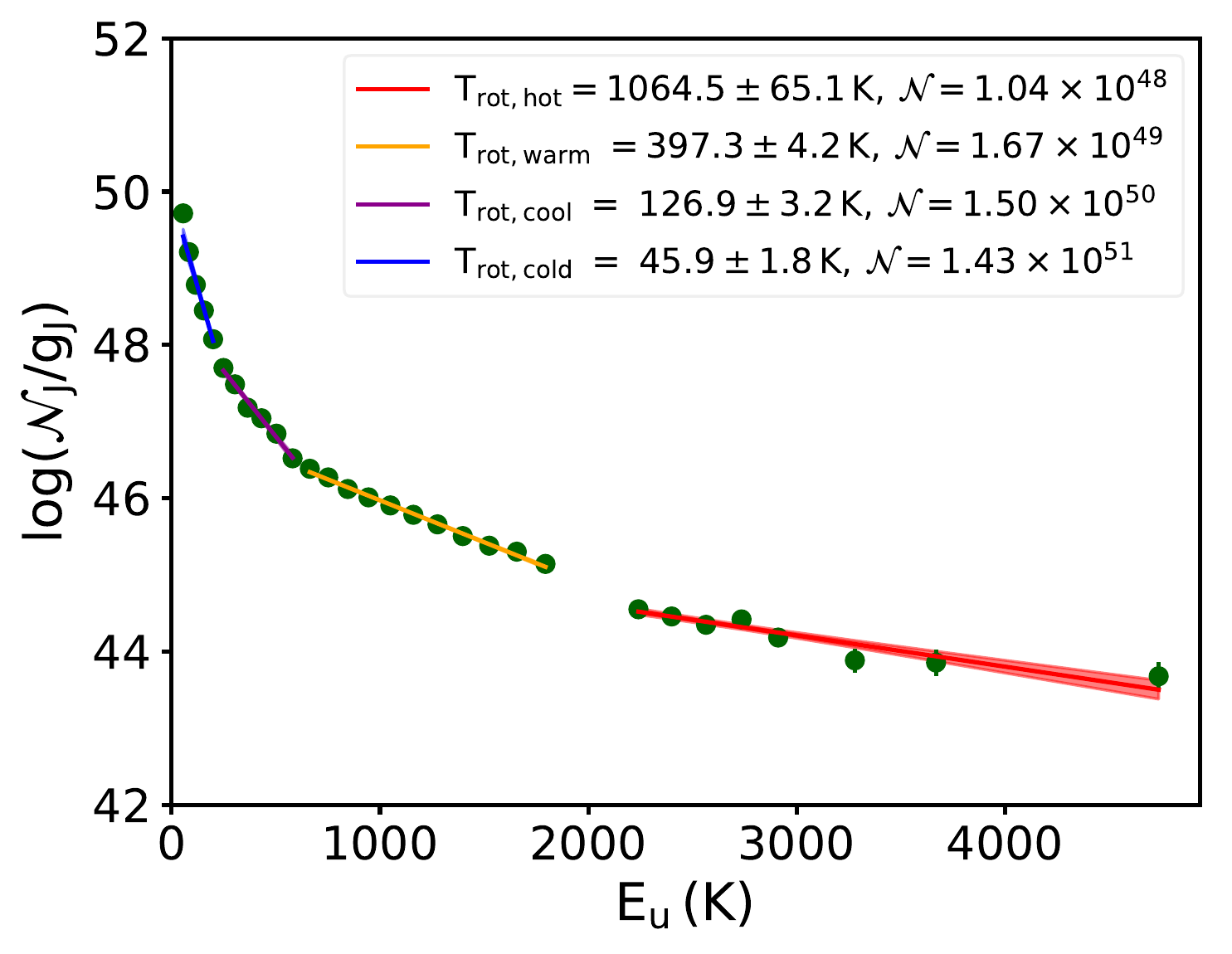}
    \includegraphics[width=0.45\textwidth]{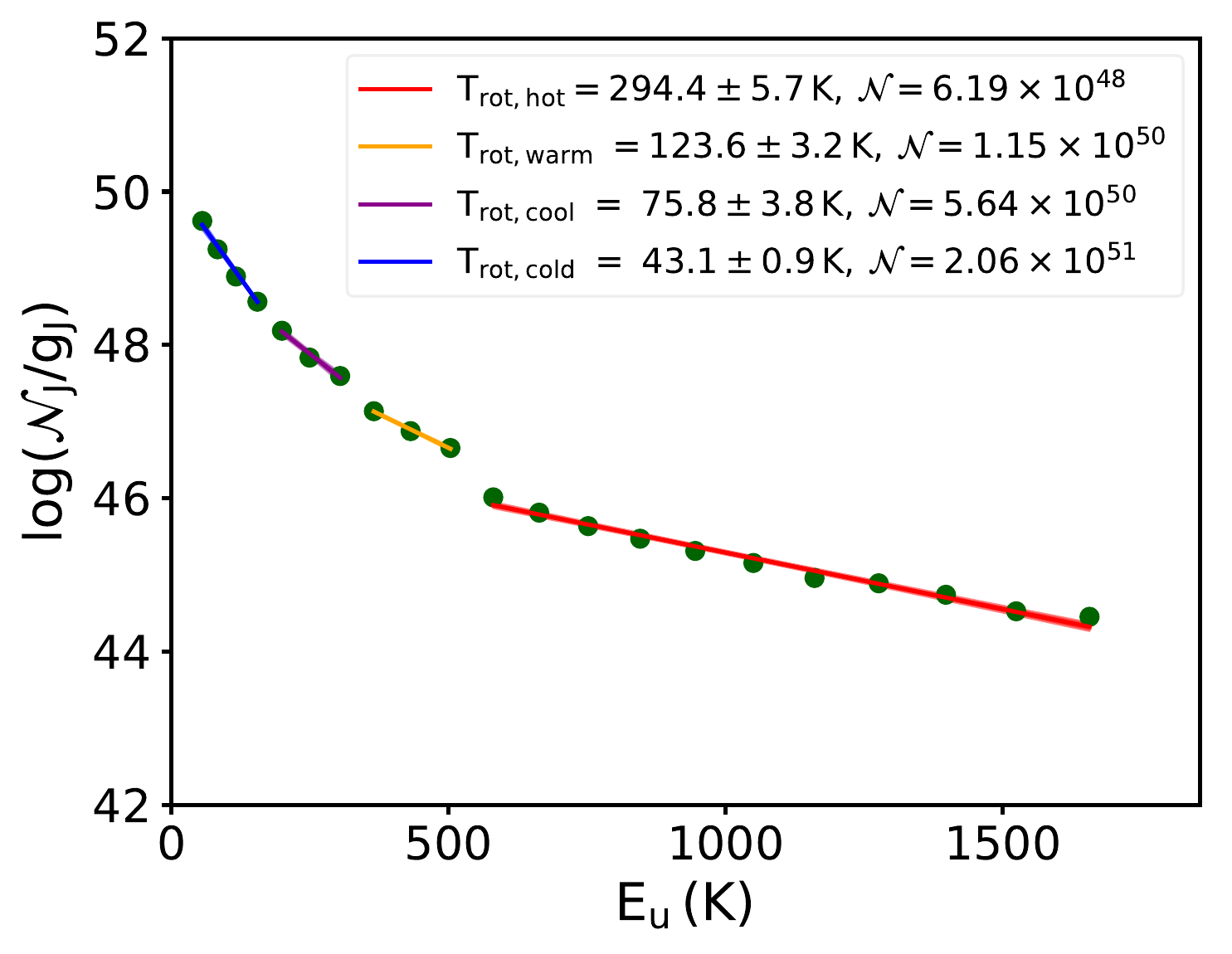}
    \caption{The CO rotational diagrams of BHR~71 (left) and VLA~1623$-$243 (right).  The solid lines indicate the fitted component of rotational temperature, while the shaded areas illustrate the corresponding uncertainties.}
    \label{fig:co_rot}
\end{figure*}

\begin{figure*}[htbp!]
    \centering
    \includegraphics[width=0.45\textwidth]{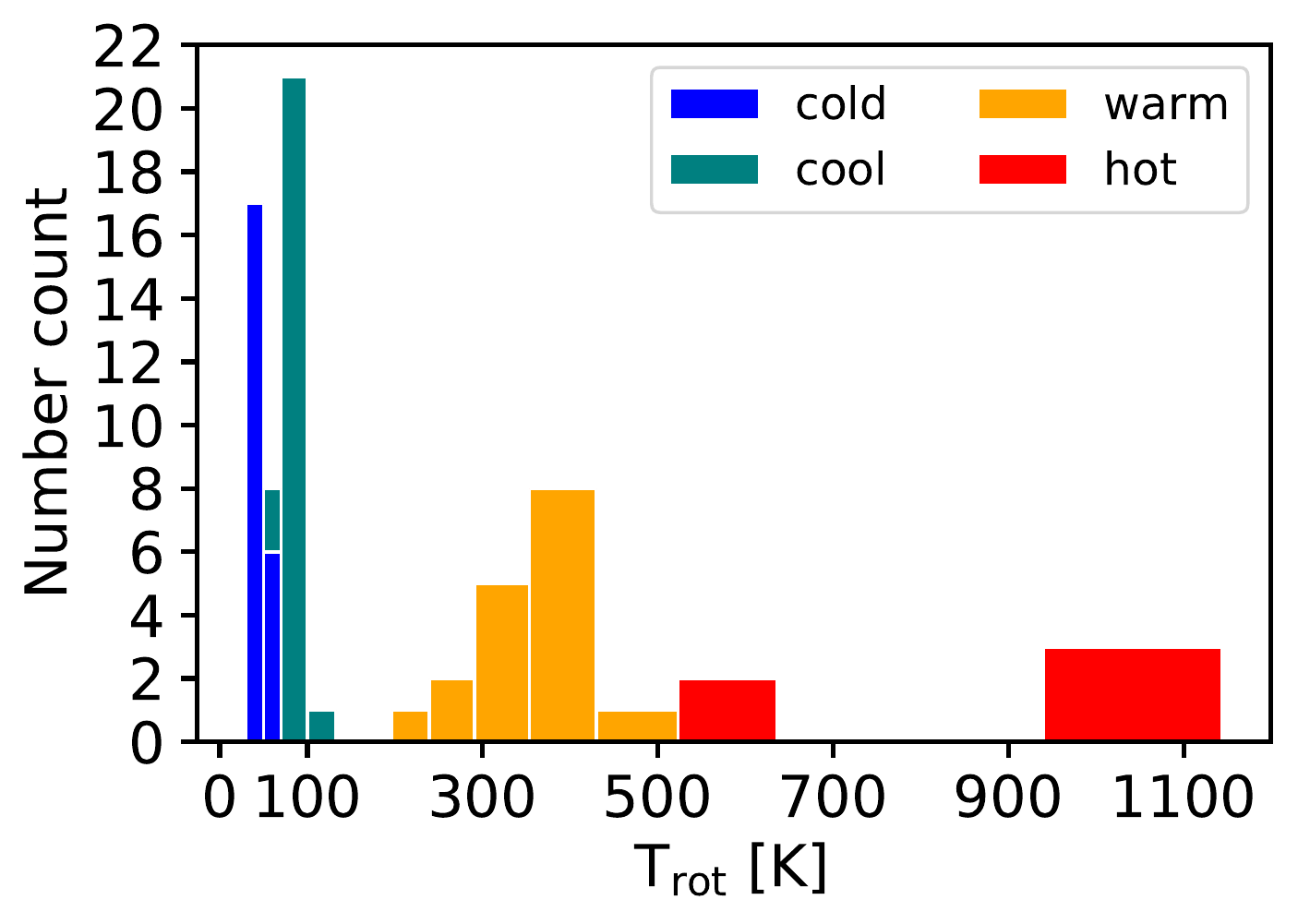}
    \includegraphics[width=0.45\textwidth]{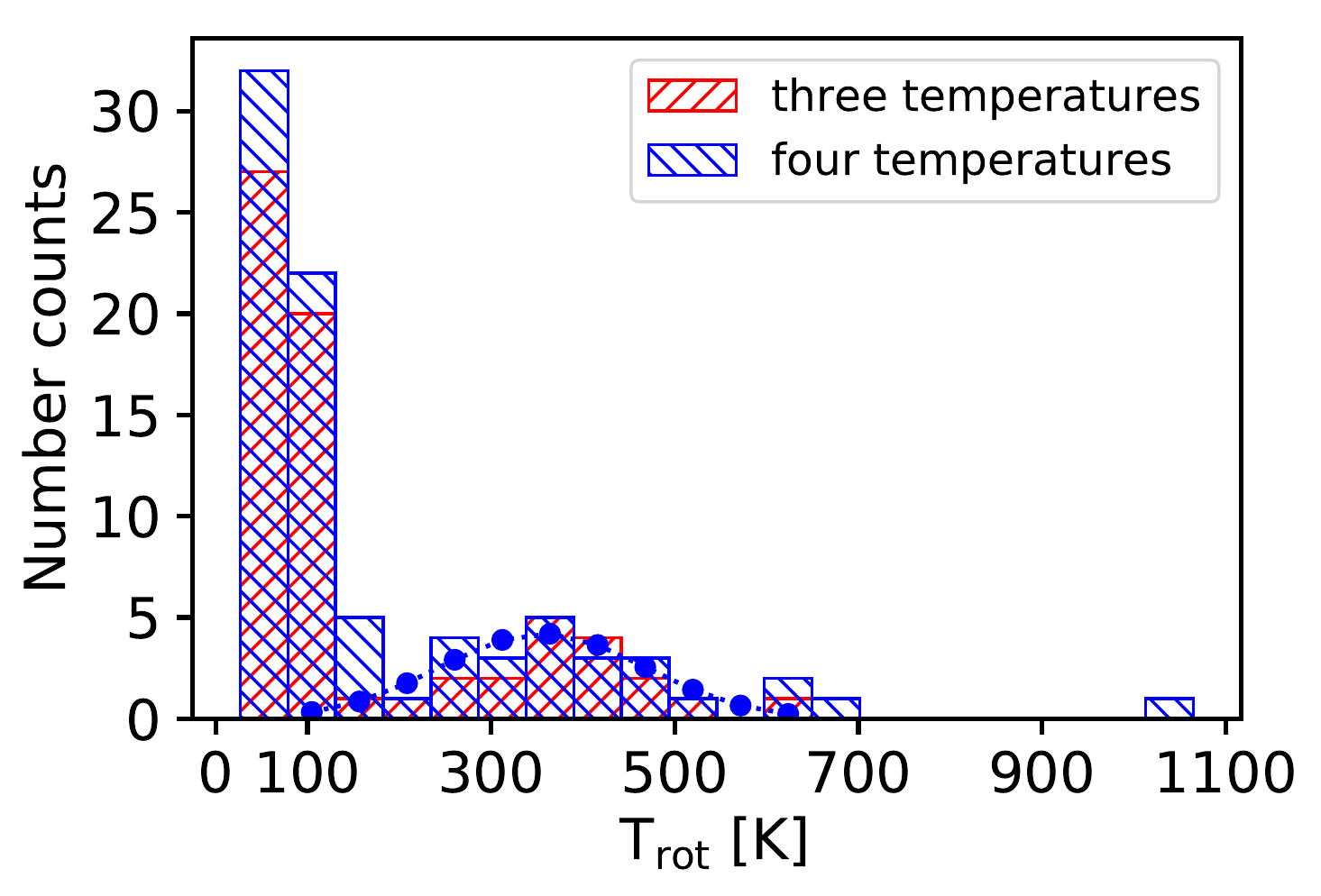}
    \caption{The distribution of the fitted rotational temperatures from all sources with fixed break points (left) and flexible break points (right).  The blue histogram shows the results of using a maximum of four temperature components, while the red histogram shows the results of using up to three temperature components.  The dotted blue line indicates the secondary population fitted with a Gaussian distribution from $T_{\rm rot}$=200--700~K, which centers at 356~K with a 1-$\sigma$ width of \replaced{126}{112}~K.}
    \label{fig:co_rot_distribution}
\end{figure*}

Table~\ref{rot_temp} shows the fitted rotational temperatures with both fixed and flexible break points as well as the corresponding number of molecules in each temperature component.  We label the temperature components by the number of components found in each source.  Figure~\ref{fig:co_rot} shows two examples of the rotational diagrams used for this study.
High rotational temperatures characterize the high-$J$ CO emission, which has high critical density, $\sim10^{7} cm^{-3}$ at \Jup=30--40. The high-$J$ CO emission comes from recently shocked gas which has been compressed to even higher density.  Alternatively, the high-$J$ emission may NOT be in LTE.

We calculate the effect of optical depth on rotational diagram analysis.  The correction mostly affects the two coldest components, where the correction of optical depth is directly applied.  For the coldest component, the temperature decreases by 14~K on average after the correction, while the temperature drops by 7~K on average after the correction for the second coldest component.  The opacity  correction also increases the total molecules found in each component.  The mean number of the CO molecules in the coldest component increases by a factor of about 15.3 from 1.8\ee{50} to 2.8\ee{51}, and the number of the CO molecules in the second coldest component increases by a factor of about 8 from 7.4\ee{49} to 5.9\ee{50}.

\startlongtable
\begin{deluxetable*}{rcccccccc}
\tabletypesize{\scriptsize}
\tablecaption{Rotational Temperatures \label{rot_temp}}
\tablewidth{0pt}
\setlength\tabcolsep{3pt}
\tablehead{
\colhead{Source} & \colhead{T$_{\rm 4}$ (K)} & \colhead{T$_{\rm 3}$ (K)} & \colhead{T$_{\rm 2}$ (K)} & \colhead{T$_{\rm 1}$ (K)} &
\colhead{$\mathcal{N}_{\rm 4}$} & \colhead{$\mathcal{N}_{\rm 3}$} & \colhead{$\mathcal{N}_{\rm 2}$} & \colhead{$\mathcal{N}_{\rm 1}$}
}

\startdata
B1-a        & \nodata           & 266.5$\pm$12.1 & 83.3$\pm$3.3  & 45.3$\pm$0.7 & \nodata                              & 6.3(48) $^{+ 2.4(48) }_{- 1.8(48) }$ & 5.5(50) $^{+ 2.9(50) }_{- 1.9(50) }$ &  2.6(51) $^{+ 2.5(51)} _{- 3.2(50)} $ \\
        ~   & 266.5$\pm$12.1    & 113.2$\pm$6.3  & 71.6$\pm$1.9  & 40.8$\pm$0.5 & 6.3(48) $^{+ 2.4(48) }_{- 1.8(48) }$ & 1.8(50) $^{+ 1.2(50) }_{- 7.1(49) }$ & 8.5(50) $^{+ 1.8(50) }_{- 1.5(50) }$ &  3.4(51) $^{+ 3.0(51)} _{- 3.6(50)} $ \\
        ~   & 14--48              & 11--13          & 8--10        & 4--7       & \nodata & \nodata & \nodata & \nodata \\
B1-c        & \nodata           & 234.4$\pm$4.1  & 90.4$\pm$3.2  & 47.7$\pm$1.2 & \nodata                              & 1.1(49) $^{+ 1.8(48) }_{- 1.6(48) }$ & 1.6(50) $^{+ 6.3(49) }_{- 4.5(49) }$ &  1.1(51) $^{+ 1.2(51)} _{- 2.0(50)} $ \\
        ~   & 238.5$\pm$5.4     & 138.7$\pm$4.1  & 73.2$\pm$0.3  & 47.7$\pm$1.2 & 1.1(49) $^{+ 2.4(48) }_{- 2.0(48) }$ & 4.0(49) $^{+ 1.1(49) }_{- 8.6(48) }$ & 2.8(50) $^{+ 9.7(48) }_{- 9.4(48) }$ &  1.1(51) $^{+ 1.0(51)} _{- 2.0(50)} $ \\
        ~   & 15--48              & 12--14          & 9--11        & 4--8       & \nodata & \nodata & \nodata & \nodata \\
B335        & \nodata           & 322.5$\pm$5.5  & 97.9$\pm$2.9  & 47.2$\pm$1.3 & \nodata                              & 1.2(48) $^{+ 1.6(47) }_{- 1.4(47) }$ & 2.5(49) $^{+ 7.5(48) }_{- 5.8(48) }$ &  1.3(50) $^{+ 1.4(50)} _{- 2.8(49)} $ \\
        ~   & 371.7$\pm$8.2     & 125.1$\pm$3.1  & 80.4$\pm$1.4  & 40.0$\pm$1.2 & 8.6(47) $^{+ 1.5(47) }_{- 1.3(47) }$ & 1.2(49) $^{+ 2.9(48) }_{- 2.3(48) }$ & 4.1(49) $^{+ 4.8(48) }_{- 4.3(48) }$ &  2.0(50) $^{+ 2.1(50)} _{- 4.6(49)} $ \\
        ~   & 16--48              & 11--15          & 8--10        & 4--7       & \nodata & \nodata & \nodata & \nodata \\
BHR~71      & 1064.5$\pm$65.1   & 387.0$\pm$4.4  & 128.3$\pm$5.0 & 45.9$\pm$1.8 & 1.0(48) $^{+ 4.5(47) }_{- 3.2(47) }$ & 1.8(49) $^{+ 1.5(48) }_{- 1.4(48) }$ & 1.5(50) $^{+ 4.4(49) }_{- 3.4(49) }$ &  1.4(51) $^{+ 1.9(51)} _{- 4.1(50)} $ \\
        ~   & 1064.5$\pm$65.1   & 397.3$\pm$4.2  & 126.9$\pm$3.2 & 45.9$\pm$1.8 & 1.0(48) $^{+ 4.5(47) }_{- 3.2(47) }$ & 1.7(49) $^{+ 1.3(48) }_{- 1.2(48) }$ & 1.5(50) $^{+ 3.1(49) }_{- 2.6(49) }$ &  1.4(51) $^{+ 1.9(51)} _{- 4.1(50)} $ \\
        ~   & 26--48              & 15--25          & 9--14        & 4--8       & \nodata & \nodata & \nodata & \nodata \\
Ced110~IRS4 & \nodata           & \nodata        & 68.0$\pm$3.0  & 47.4$\pm$0.8 & \nodata                              & \nodata                              & 1.3(50) $^{+ 7.8(49) }_{- 4.8(49) }$ &  4.0(50) $^{+ 3.3(50)} _{- 5.3(49)} $ \\
        ~   & \nodata           & \nodata        & 68.0$\pm$3.0  & 47.4$\pm$0.8 & \nodata                              & \nodata                              & 1.3(50) $^{+ 7.8(49) }_{- 4.8(49) }$ &  4.0(50) $^{+ 6.2(49)} _{- 5.3(49)} $ \\
        ~   & \nodata           & \nodata        & 9--48          & 4--8        & \nodata & \nodata & \nodata & \nodata \\
DK~Cha      & 993.4$\pm$109.4   & 431.1$\pm$4.8  & 117.7$\pm$4.6 & 48.6$\pm$1.0 & 8.3(47) $^{+ 9.7(47) }_{- 4.5(47) }$ & 6.9(48) $^{+ 5.0(47) }_{- 4.7(47) }$ & 8.0(49) $^{+ 2.7(49) }_{- 2.0(49) }$ &  5.1(50) $^{+ 5.4(50)} _{- 8.5(49)} $ \\
        ~   & 637.2$\pm$63.2    & 428.2$\pm$4.4  & 115.4$\pm$3.2 & 48.6$\pm$1.0 & 3.8(48) $^{+ 8.4(48) }_{- 2.6(48) }$ & 7.0(48) $^{+ 5.1(47) }_{- 4.7(47) }$ & 8.3(49) $^{+ 2.1(49) }_{- 1.7(49) }$ &  5.1(50) $^{+ 5.3(50)} _{- 8.5(49)} $ \\
        ~   & 31--48              & 15--30          & 9--14        & 4--8       & \nodata & \nodata & \nodata & \nodata \\
GSS~30~IRS1 & 1061.7$\pm$49.6   & 322.7$\pm$6.3  & 106.1$\pm$1.8 & 52.6$\pm$1.1 & 6.8(47) $^{+ 2.0(47) }_{- 1.6(47) }$ & 1.2(49) $^{+ 2.1(48) }_{- 1.8(48) }$ & 4.7(50) $^{+ 6.9(49) }_{- 6.0(49) }$ &  2.4(51) $^{+ 2.3(51)} _{- 3.5(50)} $ \\
        ~   & 533.7$\pm$18.4    & 246.3$\pm$5.4  & 106.1$\pm$1.8 & 52.6$\pm$1.1 & 3.5(48) $^{+ 9.7(47) }_{- 7.6(47) }$ & 2.0(49) $^{+ 3.6(48) }_{- 3.0(48) }$ & 4.7(50) $^{+ 6.9(49) }_{- 6.0(49) }$ &  2.4(51) $^{+ 2.3(51)} _{- 3.5(50)} $ \\
        ~   & 20--48              & 14--19          & 9--13        & 4--8       & \nodata & \nodata & \nodata & \nodata \\
HH~46       & \nodata           & \nodata        & 75.9$\pm$2.7  & 44.6$\pm$1.1 & \nodata                              & \nodata                              & 7.7(50) $^{+ 3.6(50) }_{- 2.4(50) }$ &  4.9(51) $^{+ 5.4(51)} _{- 1.0(51)} $ \\
        ~   & \nodata           & \nodata        & 83.5$\pm$4.0  & 45.1$\pm$0.8 & \nodata                              & \nodata                              & 5.1(50) $^{+ 3.4(50) }_{- 2.1(50) }$ &  4.8(51) $^{+ 8.6(50)} _{- 7.3(50)} $ \\
        ~   & \nodata           & \nodata        & 10--48          & 4--9       & \nodata & \nodata & \nodata & \nodata \\
IRAS~03245+3002 & \nodata       & 388.8$\pm$19.1 & 88.4$\pm$3.1  & 41.7$\pm$1.7 & \nodata                              & 2.2(48) $^{+ 7.5(47) }_{- 5.6(47) }$ & 8.9(49) $^{+ 3.6(49) }_{- 2.6(49) }$ &  7.4(50) $^{+ 1.0(51)} _{- 2.4(50)} $ \\
        ~   & 448.1$\pm$27.9    & 113.4$\pm$2.2  & 65.7$\pm$0.7  & 33.5$\pm$1.4 & 1.7(48) $^{+ 7.1(47) }_{- 5.0(47) }$ & 3.7(49) $^{+ 8.0(48) }_{- 6.5(48) }$ & 2.1(50) $^{+ 1.9(49) }_{- 1.8(49) }$ &  1.4(51) $^{+ 2.0(51)} _{- 5.0(50)} $ \\
        ~   & 16--48              & 12--15          & 8--11        & 4--7       & \nodata & \nodata & \nodata & \nodata \\
IRAS~03301+3111 & \nodata       & \nodata        & 111.5$\pm$5.1 & 29.5$\pm$1.2 & \nodata                              & \nodata                              & 6.3(49) $^{+ 2.8(49) }_{- 2.0(49) }$ &  1.2(51) $^{+ 1.8(51)} _{- 4.2(50)} $ \\
        ~   & \nodata           & \nodata        & 105.9$\pm$3.8 & 29.5$\pm$1.2 & \nodata                              & \nodata                              & 7.1(49) $^{+ 2.7(49) }_{- 1.9(49) }$ &  1.2(51) $^{+ 6.4(50)} _{- 4.2(50)} $ \\
        ~   & \nodata           & \nodata        & 8--48          & 4--7        & \nodata & \nodata & \nodata & \nodata \\
IRAS~15398$-$3359 & \nodata     & \nodata        & 115.1$\pm$3.8 & 60.5$\pm$2.8 & \nodata                              & \nodata                              & 1.1(50) $^{+ 3.2(49) }_{- 2.5(49) }$ &  2.0(50) $^{+ 1.6(50)} _{- 5.1(49)} $ \\
        ~   & \nodata           & \nodata        & 115.1$\pm$3.8 & 60.5$\pm$2.8 & \nodata                              & \nodata                              & 1.1(50) $^{+ 3.2(49) }_{- 2.5(49) }$ &  2.0(50) $^{+ 6.8(49)} _{- 5.1(49)} $ \\
        ~   & \nodata           & \nodata        & 9--48          & 4--8        & \nodata & \nodata & \nodata & \nodata \\
L1157       & -2439.8$\pm$413.3 & 367.4$\pm$5.1  & 105.8$\pm$3.6 & 44.1$\pm$1.3 & \nodata                              & 1.0(49) $^{+ 1.0(48) }_{- 9.3(47) }$ & 1.5(50) $^{+ 5.0(49) }_{- 3.8(49) }$ &  1.5(51) $^{+ 1.8(51)} _{- 3.6(50)} $ \\
        ~   & 680.7$\pm$46.1    & 357.2$\pm$5.5  & 116.6$\pm$2.2 & 47.0$\pm$1.3 & 2.5(48) $^{+ 1.3(48) }_{- 8.6(47) }$ & 1.0(49) $^{+ 1.1(48) }_{- 9.9(47) }$ & 1.1(50) $^{+ 1.9(49) }_{- 1.7(49) }$ &  1.3(51) $^{+ 1.5(51)} _{- 2.8(50)} $ \\
        ~   & 23--48              & 15--22         & 10--14        & 4--9       & \nodata & \nodata & \nodata & \nodata \\
L1455~IRS3  & \nodata & \nodata & \nodata & \nodata & \nodata & \nodata & \nodata & \nodata \\
        ~   & \nodata & \nodata & \nodata & 113.2$\pm$10.4 & \nodata & \nodata & \nodata &  5.1(49) $^{+ 4.0(49)} _{- 2.2(49)} $ \\
        ~   & \nodata & \nodata & \nodata & 4--48          & \nodata & \nodata & \nodata & \nodata \\
L1551~IRS5  & \nodata           & 378.7$\pm$9.1  & 98.3$\pm$5.0  & 46.2$\pm$2.1 & \nodata                              & 2.6(48) $^{+ 4.3(47) }_{- 3.7(47) }$ & 6.4(49) $^{+ 3.3(49) }_{- 2.2(49) }$ &  3.8(50) $^{+ 4.9(50)} _{- 1.2(50)} $ \\
        ~   & 479.5$\pm$10.3    & 127.8$\pm$4.7  & 78.9$\pm$4.4  & 38.1$\pm$1.4 & 1.7(48) $^{+ 2.4(47) }_{- 2.1(47) }$ & 2.7(49) $^{+ 1.1(49) }_{- 7.8(48) }$ & 1.1(50) $^{+ 5.5(49) }_{- 3.7(49) }$ &  5.9(50) $^{+ 7.1(50)} _{- 1.7(50)} $ \\
        ~   & 16--48              & 11--15          & 8--10        & 4--7       & \nodata & \nodata & \nodata & \nodata \\
L483        & \nodata           & \nodata        & 99.1$\pm$6.1  & 45.0$\pm$0.9 & \nodata                              & \nodata                              & 1.1(50) $^{+ 7.4(49) }_{- 4.5(49) }$ &  9.1(50) $^{+ 9.6(50)} _{- 1.4(50)} $ \\
        ~   & \nodata           & \nodata        & 99.1$\pm$6.1  & 45.0$\pm$0.9 & \nodata                              & \nodata                              & 1.1(50) $^{+ 7.4(49) }_{- 4.5(49) }$ &  9.1(50) $^{+ 1.7(50)} _{- 1.4(50)} $ \\
        ~   & \nodata           & \nodata        & 9--48          & 4--8        & \nodata & \nodata & \nodata & \nodata \\
L723~MM     & \nodata           & \nodata        & 100.2$\pm$4.6 & 44.8$\pm$1.7 & \nodata                              & \nodata                              & 1.0(50) $^{+ 4.9(49) }_{- 3.3(49) }$ &  8.9(50) $^{+ 1.1(51)} _{- 2.6(50)} $ \\
        ~   & \nodata           & \nodata        & 114.3$\pm$6.7 & 47.8$\pm$1.5 & \nodata                              & \nodata                              & 6.7(49) $^{+ 4.1(49) }_{- 2.5(49) }$ &  7.6(50) $^{+ 2.4(50)} _{- 1.8(50)} $ \\
        ~   & \nodata           & \nodata        & 10--48          & 4--9       & \nodata & \nodata & \nodata & \nodata \\
RCrA~IRS5A  & \nodata           & 273.4$\pm$5.2  & 88.2$\pm$3.2  & 51.8$\pm$1.9 & \nodata                              & 5.4(48) $^{+ 9.1(47) }_{- 7.8(47) }$ & 4.1(50) $^{+ 1.7(50) }_{- 1.2(50) }$ &  1.6(51) $^{+ 1.7(51)} _{- 3.9(50)} $ \\
        ~   & 273.4$\pm$5.2     & 119.2$\pm$2.0  & 68.9$\pm$1.3  & 40.7$\pm$1.0 & 5.4(48) $^{+ 9.1(47) }_{- 7.8(47) }$ & 1.5(50) $^{+ 2.2(49) }_{- 1.9(49) }$ & 8.3(50) $^{+ 1.3(50) }_{- 1.1(50) }$ &  2.7(51) $^{+ 2.5(51)} _{- 5.1(50)} $ \\
        ~   & 14--48              & 11--13          & 8--10        & 4--7       & \nodata & \nodata & \nodata & \nodata \\
RCrA~IRS7B/C & 598.5$\pm$16.4   & 303.0$\pm$4.7  & 103.7$\pm$2.5 & 61.8$\pm$2.3 & 4.5(48) $^{+ 1.4(48) }_{- 1.1(48) }$ & 5.3(49) $^{+ 8.3(48) }_{- 7.2(48) }$ & 1.3(51) $^{+ 3.0(50) }_{- 2.4(50) }$ &  4.1(51) $^{+ 3.9(51)} _{- 9.0(50)} $ \\
        ~   & 598.5$\pm$16.4    & 323.7$\pm$4.7  & 101.6$\pm$2.2 & 62.8$\pm$1.7 & 4.5(48) $^{+ 1.4(48) }_{- 1.1(48) }$ & 4.2(49) $^{+ 6.1(48) }_{- 5.3(48) }$ & 1.3(51) $^{+ 3.5(50) }_{- 2.8(50) }$ &  4.0(51) $^{+ 3.4(51)} _{- 6.5(50)} $ \\
        ~   & 26--48              & 16--25         & 10--15        & 4--9       & \nodata & \nodata & \nodata & \nodata \\
RNO~91      & \nodata           & \nodata        & 79.7$\pm$4.2  & 40.5$\pm$1.0 & \nodata                              & \nodata                              & 3.5(49) $^{+ 2.3(49) }_{- 1.4(49) }$ &  2.2(50) $^{+ 2.4(50)} _{- 4.4(49)} $ \\
        ~   & \nodata           & \nodata        & 75.9$\pm$2.4  & 37.6$\pm$1.4 & \nodata                              & \nodata                              & 4.2(49) $^{+ 1.3(49) }_{- 1.0(49) }$ &  2.7(50) $^{+ 1.1(50)} _{- 7.8(49)} $ \\
        ~   & \nodata           & \nodata        & 8--48          & 4--7        & \nodata & \nodata & \nodata & \nodata \\
TMC~1       & \nodata           & 359.6$\pm$19.9 & 107.2$\pm$3.6 & 45.9$\pm$0.3 & \nodata                              & 1.3(48) $^{+ 5.3(47) }_{- 3.8(47) }$ & 3.8(49) $^{+ 1.2(49) }_{- 9.0(48) }$ &  1.7(50) $^{+ 1.4(50)} _{- 9.6(48)} $ \\
        ~   & 359.6$\pm$19.9    & 138.2$\pm$5.8  & 133.4$\pm$9.9 & 44.5$\pm$0.4 & 1.3(48) $^{+ 5.3(47) }_{- 3.8(47) }$ & 1.9(49) $^{+ 6.9(48) }_{- 5.1(48) }$ & 3.0(49) $^{+ 1.1(49) }_{- 8.1(48) }$ &  1.8(50) $^{+ 1.6(50)} _{- 1.1(49)} $ \\
        ~   & 14--48              & 11--13          & 8--10        & 4--7       & \nodata & \nodata & \nodata & \nodata \\
TMC~1A      & \nodata           & 374.4$\pm$13.3 & 96.4$\pm$6.1  & 38.8$\pm$1.0 & \nodata                              & 8.2(47) $^{+ 1.9(47) }_{- 1.6(47) }$ & 1.8(49) $^{+ 1.2(49) }_{- 7.3(48) }$ &  1.5(50) $^{+ 1.8(50)} _{- 3.4(49)} $ \\
        ~   & 374.4$\pm$13.3    & 191.3$\pm$0.9  & 99.6$\pm$2.7  & 36.0$\pm$1.4 & 8.2(47) $^{+ 1.9(47) }_{- 1.6(47) }$ & 3.5(48) $^{+ 8.6(46) }_{- 8.4(46) }$ & 1.8(49) $^{+ 3.1(48) }_{- 2.6(48) }$ &  1.8(50) $^{+ 2.4(50)} _{- 5.5(49)} $ \\
        ~   & 14--48              & 11--13         & 8--10        & 4--7       & \nodata & \nodata & \nodata & \nodata \\
TMR~1       & \nodata           & 392.9$\pm$8.7  & 112.4$\pm$3.7 & 51.6$\pm$1.1 & \nodata                              & 2.0(48) $^{+ 3.0(47) }_{- 2.6(47) }$ & 4.8(49) $^{+ 1.4(49) }_{- 1.1(49) }$ &  2.3(50) $^{+ 2.2(50)} _{- 3.3(49)} $ \\
        ~   & 392.9$\pm$8.7     & 146.1$\pm$2.9  & 90.1$\pm$0.6  & 57.3$\pm$2.7 & 2.0(48) $^{+ 3.0(47) }_{- 2.6(47) }$ & 2.4(49) $^{+ 3.6(48) }_{- 3.2(48) }$ & 7.4(49) $^{+ 3.1(48) }_{- 3.0(48) }$ &  1.9(50) $^{+ 1.8(50)} _{- 4.7(49)} $ \\
        ~   & 14--48              & 11--13          & 8--10        & 4--7       & \nodata & \nodata & \nodata & \nodata \\
VLA~1623$-$243 & \nodata        & 294.4$\pm$5.7  & 86.7$\pm$3.8  & 47.4$\pm$1.0 & \nodata                              & 6.2(48) $^{+ 1.0(48) }_{- 9.0(47) }$ & 3.7(50) $^{+ 1.8(50) }_{- 1.2(50) }$ &  1.7(51) $^{+ 1.6(51)} _{- 2.6(50)} $ \\
        ~   & 294.4$\pm$5.7     & 123.6$\pm$3.2  & 75.8$\pm$3.8  & 43.1$\pm$0.9 & 6.2(48) $^{+ 1.0(48) }_{- 9.0(47) }$ & 1.2(50) $^{+ 2.5(49) }_{- 2.1(49) }$ & 5.6(50) $^{+ 2.5(50) }_{- 1.7(50) }$ &  2.1(51) $^{+ 1.9(51)} _{- 3.3(50)} $ \\
        ~   & 14--48              & 11--13          & 8--10        & 4--7       & \nodata & \nodata & \nodata & \nodata \\
WL~12       & \nodata           & 371.6$\pm$10.0 & 113.3$\pm$5.1 & 47.9$\pm$3.7 & \nodata                              & 1.5(48) $^{+ 2.7(47) }_{- 2.3(47) }$ & 6.8(49) $^{+ 2.7(49) }_{- 1.9(49) }$ &  3.9(50) $^{+ 6.5(50)} _{- 1.8(50)} $ \\
        ~   & 371.6$\pm$10.0    & 165.9$\pm$0.0  & 75.7$\pm$1.4  & 26.8$\pm$1.0 & 1.5(48) $^{+ 2.7(47) }_{- 2.3(47) }$ & 2.8(49) $^{+ 2.4(46) }_{- 2.4(46) }$ & 1.6(50) $^{+ 2.0(49) }_{- 1.8(49) }$ &  1.2(51) $^{+ 1.5(51)} _{- 3.0(50)} $ \\
        ~   & 14--48              & 11--13          & 7--10        & 4--6       & \nodata & \nodata & \nodata & \nodata \\
\enddata
\tablecomments{There are two sets of rotational temperatures and the numbers of molecules associated with each
            source.  The top row shows the temperatures and the number of molecule fitted from data with \textit{fixed break points}, and the second row shows the temperatures and the number of
            molecules fitted from data with \textit{flexible break points}.  The third row indicates the ranges of $J$ levels where the temperature and number density are fitted.  L1014 does not have enough
            detections of CO for rotational temperature analysis and is not shown in this table.
            The negative temperature in L1157 are fitted from four CO lines appearing nearly flat in the rotational diagram.}
\end{deluxetable*}

\subsection{HCO$^{+}$}
\begin{figure}[htbp!]
    \centering
    \includegraphics[width=0.45\textwidth]{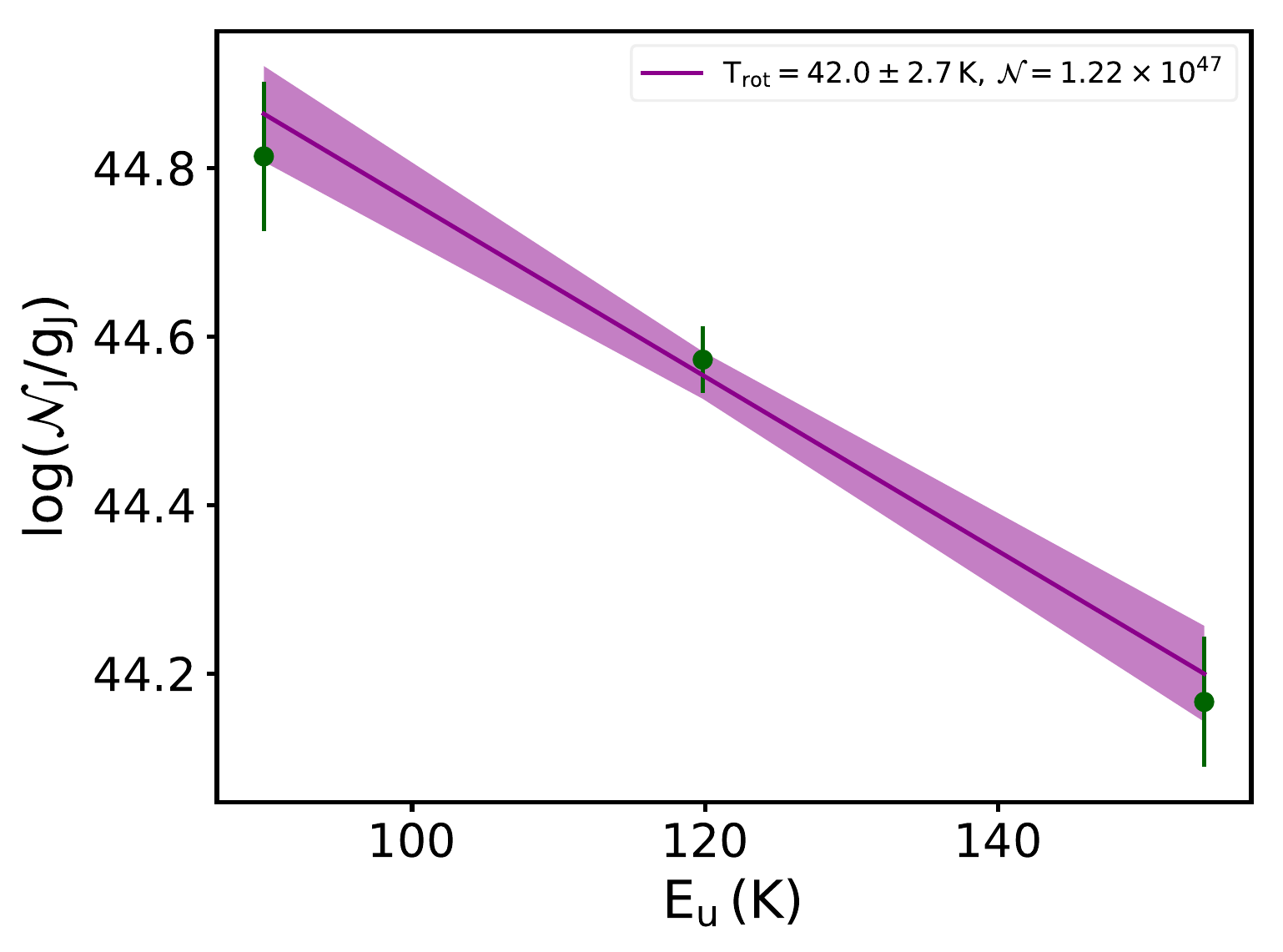}
    \caption{The rotational diagram of HCO$^{+}$ detected toward RCrA~IRS7B/C.  The solid line indicates the fitted temperature, and the filled area shows the uncertainty.}
    \label{fig:rotdia_hco+}
\end{figure}

HCO$^{+}$ has a higher critical density, $n_{\rm crit}=$(3.3--9.9)\ee{7}~cm$^{-3}$ for the lines relevant here \citep{schoier05}, which is often used for probing dense molecular gas around protostars.  Its effective density can be 1--2 orders of magnitudes lower than the critical density \citep{1999ARAA..37..311E,2015PASP..127..299S}.  We found GSS~30~IRS1 and the combined RCrA~IRS7B/C have significant detections of HCO$^{+}$.
RCrA~IRS7B/C shows the emission of HCO$^{+}$~\jj{6}{5}, \jj{7}{6}, and \jj{8}{7}, while GSS~30~IRS1 shows only the emission of HCO$^{+}$~\jj{7}{6}.
RCrA~IRS7B/C has sufficient amount of HCO$^{+}$ detections for rotational temperature analysis.  \citet{2016AA...590A.105B} detected HCO$^{+}$ toward 12 YSOs including both low-mass and high-mass sources, suggesting that our low detection rate of HCO$^{+}$~\jj{6}{5} may be due to the lack of spectral resolution, similar to the problem we have on the emission of  o-\water~$1_{10}\rightarrow1_{01}$.
Figure~\ref{fig:rotdia_hco+} shows the constructed rotational diagram along with the fitted rotational temperature of 42.0$\pm$2.7~K, which is slightly lower than the ``cold'' component (51.5$\pm$2.5~K) fitted from the CO rotational diagram, hinting that the HCO$^{+}$ may be sub-thermally excited or it is simply colder than CO.

\subsection{Beam Filling Factor}
\label{sec:filling_factor}
The optical depth and the rotational temperature provide constraints on the beam filling factor of CO emission.  We explore the filling factors derived from unresolved CO lines and compare with the maps of low-$J$ CO lines resolved both spectrally and spatially \citep[e.g.][]{2015AA...576A.109Y}. To reduce uncertainty, we only calculate the filling factors for the transitions where both $^{12}$CO and $^{13}$CO emissions are detected.

The brightness temperature of CO emission can be related to the filling factor, rotational temperature, and the optical depth with Equation~\ref{eq:t_b}.
\begin{equation}
    I_{\nu} = B_{\nu}(T_{\rm B}) = f B_{\nu}(T_{\rm rot})(1-e^{-\tau}),
    \label{eq:t_b}
\end{equation}
where B$_{\nu}(T)$ is the Planck function for a given temperature, and $f$ is the beam filling factor.  The $T_{\rm rot}$ and $\tau$ are derived from our analyses above (see Section~\ref{sec:co_opt_depth} and \ref{sec:co_rot}).  The rotational temperature is selected from the rotational diagram based on the component of temperature which dominates the transition.  The remaining variable is $T_{\rm B}$, which can be derived from the integrated fluxes if we know the line width, which is much smaller than the spectral resolution of SPIRE.
\citet{yildiz13} presented the observations of CO~\jj{6}{5} toward 26 Class 0+I protostars, 14 of which are included in the COPS sample.  We calculated an average FWHM of 4.3$\pm$1.4~\kms\ for the CO~\jj{6}{5} line from $T_{\rm peak}$ and the integrated fluxes presented in the Appendix of \citet{yildiz13}, assuming Gaussian profiles.  We take the 1-$\sigma$ value of 5.7~\kms\ as the line width for the calculation of the filling factor since the CO line often become broader at higher $J$-levels \citep{yildiz13,2015AA...576A.109Y,2017arXiv170510269K}.

Table~\ref{filling} shows the filling factors calculated with Equation~\ref{eq:t_b}.  The filling factors are all smaller than 0.53, ranging from 0.035 to 0.53.  We compared our result with the maps shown in \citet{2015AA...576A.109Y}, where 6 COPS sources (L1551~IRS5, Ced110~IRS4, GSS~30~IRS1, HH~46, DK~Cha, and TMR~1) were observed in either CO~\jj{3}{2} or CO~\jj{6}{5} lines.  In general, the relative values of our filling factors are consistent with the maps, showing that GSS30~IRS1 and HH~46 have the densest emission at the central $\sim$30\arcsec\ regions.  However, the maps of TMR~1 and DK~Cha show the emission extending in the central $\sim$20\arcsec\ regions, suggesting a higher filling factor than 0.074 (or 0.035).  Thus, we are likely to underestimate the filling factor, perhaps because we took a conservative estimate of the line width.  The uncertainty on the line width leads to a factor of two difference in the filling factor.
Another source of uncertainty comes from the correction of optical depth.   We  overestimate the optical depth by using the unresolved SPIRE spectra (see Section~\ref{sec:co_opt_depth}).  A lower optical depth results in a lower $(1-e^{-\tau})$ in Equation~\ref{eq:t_b} but a higher $T_{\rm rot}$ at the low-$J$ CO lines.  However, we found that if we artificially decrease the optical depth by 50\%, $T_{\rm rot}$ only increases 1\%\ on average.  Thus, an overestimation of optical depth makes us underestimate the derived filling factor.
Outflows often show clumpy morphology \citep{2013ApJ...774...39A,2015ApJ...805..186L}, which may remain unresolved in the observations presented by \citet{yildiz13}.  Thus, the filling factor is prone to be underestimated.

The beam filling factor decline as the $J$-level increases.  As we learned from \citet{2017arXiv170510269K}, the CO~\jj{16}{15} is dominated by broad components coming from the cavity shocks at the inner regions of outflow/envelope.  If the CO emission from the cavity shocks becomes more significant as the $J$-level increases, one should expect a decrease of the spatial extent of CO emission, closer to the small values found for water by \citet{2014AA...572A..21M}.  Thus, the filling factors extracted from the 1D spectra, which cover apertures $>$20\arcsec, naturally decrease with the increase of $J$-level due to the unresolved spatial distribution.

\begin{deluxetable*}{l r r r r r r r r r r}
\tabletypesize{\scriptsize}
\tablecaption{Beam filling factor of CO lines \label{filling}}
\tablewidth{0pt}

\tablehead{
    \colhead{Line} & \colhead{\begin{turn}{70}L1551~IRS5\end{turn} }&
    \colhead{\begin{turn}{70}Ced110~IRS4\end{turn}} &
    \colhead{\begin{turn}{70}GSS~30~IRS1\end{turn}} &
    \colhead{\begin{turn}{70}VLA~1623$-$243\end{turn}} &
    \colhead{\begin{turn}{70}WL~12\end{turn}} &
    \colhead{\begin{turn}{70}RCrA~IRS5A\end{turn}} &
    \colhead{\begin{turn}{70}RCrA~IRS7B/C\end{turn}} &
    \colhead{\begin{turn}{70}HH~46\end{turn}} &
    \colhead{\begin{turn}{70}DK~Cha\end{turn}} &
    \colhead{\begin{turn}{70}TMR~1\end{turn}}
    }
\startdata
CO~\jj{5}{4}    & 0.14      & 0.043      & 0.41       & 0.30       & 0.061      & 0.42       & 0.75 & 0.16 & 0.088 & \nodata \\
CO~\jj{6}{5}    & \nodata   & 0.061      & 0.39       & 0.29       & 0.043      & 0.39       & 0.73 & 0.15 & \nodata & 0.044 \\
CO~\jj{7}{6}    & \nodata   & \nodata   & 0.30       & \nodata   & \nodata   & \nodata   & 0.60 & \nodata & \nodata & \nodata \\
CO~\jj{8}{7}    & \nodata   & \nodata   & \nodata   & \nodata   & \nodata   & \nodata   & 0.48 & \nodata & \nodata & \nodata \\
CO~\jj{9}{8}    & 0.054      & \nodata   & 0.11       & 0.088      & \nodata   & \nodata   & 0.71 & \nodata & \nodata & \nodata \\
\enddata
\end{deluxetable*}

\subsection{The Correlations of Line Emission}
\label{sec:correlation}
Line emission from different molecules and atoms in the embedding envelope carries information of the physical properties of the gas.  However, a complete modeling of the line emission is complicated by the complexity of source-to-source difference and the combination of emission from different components (envelope, outflow, and shocks).  Thus, we took a different approach to investigate the line-to-line correlations from all of the lines detected from the 1D spectra.  We use the generalized Spearman's rank correlation to estimate the correlations of all line pairs detected toward all COPS sources, using \texttt{ASURV} Rev. 1.2 \citep{isobe90,lavalley92}, which implements the methods presented in \citet{isobe86}.  The Spearman's rank correlation estimates how well the fluxes from two transitions can be described by a monotonic function.  We further compute the significance of the Spearman's correlation coefficient ($r$) using the Fisher transformation (Equation~\ref{eq:spr.z})
\begin{equation}
    z = \sqrt{\frac{n-3}{1.06}}\frac{1}{2}ln\left(\frac{1+r}{1-r}\right),
    \label{eq:spr.z}
\end{equation}
where $z$ is the significance, $n=20$ is the sample size, and $r$ is the Spearman's rank correlation coefficient.
\texttt{ASURV} allows us to consider both the detections and the upper limits, so that the correlations remain unbiased when only a few detections are found for high energy emission lines.  We restricted our analysis to 20 sources, where both SPIRE and PACS spectra are available, to have the same number of sources for all lines.

\subsubsection{Correlation: CO}
\label{sec:co_cor}
We found that the correlation strengths of all CO line pairs show systematic behaviors within three square regions (Figure~\ref{fig:co.spr.rho}).  Those regions coincide with the wavelength ranges covered by SPIRE-SLW, SPIRE-SSW, and PACS R1 modules, suggesting an instrumental effect.  The regions covered by the PACS B2A and B2B modules show less obvious systematic effect due to fewer detections of high-$J$ CO lines.  Therefore, we only focus the correlations \textit{within} the instrumental boundaries in the following discussion.  We discuss the characteristics of the distribution of correlations of different species in the following sections.
We first discuss the correlations of CO shown in Figure~\ref{fig:co.spr.rho}; then we describe the correlations of other species, shown in Figure~\ref{fig:all.spr.rho}.

Within each module, the correlation strength smoothly decreases as the difference between the \Jup-levels of two lines increases.  No CO lines show significant offset of correlation strength other than the smooth decrease over $J$-levels.  For each CO line, the correlation strength gradually decreases when comparing either with a higher-$J$ CO line or a lower-$J$ CO line, showing no clear asymmetry on the distribution of correlations.  The smooth variation of the correlation strength together with the discrete origin of CO emission evident in the velocity-resolved profiles \citep{2017arXiv170510269K} suggests that each component of the CO emission contributes to a wide range on energy levels, and the contributing ranges of $J$-levels overlap well with each other, so that there is no distinct discontinuity seen in the distribution of correlations.

\subsubsection{Correlation: $^{13}$CO}
\label{sec:13co_cor}
The $^{13}$CO lines correlate with each other, but show much more variation from pair to pair than the distribution of correlations for CO lines (Figure~\ref{fig:all.spr.rho}), reflecting the lower detection rate of $^{13}$CO lines compared to CO lines.  Weaker but significant correlations are also found with low-$J$ $^{13}$CO lines.
The $^{13}$CO~\jj{7}{6} line shows relatively low strengths of correlation with other $^{13}$CO lines, mainly due to the fact that it is only detected toward GSS~30~IRS1 and RCrA~IRS7B/C (see more discussion of $^{13}$CO emission in Section~\ref{sec:co_opt_depth}).

\subsubsection{Correlation: \water}
\label{sec:water_cor}
\begin{figure}[htbp!]
    \includegraphics[width=0.45\textwidth]{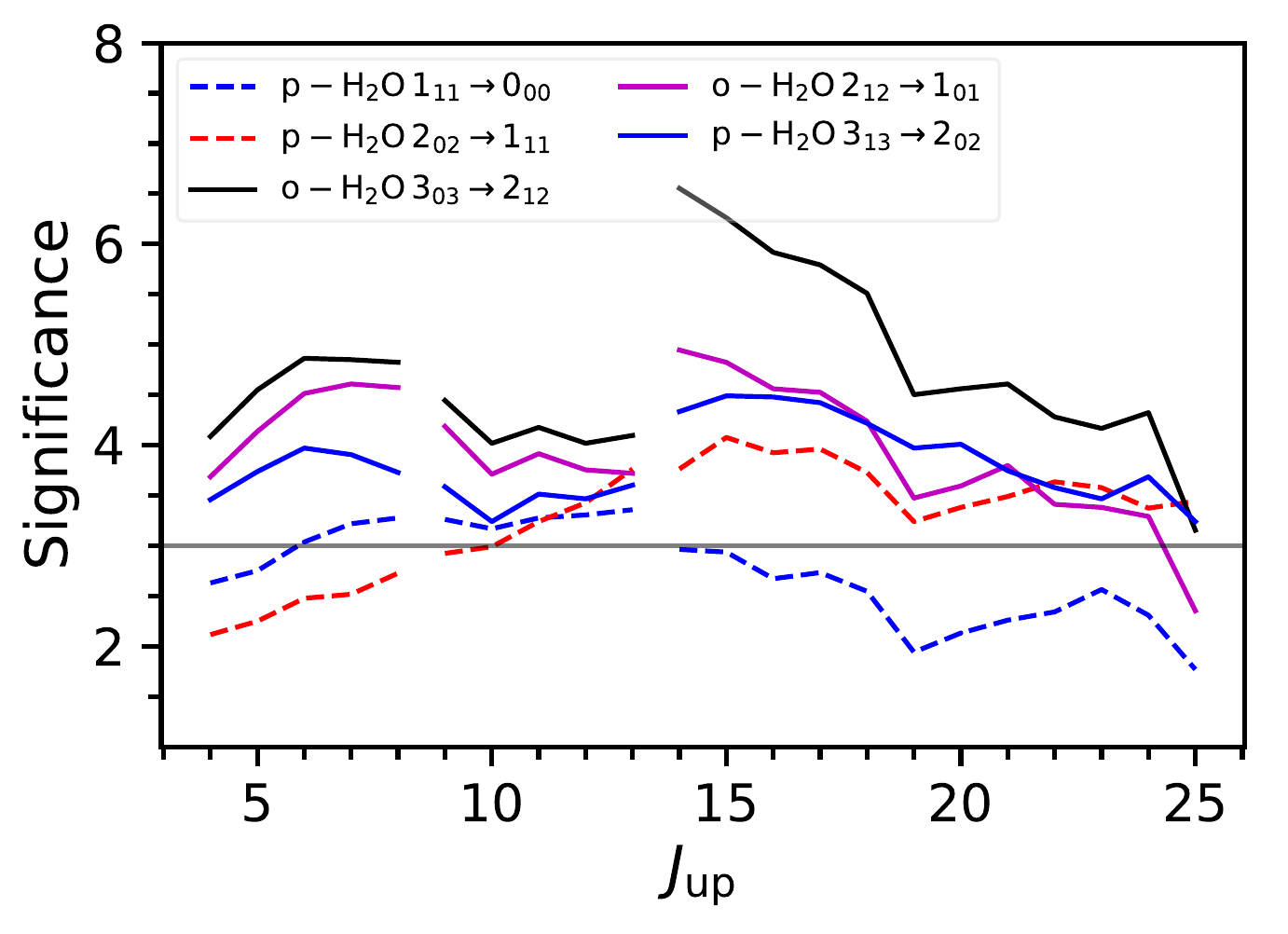}
    \caption{The 1D distributions of the significance of correlation with the CO lines for five water lines observed with PACS (p-\water~$3_{13}\rightarrow2_{02}$, o-\water~$3_{03}\rightarrow2_{12}$, and o-\water~$2_{12}\rightarrow1_{01}$) and SPIRE (p-\water~$1_{11}\rightarrow0_{00}$ and p-\water~$2_{02}\rightarrow1_{11}$).  }
    \label{fig:water_1d_correlation}
\end{figure}

Figure~\ref{fig:all.spr.rho} presents the distribution of correlation strength of selected \water\ lines, which have multiple detections among the entire sample.  All \water\ lines best correlate with other water lines.
Two selected \water\ lines are observed by SPIRE, including
p-\water~$1_{11}\rightarrow0_{00}$ (269~\micron) and p-\water~$2_{02}\rightarrow1_{11}$ (303~\micron).  The o-\water~1$_{10}\rightarrow1_{01}$ line at 557 GHz (538 \micron; $E_{\rm{up}}/k$ = 61~K), predominantly observed with HIFI, is not detected toward most of COPS sources with SPIRE (see the discussion in Section~\ref{sec:linestats}).
Three \water\ lines are selected from the PACS spectra, including
p-\water~$3_{13}\rightarrow2_{02}$ (139~\micron), o-\water~$3_{03}\rightarrow2_{12}$ (174~\micron), and o-\water~$2_{12}\rightarrow1_{01}$ (179~\micron).
Figure~\ref{fig:water_1d_correlation} shows the correlations of the selected water lines with the CO lines.  All water lines show better correlations with the CO lines in the middle of the range between \Jup=4--25.  The instrumental effect causes the discontinuity in the distribution of correlations.  The water lines observed with PACS typically have a higher correlation with the CO lines compared to the water lines observed with SPIRE, however, they all have a similar behavior for the correlations with CO lines.

Other studies have shown that the low-$J$ CO lines have weaker correlations with water.  \citet{kristensen12} found a weak correlation between o-\water~$1_{10}\rightarrow1_{01}$ line and CO~\jj{3}{2} from 29 protostars, eight of which are included in this analysis.  \citet{2017arXiv170510269K} presented a similar analysis for a larger sample.
The water line also show inconsistent spatial extent with CO~\jj{3}{2} line \citep{santangelo12,2014AA...568A.125S}, whereas the high-$J$ CO emission spatially coincides with the water emission.  The higher correlation of water lines with CO ranging from \Jup=10--15 suggests hidden components associated with both species, possibly the cavity shocks in the outflow due to the similar line profile between the emission of CO~\jj{16}{15} and water \citep{kristensen12,2017arXiv170510269K}.

\subsubsection{Correlation: OH}
\label{sec:oh_cor}
We select the strongest OH doublet lines around 119~$\mu$m to investigate the correlation of OH lines with other species.  The OH lines correlate with CO lines ($J_{\rm up} \leq$ 24) and \water\ lines.  The correlations between OH and high-$J$ CO lines (\Jup\ $>$ 24) are relatively weak despite the fact that high-$J$ CO lines are detected toward one-third of the entire sample.  \citet{wampfler13} found that the OH from a sample of Class 0+I sources, where four of them are included in this study, and intermediate-mass protostars shows correlations with water and \OI.
An outflow origin was suggested by the broad component found in the line profiles of both high-mass and low-mass protostars, W3~IRS5 \citep{wampfler11} and Serpens~SMM1 \citep{2013AA...557A..23K}.  The broad component contributes significantly to the high-$J$ CO lines, such as CO~\jj{16}{15} \citep{2017arXiv170510269K}.  Thus, it is reasonable to have significant correlations between OH and high-$J$ CO lines given their similar origins.
While our result suggests that OH correlates with water, the COPS sources show only 2.6$\sigma$ and 2.4$\sigma$ correlations between OH and \OI.

Since the water emission mostly traces the outflows \citep{kristensen12,2014AA...568A.125S}, the correlation between OH and water suggests that \water\ and OH may co-exist in the outflow (\citealt{2014AA...572A..21M,2017arXiv170510269K}, Kristensen \&\ Wampfler in prep.).
\citet{karska13} and Karska et al. (to be accepted) suggest the presence of OH from the irradiated shocks that best describe the emission of water and high-$J$ CO lines.

\subsubsection{Correlation: Atomic Species}
\label{sec:atomic_cor}
The \CI\ lines at 370~$\mu$m and 609~$\mu$m are widely detected toward our sample.  The two lines show correlations with low-$J$ $^{13}$CO lines, whereas some correlation is found with the low-$J$ $^{12}$CO lines and the \CI\ line at 370~\micron.  The $^{13}$CO lines (e.g. $^{13}$CO~\jj{6}{5}) are found to trace the UV-heated outflow cavity wall as well as the quiescent envelope \citep{yildiz12,2015AA...576A.109Y}, and the same origins were suggested for the \CI\ 370~\micron\ line \citep{2009AA...507.1425V}.

The forbidden transition of atomic oxygen, which is predominantly detected at 63~$\mu$m, shows correlations with CO lines, and correlates better with the CO lines ranging from \Jup=10 to 24.  \OI\ is thought to be a tracer of outflow \citep{hollenbach85,2017AA...597A..64D}, at least for the short term activity of outflows \citep{2017AA...600A..99M}.  The emission of outflows contributes significantly to the high-$J$ CO line, so greater strengths of correlation are to be expected between the \OI\ line and high energy CO lines.
Toward NGC1333~IRAS4A, \citet{2017AA...601L...4K} found that part of the \OI\ line at 63~\micron\ exhibits the same line profile as the CO~\jj{16}{15} line and few water lines at the same velocity channels.

\begin{figure*}[htbp!]
    \centering
    \includegraphics[width=0.9\textwidth]{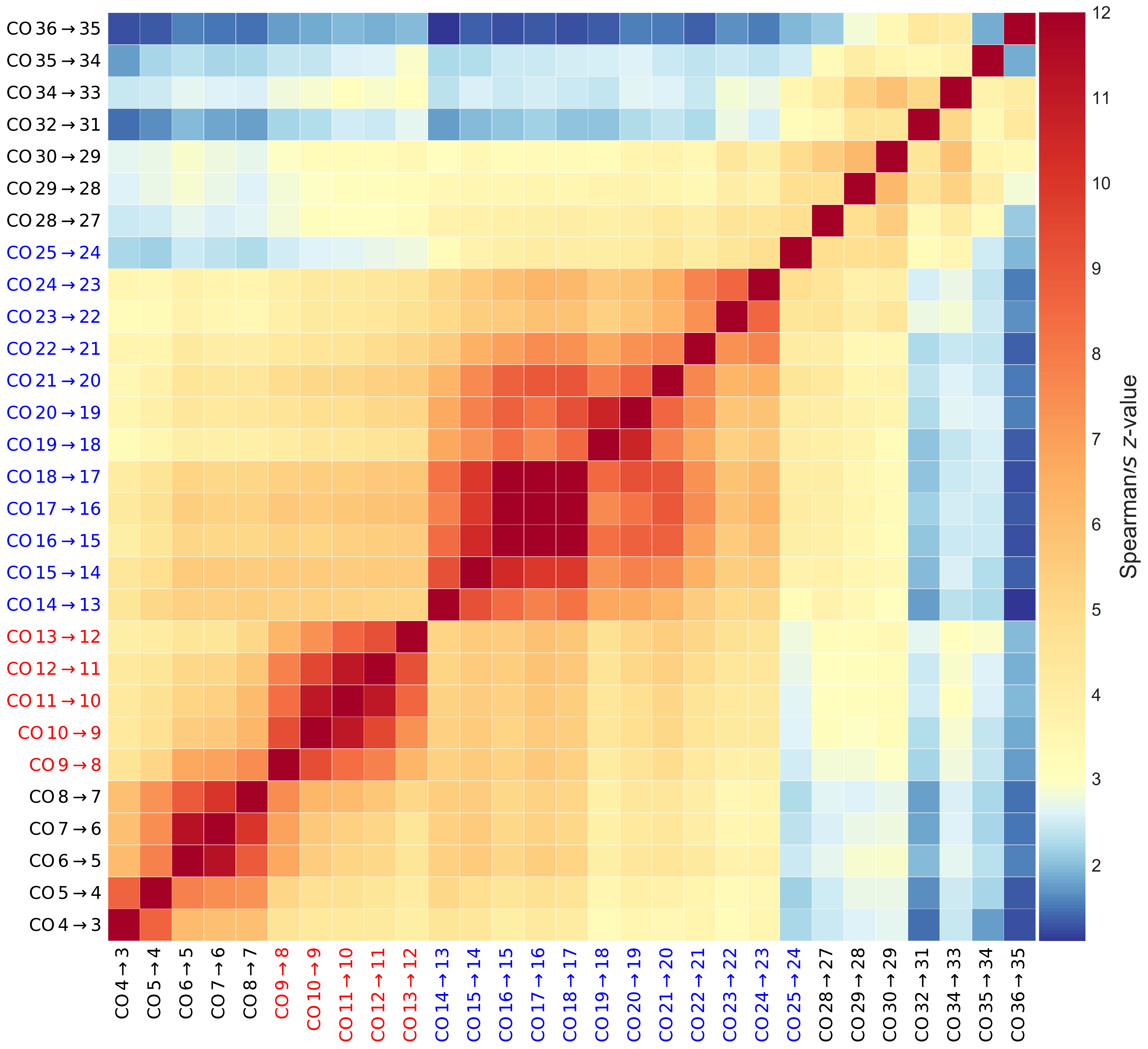}
    \caption{The Spearman's z-value for all CO line pairs.  The colors of labels indicate the lines covered by the same module/instrument.}
    \label{fig:co.spr.rho}
\end{figure*}

\begin{figure*}[htbp!]
    \centering
    \includegraphics[width=0.9\textwidth]{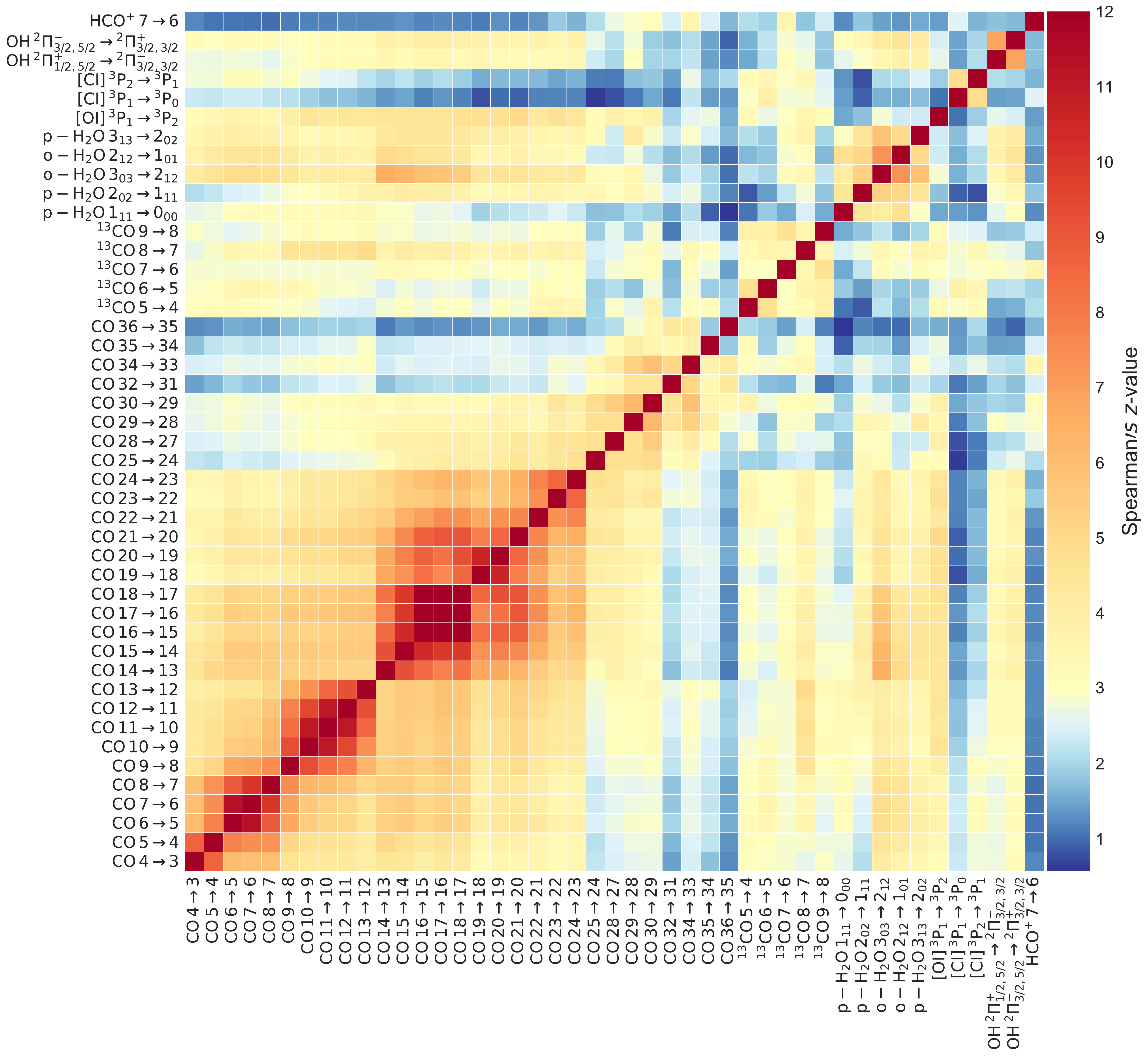}
    \caption{The Spearman's z-value for all line pairs.}
    \label{fig:all.spr.rho}
\end{figure*}

\subsection{Spatial Extent of Line and Continuum Emission}
\label{sec:lineextent}

\begin{figure*}[htbp!]
    \centering
    \includegraphics[width=0.45\textwidth]{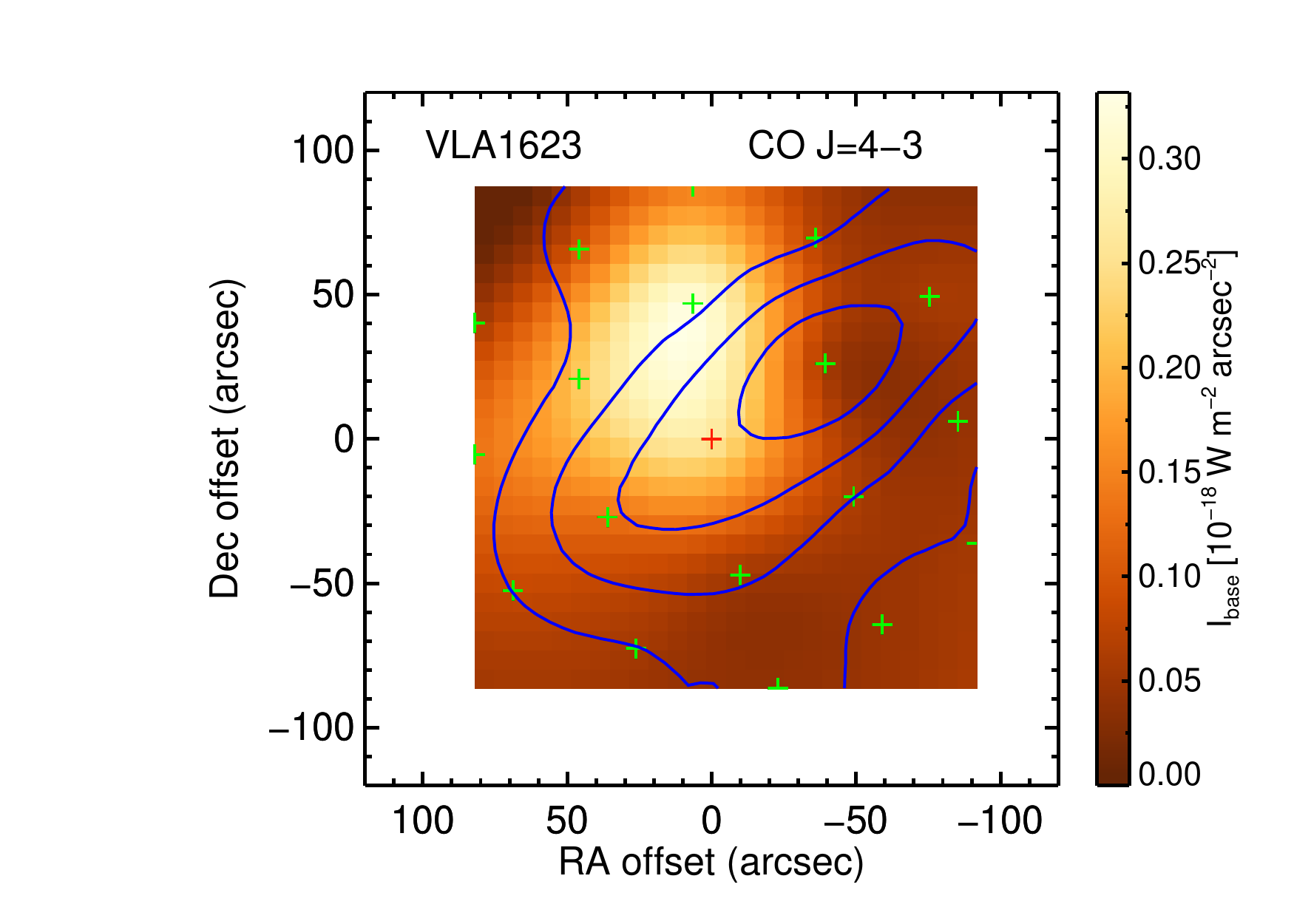}
    \includegraphics[width=0.45\textwidth]{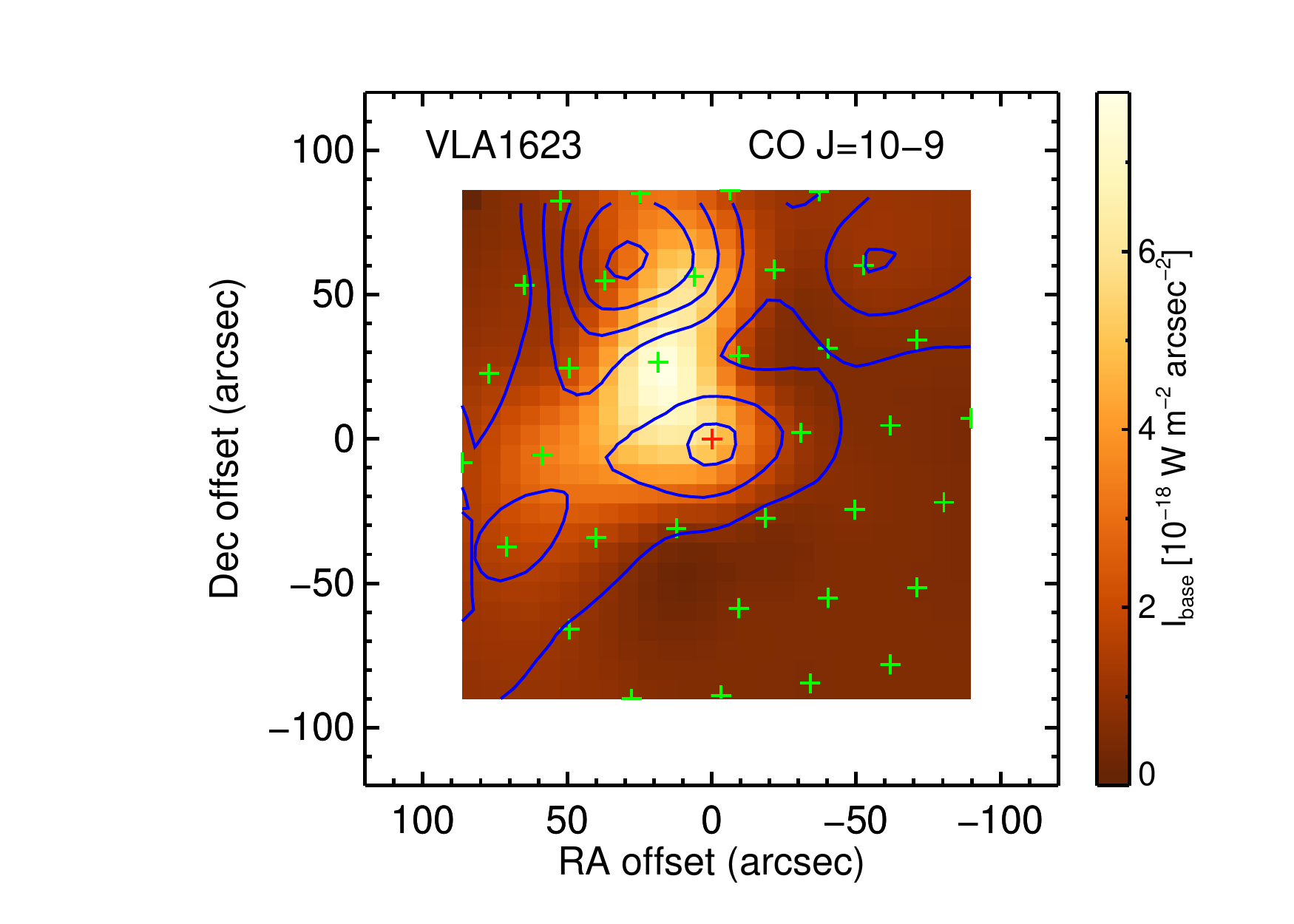}
    \caption{The smoothed contours of CO~\jj{4}{3} and CO~\jj{10}{9} emission detected toward VLA~1623$-$243.  The blue contours indicate the line fluxes down to 20\%\ of the peak flux, and the flux level increases by 20\%\ between contours.  The ``plus'' signs mark the location of spaxels in green for detections.  The central position is plotted in red ``plus.''  The background image is the smoothed distribution of continuum at the line centroid.}
    \label{fig:contours}
\end{figure*}

In the previous analyses, we focused on the 1D spectra for simplicity.  In fact, the SLW and SSW modules in SPIRE have 19 and 35 spaxels in a hexagonal layout with separations of 19\arcsec\ and 33\arcsec\ (see the ``plus'' markers in Figure~\ref{fig:contours}), respectively, providing information on the spatial morphology of emission lines.  Figure~\ref{fig:contours} shows two contours of CO~\jj{4}{3} and CO~\jj{10}{9} detected toward VLA~1623$-$243 as examples.  The CO emission is detected, not only at the central position but also in the outer spaxels, and the morphology also varies from line to line.  Although it is easier to show the morphology of emission lines as contours, the information on spatial extent needs to be simplified when we investigate the morphology of each line toward all sources at once.

\subsubsection{1D profiles of Morphology}
We take advantage of the hexagonal spaxel layout to project the 2D morphology into a 1D profile as a function of azimuthal angle.  For each non-central spaxel, we first calculate its distance to the central spaxel and the azimuthal angle with respect to north.  Then we take the ratio of the flux at the non-central spaxel to the flux at the central spaxel to construct a 1D profile of the flux ratios as a function of the azimuthal angle.
We restrict our analysis to the outer spaxels within a radius of 75\arcsec\ from the central spaxel to exclude the low SNR outermost spaxels.  The 75\arcsec\ radius allows us to include the nearest hexagon in the SLW module (6 spaxels), and the two nearest hexagons in the SSW module (16 spaxels).
As the distance to the central spaxel increases, the detected line flux generally decreases as well; however, in some cases, including B1-c, BHR~71, L1157, L1551~IRS5, VLA~1623$-$243, and WL~12, the strongest CO emission is found at off-center positions.  We reduce the radius to 50\arcsec\ for RCrA~IRS5A to avoid source confusion.

We apply a Gaussian smoothing kernel to the profile of flux ratios to produce a smooth profile for investigating the large scale morphology.  The Gaussian kernel has a 1-$\sigma$ width of 30 degree for a 60 degree spacing between two adjacent spaxels in the inner ring.

With the 1D profiles of all CO emission lines derived for each source, we investigate the variation of morphology of CO emission from line to line in two ways, the smoothed 1D profile and the difference of the maximum and minimum ratios of the 1D profile.

\subsubsection{Morphology with the 1D Profiles}
\begin{figure*}[htbp!]
    \centering
    \includegraphics[width=0.9\textwidth]{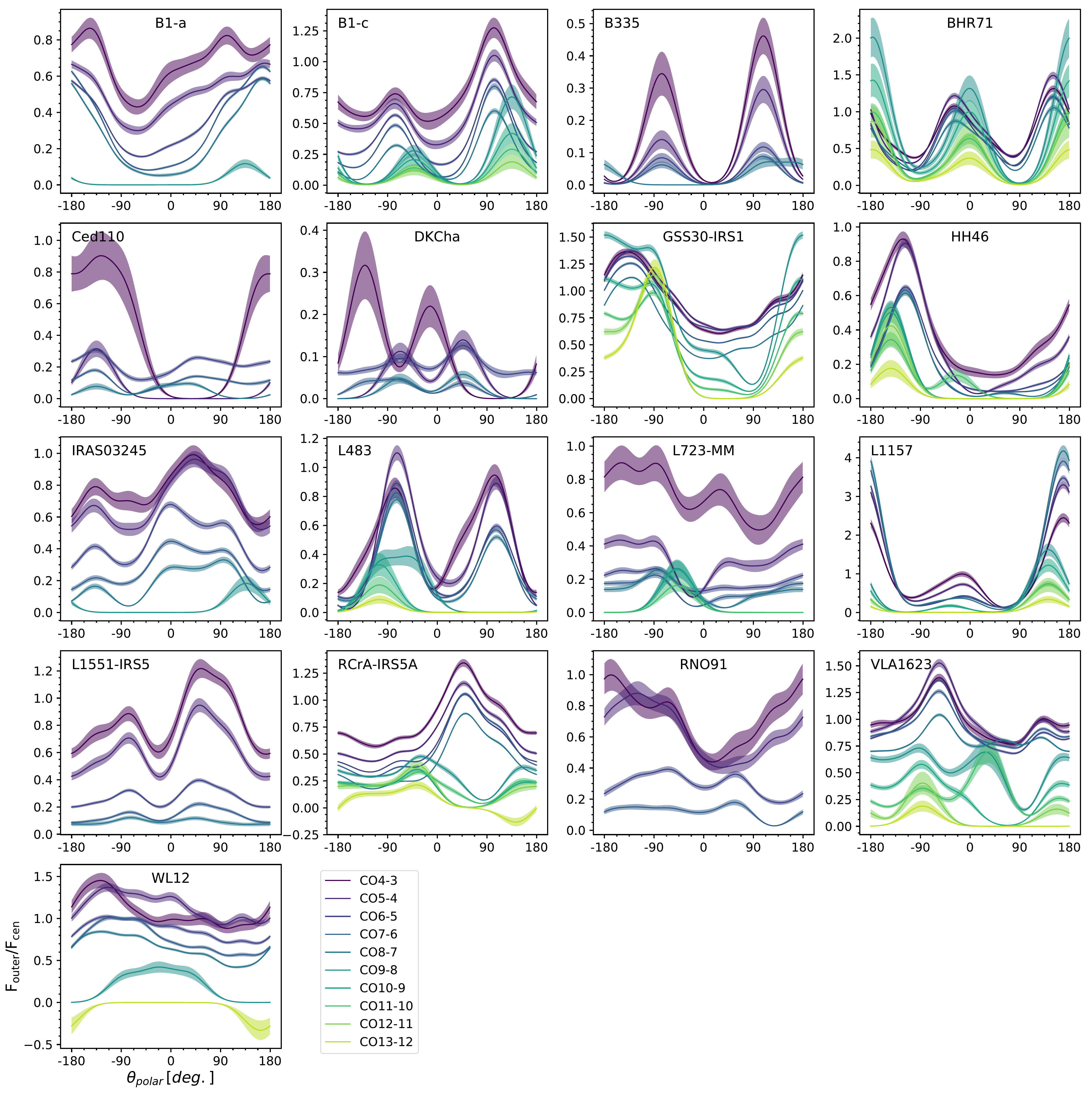}
    \caption{The distribution of flux ratios with the flux at the central spaxel as a function of the azimuthal angle.  The colors of the lines indicate the different \Jup-levels of the CO lines, and the shaded area represents the uncertainty of the smooth profile.
    We only consider the outer spaxels within 75\arcsec\ from the central spaxel, yielding 6 and 16 outer spaxels for the SLW and SSW modules, which have spaxels in hexagonal layout with separations of 19\arcsec\ and 33\arcsec, respectively.  The 75\arcsec\ radius is reduced to 50\arcsec\ for RCrA~IRS5A to avoid source confusion.}
    \label{fig:co_extent}
\end{figure*}

Figure~\ref{fig:co_extent} shows the smoothed profiles of CO emission lines.  Six sources are excluded due to the low detection rate at the outer spaxels; therefore, not enough data are available for constructing the 1D profile.
In principle, if the emission centers at the central spaxel and decreases outward in all directions uniformly, the 1D profile will appear to be a flat line (e.g. WL~12).  If the distribution of the emission is concentrated but offset from the center, the 1D profile will show a single peak at the angle toward the offset spaxel (e.g. HH~46).  If the distribution shows a bipolar feature, the 1D profile will have two peaks at two azimuthal angles separated by the angle between two lobes (e.g. BHR~71).  All three cases are seen in Figure~\ref{fig:co_extent}.

We found that all bipolar features seen in the 1D profiles correspond to the observed outflows in the spatially-resolved maps.  The single peak profiles also agree with the maps, suggesting a single outflow.  We discuss the detailed comparison in Appendix~\ref{sec:a_outflow_classification}.
For the sources with two peaks, we noted that the flux ratio decreases as the \Jup-level increases.  In addition to the variation of the flux ratios, we sometimes found the positions of the peaks also vary with the \Jup-level (B1-c, BHR~71, and L1157); however, this variation may result from the difference of spaxel layout of the SLW and SSW modules, which changes between \Jup=8 and 9.  Also, in the cases of L483, the lower-$J$ CO lines show two-peak profiles, but only a dominant peak exists in the profiles of the higher-$J$ CO lines due to the lack of detections at the higher-$J$ CO lines, making one of the lobes disappear from the observed morphology.

\subsubsection{Variations of Bipolarity}
\label{sec:varitation_bipolarity}
Beside the visual inspection of the 1D profiles, we explore the difference between the maximum and the minimum values (hereafter the peak-to-valley difference) of each profile as a probe of morphology.  A flat profile will result in a difference close to zero, while a strong double-peak profile leads to a significant difference.  However, a double-peak profile produces the same difference as a equally strong single-peak profile.  Thus, we only compare the peak-to-valley difference for the sources with bipolar features.

\begin{figure}[htbp!]
    \centering
    \includegraphics[width=0.45\textwidth]{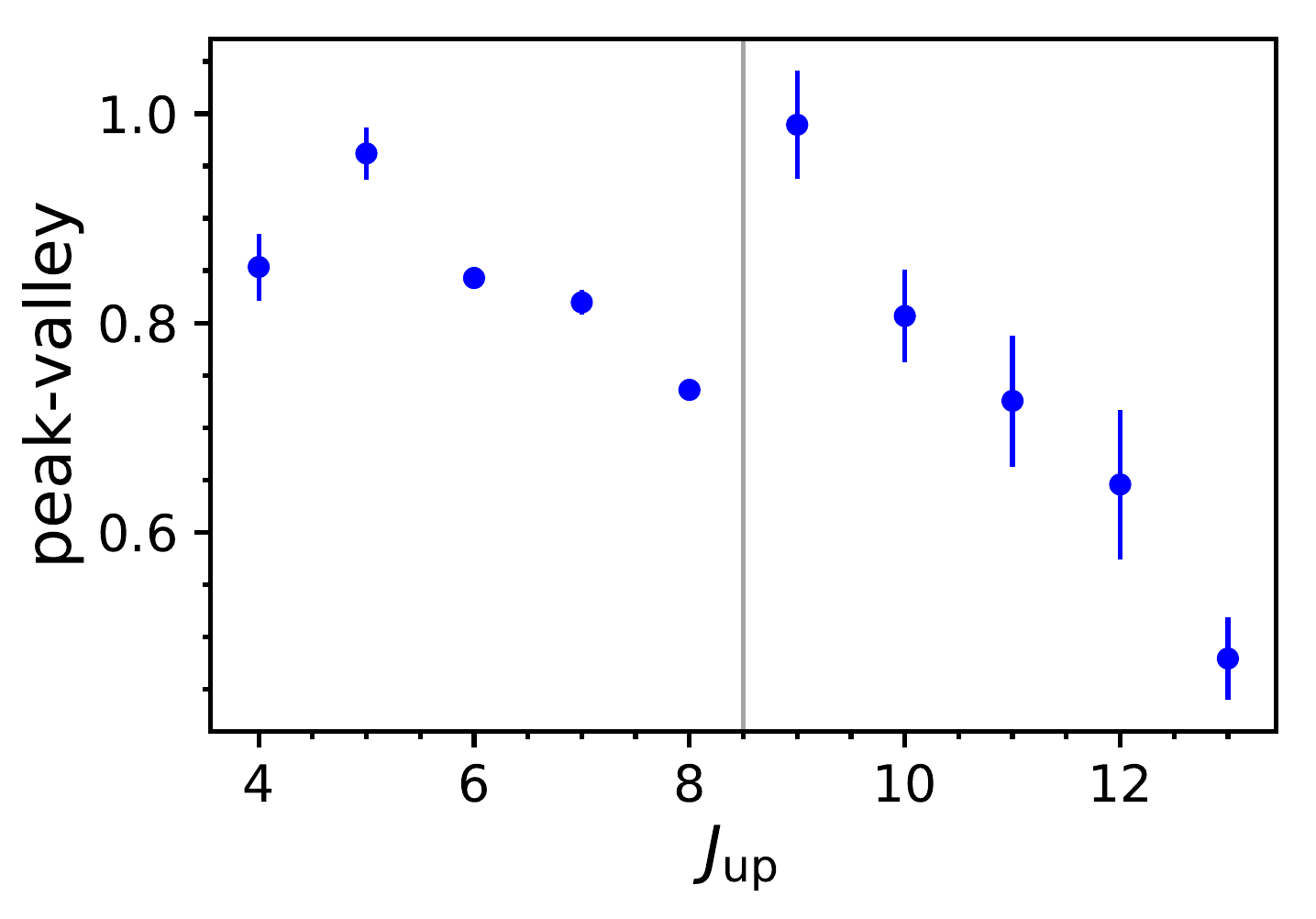}
    \caption{The averaged peak-to-valley differences from the 1D profiles of the sources with more than three detections at the outer spaxels.  The peak-to-valley differences of each source were normalized separately for the SLW and SSW modules before we took the average of all sources.  The vertical line separates the SLW and SSW modules.}
    \label{fig:co_extent_value}
\end{figure}

Figure~\ref{fig:co_extent_value} shows the distribution of the peak-to-valley difference derived from the sources with more than three detections in the outer spaxels.  The sources include B1-c, L483, L1157, HH~46, IRAS03245+3002, BHR~71, GSS~30~IRS1, and VLA~1623$-$243, where B1-c, L483, L1157, HH~46, and IRAS~03245+3002 are only considered for the SLW module.  Selecting those sources reduces the effect of sensitivity (see Appendix~\ref{sec:a_outflow_classification}).  The values of each source were normalized by the SSW and SLW modules before we averaged the peak-to-valley differences of all COPS sources.  The peak-to-valley difference shows a shallow decrease at \Jup=4--8, and exhibits a steady decrease at \Jup=9--13, suggesting that the bipolar feature is diminishing as the $J$-level increases at higher-$J$ regime.

The change of beam size between the two SPIRE modules hinders a direct comparison of the peak-to-valley differences from the SLW and SSW modules.  We should cautiously conclude that the diminishment of the bipolar feature is seen in each module.

To summarize, the flux ratios (outer to center) as a function of azimuthal angles simplifies the 2D morphology to the 1D profiles.  Among the 17 sources, 6 sources show bipolar distributions of CO emission lines, while single lobes and irregular profiles are seen in the 1D smoothed profiles of a further 11 sources.  The morphology identified from the 1D profiles is consistent with the velocity-resolved observations, suggesting that the extended CO emission seen in the SPIRE data can be related to outflows.
The CO emission exhibits bipolarity between \Jup=4--13.  We use the peak-to-valley difference derived from the smoothed 1D profiles to quantify the strength of bipolarity, and find that the bipolarity decreases as the $J$-level increases.

\section{Discussion}

\subsection{SED Classification}
\label{sec:sed_classification}
Several methods of classification have been established in the past few decades, such as the spectral index at near-infrared wavelengths \citep{adams87}, bolometric temperature \citep{myers93,chen95}, and the ratio of bolometric luminosity to the submillimeter luminosity \citep{2000prpl.conf...59A}.  \citet{evans09} summarized the history of classification methods.  SED classes were assigned for sources in the combined
c2d and Gould Belt \textit{Spitzer} Legacy programs by
\citet{2015ApJS..220...11D}, who demonstrated that the main limitation was the
lack of well-sampled data through the far-infrared to submillimeter.
Those data are now available, albeit for only a small sample, so we
focus in this section on whether the previous conclusions are changed
by such data.

Table~\ref{evolutionary} shows the $\alpha_{\rm NIR}$, \tbol, \lbol, \lsmm, and the corresponding classifications based on each method for all sources.  Our sources are all Class 0 or I with 23 of 26 protostars classified as Class 0 by the \lbol/\lsmm\ criterion.  In the following paragraphs, we follow the original definition \citep{andre93} of the Class 0 protostars, $\lbol/\lsmm > 200$, and ask if other classification methods agree with that classification. This parameter was found to be the best indicator of age for very young systems in models that followed the SED evolution through the infall phase \citep{2005ApJ...627..293Y,dunham10,dunham14a}.

\begin{figure*}[htbp!]
    \centering
    \includegraphics[width=0.32\textwidth]{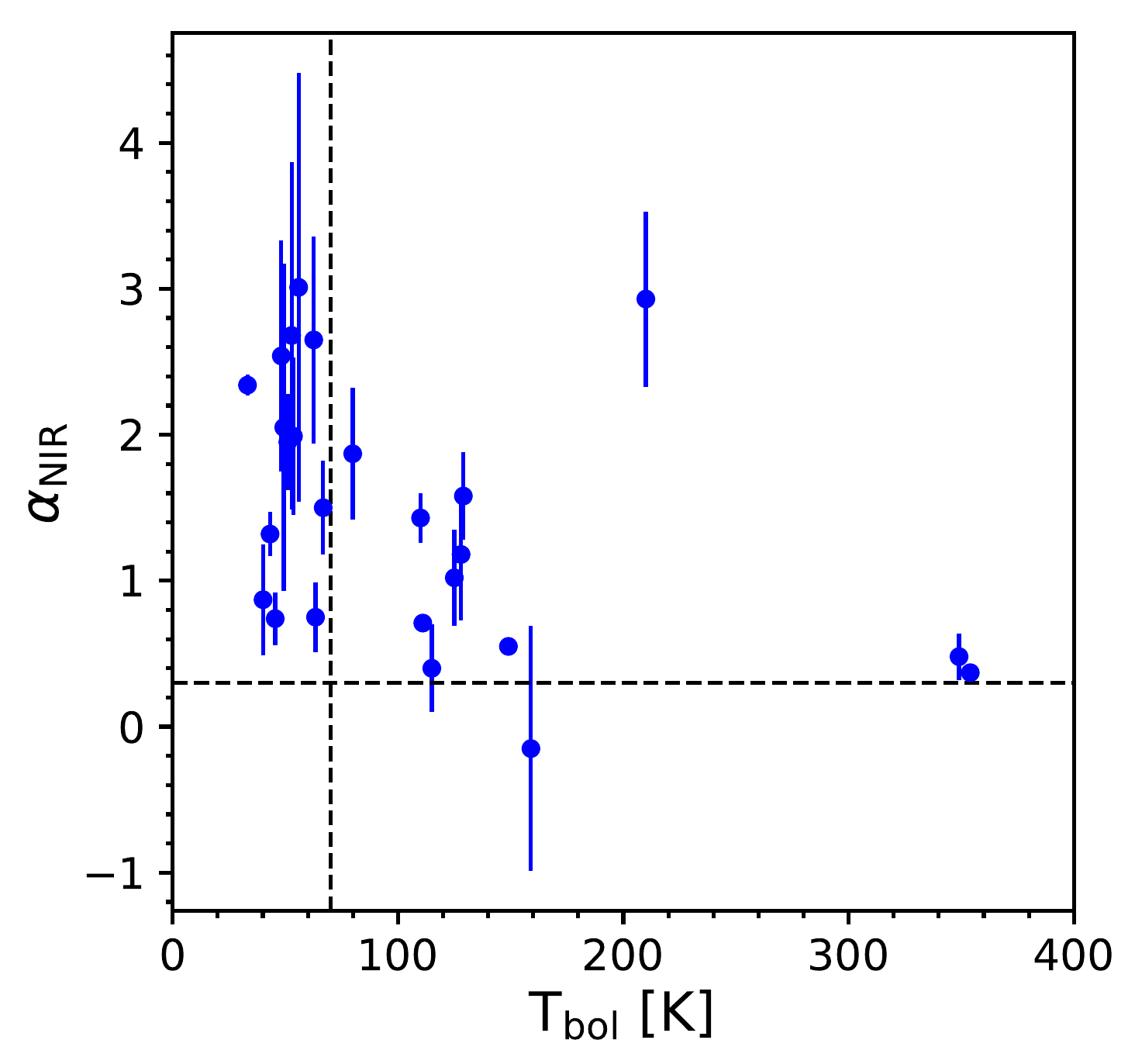}
    \includegraphics[width=0.32\textwidth]{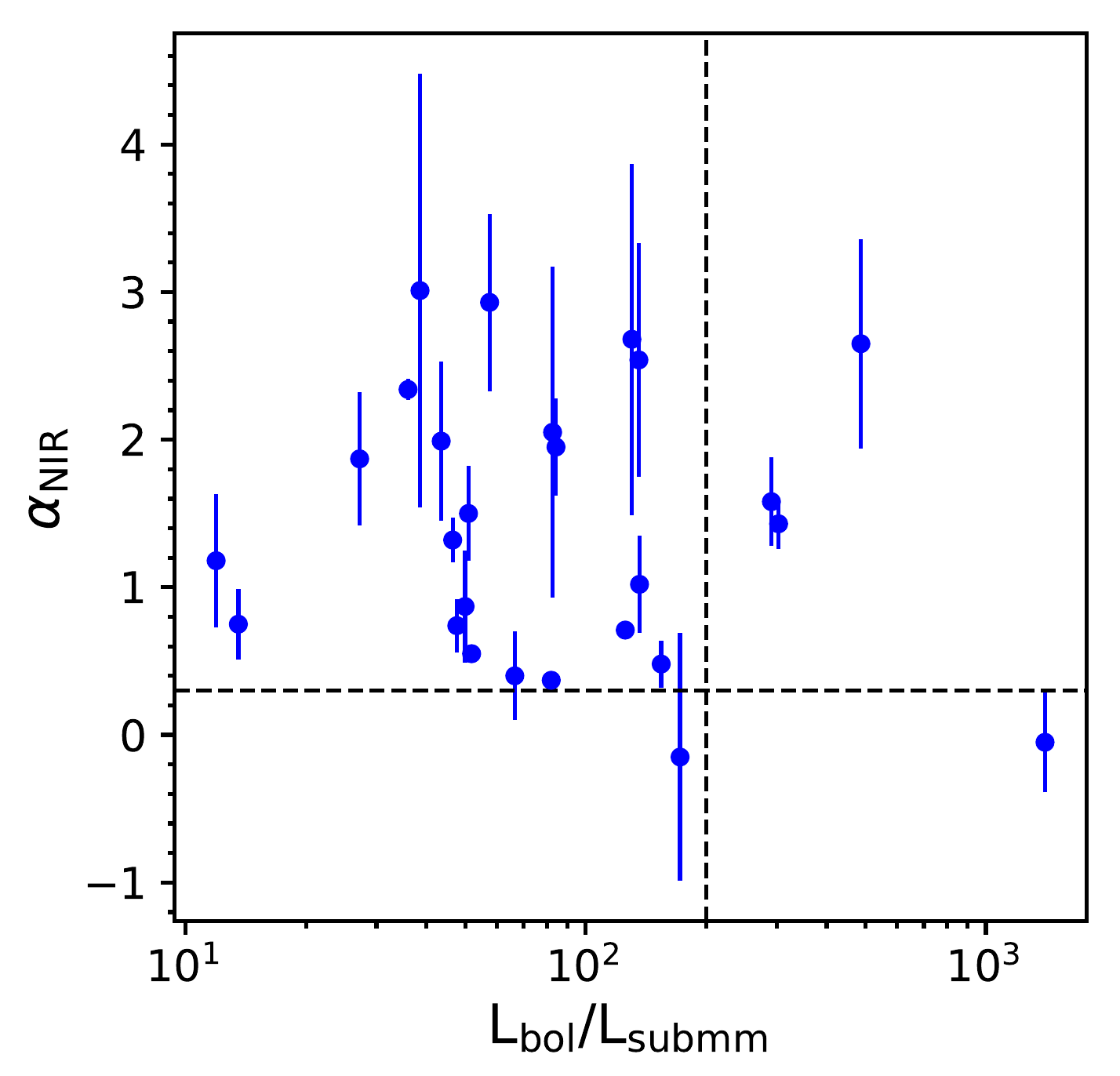}
    \includegraphics[width=0.32\textwidth]{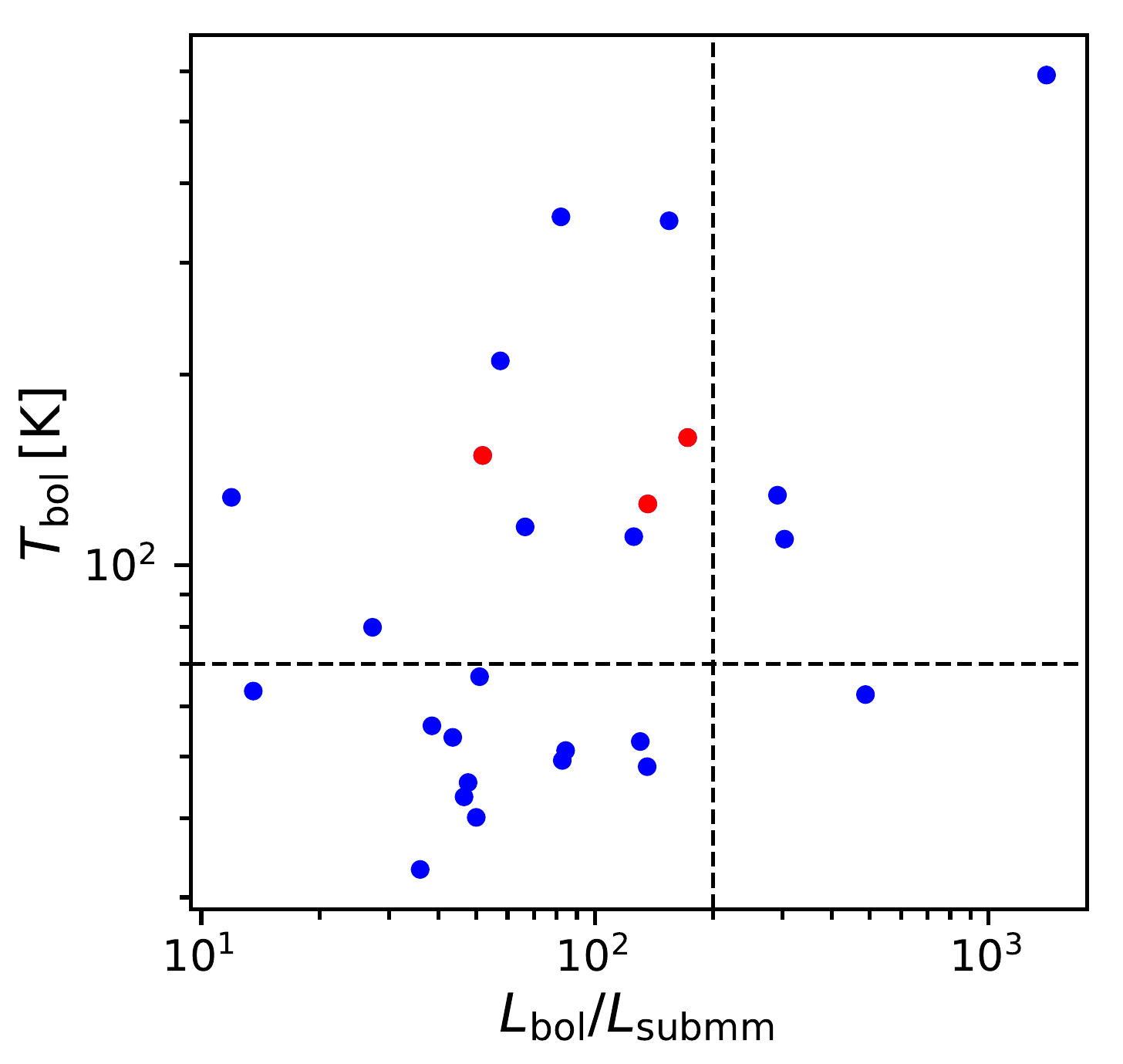}
    \caption{
        \textbf{Left:} The relation between $\alpha_{\rm NIR}$ and \tbol.
        \textbf{Middle:} The relation between $\alpha_{\rm NIR}$ and the ratio of \lbol\ to \lsmm\ for all sources in logarithm scale.
        \textbf{Right:} The relation between \tbol\ and the ratio of \lbol\ to \lsmm\ for all sources in logarithm scale.  The three Taurus sources (TMR~1, TMC~1, and TMC~1A) that are suggested as Stage I sources \citep{2014AA...562A..77H} are shown in red.
        }
    \label{fig:evo_relation1}
\end{figure*}

The original definition of SED classes was based on the slope of the relation between log($\lambda F_{\lambda}$) and log($\lambda$) between 2 and 25 \micron\ \citep{adams87,andre93,greene94}, the spectral index $\alpha$ (denoted as $\alpha_{\rm NIR}$ in this paper). While useful for distinguishing Classes
II and III from earlier classes, many papers have shown that it cannot
separate Class 0 from Class I sources (e.g., \citealt{2015ApJS..220...11D})
and this conclusion is borne out in Figure \ref{fig:evo_relation1}. The
combined Class, 0+I, defined by \citet{2015ApJS..220...11D}, is separated from Class II by $\alpha_{\rm NIR}$, with only three exceptions: IRAS~03301+3111 and TMC~1A, and DK~Cha, which are classified as flat spectrum sources.

The bolometric temperature, \tbol, is also commonly used to separate
Class 0 from Class I.  In this framework, Class 0 sources have \tbol\ less than 70~K. However, \tbol\ is quite sensitive to the near-IR flux due to inclination effects.  With inclination correction, protostars with measured \tbol\ $<$ 70~K are relatively rare, making up 3\% of the distribution of protostars in Orion \citep{fischer13}.
Figure \ref{fig:evo_relation1} shows that our well-sampled data confirm previous
results (e.g. Figure 2 of
\citealt{dunham14a}) showing that many Class 0 sources (using \lbol/\lsmm)
fall into Class I according to \tbol.
In particular, 10 of 26 protostars are classified as  Class I (or flat) sources by \tbol, but as  Class 0 sources by \lbol/\lsmm.
No simple adjustment of the boundaries can bring the two tracers into
substantially better agreement.

The original idea for defining SED classes and variables like \tbol\ was that
they would reflect physical stages. In particular,
a Class 0 source would ideally correspond to a Stage 0 source, with more mass
in the envelope than in the star. While there is general agreement between classes and stages in simulations \citep{robitaille06}, variations in original core mass (affects \lsmm),
orientation relative to the observer (affects \tbol), and evolutionary path
(e.g., episodic accretion) make this connection between class and stage
difficult \citep{dunham10,2017ApJ...840...69F}.  \citet{2014AA...562A..77H} found three Taurus sources in the COPS sample, TMC~1, TMC~1A, and TMR~1, have more than 70\%\ of the total mass in the central protostars, suggesting Stage I sources; however, all three sources are classified as Class 0 sources by \lbol/\lsmm.

Without even a decent estimate of the stellar mass in deeply embedded stages, only indirect proxies are available.  \citet{2017AA...600A..99M} found that the force in the entrained CO gas in the outflows ($F_{\rm CO}$ derived from CO~\jj{3}{2} maps) correlates weakly with \lbol\ (p=3.5$\sigma$), whereas $F_{\rm CO}$ are systematically different for Class 0 and I sources classified by \tbol.

\begin{figure*}[htbp!]
    \centering
    \includegraphics[width=0.32\textwidth]{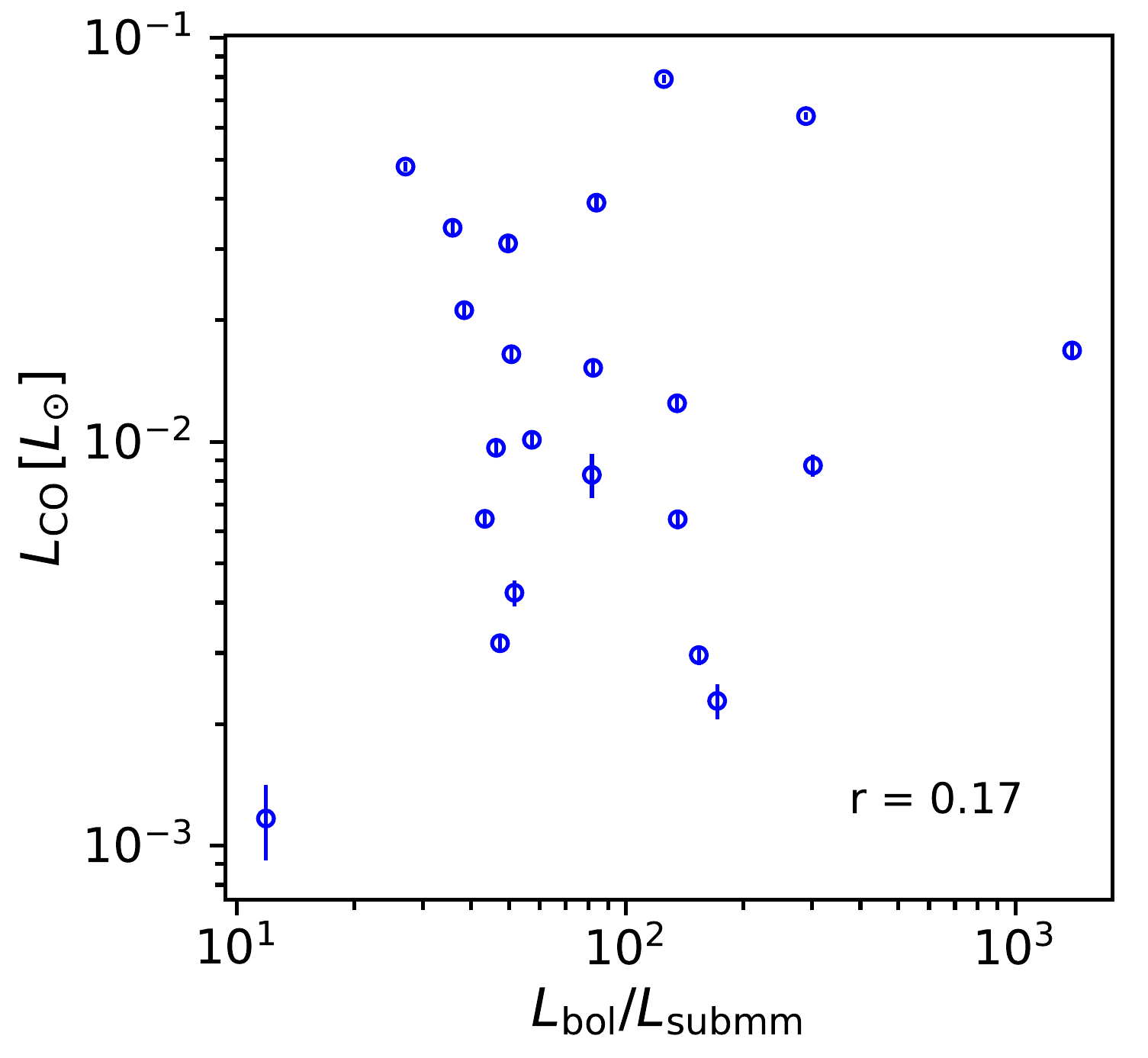}
    \includegraphics[width=0.32\textwidth]{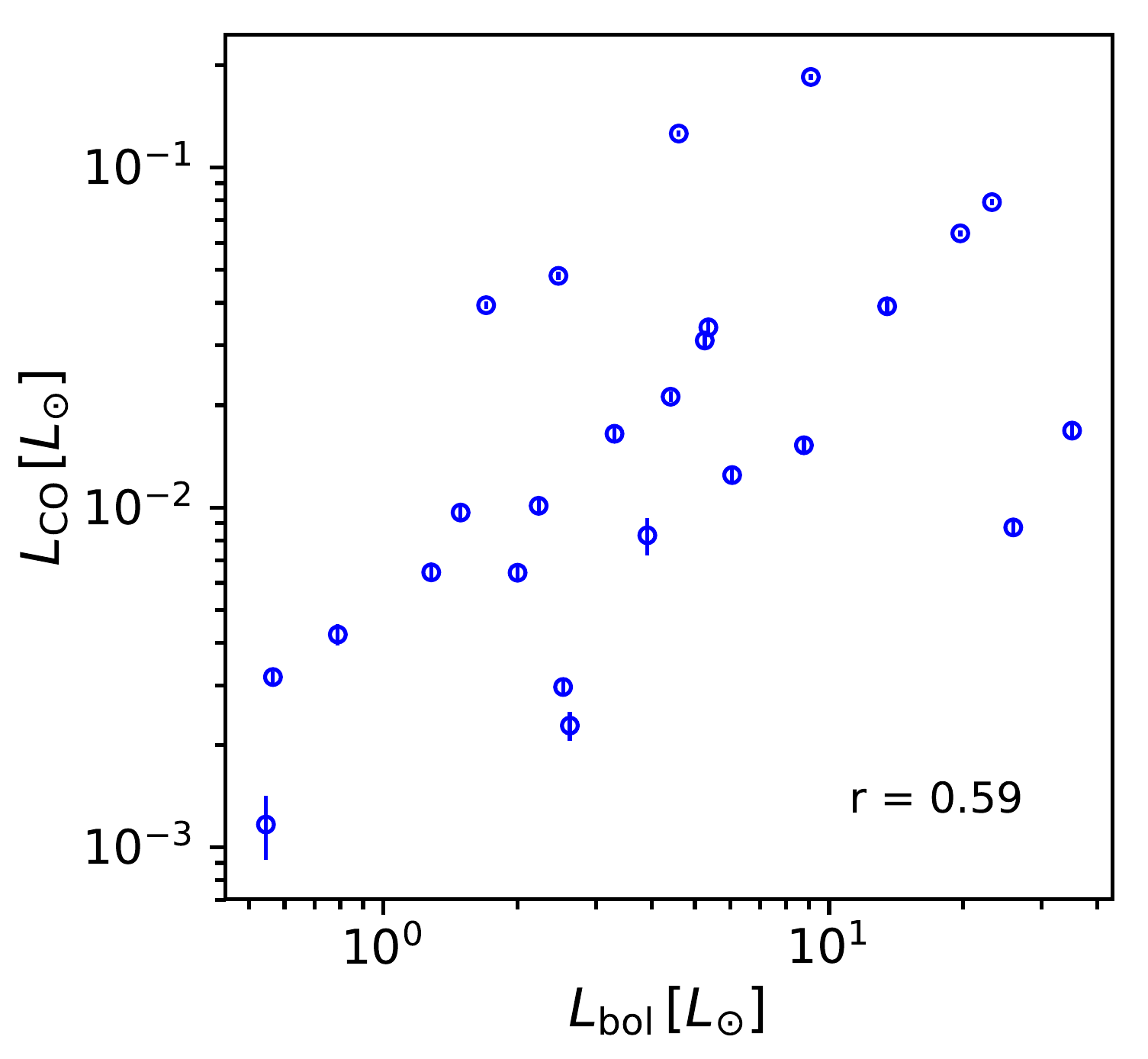}
    \includegraphics[width=0.32\textwidth]{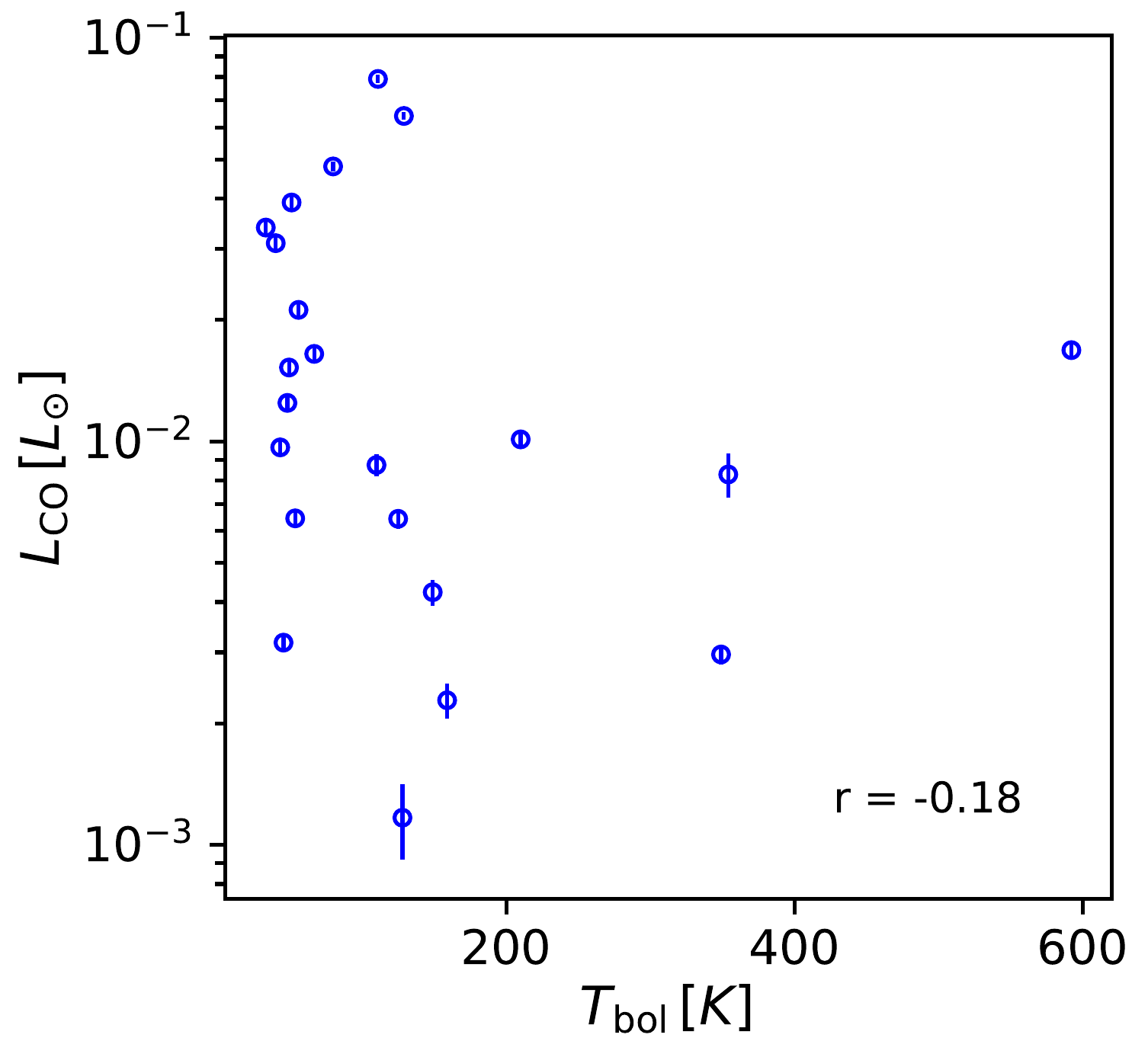}
    \caption{The relation of $L_{\rm CO}$ verses \lbol/\lsmm, \lbol, and \tbol\ from left to right.  The Pearson's correlation coefficient is shown in the bottom right for each relation.}
    \label{fig:Lco_relation}
\end{figure*}

We explore the correlation of the total CO luminosity ($L_{\rm CO}$) with different classification indicators.  We sum up the luminosities of all CO lines with the optical depth correction for calculating $L_{\rm CO}$.
Figure~\ref{fig:Lco_relation} shows the relation of $L_{\rm CO}$ versus \lbol/\lsmm, \lbol, and \tbol, where $L_{\rm CO}$ is the total CO luminosity from the CO lines observed with PACS and SPIRE.  We found a strong linear relation between $L_{\rm CO}$ and \lbol, which can be expressed as
\begin{equation}
    log(L_{\rm CO}) = (0.62\pm0.14)log(L_{\rm bol})-2.10\pm0.13,
    \label{eq:Lco_Lbol}
\end{equation}
but no systematic offset of $L_{\rm CO}$ distinguished by the \tbol\ of the sources.  The same scenario is also found for the HOPS and DIGIT programs \citep{manoj12,2016ApJ...831...69M}.

We found that the classification of embedded protostars still remains uncertain with the addition of \textit{Herschel}-SPIRE spectra.  The two main classification methods, \lbol/\lsmm\ and \tbol, only agrees for 15 of 26 COPS sources.  Studies that modeled the infall process of embedded protostars found the \lbol/\lsmm\ better described the evolution of protostars \citep[e.g.][]{2005ApJ...627..293Y,dunham10,dunham14a}, whereas some studies found that the classification based on \tbol\ is more consistent with the evolution of outflows \citep[e.g.][]{2017AA...600A..99M}.  Direct investigation on the properties of central protostars will provide an accurate estimate on the evolutionary of embedded protostars.

\subsection{The Origin of CO Emission}
\label{sec:co_origin}
It has been proposed by both observational and theoretical studies that different structure, such as UV-irradiated cavity walls, shocks, and the molecular envelope, may be responsible for CO emission over different ranges of $J$-levels. \citep[e.g.][]{vankempen10,visser12,manoj12,2014AA...572A..21M,2015ApJS..217...30L,2017arXiv170510269K}.  Combined with the PACS data, the COPS program covers a wide range of CO lines and the spatial distribution of CO emission, providing us a unique opportunity to review this issue with a more comprehensive view.  We discuss the origin of CO emission in the context of spatial extent, the comparison with kinematic studies, the rotational diagrams, and the distribution of correlations.

In Section~\ref{sec:lineextent}, we have concluded that the extended CO emission observed toward the COPS sample has an outflow-related origin.  \citet{2014AA...568A.125S} found that the morphology of CO~\jj{6}{5} line is much more consistent with the spatial extent of water emission, which traces the shocked gas, than the morphology of CO~\jj{3}{2} line.  This comparison shows that the higher-$J$ CO line traces the shocked gas, while the lower-$J$ CO line traces the gas entrained by outflows.
Velocity-resolved studies with HIFI also show a similar line profile between water and CO lines \citep{kristensen12,yildiz13,2014AA...572A..21M}.
\citet{yildiz13} showed that the CO~\jj{10}{9} line is broader than the CO~\jj{3}{2} line, where the water emission has a broad line width.  For higher-$J$ CO lines, \citet{2017arXiv170510269K} found two components in the line profile of CO~\jj{16}{15} emission, a narrow offset component and a broad centered component.  They suggested that the narrow offset component originates from the spot shocks, while the broad centered component comes from cavity shocks or disk wind.  The profile of CO emission seems to change from centered narrow+broad components from the quiescent envelope and outflowing material at low-$J$ (\jj{3}{2}) to a broader component along with a narrow offset component dominated by cavity shocks (or disk wind) and spot shocks, respectively.  It is clear that there are multiple origins contributing to the CO emission, likely dominating in different ranges of energy levels of CO.

We also see multiple origins of CO emission from our analysis of the rotational temperatures.  Figure~\ref{fig:co_rot_distribution} (right) shows the distribution of the rotational temperatures fitted with flexible break points, suggesting a primary population around 100~K (cold), and a secondary population around 400~K (warm).  The cold population represents the entrained gas in the outflows; the warm population corresponds to the cavity shocks (or disk wind), which contributes to the broad component seen in the line profiles of CO~\jj{16}{15} \citep{2017arXiv170510269K}.
Our spectrally unresolved data are insensitive for identifying spot shocks.  However, we do not find the proposed rotational temperature ($\sim$700~K) for spot shocks in the distribution of rotational temperatures, unless we use fixed break points.  We detect CO lines at \Jup=24--38, excluding \Jup=25 and 26, which are in the gap between two PACS modules, toward 5 sources on average.  Thus, we conservatively argue that spot shocks are missing from our analysis of rotational temperatures of CO, but detection limits for high-$J$ CO lines may prevent the discovery of spot shocks.

The distribution of correlation strength identifies any offset of correlations from any group of lines.  The correlations of all CO lines show no systematic offset of correlations from any group of CO lines, other than the instrumental effects, despite the fact that the distribution of the rotational temperatures clearly shows two distinct populations, suggesting that multiple origins contribute to the entire CO ladder.  If some CO lines are dominated by a certain origin, the correlations among those lines would be much higher than others, appearing to be a regional clustering in the distribution of correlation strength.  On the other hand, if the ranges of $J$-level contributed by different origins overlap with each other significantly, such regional clustering effect may not be seen.

Figure~\ref{fig:co.spr.rho} shows a rather smooth variation of correlations instead of clustering for the CO lines observed in each module.  The smooth variation suggests that each origin contributes to a wide range of CO lines, and the range of $J$-levels contributed by each origin significant overlaps with each other.  To summarize, both the distribution of rotational temperatures and the observations of resolved CO lines show multiple components of CO gas; however, the discrete origin of CO gas produces a rather smooth distribution of correlation strengths, suggesting that each origin contributes to a wide range of $J$-levels, and the dominance of each origin is smoothly varying across the CO ladder.  This illustrates the importance of spectrally resolved data to find the ``hidden'' components.

\subsubsection{A Single CO Line as a Probe of the Total CO Luminosity}
\label{sec:lco_proxy}
\begin{figure*}[htbp!]
    \centering
    \includegraphics[width=0.45\textwidth]{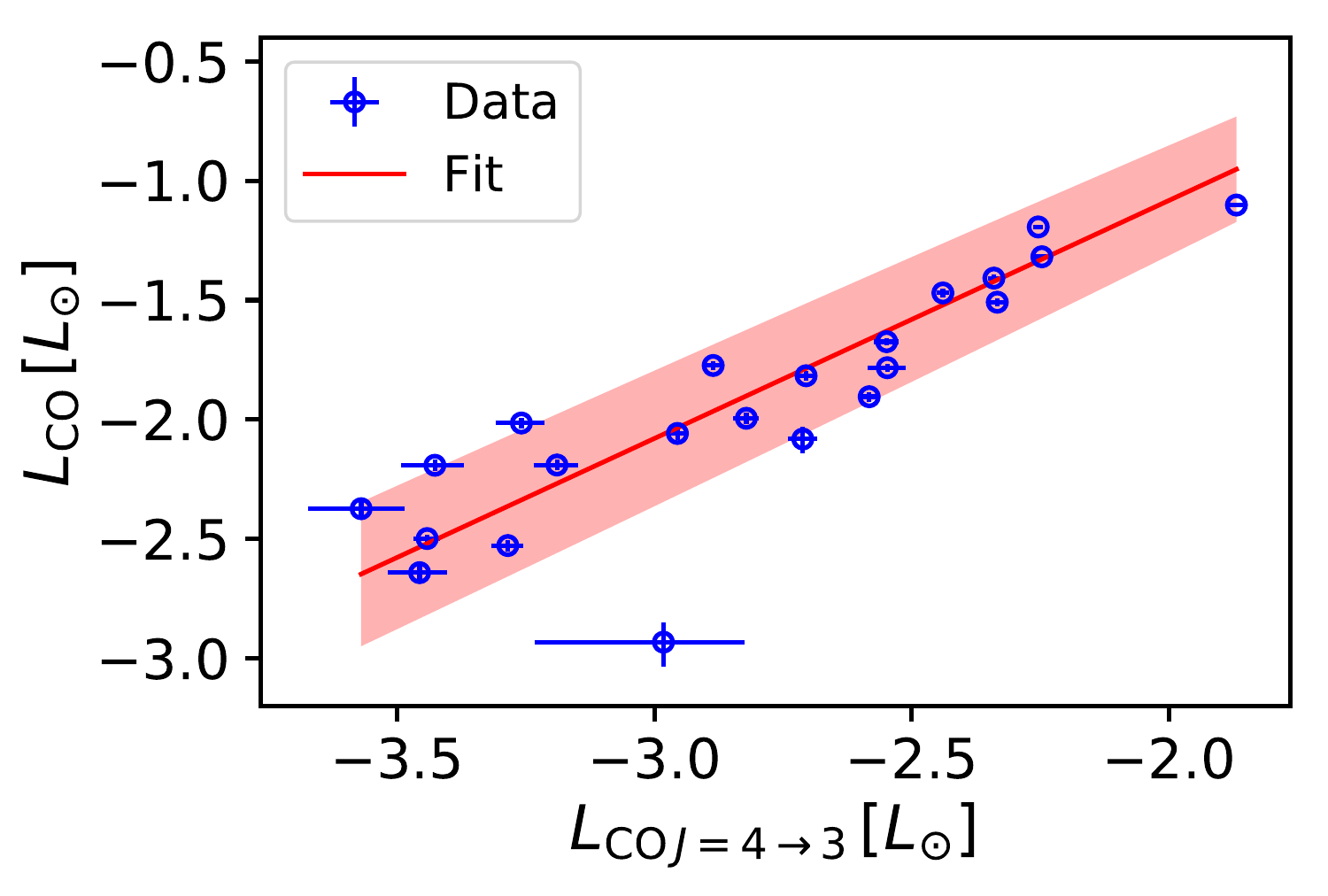}
    \includegraphics[width=0.45\textwidth]{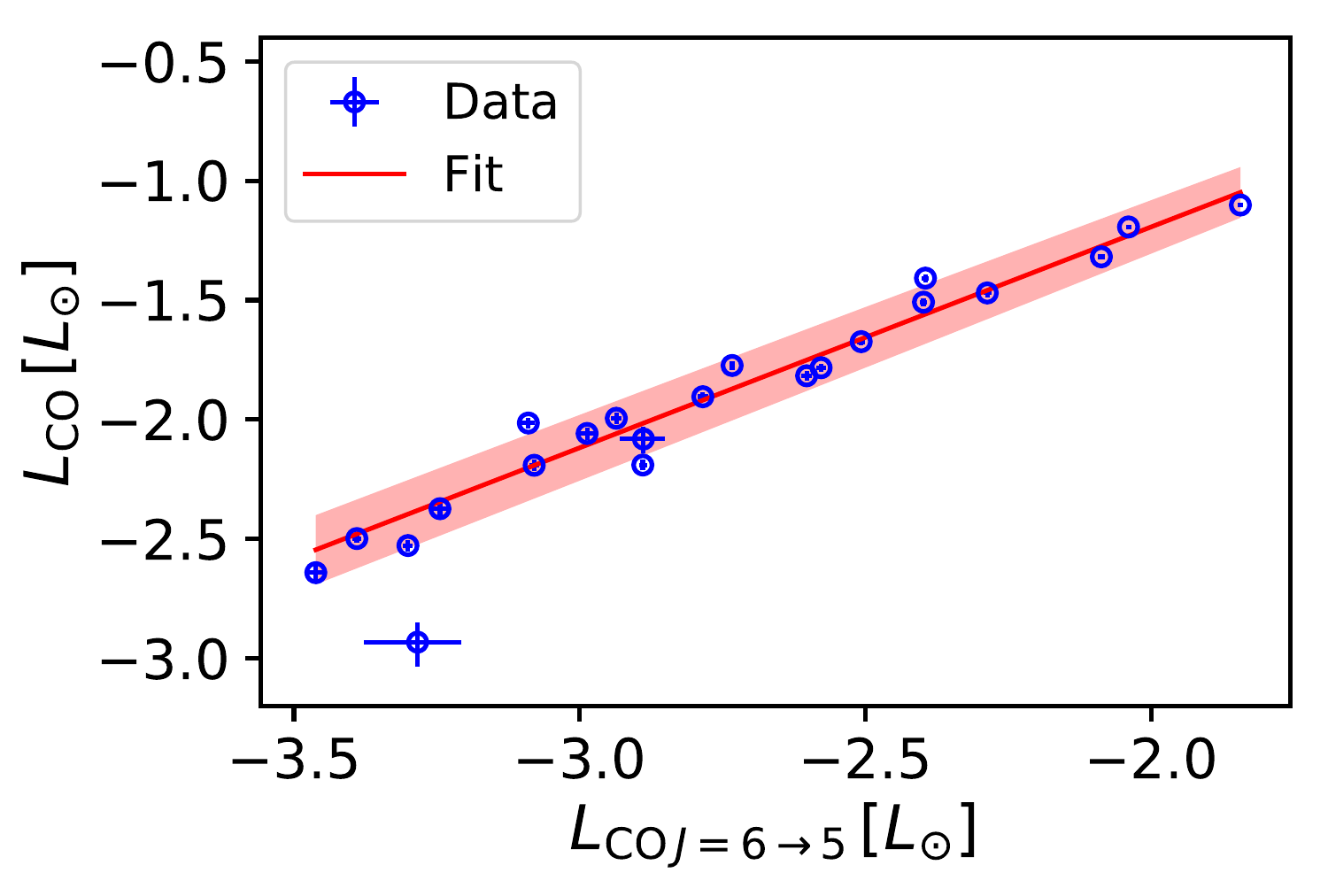}
    \caption{The correlation between the total CO luminosity and the luminosity of CO~\jj{4}{3} (left) and CO~\jj{6}{5} (right) with a fitted linear relation.  The fitted lines can be described as log($L_{{\rm CO}~J=4\rightarrow3\,[L_{\odot}]}$) = a$\times {\rm log}(L_{\rm CO}\,[L_{\odot}])$+b, where a = 1.00$\pm$0.07 and b = 0.91$\pm$0.19;
    and log($L_{{\rm CO}~J=6\rightarrow5\,[L_{\odot}]}$) = a$\times {\rm log}(L_{\rm CO}\,[L_{\odot}])$+b, where a = 0.93$\pm$0.04 and b = 0.66$\pm$0.09.}
    \label{fig:co65_co}
\end{figure*}

We explore the potential to infer $L_{\rm CO}$ with a single low-$J$ CO line.  We only focus on CO~\jj{6}{5} and \jj{4}{3} lines for their accessibility from ground-based observations.  The optical depth correction applies to the luminosities of CO~\jj{6}{5} and CO~\jj{4}{3} as well.  Both lines show strong correlation with $L_{\rm CO}$, with Spearman's $\rho$ of 0.878 and 0.926 for CO~\jj{4}{3} and \jj{6}{5}, respectively.  The data are more scattered for CO~\jj{4}{3} at the lower $L_{\rm CO}$ regime, probably due to increased contribution from the envelope.
The different spatial extent of CO~\jj{3}{2} and \jj{6}{5} lines \citep{2014AA...568A.125S} also suggests different combinations of origins for the CO~\jj{4}{3} and \jj{6}{5} emission.
We fit a power-law for the relations between the $L_{\rm CO}$ and both the luminosity of CO~\jj{4}{3} and CO~\jj{6}{5}, with the orthogonal distance regression offered in \texttt{SciPy} (Figure~\ref{fig:co65_co}).
The fitted relation can be described as
\begin{equation}
    log(L_{\rm CO}\,[L_{\odot}]) = a\times {\rm log}(L_{ \rm CO}\,[L_{\odot}])+b.
\end{equation}
For CO~\jj{6}{5}, a = 0.93$\pm$0.04 and b = 0.66$\pm$0.09; for CO~\jj{4}{3}, a = 1.00$\pm$0.07 and b = 0.91$\pm$0.19.
The tight relation between the $L_{\rm CO}$ and the luminosity of CO~\jj{6}{5} provides a pathway to assess the total CO luminosity from the ground-based observations.

\subsubsection{Comparison with FU Orionis Objects, T Tauri stars, and Herbig Ae/Be stars}
\label{sec:herbig_fuor_ttauri}
In principle, the molecular gas eventually infalls onto the protostars or dissipates into the environment, suggesting that the amount of CO should be lower for T Tauri stars and Herbig Ae/Be stars compared to embedded protostars, such as the COPS sources.  The FU Orionis objects (FUors) are often considered as the protostars which have undergone a sudden rise in accretion rate \citep{herbig77}, resulting in a rapid increase in source bolometric luminosity followed by a slow decay over 10--100 years \citep[e.g.][]{bell94,2016ApJ...832....4G}.  See \citet{audard14} for a recent review.  If the FUors just experienced an outburst of energy, their molecular gas may have been heated up, leading to an increase of the 400--500~K component in the CO rotational diagram.
We utilize the PACS spectra from the CDF archive to examine if the differences of Herbig Ae/Be stars, T~Tauri stars, and FUors appeared in the rotational diagrams of CO.  Table~\ref{tbl:source_other} shows the lists of sources.

\begin{table*}[htbp!]
    \centering
    \small
    \caption{List of Herbig Ae/Be stars, FUors, and T Tauri stars}
    \label{tbl:source_other}
    \begin{tabular}{rl}
        \toprule
        Herbig Ae/Be & AB~Aur, AS~205, HD~100453, HD~100546, HD~104237, HD~135344B, HD~139614, HD~141569 \\
        & HD~142527, HD~142666, HD~144432, HD~144668, HD~150193, HD~163296, HD~169142 \\
        & HD~179218, HD~203024, HD~245906, HD~35187, HD~36112, HD~38120, HD~50138 \\
        & HD~97048, HD~98922 \\
        \hline
        FUors & HBC~722, V1735~Cyg, V1057~Cyg, V1331~Cyg, V1515~Cyg, FU~Ori \\
        \hline
        T Tauri & S~CrA, HT~Lup, RU~Lup, RY~Lup, SR~21 \\
        \bottomrule
    \end{tabular}
\end{table*}

\begin{figure*}[htbp!]
    \centering
    \includegraphics[width=0.9\textwidth]{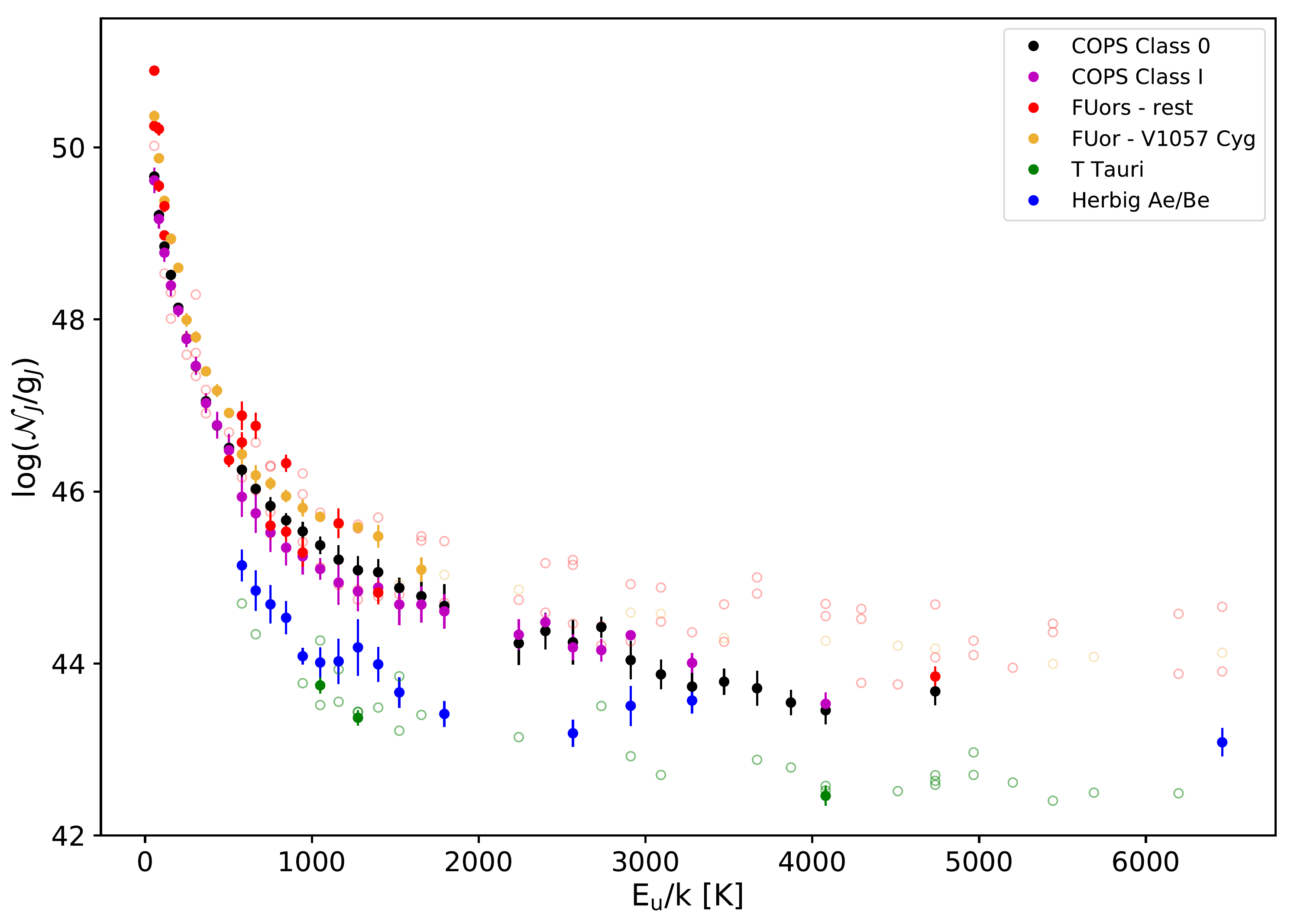}
    \caption{The CO rotational diagram of the COPS sources (black and magenta for the Class 0 and the Class I sources classified by $T_{\rm bol}$) and other types of sources from the CDF archive \citep{green16a}, including FUors (orange and red for V1057~Cyg and other FUors, respectively), T Tauri stars (green), and Herbig Ae/Be stars (blue).  The data of the COPS sources and the Herbig Ae/Be stars are shown as averaged by their upper energy, while the data are all shown for the other sources.  The upper limits are shown as open circles.}
    \label{fig:co_rot_comparison}
\end{figure*}

Figure~\ref{fig:co_rot_comparison} shows the CO rotational diagram of different types of sources as rotational diagrams.  The data for Class 0 and I sources are averaged separately.  The CO rotational diagrams of Class 0 and I sources have similar characteristics.  There are few detections of CO for FUors except for V1057~Cyg.  We found that the CO rotational diagram of V1057~Cyg is indistinguishable from the COPS sources, although there is a hint that the data of V1057~Cyg lie at the higher-end of the region populated by the embedded sources, especially at the high upper energy regime.  The CO line profiles of FUors are in fact similar to the line profiles found toward embedded protostars \citep{green13c}, which supports the similarity of the rotational diagrams between V1057~Cyg and embedded protostars.

The CO rotational diagrams of the Herbig Ae/Be stars are consistently lower than the data of the COPS sources by about 1 in log($\mathcal{N}_{J}/{\rm g}_{J}$), and they show a much more irregular decrease with $E_{\rm u}/k$ compared to the COPS protostars.  However, the detailed study of two Herbig Ae/Be stars shows a much smoother decrease similar to the protostars \citep{2017AA...605A..62J}.  Double-peaked line profiles were observed in CO~\jj{10}{9} and CO~\jj{16}{15} for disk sources \citep{2013AA...559A..77F}, which are quite different from the line profiles seen by \citet{2017arXiv170510269K}, confirming a disk origin for the CO emission.  However, how the different origins contribute to the shape of rotational diagrams is unclear without further investigation, but it certainly makes the difference of rotational diagram expected.  There are only a few detections of CO toward the T Tauri stars, which lie at the bottom-end of the diagram among the lower-end of the data from the Herbig Ae/Be stars.
In summary, V1057~Cyg has similar CO rotational diagrams as the embedded protostars, whereas the Herbig Ae/Be stars and T~Tauri stars have lower values in $\mathcal(N_{J})/g_{J}$ but similar shapes to the diagrams of embedded sources.  The similarity between embedded protostars and FUors and difference between embedded protostars and Herbig Ae/Be stars and T~Tauri stars are also seen in their line profiles.

\section{Conclusions}

We have presented a statistical analysis of the full sample of 26 Class 0+I protostars from the COPS-SPIRE Open Time Program using \textit{Herschel}-SPIRE 200--670 \micron\ spectroscopy, utilizing improvements in calibration and sensitivity from the data pipeline over the mission lifetime.

\begin{itemize}

\item We present 1--1000 \micron\ SEDs of 26 COPS sources with excellent spectral coverage in submillimeter wavelengths.  We update the reduction method used in the CDF archive \citep{green16a} to extract the 1D PACS spectra that are consistent with the SPIRE 1D spectra, which have less than 5\%\ difference of flux at the conjunction between the spectra of PACS and SPIRE for all sources, except for TMC~1, WL~12, L1014, and IRAS~03301+3111 due to excessive background emission in SPIRE.  We further investigated the change of \lbol, \tbol, and \lsmm\ because of the addition of the SPIRE spectra.  With \textit{Herschel} data, the \lbol\ increases by 50\%\ on average compared to the pre-\textit{Herschel} observations, while the \tbol\ decreases by 10\%\ on average.  (Section~\ref{sec:pre_post_herschel} and \ref{sec:seds})

\item All spectra were processed with an automatic line fitting pipeline to extract emission lines, producing the line-free continuum (Figure~\ref{fig:sed1}) and continuum-free spectra (Figure~\ref{fig:spire_1d}).  The line fitting result is provided as a machine readable table (Table~\ref{table:report}).  We found that the SPIRE photometric flux densities decrease by only $\sim$1\%\ after removing the emission lines.  (Section~\ref{sec:fitting_results} and \ref{sec:effect_photometry})

\item The $^{12}$CO lines typically have high optical depth at low-$J$, which affects the total CO luminosity as well as the comparison with the emission of other species.  With the emission of $^{13}$CO detected in a few sources, we characterize the optical depth of $^{12}$CO and discuss the uncertainties due to the lack of spectral resolution and sensitivity.  We fitted the optical depth as a function of the upper energy levels (Equation~\ref{eq:fitted_tau}), and found that the $^{12}$CO lines remain optically thick until \Jup~$>$~13.  (Section~\ref{sec:co_opt_depth})

\item We fitted multiple rotational temperatures for each source with the correction of optical depth.  We adopted the conventional break points to separate the CO ladder, and produce four components of rotational temperature, $<$100~K, $\sim$100~K, $\sim$200--300~K, and $\gtrsim$500~K, which is similar to the scenario concluded by many studies \citep[e.g.][]{yildiz13,green13b,manoj12,karska14,2015AA...578A..20M}.
However, the separation between components becomes indistinct when the break points remain flexible during the fitting of rotational temperature.  The distribution of rotational temperatures (Figure~\ref{fig:co_rot_distribution}) shows two populations, the dominant cold components (50--120~K) and a secondary population centered at $\sim$400~K, which trace the entrained gas in the outflows and possibly the cavity shocks/disk wind, respectively.  (Section~\ref{sec:co_rot})

\item We calculate the beam filling factor for the sources where we detect $^{13}$CO emission, which constrains the optical depth directly.  The beam filling factors range from 0.035 to 0.53, which are lower limit of the filling factor due to the uncertainties on the line width and optical depth.  A clumpy nature of the CO gas at embedded protostars learned from ground-based observations with higher spatial resolution (9\arcsec\ beam) \citep{2013ApJ...774...39A,2015ApJ...805..186L,2015AA...576A.109Y}, suggesting a smaller beam filling factor, which is consistent with our result.  (Section~\ref{sec:filling_factor})

\item We constrain the correlations of all pairs of lines drawn from the line fitting results with the generalized Spearman's rank correlation.  The distribution of correlation strength (Figure~\ref{fig:all.spr.rho}) show systematic offsets of correlation strength from several groups of lines, where the lines of each group are observed with the same module/instrument, suggesting that the bias is due to the instrumental effect.  Therefore, any comparison of the lines observed in different modules needs to be done with caution.  (Section~\ref{sec:correlation})

\item Within each module, the correlations between each CO line pair are smoothly decreasing as the difference between $J$-level between two lines increases.
The emission of $^{13}$CO is found to have weak correlations with $^{12}$CO, and stronger correlations with \CI~$^{3}P_{2}\rightarrow^{3}P_{1}$.
Both $^{13}$CO and \CI\ were associated with the UV-heated outflow cavity wall, which explains the correlation.
The \water\ lines show significant correlations with CO and OH.  The water lines correlate better with the CO lines ranging from \Jup=10--15 in the middle of the range of $J$-levels shown in Figure~\ref{fig:water_1d_correlation}.
The \OI~$^{3}P_{1}\rightarrow^{3}P_{2}$ at 63~$\mu$m has weak correlations with low-$J$ CO lines observed with SPIRE, and shows no correlation with \CI.  (Section~\ref{sec:co_cor} to \ref{sec:atomic_cor})

\item The spatial distribution of CO emission shows an extended feature in low-$J$ CO lines.  We develop a method to project the 2D distribution into an 1D profile as a function of the azimuthal angles from the center of the sources.  The 1D profile shows the ratios of the flux at outer spaxels to the flux at the central spaxel.  The 1D profiles show consistent morphology with the spatially resolved map.  The bipolar CO distribution indeed has an outflow-related origin.  The bipolar distribution of CO emission is more significant for lower-$J$ CO lines, and typically show a separation of 180 degree between the two peaks.

We use the difference of the maximum and the minimum values of the extracted 1D profiles as a probe of the extent of bipolar feature.  The averaged peak-to-valley differences show a general decline as the $J$-level increases, suggesting a less predominant bipolar feature at higher-\Jup\ CO lines.  (Section~\ref{sec:lineextent})

\item The classifications using \lbol/\lsmm, \tbol, and the spectral index at near-infrared ($\alpha_{\rm NIR}$) as the indicators all conclude that our sample consists of Class 0+I protostars.  The classifications using \tbol\ and \lbol/\lsmm\ are not very consistent in separating 0 from I: 10 of 26 sources have different classifications under these two methods.  (Section~\ref{sec:sed_classification})

\item The origin of the CO emission can be constrained with its spatial extent, rotational temperature, and correlations with other CO lines.  We show that the extended CO emission traces the outflows.  Also, two populations of the CO rotational temperatures are found in the distribution of rotational temperatures, which is consistent with multiple origins of CO emission inferred from the velocity-resolved line profiles \citep{kristensen12,yildiz13,2013AA...553A.125S,2017arXiv170510269K}.  The cold population ($\lesssim$100~K) indicates the entrained gas in the outflows.  While the origin of the broad component of the resolved CO line profiles is consistent with the warm population ($\sim$400~K), we do not find the hot component ($\sim$700~K) in the distribution of rotational temperatures.  The distributions of correlation strength do not exhibit multiple distinct patterns, suggesting that either the CO emission comes from numerous components, or the few mechanisms of CO excitation contribute to a wide range of CO ladder and well-overlap with each other.  The kinematics of the resolved line profiles rule out the former scenario \citep{2017arXiv170510269K}.  Thus, the discrete origins of CO emission contribute to a wide range of CO ladder, and their contributions significantly overlap with each other on the CO ladder.  (Section~\ref{sec:co_origin})

\item The comparison of the CO rotational diagrams of different types of sources, such as embedded protostars, FUors, T~Tauri stars, and Herbig Ae/Be stars, collected from the CDF archive \citep{green16a}, shows that the rotational diagrams of one FUor (V1057~Cyg) is indistinguishable from that of an emvedded protostar, except for the higher $\mathcal{N_{J}}/g_{J}$ at high upper energy transitions(Figure~\ref{fig:co_rot_comparison}).
The other FUors exhibit few CO detections.  The data from the embedded sources form a coherent pattern in the rotational diagram, where the data of the Herbig Ae/Be stars lie at the lower end of log($\mathcal{N}_{J}/g_{J}$) compared to the embedded sources.  Only a few detections are found toward the T Tauri stars, which have weaker emission than the Herbig Ae/Be stars.  (Section~\ref{sec:herbig_fuor_ttauri})

\end{itemize}

\acknowledgements
Support for this work, part of the \textit{Herschel} Open Time Key Project Program, was provided by NASA through an award issued by the Jet Propulsion Laboratory, California Institute of Technology.  Y.-L.Y. acknowledges the support of the University Graduate Continuing Fellowship from University of Texas at Austin.  Y.-L.Y. and J.G. acknowledge support in part from a 2016 DDRF award at the Space Telescope Science Institute (STScI/AURA) and the support from NASA Herschel Science Center Cycle 2 grants.  Some financial support for this work was provided by NASA through awards \#\ SOF 04-0073 and 04-0146 issued by USRA. SOFIA is jointly operated by the Universities Space Research Association, Inc. (USRA), under NASA contract NAS2-97001, and the Deutsches SOFIA Institut (DSI) under DLR contract 50 OK 0901 to the University of Stuttgart.  JEL is supported by the Basic Science Research Program through the National Research Foundation of Korea (grant No. NRF-2018R1A2B6003423) and the Korea Astronomy and Space Science Institute under the R\&D program supervised by the Ministry of Science, ICT and Future Planning.  LEK and JKJ acknowledge support from the European Research Council (ERC) under the European Union???s Horizon 2020 research and innovation programme (grant agreement No 646908) through ERC Consolidator Grant ``S4F''.  The research at the Centre for Star and Planet Formation is supported by the Danish National Research Foundation and the University of Copenhagen's programme of excellence.  JCM acknowledges support from the European Research Council under the European Community???s Horizon 2020 framework program (2014-2020) via the ERC Consolidator grant `From Cloud to Star Formation (CSF)' (project number 648505).  The research of OD was supported by the Austrian Research Foundation Agency (FFG) under the Austrian Space Applications Program (ASAP) projects JetPro* and PROTEUS (FFG-854025, FFG-866005).  AK acknowledges support from the Polish National Science Center grants 2013/11/N/ST9/00400 and 2016/21/D/ST9/01098.  Y.-L.Y. would like to thank Trey Heinen for assistance with PACS and SPIRE data reduction, and acknowledge numerous helpful discussions with Emma Yu, Isa Oliveira, Manoj Puravankara, Dan Watson, Amy Stutz, John Lacy, and Dan Jaffe.  This research made use of Astropy, a community-developed core Python package for Astronomy (Astropy Collaboration, 2013).  This research has made use of the VizieR catalogue access tool, CDS, Strasbourg, France. The original description of the VizieR service was published in A\&AS 143, 23.  This publication makes use of data products from the Two Micron All Sky Survey, which is a joint project of the University of Massachusetts and the Infrared Processing and Analysis Center/California Institute of Technology, funded by the National Aeronautics and Space Administration and the National Science Foundation.

\facilities{\textit{Spitzer} (IRS), \textit{Herschel} (PACS), \textit{Herschel} (SPIRE)}

\software{astropy \citep{2013AA...558A..33A}, scipy \citep{scipy}, seaborn \citep{michael_waskom_2014_12710}, HIPE v.14 \citep{ott10}, ASURV Rev. 1.2 \citep{isobe90,lavalley92}}
\textbf{}

\begin{turnpage}
\begin{deluxetable*}{l ccccccccccccccccccccccccccc}
\tabletypesize{\scriptsize}
\tablecaption{Line Detection Summary \label{detection}}
\tablewidth{0pt}
\setlength\tabcolsep{3pt}

\tablehead{
	 \colhead{Source} &
	 \colhead{\begin{turn}{90}{$^{13}$CO~\jj{5}{4}}\end{turn}} &
	 \colhead{\begin{turn}{90}{$^{13}$CO~\jj{6}{5}}\end{turn}} &
	 \colhead{\begin{turn}{90}{$^{13}$CO~\jj{7}{6}}\end{turn}} &
	 \colhead{\begin{turn}{90}{$^{13}$CO~\jj{8}{7}}\end{turn}} &
	 \colhead{\begin{turn}{90}{$^{13}$CO~\jj{9}{8}}\end{turn}} &
	 \colhead{\begin{turn}{90}{CO~\jj{4}{3}}\end{turn}} &
	 \colhead{\begin{turn}{90}{CO~\jj{5}{4}}\end{turn}} &
	 \colhead{\begin{turn}{90}{CO~\jj{6}{5}}\end{turn}} &
	 \colhead{\begin{turn}{90}{CO~\jj{7}{6}}\end{turn}} &
	 \colhead{\begin{turn}{90}{CO~\jj{8}{7}}\end{turn}} &
	 \colhead{\begin{turn}{90}{CO~\jj{9}{8}}\end{turn}} &
	 \colhead{\begin{turn}{90}{CO~\jj{10}{9}}\end{turn}} &
	 \colhead{\begin{turn}{90}{CO~\jj{11}{10}}\end{turn}} &
	 \colhead{\begin{turn}{90}{CO~\jj{12}{11}}\end{turn}} &
	 \colhead{\begin{turn}{90}{CO~\jj{13}{12}}\end{turn}} &
	 \colhead{\begin{turn}{90}{\CI~$^{3}P_{1}\rightarrow ^{3}P_{0}$}\end{turn}} &
	 \colhead{\begin{turn}{90}{\CI~$^{3}P_{2}\rightarrow ^{3}P_{1}$}\end{turn}} &
	 \colhead{\begin{turn}{90}{HCO$^{+}$~\jj{6}{5}}\end{turn}} &
	 \colhead{\begin{turn}{90}{HCO$^{+}$~\jj{7}{6}}\end{turn}} &
	 \colhead{\begin{turn}{90}{HCO$^{+}$~\jj{8}{7}}\end{turn}} &
	 \colhead{\begin{turn}{90}{o-H$_{2}$O$~1_{10} \rightarrow 1_{01}$}\end{turn}} &
	 \colhead{\begin{turn}{90}{o-H$_{2}$O$~3_{12} \rightarrow 2_{21}$}\end{turn}} &
	 \colhead{\begin{turn}{90}{o-H$_{2}$O$~3_{12} \rightarrow 3_{03}$}\end{turn}} &
	 \colhead{\begin{turn}{90}{o-H$_{2}$O$~3_{21} \rightarrow 3_{12}$}\end{turn}} &
	 \colhead{\begin{turn}{90}{p-H$_{2}$O$~1_{11} \rightarrow 0_{00}$}\end{turn}} &
	 \colhead{\begin{turn}{90}{p-H$_{2}$O$~2_{02} \rightarrow 1_{11}$}\end{turn}} &
	 \colhead{\begin{turn}{90}{p-H$_{2}$O$~2_{11} \rightarrow 2_{02}$}\end{turn}} }
\startdata
B1-a & \nodata  & \nodata  & \nodata  & \nodata  & \nodata  & Y  & Y  & Y  & Y  & Y  & Y  & Y  & Y  & Y  & Y  & Y  & Y  & \nodata  & \nodata  & \nodata  & \nodata  & Y  & \nodata  & \nodata  & Y  & \nodata  & \nodata  \\
B1-c & \nodata  & \nodata  & \nodata  & \nodata  & \nodata  & Y  & Y  & Y  & Y  & Y  & Y  & Y  & Y  & Y  & Y  & Y  & Y  & \nodata  & \nodata  & \nodata  & Y  & Y  & Y  & Y  & Y  & Y  & Y  \\
B335 & \nodata  & \nodata  & \nodata  & \nodata  & \nodata  & Y  & Y  & Y  & Y  & Y  & Y  & Y  & Y  & Y  & Y  & \nodata  & \nodata  & \nodata  & \nodata  & \nodata  & \nodata  & \nodata  & \nodata  & \nodata  & Y  & Y  & \nodata  \\
BHR~71 & \nodata  & \nodata  & \nodata  & \nodata  & \nodata  & Y  & Y  & Y  & Y  & Y  & Y  & Y  & Y  & Y  & Y  & \nodata  & \nodata  & \nodata  & \nodata  & \nodata  & Y  & Y  & \nodata  & \nodata  & Y  & Y  & Y  \\
Ced110~IRS4 & Y  & Y  & \nodata  & \nodata  & \nodata  & Y  & Y  & Y  & Y  & Y  & Y  & Y  & Y  & Y  & Y  & Y  & Y  & \nodata  & \nodata  & \nodata  & \nodata  & \nodata  & \nodata  & \nodata  & \nodata  & \nodata  & \nodata  \\
DK~Cha & Y  & \nodata  & \nodata  & \nodata  & \nodata  & Y  & Y  & Y  & Y  & Y  & Y  & Y  & Y  & Y  & Y  & \nodata  & \nodata  & \nodata  & \nodata  & \nodata  & \nodata  & Y  & \nodata  & \nodata  & \nodata  & Y  & \nodata  \\
GSS~30~IRS1 & Y  & Y  & Y  & \nodata  & Y  & Y  & Y  & Y  & Y  & Y  & Y  & Y  & Y  & Y  & Y  & Y  & Y  & \nodata  & Y  & \nodata  & \nodata  & Y  & \nodata  & \nodata  & Y  & Y  & \nodata  \\
HH~46 & Y  & Y  & \nodata  & \nodata  & \nodata  & Y  & Y  & Y  & Y  & Y  & Y  & Y  & Y  & Y  & Y  & Y  & Y  & \nodata  & \nodata  & \nodata  & \nodata  & \nodata  & \nodata  & \nodata  & \nodata  & \nodata  & \nodata  \\
IRAS~03245+3002 & \nodata  & \nodata  & \nodata  & \nodata  & \nodata  & Y  & Y  & Y  & Y  & Y  & Y  & Y  & Y  & Y  & Y  & Y  & Y  & \nodata  & \nodata  & \nodata  & \nodata  & Y  & \nodata  & \nodata  & \nodata  & \nodata  & \nodata  \\
IRAS~03301+3111 & \nodata  & \nodata  & \nodata  & \nodata  & \nodata  & Y  & Y  & Y  & Y  & \nodata  & Y  & Y  & Y  & Y  & Y  & Y  & Y  & \nodata  & \nodata  & \nodata  & \nodata  & \nodata  & \nodata  & \nodata  & \nodata  & \nodata  & \nodata  \\
IRAS~15398$-$3359 & \nodata  & \nodata  & \nodata  & \nodata  & \nodata  & Y  & Y  & Y  & Y  & Y  & Y  & Y  & Y  & Y  & Y  & \nodata  & \nodata  & \nodata  & \nodata  & \nodata  & \nodata  & \nodata  & \nodata  & \nodata  & Y  & Y  & \nodata  \\
L1014 & \nodata  & \nodata  & \nodata  & \nodata  & \nodata  & \nodata  & \nodata  & \nodata  & \nodata  & \nodata  & \nodata  & \nodata  & \nodata  & \nodata  & \nodata  & Y  & \nodata  & \nodata  & \nodata  & \nodata  & \nodata  & \nodata  & \nodata  & \nodata  & \nodata  & \nodata  & \nodata  \\
L1157 & \nodata  & \nodata  & \nodata  & \nodata  & \nodata  & Y  & Y  & Y  & Y  & Y  & Y  & Y  & Y  & Y  & Y  & \nodata  & \nodata  & \nodata  & \nodata  & \nodata  & \nodata  & Y  & Y  & \nodata  & Y  & Y  & Y  \\
L1455~IRS3 & \nodata  & \nodata  & \nodata  & \nodata  & \nodata  & \nodata  & \nodata  & Y  & Y  & \nodata  & \nodata  & \nodata  & \nodata  & \nodata  & Y  & \nodata  & \nodata  & \nodata  & \nodata  & \nodata  & \nodata  & \nodata  & \nodata  & \nodata  & \nodata  & \nodata  & \nodata  \\
L1551~IRS5 & Y  & \nodata  & \nodata  & \nodata  & Y  & Y  & Y  & Y  & Y  & Y  & Y  & Y  & Y  & Y  & Y  & Y  & \nodata  & \nodata  & \nodata  & \nodata  & \nodata  & \nodata  & \nodata  & \nodata  & \nodata  & \nodata  & \nodata  \\
L483 & \nodata  & \nodata  & \nodata  & \nodata  & \nodata  & Y  & Y  & Y  & Y  & Y  & Y  & Y  & Y  & Y  & Y  & Y  & \nodata  & \nodata  & \nodata  & \nodata  & \nodata  & Y  & \nodata  & \nodata  & Y  & Y  & \nodata  \\
L723~MM & \nodata  & \nodata  & \nodata  & \nodata  & \nodata  & Y  & Y  & Y  & Y  & Y  & Y  & Y  & Y  & Y  & Y  & Y  & \nodata  & \nodata  & \nodata  & \nodata  & \nodata  & \nodata  & \nodata  & \nodata  & \nodata  & Y  & \nodata  \\
RCrA~IRS5A & Y  & Y  & \nodata  & \nodata  & \nodata  & Y  & Y  & Y  & Y  & Y  & Y  & Y  & Y  & Y  & Y  & Y  & Y  & \nodata  & \nodata  & \nodata  & \nodata  & \nodata  & \nodata  & \nodata  & Y  & \nodata  & \nodata  \\
RCrA~IRS7B & Y  & Y  & Y  & Y  & Y  & Y  & Y  & Y  & Y  & Y  & Y  & Y  & Y  & Y  & Y  & Y  & Y  & Y  & Y  & Y  & \nodata  & Y  & \nodata  & \nodata  & \nodata  & Y  & \nodata  \\
RCrA~IRS7C & Y  & Y  & Y  & Y  & Y  & Y  & Y  & Y  & Y  & Y  & Y  & Y  & Y  & Y  & Y  & Y  & Y  & Y  & Y  & Y  & \nodata  & \nodata  & \nodata  & \nodata  & Y  & Y  & \nodata  \\
RNO~91 & \nodata  & \nodata  & \nodata  & \nodata  & \nodata  & Y  & Y  & Y  & Y  & Y  & Y  & Y  & Y  & Y  & Y  & Y  & Y  & \nodata  & \nodata  & \nodata  & \nodata  & \nodata  & \nodata  & \nodata  & \nodata  & \nodata  & \nodata  \\
TMC~1 & \nodata  & \nodata  & \nodata  & \nodata  & \nodata  & Y  & Y  & Y  & Y  & Y  & Y  & Y  & Y  & Y  & Y  & \nodata  & \nodata  & \nodata  & \nodata  & \nodata  & \nodata  & \nodata  & \nodata  & \nodata  & \nodata  & \nodata  & \nodata  \\
TMC~1A & \nodata  & \nodata  & \nodata  & \nodata  & \nodata  & Y  & Y  & Y  & Y  & Y  & Y  & Y  & Y  & Y  & Y  & \nodata  & Y  & \nodata  & \nodata  & \nodata  & \nodata  & \nodata  & \nodata  & \nodata  & \nodata  & \nodata  & \nodata  \\
TMR~1 & \nodata  & Y  & \nodata  & \nodata  & \nodata  & Y  & Y  & Y  & Y  & Y  & Y  & Y  & Y  & Y  & Y  & Y  & Y  & \nodata  & \nodata  & \nodata  & \nodata  & Y  & \nodata  & \nodata  & Y  & Y  & \nodata  \\
VLA~1623$-$243 & Y  & Y  & \nodata  & \nodata  & Y  & Y  & Y  & Y  & Y  & Y  & Y  & Y  & Y  & Y  & Y  & Y  & Y  & \nodata  & \nodata  & \nodata  & \nodata  & \nodata  & \nodata  & \nodata  & \nodata  & \nodata  & \nodata  \\
WL~12 & Y  & Y  & \nodata  & \nodata  & \nodata  & Y  & Y  & Y  & Y  & Y  & Y  & Y  & Y  & Y  & Y  & Y  & Y  & \nodata  & \nodata  & \nodata  & \nodata  & \nodata  & \nodata  & \nodata  & \nodata  & \nodata  & \nodata  \\
\enddata
\end{deluxetable*}
\end{turnpage}

\clearpage
\appendix

\section{List of OBSID of photometry}
\label{photref}

Table~\ref{obsid_phot} lists the OBSIDs used for calculating the PACS and SPIRE photometry in this study.

\begin{deluxetable*}{r l | l}
\tabletypesize{\scriptsize}
\tablecaption{OBSID of Photometry \label{obsid_phot}}
\tablewidth{\textwidth}
\tablehead{
    \colhead{Source} & \colhead{PACS} & \colhead{SPIRE} }
\startdata
IRAS 03245+3002  & 1342227103, 1342227104 & 1342190327, 1342190326 \\
L1455~IRS3       & 1342227103, 1342227104 & 1342190327, 1342190326 \\
IRAS 03301+3111  & 1342227103, 1342227104 & 1342190327, 1342190326 \\
B1-a             & 1342227103, 1342227104 & 1342190327, 1342190326 \\
B1-c             & 1342267246, 1342267247 & 1342190327, 1342190326 \\
L1551~IRS5       & 1342202251 & 1342202250, 1342202251 \\
TMR 1            & 1342228175, 1342228174 & 1342202252, 1342202253 \\
TMC 1A           & 1342202252 & 1342202252, 1342202253 \\
TMC 1            & 1342202252 & 1342202252, 1342202253 \\
HH~46            & \nodata & \nodata \\
Ced110~IRS4      & 1342223480, 1342224782, 1342224783 & 1342213179, 1342213178 \\
BHR 71           & 1342224922, 1342224925, 1342224924, 1342224923 & 1342226633 \\
DK Cha           & 1342212709, 1342212708, 1342213180 & 1342213180, 1342213181 \\
IRAS 15398--3359 & 1342226705 & 1342213182, 1342213183 \\
GSS~30~IRS1      & 1342227148, 1342227149, 1342205093, 1342205094 & 1342203074, 1342205093, 1342205094 \\
VLA~1623$-$243     & 1342205093, 1342205094 & 1342203074, 1342205093, 1342205094 \\
WL~12            & 1342238817, 1342238816 & 1342205093, 1342205094 \\
RNO~91           & 1342263844, 1342263845 & 1342263844, 1342263845 \\
L483             & 1342228397, 1342228398, 1342228395, 1342228396 & 1342229186 \\
RCrA~IRS5A       & 1342267429, 1342267427, 1342267428, 1342267430, 1342242076 & 1342206677, 1342206678, 1342216002 \\
                 & 1342241402, 1342241519, 1342241403, 1342242555, 1342242077 & \\
                 & 1342241314, 1342241520, 1342241313, 1342242554 &  \\
RCrA~IRS7B/C       & 1342184510, 1342184511 & 1342206677, 1342206678, 1342216002 \\
L723~MM          & 1342231917, 1342231918 & 1342229605 \\
B335             & 1342196030, 1342196031 & 1342192685 \\
L1157            & 1342224778, 1342224779, 1342189845 & 1342189844, 1342189843 \\
L1014            & 1342225450, 1342225449 & 1342219974, 1342220631 \\
\enddata
\tablecomments{We exclude HH~100 from the calibration with photometry, because the observation of HH~100 is heavily contaminated by a nearby source, RCrA~IRS7B.}
\end{deluxetable*}

\section{The 1D SPIRE spectra for the COPS-SPIRE sources}
\label{sec:a_spire_1d}

\begin{figure*}[htbp!]
    \includegraphics[width=\textwidth]{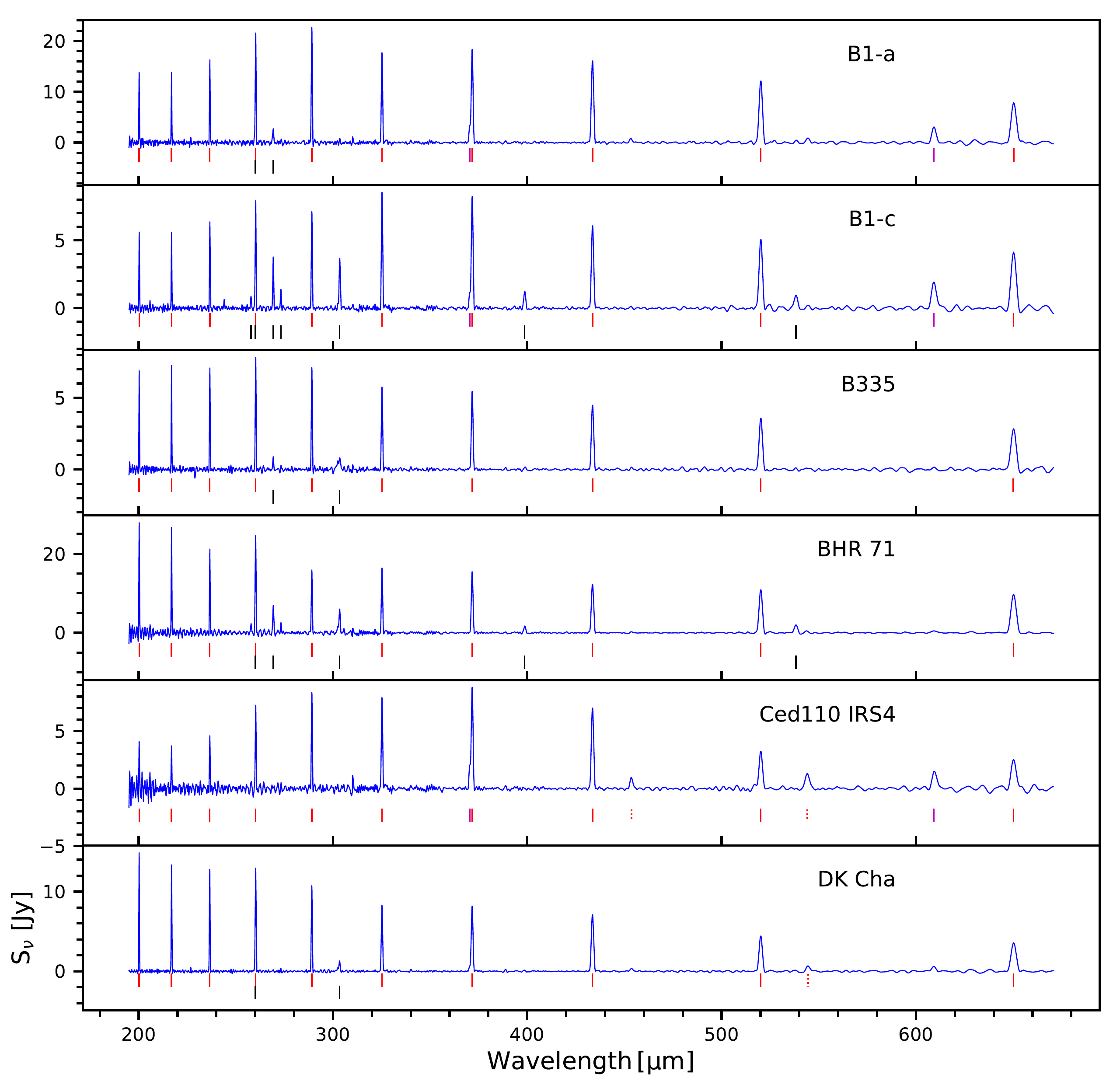}
    \caption{The continuum-free SPIRE 1D spectra for the COPS sources.  The vertical bars indicate the significant detections on emission lines.  The red solid bars represent the CO emission, while the red dotted bars represent the $^{13}$CO emission.  The water lines are shown in black solid bars, while the OH lines are shown in black dotted bars.  The orange bars indicate the emission of HCO$^{+}$, while the \CI\ lines are shown in magenta bars.  Note that the figure of RCrA~IRS7B/C shows the spectrum of RCrA~IRS7C due to the blending between of two sources.}
    \label{fig:spire_1d}
\end{figure*}

Figure~\ref{fig:spire_1d} shows the continuum-subtracted SPIRE 1-D spectra for the COPS sources.  Three versions of spectra are available online, the files named by \texttt{[object]\_spire\_corrected} contain the SPIRE 1-D spectra used in this study, \texttt{flat\_spectrum} in the file name indicates the continuum-subtracted spectra, \texttt{residual\_spectrum} in the file name indicates the residual spectra after subtracting emission lines and continuum.

\renewcommand{\thefigure}{\arabic{figure} (Cont.)}
\addtocounter{figure}{-1}

\begin{figure*}[htbp!]
    \includegraphics[width=\textwidth]{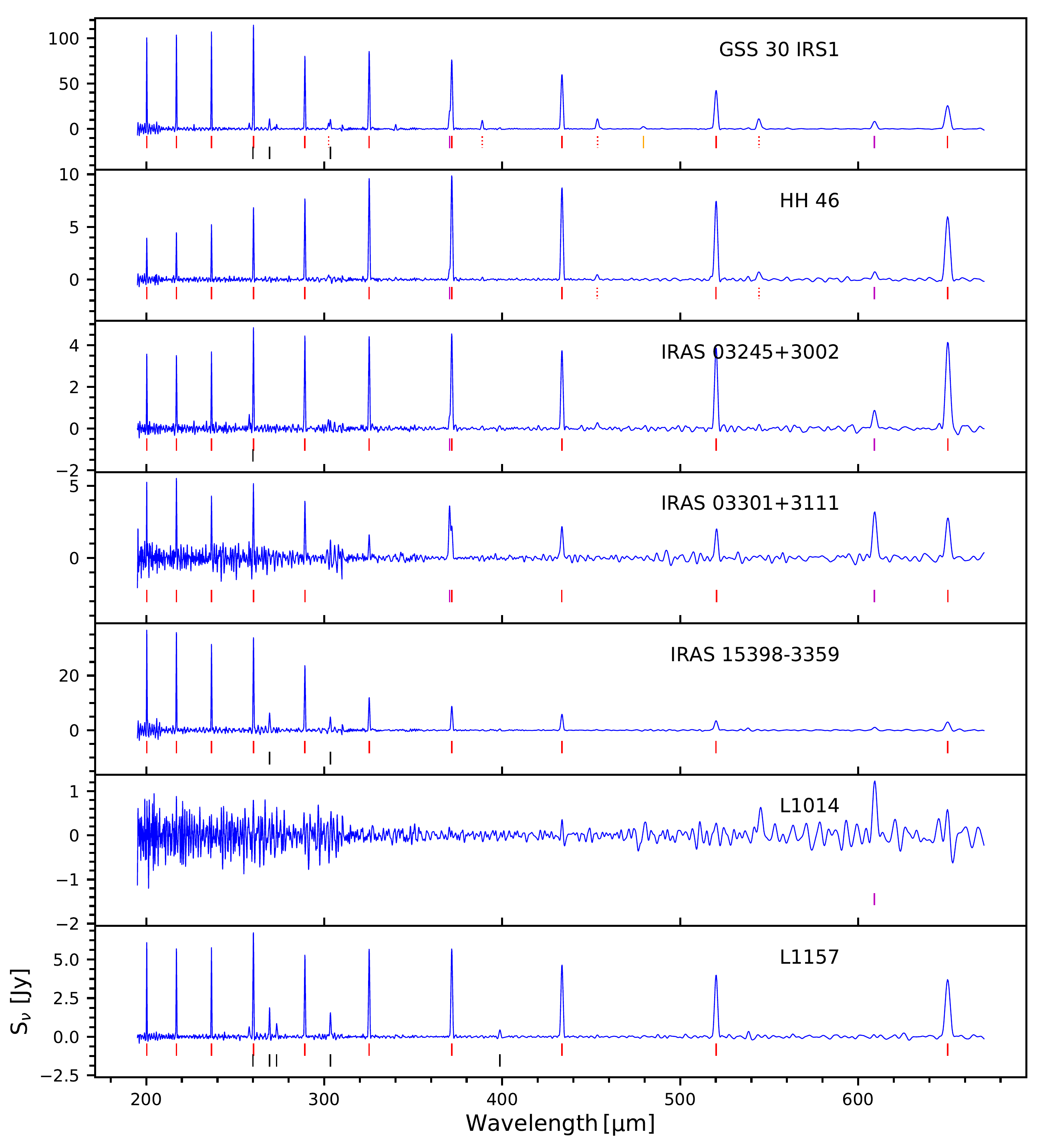}
    \caption{}
\end{figure*}
\addtocounter{figure}{-1}

\begin{figure*}[htbp!]
    \includegraphics[width=\textwidth]{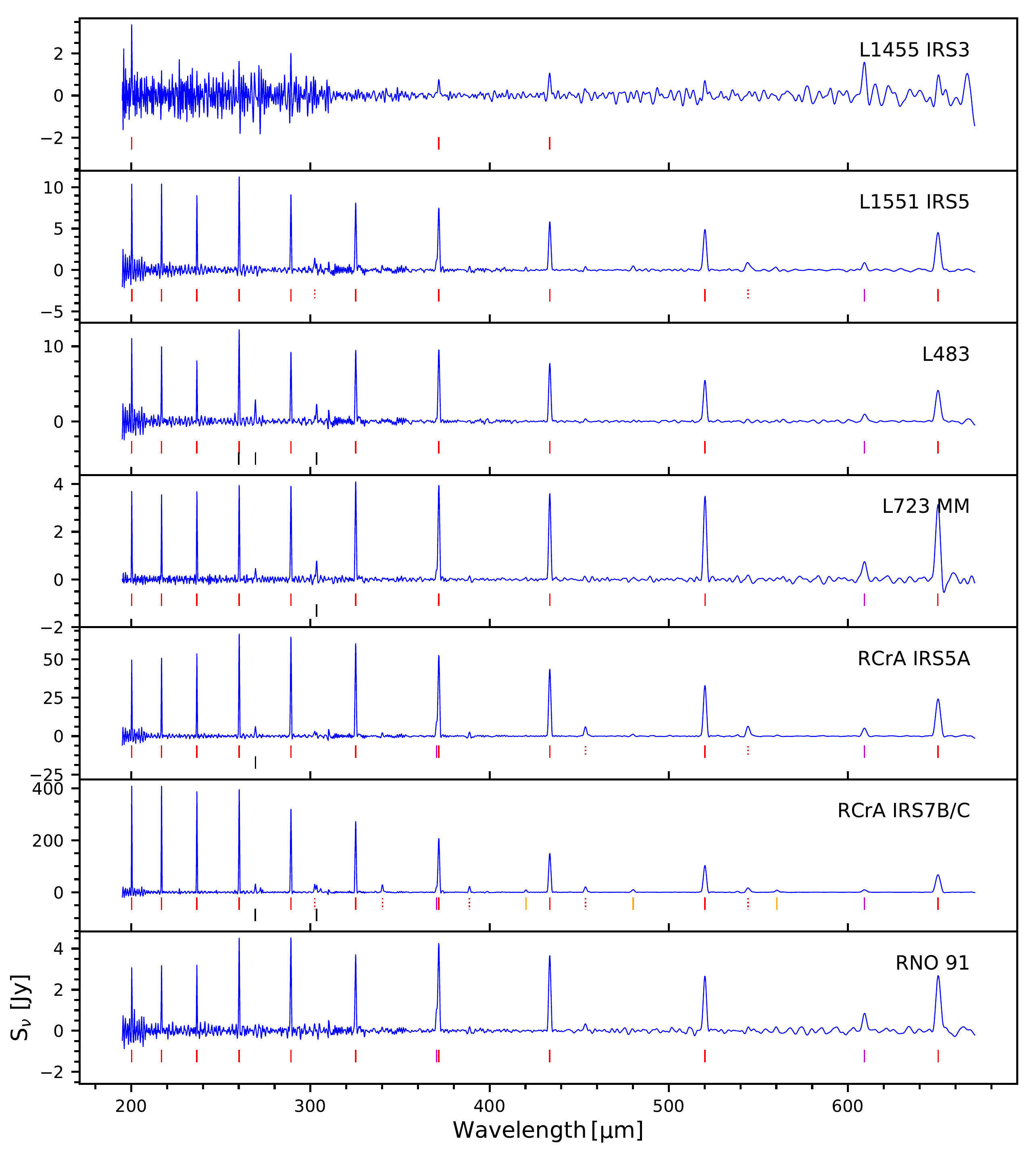}
    \caption{}
\end{figure*}
\addtocounter{figure}{-1}

\begin{figure*}[htbp!]
    \includegraphics[width=\textwidth]{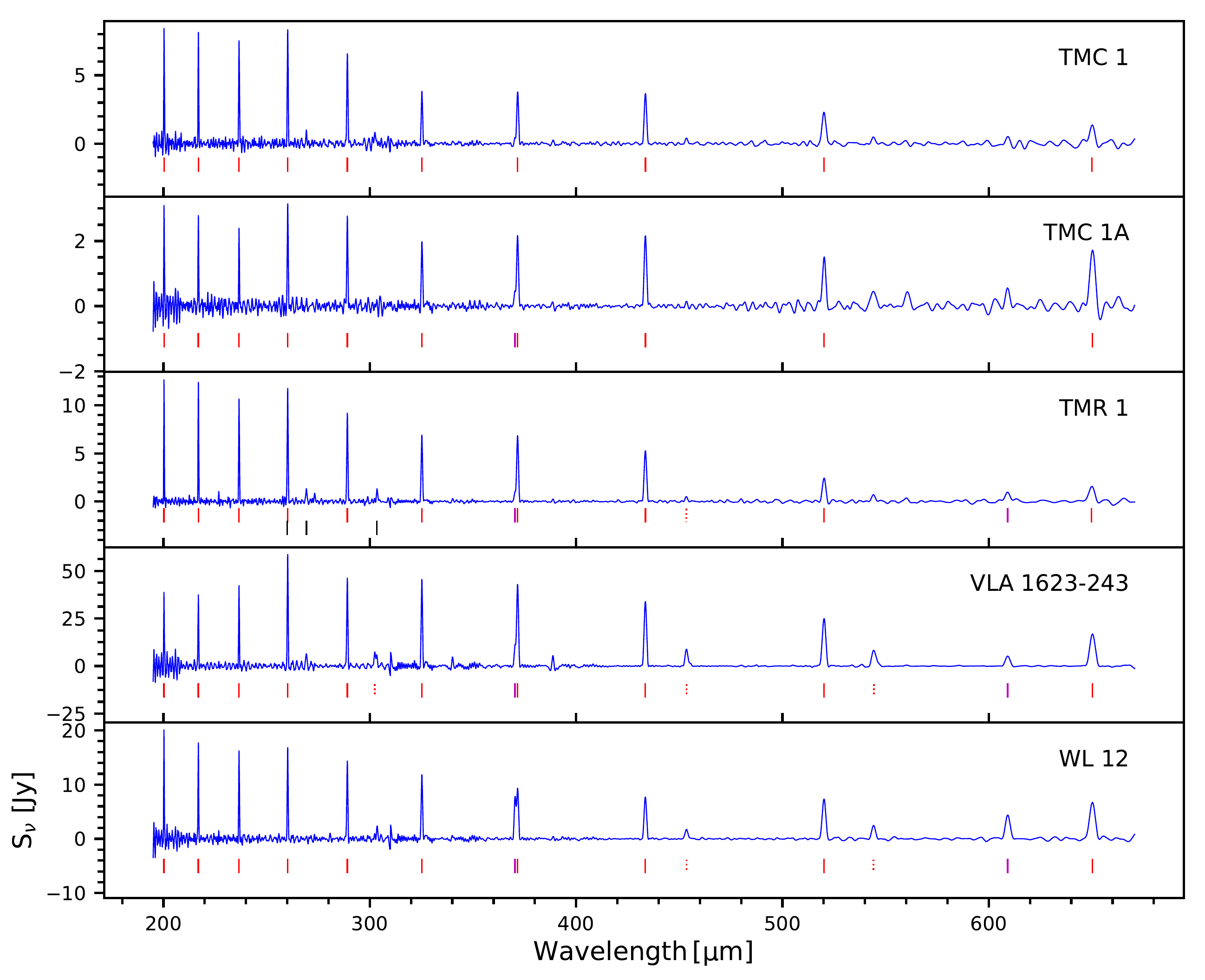}
    \caption{}
\end{figure*}
\renewcommand{\thefigure}{\arabic{figure}}

\section{Classification of the 1D Profiles for Morphology}
\label{sec:a_outflow_classification}
\begin{deluxetable*}{lrrrr}
    \tabletypesize{\scriptsize}
    \tablecaption{Summary of the morphology of CO emission \label{co_morphology}}
    \tablewidth{\textwidth}
    \tablehead{
        \colhead{Source} & \colhead{1D profile} & \colhead{Presence of outflow} & \colhead{Ref.} & \colhead{Type$^{a}$}
        }
    \startdata
    IRAS 03245+3002  & Ir & B/R & \citet{2016ApJ...823..151C} & 4 \\
    L1455~IRS3       & ND & N & Kang et al. (in prep.)        & 2 \\
    IRAS 03301+3111  & ND & N & Kang et al. (in prep.)        & 2 \\
    B1-a             & Si & B & Kang et al. (in prep.)        & 1 \\
    B1-c             & Bi & B/R & \citet{2014ApJ...794..165S} & 1 \\
    L1551~IRS5       & Bi & B/R & \citet{2009ApJ...698..184W} & 1 \\
    TMR 1            & ND & B & Kang et al. (in prep.)        & 3 \\
    TMC 1A           & ND & B & \citet{2015AA...576A.109Y}    & 3 \\
    TMC 1            & ND & B/R & \citet{2015AA...576A.109Y}  & 3 \\
    HH~46            & Si & R & \citet{2015AA...576A.109Y}    & 1 \\
    Ced110~IRS4      & Ir & N & \citet{2015AA...576A.109Y}    & 2 \\
    BHR 71           & Bi & B/R & \citet{2006AA...454L..79P}  & 1 \\
    DK Cha           & Ir & N & \citet{2015AA...576A.109Y}    & 2 \\
    IRAS 15398--3359 & ND & B/R & \citet{2015AA...576A.109Y}  & 3 \\
    GSS~30~IRS1      & Si & B & \citet{2015AA...576A.109Y}    & 1 \\
    VLA~1623$-$243     & Ir & B/R & \citet{2015AA...581A..91S}  & 4 \\
    WL~12            & Ir & B/R & \citet{1996AA...311..858B}  & 3 \\
    RNO~91           & Ir & N & \citet{2015AA...576A.109Y}    & 2 \\
    L483             & Bi & B/R & \citet{2014ApJ...783...29D} & 1 \\
    RCrA~IRS5A       & Si & N & \citet{2011ApJS..194...43P} & \nodata$^{b}$ \\
    L723~MM          & Ir & B/R & \citet{1996AA...311..858B}  & 3 \\
    B335             & Bi & B/R & \citet{2008ApJ...687..389S} & 1 \\
    L1157            & Bi & B/R & \citet{2015AA...573L...2T}  & 1 \\
    L1014            & ND & B/R & \citet{2005ApJ...633L.129B} & 3 \\
    \enddata
    \tablecomments{The notation for the classification of 1D smooth profile: Bi: bipolar lobes; Si: single lobe; Ir: irregular profile; ND: not enough detection for constructing the 1D profile.  The presence of outflow is labelled by B, R, and N for blue, red, and no outflow, respectively. \\ $^{a}$Please see Appendix~\ref{sec:a_outflow_classification} for the discussion of the types. \\
    $^{b}$No clear outflow detection for RCrA~IRS5 in other studies; therefore, it is unclear whether the SIPRE morphology tracing outflows.}
\end{deluxetable*}

We classified the 1D smooth profile into three classes, bipolar, single, and irregular, depending on whether they have the double peaks, single peak, and flat profile, respectively.  The irregular profile can be a result of complex source structure or the absence of outflows.  The sources that have a bipolar feature identified in the 1D profile all exhibit outflows in the velocity-resolved observations.  However, not all outflows observed in the velocity-resolved observations are identified from the 1D profiles due to the weak emission of the outflows or the complexity of the source.  Thus, we further classified the 1D profiles into three types: (1) bipolar outflows (or single outflow) are identified in the 1D profile, which is also confirmed with the velocity-resolved observations; (2) no bipolar feature seen in the 1D profile, and the velocity-resolved observations show little CO emission in the high velocity wings; (3) outflows are seen from the velocity-resolved observations, whereas few off-center detection is found in our SPIRE data due to insufficient sensitivity; (4) the 1D profile is irregular, while the irregularity is consistent with the velocity-resolved observations, mostly due to the presence of multiple sources or broad outflows.  Table~\ref{co_morphology} lists the class and type for the 1D profile of each source.  We found that spatially extended CO emission observed with SPIRE always has an outflow-related origin based on its morphology.

\section{Effect of Sensitivity on the Variation of Bipolarity}
\label{sec:a_sensitivity}
\begin{figure*}[htbp!]
    \centering
    \includegraphics[width=0.45\textwidth]{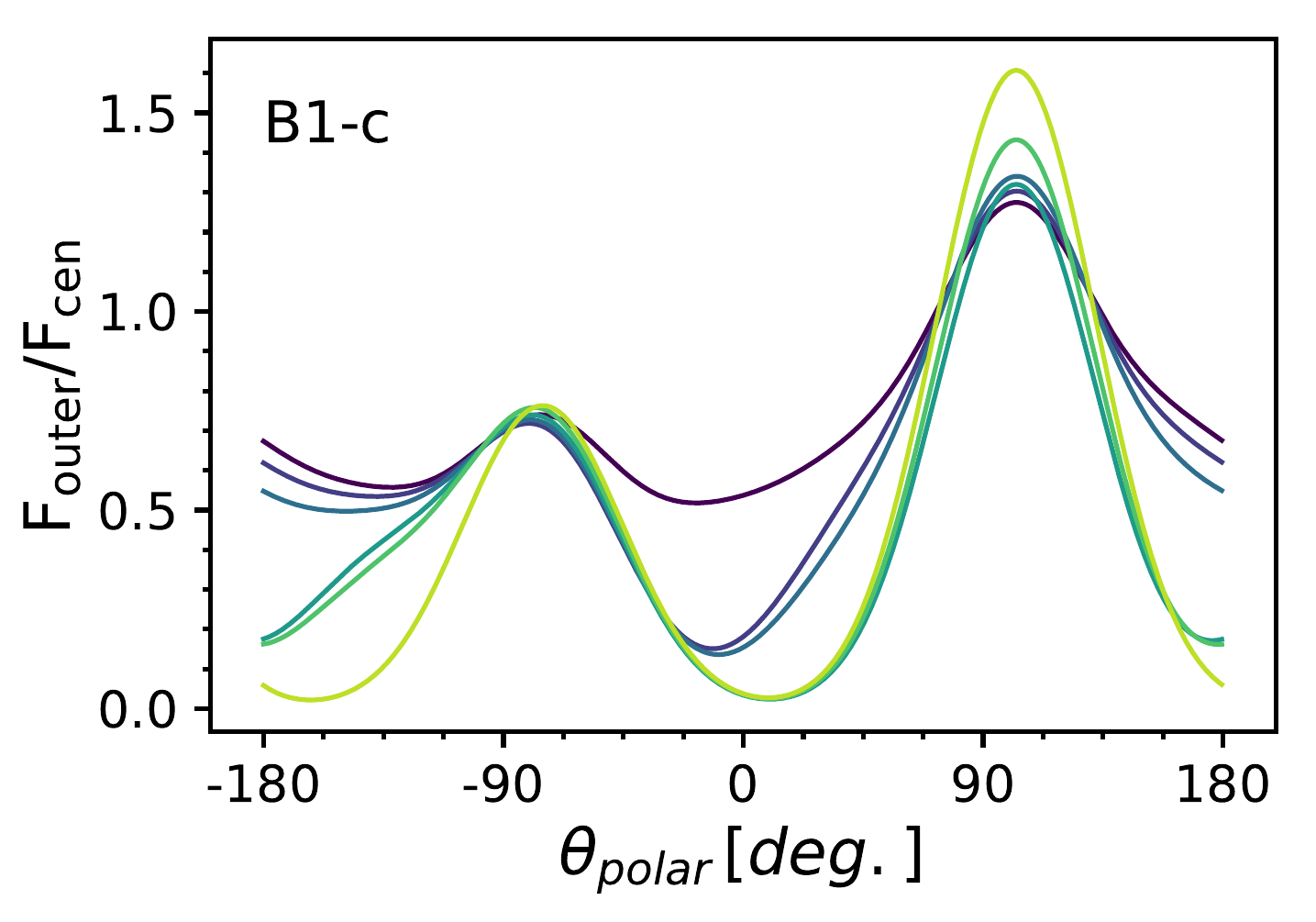}
    \includegraphics[width=0.45\textwidth]{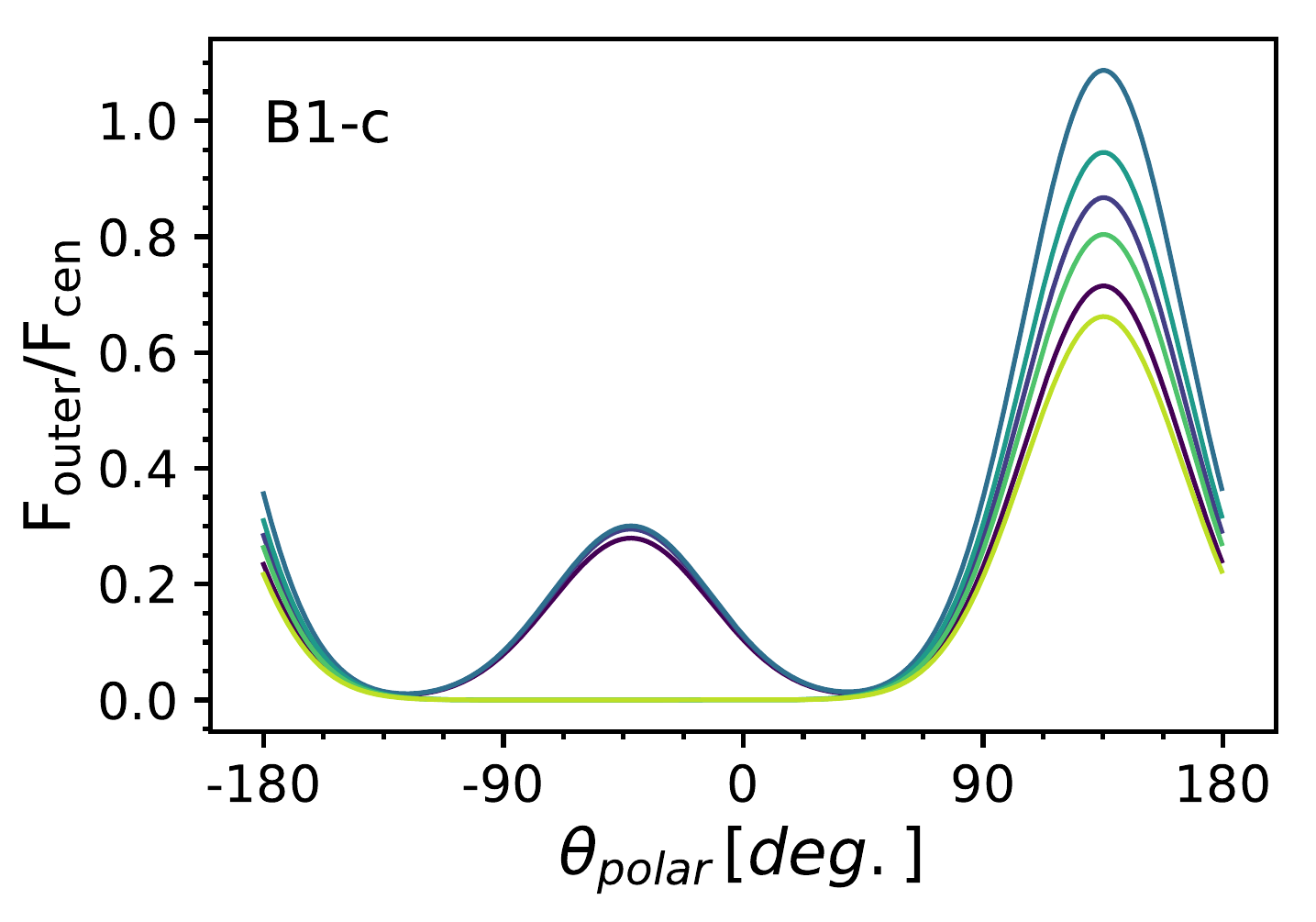}
    \includegraphics[width=0.45\textwidth]{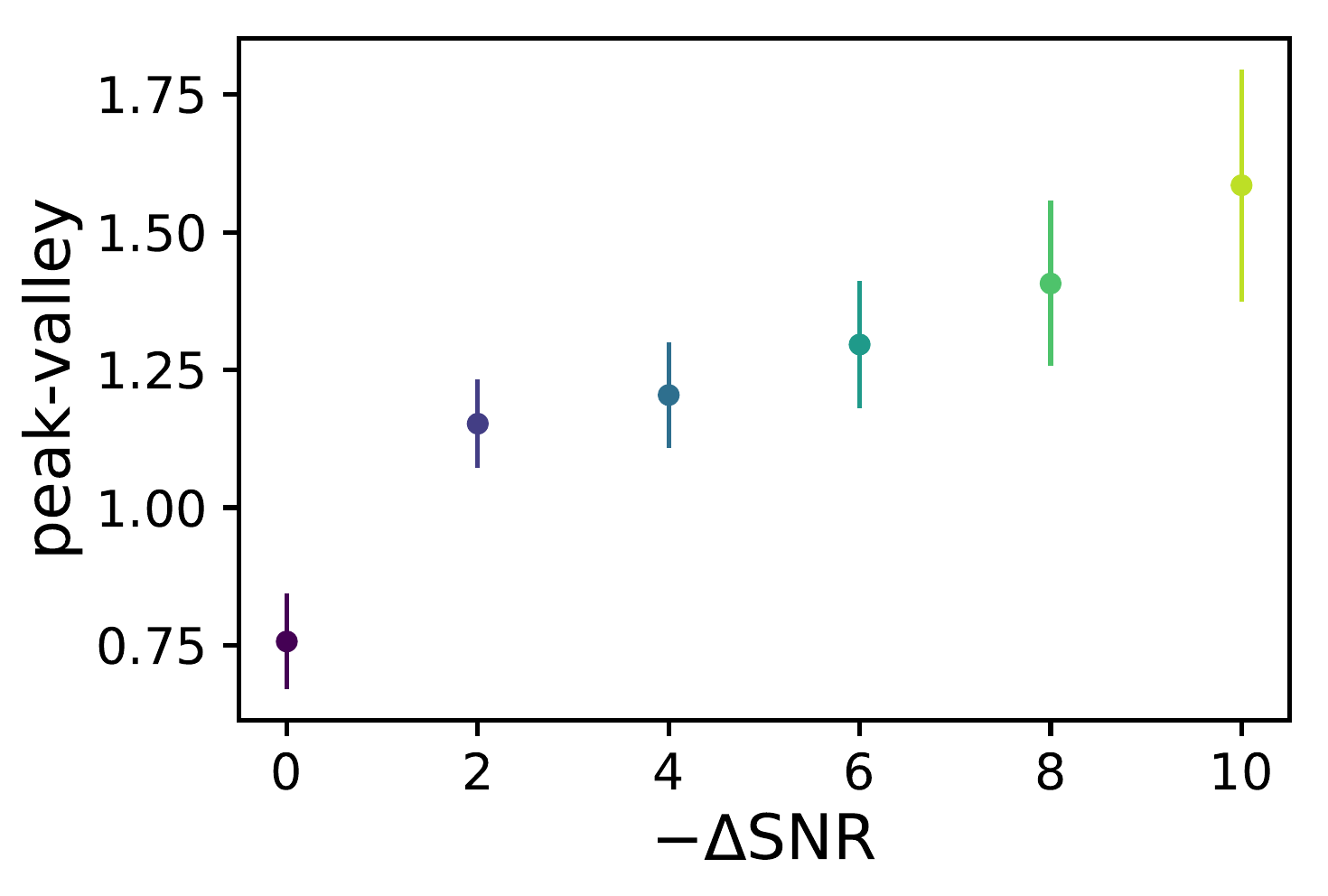}
    \includegraphics[width=0.45\textwidth]{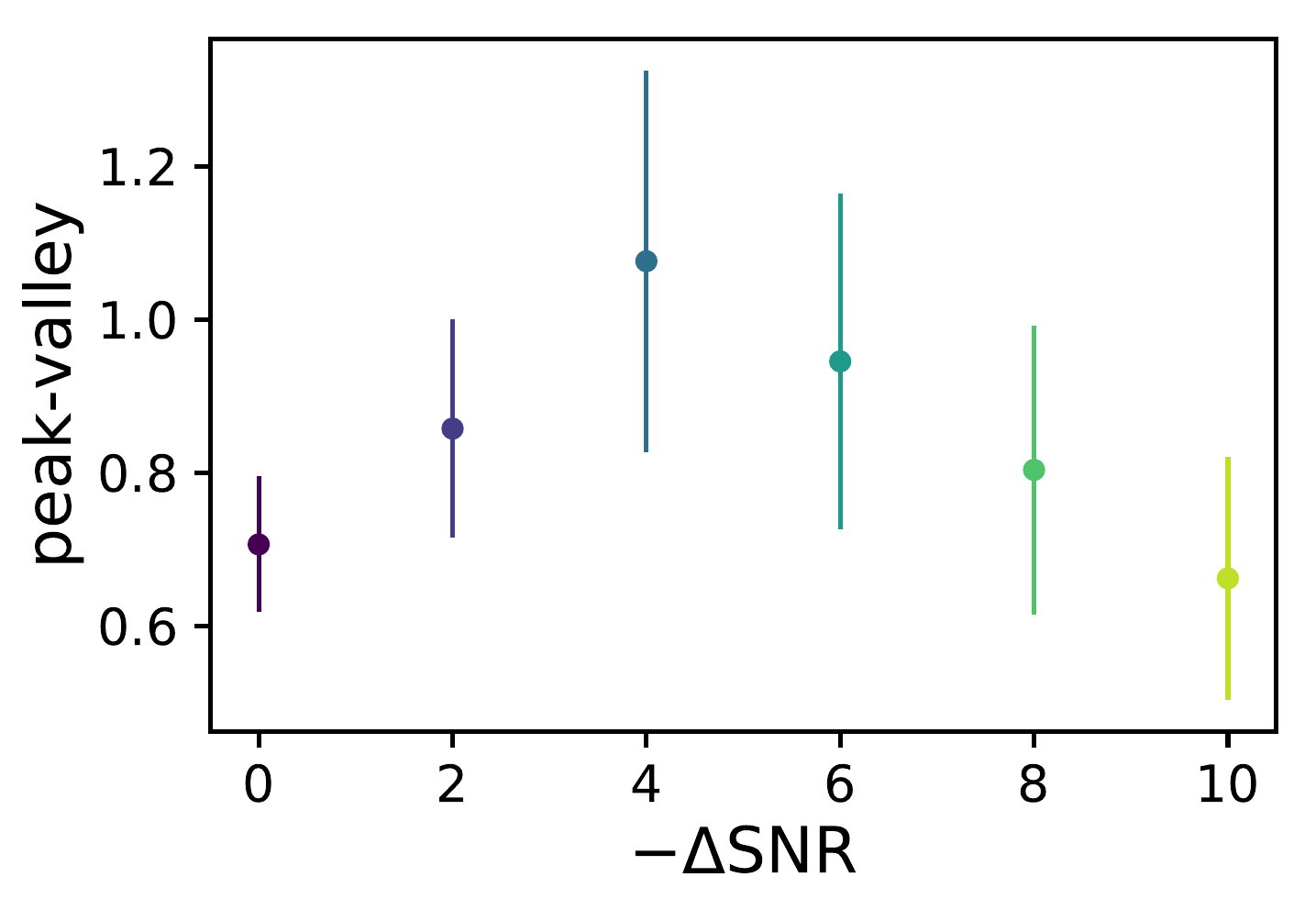}
    \caption{\textbf{Top:} The 1D smoothed profiles of CO~\jj{4}{3} (left) and CO~\jj{9}{8} (right) with SNR reductions applied.  The applied SNR reduction is indicated by the color of points in the figures shown in the bottom.  \textbf{Bottom: } The peak-to-valley differences derived from the profiles with different SNR reductions applied.}
    \label{fig:co_extent_sensitivity}
\end{figure*}

In Section~\ref{sec:varitation_bipolarity}, we use the peak-to-valley difference to quantify the strength of bipolarity and found a decrease of the difference suggesting the bipolarity diminishing at higher-$J$ CO lines.  However, the cause of the decrease of the peak-to-valley difference can be either the bipolar feature diminishing toward the higher $J$-levels or simply lower sensitivities at the higher $J$-levels.  Thus, we investigate the effect of sensitivity by artificially decreasing the fluxes seen in CO~\jj{4}{3} (CO~\jj{9}{8}) for the SLW (SSW) module and calculate the peak-to-valley differences.  These two lines have the strongest strength at outer spaxels in each module.
We use B1-c as an example.  Figure~\ref{fig:co_extent_sensitivity} shows two sets of 1D smoothed profiles applied with various SNR reductions for the SLW and SSW modules, respectively.  Ideally, if all signals remain significant after the SNR reduction is applied, the 1D profile should be the same, therefore the same peak-to-valley difference.  It is only when the signal becomes non-detectable (i.e. SNR $<$ 4), then the 1D profile and the peak-to-valley difference change correspondingly.

When the SNR reduction is small ($-\Delta$SNR=2 or 4), the low-SNR detections are reduced to the noise level, while the high-SNR detections remain significant but weaker.  Therefore, the valley of the 1D profile becomes lower, resulting in a higher peak-to-valley difference (Figure~\ref{fig:co_extent_sensitivity}).  As the SNR reduction becomes larger, we start to see the peak-to-valley difference decrease (Figure~\ref{fig:co_extent_sensitivity}, right).  When most of the original detections become non-detections, the 1D profile is dominated by a few high-SNR detections preferentially found in the bipolar feature.  Thus, a further SNR reduction simply decreases the height of the peak in the 1D profile, reducing the derived peak-to-valley difference.  In summary, the most robust analysis with the peak-to-valley difference should be restricted to the sources with significant detections at the outer spaxels over the $J$-levels (\Jup=4--13).

\section{Archival Photometry for COPS sources}
Table~\ref{phot_reference} shows part of a table that contains all PACS and SPIRE photometry derived in this study as well as the photometry collected from literatures.  The entire table is available as machine readable table online.

\begin{deluxetable*}{rllrllrll}
\tabletypesize{\scriptsize}
\tablecaption{Reference for Photometry \label{phot_reference}}
\tablewidth{0pt}
\setlength\tabcolsep{3pt}
\tablehead{ \colhead{Wave.} & \colhead{Flux} & \colhead{Ref.} &
\colhead{Wave.} & \colhead{Flux} & \colhead{Ref.} &
\colhead{Wave.} & \colhead{Flux} & \colhead{Ref.} \\
\colhead{[\micron]} & \colhead{[Jy]} & \colhead{} &
\colhead{[\micron]} & \colhead{[Jy]} & \colhead{} &
\colhead{[\micron]} & \colhead{[Jy]} & \colhead{}}
\startdata
\multicolumn{9}{c}{\textbf{B1-a}} \\
\hline
1.2 & 8.90($-$3) & C03 & 1.7 & 3.80($-$4) & C03 & 2.2 & 1.40($-$3) & C03 \\
3.6 & 1.64($-$2)$\pm$9.00($-$4) & c2d-DR4 & 4.5 & 4.93($-$2)$\pm$2.50($-$3) & c2d-DR4 & 5.8 & 8.45($-$2)$\pm$4.00($-$3) & c2d-DR4 \\
8.0 & 1.15($-$1)$\pm$5.00($-$3) & c2d-DR4 & 12.0 & 2.50($-$1) & IRAS2 & 24.0 & 1.69$\pm$1.60($-$1) & c2d-DR4 \\
25.0 & 1.39$\pm$1.40($-$1) & IRAS2 & 70.0 & 6.45$\pm$6.10($-$1) & c2d-DR4 & 100.0 & 7.54$\pm$3.64 & CDF-v2 \\
160.0 & 1.18(1)$\pm$4.45 & CDF-v2 & 250.0 & 2.40(1)$\pm$7.95 & CDF-v2 & 350.0 & 2.02(1)$\pm$7.20 & CDF-v2 \\
500.0 & 1.50(1)$\pm$6.41 & CDF-v2 & 850.0 & 3.69 & DF08 & & & \\
 \hline
\enddata
\tablecomments{The numbers in the prentices indicate the exponent of the power of ten.  The uncertainties of fluxes are only shown if published in literatures.  The entire table is available online as a machine readable table.}
\tablerefs{A93~=~\citet{1993ApJ...406..122A},
A94~=~\citet{1994ApJ...420..837A},
A05~=~\citet{2005ApJ...631.1134A},
B81~=~\citet{1981ApJ...245..589B},
C84~=~\citet{1984ApJ...278..671C},
C85~=~\citet{1985ApJ...296..633C},
C00~=~\citet{2000ApJ...530..851C},
C03~=~\citet{2003yCat.2246....0C},
C08~=~\citet{2008ApJ...683..862C},
C12~=~\citet{2012yCat.2311....0C},
c2d-DR4~=~\citet{2003PASP..115..965E},
CDF-v2~=~this work,
D15~=~\citet{2015ApJS..220...11D},
D98~=~\citet{1998MNRAS.301.1049D},
DF08~=~\citet{2008ApJS..175..277D},
E05~=~\citet{2005ApJ...635..396E},
E08~=~\citet{2008ApJ...684.1240E},
E09~=~\citet{2009ApJS..181..321E},
G85~=~\citet{1985MNRAS.215P..15G},
G07~=~\citet{2007ApJ...670..489G},
G09~=~\citet{2009ApJS..184...18G},
H92~=~\citet{1992AJ....104..680H},
H93~=~\citet{1993AA...276..129H},
H00~=~\citet{2000ApJ...534..880H},
H16~=~\citet{2016yCat.2336....0H},
I10~=~\citet{2010yCat.2297....0I},
IF2~=~\citet{1989ifss.book.....M},
IRAS2~=~\citet{1988iras....7.....H},
J08~=~\citet{2008ApJ...683..822J},
K83~=~\citet{1983ApJ...274L..43K},
K93~=~\citet{1993ApJ...414..676K},
K09~=~\citet{2009ApJS..185..198K},
L10~=~\citet{2010ApJS..188..139L},
M87~=~\citet{1987ApJ...319..340M},
MS94~=~\citet{1994ApJ...436..800M},
M01~=~\citet{2001AA...365..440M},
M10~=~\citet{2010ApJS..188...75M},
P11~=~\citet{2011ApJS..194...43P},
R10~=~\citet{2010ApJS..186..259R},
S96~=~\citet{1996AA...309..827S},
S00~=~\citet{2000ApJS..131..249S},
S06~=~\citet{2006AA...447..609S},
S07~=~\citet{2007AA...470..281S},
S08~=~\citet{2008ApJ...687..389S},
T90~=~\citet{1990ApJ...362L..63T},
V02~=~\citet{2002AJ....124.2756V},
W07~=~\citet{2007AJ....133.1560W},
Y05~=~\citet{2005ApJ...628..283Y},
Y10~=~\citet{2010yCat.2298....0Y},
Z04~=~\citet{2004AAS...205.4815Z}.
}\end{deluxetable*}

\clearpage
\bibliography{dissbib}

\begin{thebibliography}{}
\expandafter\ifx\csname natexlab\endcsname\relax\def\natexlab#1{#1}\fi

\bibitem[{{Adams} {et~al.}(1987){Adams}, {Lada}, \& {Shu}}]{adams87}
{Adams}, F.~C., {Lada}, C.~J., \& {Shu}, F.~H. 1987, \apj, 312, 788

\bibitem[{{Andre} \& {Montmerle}(1994)}]{1994ApJ...420..837A}
{Andre}, P., \& {Montmerle}, T. 1994, \apj, 420, 837

\bibitem[{{Andre} {et~al.}(1993{\natexlab{a}}){Andre}, {Ward-Thompson}, \&
  {Barsony}}]{andre93}
{Andre}, P., {Ward-Thompson}, D., \& {Barsony}, M. 1993{\natexlab{a}}, \apj,
  406, 122

\bibitem[{{Andre} {et~al.}(1993{\natexlab{b}}){Andre}, {Ward-Thompson}, \&
  {Barsony}}]{1993ApJ...406..122A}
---. 1993{\natexlab{b}}, \apj, 406, 122

\bibitem[{{Andre} {et~al.}(2000){Andre}, {Ward-Thompson}, \&
  {Barsony}}]{2000prpl.conf...59A}
---. 2000, Protostars and Planets IV, 59

\bibitem[{{Andr{\'e}} {et~al.}(2010){Andr{\'e}}, {Men'shchikov}, {Bontemps},
  {K{\"o}nyves}, {Motte}, {Schneider}, {Didelon}, {Minier}, {Saraceno},
  {Ward-Thompson}, {di Francesco}, {White}, {Molinari}, {Testi}, {Abergel},
  {Griffin}, {Henning}, {Royer}, {Mer{\'{\i}}n}, {Vavrek}, {Attard},
  {Arzoumanian}, {Wilson}, {Ade}, {Aussel}, {Baluteau}, {Benedettini},
  {Bernard}, {Blommaert}, {Cambr{\'e}sy}, {Cox}, {di Giorgio}, {Hargrave},
  {Hennemann}, {Huang}, {Kirk}, {Krause}, {Launhardt}, {Leeks}, {Le Pennec},
  {Li}, {Martin}, {Maury}, {Olofsson}, {Omont}, {Peretto}, {Pezzuto}, {Prusti},
  {Roussel}, {Russeil}, {Sauvage}, {Sibthorpe}, {Sicilia-Aguilar}, {Spinoglio},
  {Waelkens}, {Woodcraft}, \& {Zavagno}}]{2010AA...518L.102A}
{Andr{\'e}}, P., {Men'shchikov}, A., {Bontemps}, S., {et~al.} 2010, \aap, 518,
  L102

\bibitem[{{Andrews} \& {Williams}(2005)}]{2005ApJ...631.1134A}
{Andrews}, S.~M., \& {Williams}, J.~P. 2005, \apj, 631, 1134

\bibitem[{{Apai} {et~al.}(2005){Apai}, {T{\'o}th}, {Henning}, {Vavrek},
  {Kov{\'a}cs}, \& {Lemke}}]{apai05}
{Apai}, D., {T{\'o}th}, L.~V., {Henning}, T., {et~al.} 2005, \aap, 433, L33

\bibitem[{{Arce} \& {Goodman}(2001)}]{2001ApJ...554..132A}
{Arce}, H.~G., \& {Goodman}, A.~A. 2001, \apj, 554, 132

\bibitem[{{Arce} {et~al.}(2013){Arce}, {Mardones}, {Corder}, {Garay},
  {Noriega-Crespo}, \& {Raga}}]{2013ApJ...774...39A}
{Arce}, H.~G., {Mardones}, D., {Corder}, S.~A., {et~al.} 2013, \apj, 774, 39

\bibitem[{{Arce} {et~al.}(2007){Arce}, {Shepherd}, {Gueth}, {Lee}, {Bachiller},
  {Rosen}, \& {Beuther}}]{2007prpl.conf..245A}
{Arce}, H.~G., {Shepherd}, D., {Gueth}, F., {et~al.} 2007, Protostars and
  Planets V, 245

\bibitem[{{Astropy Collaboration} {et~al.}(2013){Astropy Collaboration},
  {Robitaille}, {Tollerud}, {Greenfield}, {Droettboom}, {Bray}, {Aldcroft},
  {Davis}, {Ginsburg}, {Price-Whelan}, {Kerzendorf}, {Conley}, {Crighton},
  {Barbary}, {Muna}, {Ferguson}, {Grollier}, {Parikh}, {Nair}, {Unther},
  {Deil}, {Woillez}, {Conseil}, {Kramer}, {Turner}, {Singer}, {Fox}, {Weaver},
  {Zabalza}, {Edwards}, {Azalee Bostroem}, {Burke}, {Casey}, {Crawford},
  {Dencheva}, {Ely}, {Jenness}, {Labrie}, {Lim}, {Pierfederici}, {Pontzen},
  {Ptak}, {Refsdal}, {Servillat}, \& {Streicher}}]{2013AA...558A..33A}
{Astropy Collaboration}, {Robitaille}, T.~P., {Tollerud}, E.~J., {et~al.} 2013,
  \aap, 558, A33

\bibitem[{{Audard} {et~al.}(2014){Audard}, {{\'A}brah{\'a}m}, {Dunham},
  {Green}, {Grosso}, {Hamaguchi}, {Kastner}, {K{\'o}sp{\'a}l}, {Lodato},
  {Romanova}, {Skinner}, {Vorobyov}, \& {Zhu}}]{audard14}
{Audard}, M., {{\'A}brah{\'a}m}, P., {Dunham}, M.~M., {et~al.} 2014, Protostars
  and Planets VI, 387

\bibitem[{{Beichman} \& {Harris}(1981)}]{1981ApJ...245..589B}
{Beichman}, C., \& {Harris}, S. 1981, \apj, 245, 589

\bibitem[{{Belikov} {et~al.}(2002){Belikov}, {Kharchenko}, {Piskunov},
  {Schilbach}, \& {Scholz}}]{2002AA...387..117B}
{Belikov}, A.~N., {Kharchenko}, N.~V., {Piskunov}, A.~E., {Schilbach}, E., \&
  {Scholz}, R.-D. 2002, \aap, 387, 117

\bibitem[{{Bell} \& {Lin}(1994)}]{bell94}
{Bell}, K.~R., \& {Lin}, D.~N.~C. 1994, \apj, 427, 987

\bibitem[{{Benz} {et~al.}(2016){Benz}, {Bruderer}, {van Dishoeck}, {Melchior},
  {Wampfler}, {van der Tak}, {Goicoechea}, {Indriolo}, {Kristensen}, {Lis},
  {Mottram}, {Bergin}, {Caselli}, {Herpin}, {Hogerheijde}, {Johnstone},
  {Liseau}, {Nisini}, {Tafalla}, {Visser}, \& {Wyrowski}}]{2016AA...590A.105B}
{Benz}, A.~O., {Bruderer}, S., {van Dishoeck}, E.~F., {et~al.} 2016, \aap, 590,
  A105

\bibitem[{{Bertout} {et~al.}(1999){Bertout}, {Robichon}, \&
  {Arenou}}]{1999AA...352..574B}
{Bertout}, C., {Robichon}, N., \& {Arenou}, F. 1999, \aap, 352, 574

\bibitem[{{Bontemps} {et~al.}(1996){Bontemps}, {Andre}, {Terebey}, \&
  {Cabrit}}]{1996AA...311..858B}
{Bontemps}, S., {Andre}, P., {Terebey}, S., \& {Cabrit}, S. 1996, \aap, 311,
  858

\bibitem[{{Bourke} {et~al.}(2005){Bourke}, {Crapsi}, {Myers}, {Evans},
  {Wilner}, {Huard}, {J{\o}rgensen}, \& {Young}}]{2005ApJ...633L.129B}
{Bourke}, T.~L., {Crapsi}, A., {Myers}, P.~C., {et~al.} 2005, \apjl, 633, L129

\bibitem[{{Brandt} {et~al.}(1971){Brandt}, {Stecher}, {Crawford}, \&
  {Maran}}]{1971ApJ...163L..99B}
{Brandt}, J.~C., {Stecher}, T.~P., {Crawford}, D.~L., \& {Maran}, S.~P. 1971,
  \apjl, 163, L99

\bibitem[{{Brinch} {et~al.}(2007){Brinch}, {Crapsi}, {Hogerheijde}, \&
  {J{\o}rgensen}}]{brinch07}
{Brinch}, C., {Crapsi}, A., {Hogerheijde}, M.~R., \& {J{\o}rgensen}, J.~K.
  2007, \aap, 461, 1037

\bibitem[{{Casey} {et~al.}(1998){Casey}, {Mathieu}, {Vaz}, {Andersen}, \&
  {Suntzeff}}]{1998AJ....115.1617C}
{Casey}, B.~W., {Mathieu}, R.~D., {Vaz}, L.~P.~R., {Andersen}, J., \&
  {Suntzeff}, N.~B. 1998, \aj, 115, 1617

\bibitem[{{Ceccarelli} {et~al.}(1999){Ceccarelli}, {Caux}, {Loinard},
  {Castets}, {Tielens}, {Molinari}, {Liseau}, {Saraceno}, {Smith}, \&
  {White}}]{ceccarelli99}
{Ceccarelli}, C., {Caux}, E., {Loinard}, L., {et~al.} 1999, \aap, 342, L21

\bibitem[{{\v{C}ernis}(1993)}]{1993BaltA...2..214C}
{\v{C}ernis}, K. 1993, Baltic Astronomy, 2, 214

\bibitem[{{\v{C}ernis} \& {Strai\v{z}ys}(2003)}]{2003BaltA..12..301C}
{\v{C}ernis}, K., \& {Strai\v{z}ys}, V. 2003, Baltic Astronomy, 12, 301

\bibitem[{{Chandler} \& {Richer}(2000)}]{2000ApJ...530..851C}
{Chandler}, C.~J., \& {Richer}, J.~S. 2000, \apj, 530, 851

\bibitem[{{Chen} {et~al.}(1995){Chen}, {Myers}, {Ladd}, \& {Wood}}]{chen95}
{Chen}, H., {Myers}, P.~C., {Ladd}, E.~F., \& {Wood}, D.~O.~S. 1995, \apj, 445,
  377

\bibitem[{{Chen} {et~al.}(2008){Chen}, {Launhardt}, {Bourke}, {Henning}, \&
  {Barnes}}]{2008ApJ...683..862C}
{Chen}, X., {Launhardt}, R., {Bourke}, T.~L., {Henning}, T., \& {Barnes}, P.~J.
  2008, \apj, 683, 862

\bibitem[{{Chou} {et~al.}(2016){Chou}, {Yen}, {Koch}, \&
  {Guilloteau}}]{2016ApJ...823..151C}
{Chou}, H.-G., {Yen}, H.-W., {Koch}, P.~M., \& {Guilloteau}, S. 2016, \apj,
  823, 151

\bibitem[{{Cohen} {et~al.}(1985){Cohen}, {Harvey}, \&
  {Schwartz}}]{1985ApJ...296..633C}
{Cohen}, M., {Harvey}, P.~M., \& {Schwartz}, R.~D. 1985, \apj, 296, 633

\bibitem[{{Cohen} {et~al.}(1984){Cohen}, {Harvey}, {Wilking}, \&
  {Schwartz}}]{1984ApJ...278..671C}
{Cohen}, M., {Harvey}, P.~M., {Wilking}, B.~A., \& {Schwartz}, R.~D. 1984,
  \apj, 278, 671

\bibitem[{{Cutri} \& {et al.}(2012)}]{2012yCat.2311....0C}
{Cutri}, R.~M., \& {et al.} 2012, VizieR Online Data Catalog, 2311

\bibitem[{{Cutri} {et~al.}(2003){Cutri}, {Skrutskie}, {van Dyk}, {Beichman},
  {Carpenter}, {Chester}, {Cambresy}, {Evans}, {Fowler}, {Gizis}, {Howard},
  {Huchra}, {Jarrett}, {Kopan}, {Kirkpatrick}, {Light}, {Marsh}, {McCallon},
  {Schneider}, {Stiening}, {Sykes}, {Weinberg}, {Wheaton}, {Wheelock}, \&
  {Zacarias}}]{2003yCat.2246....0C}
{Cutri}, R.~M., {Skrutskie}, M.~F., {van Dyk}, S., {et~al.} 2003, VizieR Online
  Data Catalog, 2246

\bibitem[{{Dame} \& {Thaddeus}(1985)}]{1985ApJ...297..751D}
{Dame}, T.~M., \& {Thaddeus}, P. 1985, \apj, 297, 751

\bibitem[{{de Geus} {et~al.}(1990){de Geus}, {Bronfman}, \&
  {Thaddeus}}]{1990AA...231..137D}
{de Geus}, E.~J., {Bronfman}, L., \& {Thaddeus}, P. 1990, \aap, 231, 137

\bibitem[{{de Zeeuw} {et~al.}(1999){de Zeeuw}, {Hoogerwerf}, {de Bruijne},
  {Brown}, \& {Blaauw}}]{1999AJ....117..354D}
{de Zeeuw}, P.~T., {Hoogerwerf}, R., {de Bruijne}, J.~H.~J., {Brown}, A.~G.~A.,
  \& {Blaauw}, A. 1999, \aj, 117, 354

\bibitem[{{Dent} {et~al.}(1998){Dent}, {Matthews}, \&
  {Ward-Thompson}}]{1998MNRAS.301.1049D}
{Dent}, W.~R.~F., {Matthews}, H.~E., \& {Ward-Thompson}, D. 1998, \mnras, 301,
  1049

\bibitem[{{Di Francesco} {et~al.}(2008){Di Francesco}, {Johnstone}, {Kirk},
  {MacKenzie}, \& {Ledwosinska}}]{2008ApJS..175..277D}
{Di Francesco}, J., {Johnstone}, D., {Kirk}, H., {MacKenzie}, T., \&
  {Ledwosinska}, E. 2008, \apjs, 175, 277

\bibitem[{{Dionatos} \& {G{\"u}del}(2017)}]{2017AA...597A..64D}
{Dionatos}, O., \& {G{\"u}del}, M. 2017, \aap, 597, A64

\bibitem[{{Dionatos} {et~al.}(2013){Dionatos}, {J{\o}rgensen}, {Green},
  {Herczeg}, {Evans}, {Kristensen}, {Lindberg}, \& {van Dishoeck}}]{dionatos13}
{Dionatos}, O., {J{\o}rgensen}, J.~K., {Green}, J.~D., {et~al.} 2013, \aap,
  558, A88

\bibitem[{{Drabek} {et~al.}(2012){Drabek}, {Hatchell}, {Friberg}, {Richer},
  {Graves}, {Buckle}, {Nutter}, {Johnstone}, \& {Di
  Francesco}}]{2012MNRAS.426...23D}
{Drabek}, E., {Hatchell}, J., {Friberg}, P., {et~al.} 2012, \mnras, 426, 23

\bibitem[{{Dunham} {et~al.}(2014{\natexlab{a}}){Dunham}, {Arce}, {Mardones},
  {Lee}, {Matthews}, {Stutz}, \& {Williams}}]{2014ApJ...783...29D}
{Dunham}, M.~M., {Arce}, H.~G., {Mardones}, D., {et~al.} 2014{\natexlab{a}},
  \apj, 783, 29

\bibitem[{{Dunham} {et~al.}(2010){Dunham}, {Evans}, {Terebey}, {Dullemond}, \&
  {Young}}]{dunham10}
{Dunham}, M.~M., {Evans}, N.~J., {Terebey}, S., {Dullemond}, C.~P., \& {Young},
  C.~H. 2010, \apj, 710, 470

\bibitem[{{Dunham} {et~al.}(2006){Dunham}, {Evans}, {Bourke}, {Dullemond},
  {Young}, {Brooke}, {Chapman}, {Myers}, {Porras}, {Spiesman}, {Teuben}, \&
  {Wahhaj}}]{dunham06}
{Dunham}, M.~M., {Evans}, II, N.~J., {Bourke}, T.~L., {et~al.} 2006, \apj, 651,
  945

\bibitem[{{Dunham} {et~al.}(2013){Dunham}, {Arce}, {Allen}, {Evans},
  {Broekhoven-Fiene}, {Chapman}, {Cieza}, {Gutermuth}, {Harvey}, {Hatchell},
  {Huard}, {Kirk}, {Matthews}, {Mer{\'{\i}}n}, {Miller}, {Peterson}, \&
  {Spezzi}}]{dunham13}
{Dunham}, M.~M., {Arce}, H.~G., {Allen}, L.~E., {et~al.} 2013, \aj, 145, 94

\bibitem[{{Dunham} {et~al.}(2014{\natexlab{b}}){Dunham}, {Stutz}, {Allen},
  {Evans}, {Fischer}, {Megeath}, {Myers}, {Offner}, {Poteet}, {Tobin}, \&
  {Vorobyov}}]{dunham14a}
{Dunham}, M.~M., {Stutz}, A.~M., {Allen}, L.~E., {et~al.} 2014{\natexlab{b}},
  Protostars and Planets VI, 195

\bibitem[{{Dunham} {et~al.}(2015){Dunham}, {Allen}, {Evans},
  {Broekhoven-Fiene}, {Cieza}, {Di Francesco}, {Gutermuth}, {Harvey},
  {Hatchell}, {Heiderman}, {Huard}, {Johnstone}, {Kirk}, {Matthews}, {Miller},
  {Peterson}, \& {Young}}]{2015ApJS..220...11D}
{Dunham}, M.~M., {Allen}, L.~E., {Evans}, II, N.~J., {et~al.} 2015, \apjs, 220,
  11

\bibitem[{{Eggen}(1980)}]{1980ApJ...238..919E}
{Eggen}, O.~J. 1980, \apj, 238, 919

\bibitem[{{Eisner} {et~al.}(2005){Eisner}, {Hillenbrand}, {Carpenter}, \&
  {Wolf}}]{2005ApJ...635..396E}
{Eisner}, J.~A., {Hillenbrand}, L.~A., {Carpenter}, J.~M., \& {Wolf}, S. 2005,
  \apj, 635, 396

\bibitem[{{Enoch} {et~al.}(2008){Enoch}, {Evans}, {Sargent}, {Glenn},
  {Rosolowsky}, \& {Myers}}]{2008ApJ...684.1240E}
{Enoch}, M.~L., {Evans}, II, N.~J., {Sargent}, A.~I., {et~al.} 2008, \apj, 684,
  1240

\bibitem[{{Enoch} {et~al.}(2006){Enoch}, {Young}, {Glenn}, {Evans}, {Golwala},
  {Sargent}, {Harvey}, {Aguirre}, {Goldin}, {Haig}, {Huard}, {Lange},
  {Laurent}, {Maloney}, {Mauskopf}, {Rossinot}, \& {Sayers}}]{enoch06}
{Enoch}, M.~L., {Young}, K.~E., {Glenn}, J., {et~al.} 2006, \apj, 638, 293

\bibitem[{{Eric Jones} {et~al.}(2001--){Eric Jones}, {Pearu Peterson},
  {et~al.}}]{scipy}
{Eric Jones}, T., {Pearu Peterson}, {et~al.} 2001--, {SciPy}: Open source
  scientific tools for {Python}, ,

\bibitem[{{Evans} {et~al.}(2009{\natexlab{a}}){Evans}, {Dunham},
  {J{\o}rgensen}, {Enoch}, {Mer{\'{\i}}n}, {van Dishoeck}, {Alcal{\'a}},
  {Myers}, {Stapelfeldt}, {Huard}, {Allen}, {Harvey}, {van Kempen}, {Blake},
  {Koerner}, {Mundy}, {Padgett}, \& {Sargent}}]{evans09}
{Evans}, N.~J., {Dunham}, M.~M., {J{\o}rgensen}, J.~K., {et~al.}
  2009{\natexlab{a}}, \apjs, 181, 321

\bibitem[{{Evans}(1999)}]{1999ARAA..37..311E}
{Evans}, II, N.~J. 1999, \araa, 37, 311

\bibitem[{{Evans} {et~al.}(2003){Evans}, {Allen}, {Blake}, {Boogert}, {Bourke},
  {Harvey}, {Kessler}, {Koerner}, {Lee}, {Mundy}, {Myers}, {Padgett},
  {Pontoppidan}, {Sargent}, {Stapelfeldt}, {van Dishoeck}, {Young}, \&
  {Young}}]{2003PASP..115..965E}
{Evans}, II, N.~J., {Allen}, L.~E., {Blake}, G.~A., {et~al.} 2003, \pasp, 115,
  965

\bibitem[{{Evans} {et~al.}(2009{\natexlab{b}}){Evans}, {Dunham},
  {J{\o}rgensen}, {Enoch}, {Mer{\'{\i}}n}, {van Dishoeck}, {Alcal{\'a}},
  {Myers}, {Stapelfeldt}, {Huard}, {Allen}, {Harvey}, {van Kempen}, {Blake},
  {Koerner}, {Mundy}, {Padgett}, \& {Sargent}}]{2009ApJS..181..321E}
{Evans}, II, N.~J., {Dunham}, M.~M., {J{\o}rgensen}, J.~K., {et~al.}
  2009{\natexlab{b}}, \apjs, 181, 321

\bibitem[{{Fedele} {et~al.}(2013){Fedele}, {Bruderer}, {van Dishoeck}, {Carr},
  {Herczeg}, {Salyk}, {Evans}, {Bouwman}, {Meeus}, {Henning}, {Green},
  {Najita}, \& {G{\"u}del}}]{2013AA...559A..77F}
{Fedele}, D., {Bruderer}, S., {van Dishoeck}, E.~F., {et~al.} 2013, \aap, 559,
  A77

\bibitem[{{Fischer} {et~al.}(2013){Fischer}, {Megeath}, {Stutz}, {Tobin},
  {Ali}, {Stanke}, {Osorio}, {Furlan}, {HOPS Team}, \& {Orion Protostar
  Survey}}]{fischer13}
{Fischer}, W.~J., {Megeath}, S.~T., {Stutz}, A.~M., {et~al.} 2013,
  Astronomische Nachrichten, 334, 53

\bibitem[{{Fischer} {et~al.}(2017){Fischer}, {Megeath}, {Furlan}, {Ali},
  {Stutz}, {Tobin}, {Osorio}, {Stanke}, {Manoj}, {Poteet}, {Booker},
  {Hartmann}, {Wilson}, {Myers}, \& {Watson}}]{2017ApJ...840...69F}
{Fischer}, W.~J., {Megeath}, S.~T., {Furlan}, E., {et~al.} 2017, \apj, 840, 69

\bibitem[{{Flower} \& {Pineau Des For{\^e}ts}(2010)}]{flower10}
{Flower}, D.~R., \& {Pineau Des For{\^e}ts}, G. 2010, \mnras, 406, 1745

\bibitem[{{Furlan} {et~al.}(2008){Furlan}, {McClure}, {Calvet}, {Hartmann},
  {D'Alessio}, {Forrest}, {Watson}, {Uchida}, {Sargent}, {Green}, \&
  {Herter}}]{2008ApJS..176..184F}
{Furlan}, E., {McClure}, M., {Calvet}, N., {et~al.} 2008, \apjs, 176, 184

\bibitem[{{Gee} {et~al.}(1985){Gee}, {Griffin}, {Cunningham}, {Emerson}, {Ade},
  \& {Caroff}}]{1985MNRAS.215P..15G}
{Gee}, G., {Griffin}, M.~J., {Cunningham}, T., {et~al.} 1985, \mnras, 215, 15P

\bibitem[{{Giannini} {et~al.}(2001){Giannini}, {Nisini}, \&
  {Lorenzetti}}]{giannini01}
{Giannini}, T., {Nisini}, B., \& {Lorenzetti}, D. 2001, \apj, 555, 40

\bibitem[{{Giannini} {et~al.}(1999){Giannini}, {Lorenzetti}, {Tommasi},
  {Nisini}, {Benedettini}, {Pezzuto}, {Strafella}, {Barlow}, {Clegg}, {Cohen},
  {di Giorgio}, {Liseau}, {Molinari}, {Palla}, {Saraceno}, {Smith},
  {Spinoglio}, \& {White}}]{giannini99}
{Giannini}, T., {Lorenzetti}, D., {Tommasi}, E., {et~al.} 1999, \aap, 346, 617

\bibitem[{{Goicoechea} {et~al.}(2012){Goicoechea}, {Cernicharo}, {Karska},
  {Herczeg}, {Polehampton}, {Wampfler}, {Kristensen}, {van Dishoeck},
  {Etxaluze}, {Bern{\'e}}, \& {Visser}}]{goi12}
{Goicoechea}, J.~R., {Cernicharo}, J., {Karska}, A., {et~al.} 2012, \aap, 548,
  A77

\bibitem[{{Goldsmith} \& {Langer}(1999)}]{goldsmith99}
{Goldsmith}, P.~F., \& {Langer}, W.~D. 1999, \apj, 517, 209

\bibitem[{{Goldsmith} {et~al.}(1984){Goldsmith}, {Snell}, {Hemeon-Heyer}, \&
  {Langer}}]{1984ApJ...286..599G}
{Goldsmith}, P.~F., {Snell}, R.~L., {Hemeon-Heyer}, M., \& {Langer}, W.~D.
  1984, \apj, 286, 599

\bibitem[{{Green} {et~al.}(2013{\natexlab{a}}){Green}, {Evans},
  {K{\'o}sp{\'a}l}, {Herczeg}, {Quanz}, {Henning}, {van Kempen}, {Lee},
  {Dunham}, {Meeus}, {Bouwman}, {Chen}, {G{\"u}del}, {Skinner}, {Liebhart}, \&
  {Merello}}]{green13c}
{Green}, J.~D., {Evans}, II, N.~J., {K{\'o}sp{\'a}l}, {\'A}., {et~al.}
  2013{\natexlab{a}}, \apj, 772, 117

\bibitem[{{Green} {et~al.}(2013{\natexlab{b}}){Green}, {Evans}, {J{\o}rgensen},
  {Herczeg}, {Kristensen}, {Lee}, {Dionatos}, {Yildiz}, {Salyk}, {Meeus},
  {Bouwman}, {Visser}, {Bergin}, {van Dishoeck}, {Rascati}, {Karska}, {van
  Kempen}, {Dunham}, {Lindberg}, {Fedele}, \& {DIGIT Team}}]{green13b}
{Green}, J.~D., {Evans}, II, N.~J., {J{\o}rgensen}, J.~K., {et~al.}
  2013{\natexlab{b}}, \apj, 770, 123

\bibitem[{{Green} {et~al.}(2016{\natexlab{a}}){Green}, {Yang}, {Evans},
  {Karska}, {Herczeg}, {van Dishoeck}, {Lee}, {Larson}, \&
  {Bouwman}}]{green16a}
{Green}, J.~D., {Yang}, Y.-L., {Evans}, II, N.~J., {et~al.} 2016{\natexlab{a}},
  \aj, 151, 75

\bibitem[{{Green} {et~al.}(2016{\natexlab{b}}){Green}, {Jones}, {Keller},
  {Poteet}, {Yang}, {Fischer}, {Evans}, {Sargent}, \&
  {Rebull}}]{2016ApJ...832....4G}
{Green}, J.~D., {Jones}, O.~C., {Keller}, L.~D., {et~al.} 2016{\natexlab{b}},
  \apj, 832, 4

\bibitem[{{Greene} {et~al.}(1994){Greene}, {Wilking}, {Andre}, {Young}, \&
  {Lada}}]{greene94}
{Greene}, T.~P., {Wilking}, B.~A., {Andre}, P., {Young}, E.~T., \& {Lada},
  C.~J. 1994, \apj, 434, 614

\bibitem[{{Griffin} {et~al.}(2010){Griffin}, {Abergel}, {Abreu}, {Ade},
  {Andr{\'e}}, {Augueres}, {Babbedge}, {Bae}, {Baillie}, {Baluteau}, {Barlow},
  {Bendo}, {Benielli}, {Bock}, {Bonhomme}, {Brisbin}, {Brockley-Blatt},
  {Caldwell}, {Cara}, {Castro-Rodriguez}, {Cerulli}, {Chanial}, {Chen},
  {Clark}, {Clements}, {Clerc}, {Coker}, {Communal}, {Conversi}, {Cox},
  {Crumb}, {Cunningham}, {Daly}, {Davis}, {de Antoni}, {Delderfield}, {Devin},
  {di Giorgio}, {Didschuns}, {Dohlen}, {Donati}, {Dowell}, {Dowell}, {Duband},
  {Dumaye}, {Emery}, {Ferlet}, {Ferrand}, {Fontignie}, {Fox}, {Franceschini},
  {Frerking}, {Fulton}, {Garcia}, {Gastaud}, {Gear}, {Glenn}, {Goizel},
  {Griffin}, {Grundy}, {Guest}, {Guillemet}, {Hargrave}, {Harwit}, {Hastings},
  {Hatziminaoglou}, {Herman}, {Hinde}, {Hristov}, {Huang}, {Imhof}, {Isaak},
  {Israelsson}, {Ivison}, {Jennings}, {Kiernan}, {King}, {Lange}, {Latter},
  {Laurent}, {Laurent}, {Leeks}, {Lellouch}, {Levenson}, {Li}, {Li},
  {Lilienthal}, {Lim}, {Liu}, {Lu}, {Madden}, {Mainetti}, {Marliani}, {McKay},
  {Mercier}, {Molinari}, {Morris}, {Moseley}, {Mulder}, {Mur}, {Naylor},
  {Nguyen}, {O'Halloran}, {Oliver}, {Olofsson}, {Olofsson}, {Orfei}, {Page},
  {Pain}, {Panuzzo}, {Papageorgiou}, {Parks}, {Parr-Burman}, {Pearce},
  {Pearson}, {P{\'e}rez-Fournon}, {Pinsard}, {Pisano}, {Podosek}, {Pohlen},
  {Polehampton}, {Pouliquen}, {Rigopoulou}, {Rizzo}, {Roseboom}, {Roussel},
  {Rowan-Robinson}, {Rownd}, {Saraceno}, {Sauvage}, {Savage}, {Savini},
  {Sawyer}, {Scharmberg}, {Schmitt}, {Schneider}, {Schulz}, {Schwartz},
  {Shafer}, {Shupe}, {Sibthorpe}, {Sidher}, {Smith}, {Smith}, {Smith},
  {Spencer}, {Stobie}, {Sudiwala}, {Sukhatme}, {Surace}, {Stevens}, {Swinyard},
  {Trichas}, {Tourette}, {Triou}, {Tseng}, {Tucker}, {Turner}, {Vaccari},
  {Valtchanov}, {Vigroux}, {Virique}, {Voellmer}, {Walker}, {Ward}, {Waskett},
  {Weilert}, {Wesson}, {White}, {Whitehouse}, {Wilson}, {Winter}, {Woodcraft},
  {Wright}, {Xu}, {Zavagno}, {Zemcov}, {Zhang}, \& {Zonca}}]{griffin10}
{Griffin}, M.~J., {Abergel}, A., {Abreu}, A., {et~al.} 2010, \aap, 518, L3+

\bibitem[{{Groppi} {et~al.}(2007){Groppi}, {Hunter}, {Blundell}, \&
  {Sandell}}]{2007ApJ...670..489G}
{Groppi}, C.~E., {Hunter}, T.~R., {Blundell}, R., \& {Sandell}, G. 2007, \apj,
  670, 489

\bibitem[{{Gutermuth} {et~al.}(2009){Gutermuth}, {Megeath}, {Myers}, {Allen},
  {Pipher}, \& {Fazio}}]{2009ApJS..184...18G}
{Gutermuth}, R.~A., {Megeath}, S.~T., {Myers}, P.~C., {et~al.} 2009, \apjs,
  184, 18

\bibitem[{{Haisch} {et~al.}(2004){Haisch}, {Greene}, {Barsony}, \&
  {Stahler}}]{haisch04}
{Haisch}, Jr., K.~E., {Greene}, T.~P., {Barsony}, M., \& {Stahler}, S.~W. 2004,
  \aj, 127, 1747

\bibitem[{{Harsono} {et~al.}(2014){Harsono}, {J{\o}rgensen}, {van Dishoeck},
  {Hogerheijde}, {Bruderer}, {Persson}, \& {Mottram}}]{2014AA...562A..77H}
{Harsono}, D., {J{\o}rgensen}, J.~K., {van Dishoeck}, E.~F., {et~al.} 2014,
  \aap, 562, A77

\bibitem[{{Helou} \& {Walker}(1988)}]{1988iras....7.....H}
{Helou}, G., \& {Walker}, D.~W., eds. 1988, {Infrared astronomical satellite
  (IRAS) catalogs and atlases. Volume 7: The small scale structure catalog},
  Vol.~7, 1--265

\bibitem[{{Henden} {et~al.}(2016){Henden}, {Templeton}, {Terrell}, {Smith},
  {Levine}, \& {Welch}}]{2016yCat.2336....0H}
{Henden}, A.~A., {Templeton}, M., {Terrell}, D., {et~al.} 2016, VizieR Online
  Data Catalog, 2336

\bibitem[{{Henning} {et~al.}(1993){Henning}, {Pfau}, {Zinnecker}, \&
  {Prusti}}]{1993AA...276..129H}
{Henning}, T., {Pfau}, W., {Zinnecker}, H., \& {Prusti}, T. 1993, \aap, 276,
  129

\bibitem[{{Herbig}(1977)}]{herbig77}
{Herbig}, G.~H. 1977, \apj, 217, 693

\bibitem[{{Herczeg} {et~al.}(2012){Herczeg}, {Karska}, {Bruderer},
  {Kristensen}, {van Dishoeck}, {J{\o}rgensen}, {Visser}, {Wampfler}, {Bergin},
  {Y{\i}ld{\i}z}, {Pontoppidan}, \& {Gracia-Carpio}}]{herczeg12}
{Herczeg}, G.~J., {Karska}, A., {Bruderer}, S., {et~al.} 2012, \aap, 540, A84

\bibitem[{{Hirota} {et~al.}(2011){Hirota}, {Honma}, {Imai}, {Sunada}, {Ueno},
  {Kobayashi}, \& {Kawaguchi}}]{hirota11}
{Hirota}, T., {Honma}, M., {Imai}, H., {et~al.} 2011, \pasj, 63, 1

\bibitem[{{Hogerheijde} \& {Sandell}(2000)}]{2000ApJ...534..880H}
{Hogerheijde}, M.~R., \& {Sandell}, G. 2000, \apj, 534, 880

\bibitem[{{Hollenbach}(1985)}]{hollenbach85}
{Hollenbach}, D. 1985, Icarus, 61, 36

\bibitem[{{Hughes} \& {Hartigan}(1992)}]{1992AJ....104..680H}
{Hughes}, J., \& {Hartigan}, P. 1992, \aj, 104, 680

\bibitem[{{Ishihara} {et~al.}(2010){Ishihara}, {Onaka}, {Kataza}, {Salama},
  {Alfageme}, {Cassatella}, {Cox}, {Garcia-Lario}, {Stephenson}, {Cohen},
  {Fujishiro}, {Fujiwara}, {Hasegawa}, {Ita}, {Kim}, {Matsuhara}, {Murakami},
  {Muller}, {Nakagawa}, {Ohyama}, {Oyabu}, {Pyo}, {Sakon}, {Shibai}, {Takita},
  {Tanab}, {Uemizu}, {Ueno}, {Usui}, {Wada}, {Watarai}, {Yamamura}, \&
  {Yamauchi}}]{2010yCat.2297....0I}
{Ishihara}, D., {Onaka}, T., {Kataza}, H., {et~al.} 2010, VizieR Online Data
  Catalog, 2297

\bibitem[{{Isobe} \& {Feigelson}(1990)}]{isobe90}
{Isobe}, T., \& {Feigelson}, E.~D. 1990, in \baas, Vol.~22, Bulletin of the
  American Astronomical Society, 917--918

\bibitem[{{Isobe} {et~al.}(1986){Isobe}, {Feigelson}, \& {Nelson}}]{isobe86}
{Isobe}, T., {Feigelson}, E.~D., \& {Nelson}, P.~I. 1986, \apj, 306, 490

\bibitem[{{Jim{\'e}nez-Donaire} {et~al.}(2017){Jim{\'e}nez-Donaire}, {Meeus},
  {Karska}, {Montesinos}, {Bouwman}, {Eiroa}, \&
  {Henning}}]{2017AA...605A..62J}
{Jim{\'e}nez-Donaire}, M.~J., {Meeus}, G., {Karska}, A., {et~al.} 2017, \aap,
  605, A62

\bibitem[{{J{\o}rgensen} {et~al.}(2008){J{\o}rgensen}, {Johnstone}, {Kirk},
  {Myers}, {Allen}, \& {Shirley}}]{2008ApJ...683..822J}
{J{\o}rgensen}, J.~K., {Johnstone}, D., {Kirk}, H., {et~al.} 2008, \apj, 683,
  822

\bibitem[{{J{\o}rgensen} {et~al.}(2002){J{\o}rgensen}, {Sch{\"o}ier}, \& {van
  Dishoeck}}]{2002AA...389..908J}
{J{\o}rgensen}, J.~K., {Sch{\"o}ier}, F.~L., \& {van Dishoeck}, E.~F. 2002,
  \aap, 389, 908

\bibitem[{{J{\o}rgensen} {et~al.}(2004){J{\o}rgensen}, {Sch{\"o}ier}, \& {van
  Dishoeck}}]{2004AA...416..603J}
---. 2004, \aap, 416, 603

\bibitem[{{J{\o}rgensen} {et~al.}(2009){J{\o}rgensen}, {van Dishoeck},
  {Visser}, {Bourke}, {Wilner}, {Lommen}, {Hogerheijde}, \&
  {Myers}}]{jorgensen09}
{J{\o}rgensen}, J.~K., {van Dishoeck}, E.~F., {Visser}, R., {et~al.} 2009,
  \aap, 507, 861

\bibitem[{{J{\o}rgensen} {et~al.}(2007){J{\o}rgensen}, {Bourke}, {Myers}, {Di
  Francesco}, {van Dishoeck}, {Lee}, {Ohashi}, {Sch{\"o}ier}, {Takakuwa},
  {Wilner}, \& {Zhang}}]{jorgensen07a}
{J{\o}rgensen}, J.~K., {Bourke}, T.~L., {Myers}, P.~C., {et~al.} 2007, \apj,
  659, 479

\bibitem[{{Karska} {et~al.}(2013){Karska}, {Herczeg}, {van Dishoeck},
  {Wampfler}, {Kristensen}, {Goicoechea}, {Visser}, {Nisini}, {San
  Jos{\'e}-Garc{\'{\i}}a}, {Bruderer}, {{\'S}niady}, {Doty}, {Fedele},
  {Y{\i}ld{\i}z}, {Benz}, {Bergin}, {Caselli}, {Herpin}, {Hogerheijde},
  {Johnstone}, {J{\o}rgensen}, {Liseau}, {Tafalla}, {van der Tak}, \&
  {Wyrowski}}]{karska13}
{Karska}, A., {Herczeg}, G.~J., {van Dishoeck}, E.~F., {et~al.} 2013, \aap,
  552, A141

\bibitem[{{Karska} {et~al.}(2014){Karska}, {Kristensen}, {van Dishoeck},
  {Drozdovskaya}, {Mottram}, {Herczeg}, {Bruderer}, {Cabrit}, {Evans},
  {Fedele}, {Gusdorf}, {J{\o}rgensen}, {Kaufman}, {Melnick}, {Neufeld},
  {Nisini}, {Santangelo}, {Tafalla}, \& {Wampfler}}]{karska14}
{Karska}, A., {Kristensen}, L.~E., {van Dishoeck}, E.~F., {et~al.} 2014, \aap,
  572, A9

\bibitem[{{Kaufman} \& {Neufeld}(1996)}]{kaufman96}
{Kaufman}, M.~J., \& {Neufeld}, D.~A. 1996, \apj, 456, 611

\bibitem[{{Keene} {et~al.}(1983){Keene}, {Davidson}, {Harper}, {Hildebrand},
  {Jaffe}, {Loewenstein}, {Low}, \& {Pernic}}]{1983ApJ...274L..43K}
{Keene}, J., {Davidson}, J.~A., {Harper}, D.~A., {et~al.} 1983, \apjl, 274, L43

\bibitem[{{Kenyon} {et~al.}(1993){Kenyon}, {Calvet}, \&
  {Hartmann}}]{1993ApJ...414..676K}
{Kenyon}, S.~J., {Calvet}, N., \& {Hartmann}, L. 1993, \apj, 414, 676

\bibitem[{{Kenyon} {et~al.}(1994){Kenyon}, {Dobrzycka}, \&
  {Hartmann}}]{1994AJ....108.1872K}
{Kenyon}, S.~J., {Dobrzycka}, D., \& {Hartmann}, L. 1994, \aj, 108, 1872

\bibitem[{{Kirk} {et~al.}(2009){Kirk}, {Ward-Thompson}, {Di Francesco},
  {Bourke}, {Evans}, {Mer{\'{\i}}n}, {Allen}, {Cieza}, {Dunham}, {Harvey},
  {Huard}, {J{\o}rgensen}, {Miller}, {Noriega-Crespo}, {Peterson}, {Ray}, \&
  {Rebull}}]{2009ApJS..185..198K}
{Kirk}, J.~M., {Ward-Thompson}, D., {Di Francesco}, J., {et~al.} 2009, \apjs,
  185, 198

\bibitem[{{Kirk} {et~al.}(2013){Kirk}, {Ward-Thompson}, {Palmeirim},
  {Andr{\'e}}, {Griffin}, {Hargrave}, {K{\"o}nyves}, {Bernard}, {Nutter},
  {Sibthorpe}, {Di Francesco}, {Abergel}, {Arzoumanian}, {Benedettini},
  {Bontemps}, {Elia}, {Hennemann}, {Hill}, {Men'shchikov}, {Motte},
  {Nguyen-Luong}, {Peretto}, {Pezzuto}, {Rygl}, {Sadavoy}, {Schisano},
  {Schneider}, {Testi}, \& {White}}]{2013MNRAS.432.1424K}
{Kirk}, J.~M., {Ward-Thompson}, D., {Palmeirim}, P., {et~al.} 2013, \mnras,
  432, 1424

\bibitem[{{Knude} \& {Hog}(1998)}]{1998AA...338..897K}
{Knude}, J., \& {Hog}, E. 1998, \aap, 338, 897

\bibitem[{{Kristensen} {et~al.}(2017{\natexlab{a}}){Kristensen}, {Gusdorf},
  {Mottram}, {Karska}, {Visser}, {Wiesemeyer}, {G{\"u}sten}, \&
  {Simon}}]{2017AA...601L...4K}
{Kristensen}, L.~E., {Gusdorf}, A., {Mottram}, J.~C., {et~al.}
  2017{\natexlab{a}}, \aap, 601, L4

\bibitem[{{Kristensen} {et~al.}(2013){Kristensen}, {van Dishoeck}, {Benz},
  {Bruderer}, {Visser}, \& {Wampfler}}]{2013AA...557A..23K}
{Kristensen}, L.~E., {van Dishoeck}, E.~F., {Benz}, A.~O., {et~al.} 2013, \aap,
  557, A23

\bibitem[{{Kristensen} {et~al.}(2010){Kristensen}, {Visser}, {van Dishoeck},
  {Y{\i}ld{\i}z}, {Doty}, {Herczeg}, {Liu}, {Parise}, {J{\o}rgensen}, {van
  Kempen}, {Brinch}, {Wampfler}, {Bruderer}, {Benz}, {Hogerheijde}, {Deul},
  {Bachiller}, {Baudry}, {Benedettini}, {Bergin}, {Bjerkeli}, {Blake},
  {Bontemps}, {Braine}, {Caselli}, {Cernicharo}, {Codella}, {Daniel}, {de
  Graauw}, {di Giorgio}, {Dominik}, {Encrenaz}, {Fich}, {Fuente}, {Giannini},
  {Goicoechea}, {Helmich}, {Herpin}, {Jacq}, {Johnstone}, {Kaufman}, {Larsson},
  {Lis}, {Liseau}, {Marseille}, {McCoey}, {Melnick}, {Neufeld}, {Nisini},
  {Olberg}, {Pearson}, {Plume}, {Risacher}, {Santiago-Garc{\'{\i}}a},
  {Saraceno}, {Shipman}, {Tafalla}, {Tielens}, {van der Tak}, {Wyrowski},
  {Beintema}, {de Jonge}, {Dieleman}, {Ossenkopf}, {Roelfsema}, {Stutzki}, \&
  {Whyborn}}]{2010AA...521L..30K}
{Kristensen}, L.~E., {Visser}, R., {van Dishoeck}, E.~F., {et~al.} 2010, \aap,
  521, L30

\bibitem[{{Kristensen} {et~al.}(2012){Kristensen}, {van Dishoeck}, {Bergin},
  {Visser}, {Y{\i}ld{\i}z}, {San Jose-Garcia}, {J{\o}rgensen}, {Herczeg},
  {Johnstone}, {Wampfler}, {Benz}, {Bruderer}, {Cabrit}, {Caselli}, {Doty},
  {Harsono}, {Herpin}, {Hogerheijde}, {Karska}, {van Kempen}, {Liseau},
  {Nisini}, {Tafalla}, {van der Tak}, \& {Wyrowski}}]{kristensen12}
{Kristensen}, L.~E., {van Dishoeck}, E.~F., {Bergin}, E.~A., {et~al.} 2012,
  \aap, 542, A8

\bibitem[{{Kristensen} {et~al.}(2017{\natexlab{b}}){Kristensen}, {van
  Dishoeck}, {Mottram}, {Karska}, {Y{\i}ld{\i}z}, {Bergin}, {Bjerkeli},
  {Cabrit}, {Doty}, {Evans}, {Gusdorf}, {Harsono}, {Herczeg}, {Johnstone},
  {J{\o}rgensen}, {van Kempen}, {Lee}, {Maret}, {Tafalla}, {Visser}, \&
  {Wampfler}}]{2017arXiv170510269K}
{Kristensen}, L.~E., {van Dishoeck}, E.~F., {Mottram}, J.~C., {et~al.}
  2017{\natexlab{b}}, \aap, 605, A93

\bibitem[{{Langer} \& {Penzias}(1993)}]{1993ApJ...408..539L}
{Langer}, W.~D., \& {Penzias}, A.~A. 1993, \apj, 408, 539

\bibitem[{{Launhardt} {et~al.}(2010){Launhardt}, {Nutter}, {Ward-Thompson},
  {Bourke}, {Henning}, {Khanzadyan}, {Schmalzl}, {Wolf}, \&
  {Zylka}}]{2010ApJS..188..139L}
{Launhardt}, R., {Nutter}, D., {Ward-Thompson}, D., {et~al.} 2010, \apjs, 188,
  139

\bibitem[{{Lavalley} {et~al.}(1992){Lavalley}, {Isobe}, \&
  {Feigelson}}]{lavalley92}
{Lavalley}, M.~P., {Isobe}, T., \& {Feigelson}, E.~D. 1992, in \baas, Vol.~24,
  Bulletin of the American Astronomical Society, 839--840

\bibitem[{{Lee} {et~al.}(2015{\natexlab{a}}){Lee}, {Hirano}, {Zhang}, {Shang},
  {Ho}, \& {Mizuno}}]{2015ApJ...805..186L}
{Lee}, C.-F., {Hirano}, N., {Zhang}, Q., {et~al.} 2015{\natexlab{a}}, \apj,
  805, 186

\bibitem[{{Lee} {et~al.}(2014){Lee}, {Lee}, {Lee}, {Evans}, \&
  {Green}}]{2014ApJS..214...21L}
{Lee}, J.-E., {Lee}, J., {Lee}, S., {Evans}, II, N.~J., \& {Green}, J.~D. 2014,
  \apjs, 214, 21

\bibitem[{{Lee} {et~al.}(2015{\natexlab{b}}){Lee}, {Lee}, \&
  {Bergin}}]{2015ApJS..217...30L}
{Lee}, S., {Lee}, J.-E., \& {Bergin}, E.~A. 2015{\natexlab{b}}, \apjs, 217, 30

\bibitem[{{Lindberg} {et~al.}(2014){Lindberg}, {J{\o}rgensen}, {Green},
  {Herczeg}, {Dionatos}, {Evans}, {Karska}, \& {Wampfler}}]{2014AA...565A..29L}
{Lindberg}, J.~E., {J{\o}rgensen}, J.~K., {Green}, J.~D., {et~al.} 2014, \aap,
  565, A29

\bibitem[{{Lindberg} {et~al.}(2011){Lindberg}, {Jorgensen}, \& {Herschel DIGIT
  Team}}]{lindberg11}
{Lindberg}, J.~E., {Jorgensen}, J.~K., \& {Herschel DIGIT Team}. 2011, in IAU
  Symposium, Vol. 280, IAU Symposium, 235P

\bibitem[{{Lorenzetti} {et~al.}(1999){Lorenzetti}, {Tommasi}, {Giannini},
  {Nisini}, {Benedettini}, {Pezzuto}, {Strafella}, {Barlow}, {Clegg}, {Cohen},
  {di Giorgio}, {Liseau}, {Molinari}, {Palla}, {Saraceno}, {Smith},
  {Spinoglio}, \& {White}}]{lorenzetti99}
{Lorenzetti}, D., {Tommasi}, E., {Giannini}, T., {et~al.} 1999, \aap, 346, 604

\bibitem[{{Lorenzetti} {et~al.}(2000){Lorenzetti}, {Giannini}, {Nisini},
  {Benedettini}, {Creech-Eakman}, {Blake}, {van Dishoeck}, {Cohen}, {Liseau},
  {Molinari}, {Pezzuto}, {Saraceno}, {Smith}, {Spinoglio}, \&
  {White}}]{lorenzetti00}
{Lorenzetti}, D., {Giannini}, T., {Nisini}, B., {et~al.} 2000, \aap, 357, 1035

\bibitem[{{Luhman}(2008)}]{2008hsf2.book..169L}
{Luhman}, K.~L. 2008, {Chamaeleon}, ed. B.~{Reipurth}, 169

\bibitem[{{Maheswar} {et~al.}(2004){Maheswar}, {Manoj}, \&
  {Bhatt}}]{2004MNRAS.355.1272M}
{Maheswar}, G., {Manoj}, P., \& {Bhatt}, H.~C. 2004, \mnras, 355, 1272

\bibitem[{{Makiwa} {et~al.}(2013){Makiwa}, {Naylor}, {Ferlet}, {Salji},
  {Swinyard}, {Polehampton}, \& {van der Wiel}}]{2013ApOpt..52.3864M}
{Makiwa}, G., {Naylor}, D.~A., {Ferlet}, M., {et~al.} 2013, \ao, 52, 3864

\bibitem[{{Makiwa} {et~al.}(2016){Makiwa}, {Naylor}, {van der Wiel},
  {Ward-Thompson}, {Kirk}, {Eyres}, {Abergel}, \& {K{\"o}hler}}]{makiwa16}
{Makiwa}, G., {Naylor}, D.~A., {van der Wiel}, M.~H.~D., {et~al.} 2016, \mnras,
  458, 2150

\bibitem[{{Manoj} {et~al.}(2013){Manoj}, {Watson}, {Neufeld}, {Megeath},
  {Vavrek}, {Yu}, {Visser}, {Bergin}, {Fischer}, {Tobin}, {Stutz}, {Ali},
  {Wilson}, {Di Francesco}, {Osorio}, {Maret}, \& {Poteet}}]{manoj12}
{Manoj}, P., {Watson}, D.~M., {Neufeld}, D.~A., {et~al.} 2013, \apj, 763, 83

\bibitem[{{Manoj} {et~al.}(2016){Manoj}, {Green}, {Megeath}, {Evans}, {Stutz},
  {Tobin}, {Watson}, {Fischer}, {Furlan}, \& {Henning}}]{2016ApJ...831...69M}
{Manoj}, P., {Green}, J.~D., {Megeath}, S.~T., {et~al.} 2016, \apj, 831, 69

\bibitem[{{Matuszak} {et~al.}(2015){Matuszak}, {Karska}, {Kristensen},
  {Herczeg}, {Tychoniec}, {van Kempen}, \& {Fuente}}]{2015AA...578A..20M}
{Matuszak}, M., {Karska}, A., {Kristensen}, L.~E., {et~al.} 2015, \aap, 578,
  A20

\bibitem[{{McClure} {et~al.}(2010){McClure}, {Furlan}, {Manoj}, {Luhman},
  {Watson}, {Forrest}, {Espaillat}, {Calvet}, {D'Alessio}, {Sargent}, {Tobin},
  \& {Chiang}}]{2010ApJS..188...75M}
{McClure}, M.~K., {Furlan}, E., {Manoj}, P., {et~al.} 2010, \apjs, 188, 75

\bibitem[{{Meeus} {et~al.}(2013){Meeus}, {Salyk}, {Bruderer}, {Fedele},
  {Maaskant}, {Evans}, {van Dishoeck}, {Montesinos}, {Herczeg}, {Bouwman},
  {Green}, {Dominik}, {Henning}, \& {Vicente}}]{2013AA...559A..84M}
{Meeus}, G., {Salyk}, C., {Bruderer}, S., {et~al.} 2013, \aap, 559, A84

\bibitem[{{Moriarty-Schieven} {et~al.}(1994){Moriarty-Schieven}, {Wannier},
  {Keene}, \& {Tamura}}]{1994ApJ...436..800M}
{Moriarty-Schieven}, G.~H., {Wannier}, P.~G., {Keene}, J., \& {Tamura}, M.
  1994, \apj, 436, 800

\bibitem[{{Moshir}(1989)}]{1989ifss.book.....M}
{Moshir}, M. 1989, {IRAS Faint Source Survey, Explanatory supplement version 1
  and tape}

\bibitem[{{Motte} \& {Andr{\'e}}(2001)}]{2001AA...365..440M}
{Motte}, F., \& {Andr{\'e}}, P. 2001, \aap, 365, 440

\bibitem[{{Mottram} {et~al.}(2014){Mottram}, {Kristensen}, {van Dishoeck},
  {Bruderer}, {San Jos{\'e}-Garc{\'{\i}}a}, {Karska}, {Visser}, {Santangelo},
  {Benz}, {Bergin}, {Caselli}, {Herpin}, {Hogerheijde}, {Johnstone}, {van
  Kempen}, {Liseau}, {Nisini}, {Tafalla}, {van der Tak}, \&
  {Wyrowski}}]{2014AA...572A..21M}
{Mottram}, J.~C., {Kristensen}, L.~E., {van Dishoeck}, E.~F., {et~al.} 2014,
  \aap, 572, A21

\bibitem[{{Mottram} {et~al.}(2017){Mottram}, {van Dishoeck}, {Kristensen},
  {Karska}, {San Jos{\'e}-Garc{\'{\i}}a}, {Khanna}, {Herczeg}, {Andr{\'e}},
  {Bontemps}, {Cabrit}, {Carney}, {Drozdovskaya}, {Dunham}, {Evans}, {Fedele},
  {Green}, {Harsono}, {Johnstone}, {J{\o}rgensen}, {K{\"o}nyves}, {Nisini},
  {Persson}, {Tafalla}, {Visser}, \& {Y{\i}ld{\i}z}}]{2017AA...600A..99M}
{Mottram}, J.~C., {van Dishoeck}, E.~F., {Kristensen}, L.~E., {et~al.} 2017,
  \aap, 600, A99

\bibitem[{{Myers} {et~al.}(1987){Myers}, {Fuller}, {Mathieu}, {Beichman},
  {Benson}, {Schild}, \& {Emerson}}]{1987ApJ...319..340M}
{Myers}, P.~C., {Fuller}, G.~A., {Mathieu}, R.~D., {et~al.} 1987, \apj, 319,
  340

\bibitem[{{Myers} \& {Ladd}(1993)}]{myers93}
{Myers}, P.~C., \& {Ladd}, E.~F. 1993, \apjl, 413, L47

\bibitem[{{Neufeld}(2012)}]{neufeld12}
{Neufeld}, D.~A. 2012, \apj, 749, 125

\bibitem[{{Neuh{\"a}user} \& {Forbrich}(2008)}]{neuhauser08}
{Neuh{\"a}user}, R., \& {Forbrich}, J. 2008, {The Corona Australis Star Forming
  Region} (Handbook of Star Forming Regions, Volume II, ed. Bo Reipurth), 735

\bibitem[{{Nisini} {et~al.}(2005){Nisini}, {Antoniucci}, {Giannini}, \&
  {Lorenzetti}}]{nisini05}
{Nisini}, B., {Antoniucci}, S., {Giannini}, T., \& {Lorenzetti}, D. 2005, \aap,
  429, 543

\bibitem[{{Nisini} {et~al.}(2002){Nisini}, {Giannini}, \&
  {Lorenzetti}}]{nisini02}
{Nisini}, B., {Giannini}, T., \& {Lorenzetti}, D. 2002, \apj, 574, 246

\bibitem[{{Nisini} {et~al.}(2010){Nisini}, {Benedettini}, {Codella},
  {Giannini}, {Liseau}, {Neufeld}, {Tafalla}, {van Dishoeck}, {Bachiller},
  {Baudry}, {Benz}, {Bergin}, {Bjerkeli}, {Blake}, {Bontemps}, {Braine},
  {Bruderer}, {Caselli}, {Cernicharo}, {Daniel}, {Encrenaz}, {di Giorgio},
  {Dominik}, {Doty}, {Fich}, {Fuente}, {Goicoechea}, {de Graauw}, {Helmich},
  {Herczeg}, {Herpin}, {Hogerheijde}, {Jacq}, {Johnstone}, {J{\o}rgensen},
  {Kaufman}, {Kristensen}, {Larsson}, {Lis}, {Marseille}, {McCoey}, {Melnick},
  {Olberg}, {Parise}, {Pearson}, {Plume}, {Risacher}, {Santiago}, {Saraceno},
  {Shipman}, {van Kempen}, {Visser}, {Viti}, {Wampfler}, {Wyrowski}, {van der
  Tak}, {Y{\i}ld{\i}z}, {Delforge}, {Desbat}, {Hatch}, {P{\'e}ron}, {Schieder},
  {Stern}, {Teyssier}, \& {Whyborn}}]{nisini10}
{Nisini}, B., {Benedettini}, M., {Codella}, C., {et~al.} 2010, \aap, 518, L120

\bibitem[{{Olofsson} {et~al.}(2009){Olofsson}, {Augereau}, {van Dishoeck},
  {Mer{\'{\i}}n}, {Lahuis}, {Kessler-Silacci}, {Dullemond}, {Oliveira},
  {Blake}, {Boogert}, {Brown}, {Evans}, {Geers}, {Knez}, {Monin}, \&
  {Pontoppidan}}]{olofsson09}
{Olofsson}, J., {Augereau}, J.-C., {van Dishoeck}, E.~F., {et~al.} 2009, \aap,
  507, 327

\bibitem[{{Ortiz-Le{\'o}n} {et~al.}(2017){Ortiz-Le{\'o}n}, {Loinard},
  {Kounkel}, {Dzib}, {Mioduszewski}, {Rodr{\'{\i}}guez}, {Torres},
  {Gonz{\'a}lez-L{\'o}pezlira}, {Pech}, {Rivera}, {Hartmann}, {Boden}, {Evans},
  {Brice{\~n}o}, {Tobin}, {Galli}, \& {Gudehus}}]{ortiz-leon17}
{Ortiz-Le{\'o}n}, G.~N., {Loinard}, L., {Kounkel}, M.~A., {et~al.} 2017, \apj,
  834, 141

\bibitem[{{Ott}(2010)}]{ott10}
{Ott}, S. 2010, in Astronomical Society of the Pacific Conference Series, Vol.
  434, Astronomical Data Analysis Software and Systems XIX, ed. Y.~{Mizumoto},
  K.-I. {Morita}, \& M.~{Ohishi}, 139

\bibitem[{{Parise} {et~al.}(2006){Parise}, {Belloche}, {Leurini}, {Schilke},
  {Wyrowski}, \& {G{\"u}sten}}]{2006AA...454L..79P}
{Parise}, B., {Belloche}, A., {Leurini}, S., {et~al.} 2006, \aap, 454, L79

\bibitem[{{Peterson} {et~al.}(2011){Peterson}, {Caratti o Garatti}, {Bourke},
  {Forbrich}, {Gutermuth}, {J{\o}rgensen}, {Allen}, {Patten}, {Dunham},
  {Harvey}, {Mer{\'{\i}}n}, {Chapman}, {Cieza}, {Huard}, {Knez}, {Prager}, \&
  {Evans}}]{2011ApJS..194...43P}
{Peterson}, D.~E., {Caratti o Garatti}, A., {Bourke}, T.~L., {et~al.} 2011,
  \apjs, 194, 43

\bibitem[{{Pilbratt} {et~al.}(2010){Pilbratt}, {Riedinger}, {Passvogel},
  {Crone}, {Doyle}, {Gageur}, {Heras}, {Jewell}, {Metcalfe}, {Ott}, \&
  {Schmidt}}]{pilbratt10}
{Pilbratt}, G.~L., {Riedinger}, J.~R., {Passvogel}, T., {et~al.} 2010, \aap,
  518, L1

\bibitem[{{Poglitsch} {et~al.}(2010){Poglitsch}, {Waelkens}, {Geis},
  {Feuchtgruber}, {Vandenbussche}, {Rodriguez}, {Krause}, {Renotte}, {van
  Hoof}, {Saraceno}, {Cepa}, {Kerschbaum}, {Agn{\`e}se}, {Ali}, {Altieri},
  {Andreani}, {Augueres}, {Balog}, {Barl}, {Bauer}, {Belbachir}, {Benedettini},
  {Billot}, {Boulade}, {Bischof}, {Blommaert}, {Callut}, {Cara}, {Cerulli},
  {Cesarsky}, {Contursi}, {Creten}, {De Meester}, {Doublier}, {Doumayrou},
  {Duband}, {Exter}, {Genzel}, {Gillis}, {Gr{\"o}zinger}, {Henning},
  {Herreros}, {Huygen}, {Inguscio}, {Jakob}, {Jamar}, {Jean}, {de Jong},
  {Katterloher}, {Kiss}, {Klaas}, {Lemke}, {Lutz}, {Madden}, {Marquet},
  {Martignac}, {Mazy}, {Merken}, {Montfort}, {Morbidelli}, {M{\"u}ller},
  {Nielbock}, {Okumura}, {Orfei}, {Ottensamer}, {Pezzuto}, {Popesso},
  {Putzeys}, {Regibo}, {Reveret}, {Royer}, {Sauvage}, {Schreiber}, {Stegmaier},
  {Schmitt}, {Schubert}, {Sturm}, {Thiel}, {Tofani}, {Vavrek}, {Wetzstein},
  {Wieprecht}, \& {Wiezorrek}}]{poglitsch10}
{Poglitsch}, A., {Waelkens}, C., {Geis}, N., {et~al.} 2010, \aap, 518, L2

\bibitem[{{Rebull} {et~al.}(2010){Rebull}, {Padgett}, {McCabe}, {Hillenbrand},
  {Stapelfeldt}, {Noriega-Crespo}, {Carey}, {Brooke}, {Huard}, {Terebey},
  {Audard}, {Monin}, {Fukagawa}, {G{\"u}del}, {Knapp}, {Menard}, {Allen},
  {Angione}, {Baldovin-Saavedra}, {Bouvier}, {Briggs}, {Dougados}, {Evans},
  {Flagey}, {Guieu}, {Grosso}, {Glauser}, {Harvey}, {Hines}, {Latter},
  {Skinner}, {Strom}, {Tromp}, \& {Wolf}}]{2010ApJS..186..259R}
{Rebull}, L.~M., {Padgett}, D.~L., {McCabe}, C.-E., {et~al.} 2010, \apjs, 186,
  259

\bibitem[{{Rebull} {et~al.}(2015){Rebull}, {Stauffer}, {Cody}, {G{\"u}nther},
  {Hillenbrand}, {Poppenhaeger}, {Wolk}, {Hora}, {Hernandez}, {Bayo}, {Covey},
  {Forbrich}, {Gutermuth}, {Morales-Calder{\'o}n}, {Plavchan}, {Song}, {Bouy},
  {Terebey}, {Cuillandre}, \& {Allen}}]{2015AJ....150..175R}
{Rebull}, L.~M., {Stauffer}, J.~R., {Cody}, A.~M., {et~al.} 2015, \aj, 150,
  doi:10.1088/0004-6256/150/6/175

\bibitem[{{Reynolds}(1976)}]{1976ApJ...203..151R}
{Reynolds}, R.~J. 1976, \apj, 203, 151

\bibitem[{{Robitaille} {et~al.}(2006){Robitaille}, {Whitney}, {Indebetouw},
  {Wood}, \& {Denzmore}}]{robitaille06}
{Robitaille}, T.~P., {Whitney}, B.~A., {Indebetouw}, R., {Wood}, K., \&
  {Denzmore}, P. 2006, \apjs, 167, 256

\bibitem[{{San Jos{\'e}-Garc{\'{\i}}a} {et~al.}(2013){San
  Jos{\'e}-Garc{\'{\i}}a}, {Mottram}, {Kristensen}, {van Dishoeck},
  {Y{\i}ld{\i}z}, {van der Tak}, {Herpin}, {Visser}, {McCoey}, {Wyrowski},
  {Braine}, \& {Johnstone}}]{2013AA...553A.125S}
{San Jos{\'e}-Garc{\'{\i}}a}, I., {Mottram}, J.~C., {Kristensen}, L.~E.,
  {et~al.} 2013, \aap, 553, A125

\bibitem[{{Santangelo} {et~al.}(2015){Santangelo}, {Murillo}, {Nisini},
  {Codella}, {Bruderer}, {Lai}, \& {van Dishoeck}}]{2015AA...581A..91S}
{Santangelo}, G., {Murillo}, N.~M., {Nisini}, B., {et~al.} 2015, \aap, 581, A91

\bibitem[{{Santangelo} {et~al.}(2012){Santangelo}, {Nisini}, {Giannini},
  {Antoniucci}, {Vasta}, {Codella}, {Lorenzani}, {Tafalla}, {Liseau}, {van
  Dishoeck}, \& {Kristensen}}]{santangelo12}
{Santangelo}, G., {Nisini}, B., {Giannini}, T., {et~al.} 2012, \aap, 538, A45

\bibitem[{{Santangelo} {et~al.}(2014){Santangelo}, {Nisini}, {Codella},
  {Lorenzani}, {Y{\i}ld{\i}z}, {Antoniucci}, {Bjerkeli}, {Cabrit}, {Giannini},
  {Kristensen}, {Liseau}, {Mottram}, {Tafalla}, \& {van
  Dishoeck}}]{2014AA...568A.125S}
{Santangelo}, G., {Nisini}, B., {Codella}, C., {et~al.} 2014, \aap, 568, A125

\bibitem[{{Saraceno} {et~al.}(1996){Saraceno}, {Andre}, {Ceccarelli},
  {Griffin}, \& {Molinari}}]{1996AA...309..827S}
{Saraceno}, P., {Andre}, P., {Ceccarelli}, C., {Griffin}, M., \& {Molinari}, S.
  1996, \aap, 309, 827

\bibitem[{{Sch{\"o}ier} {et~al.}(2005){Sch{\"o}ier}, {van der Tak}, {van
  Dishoeck}, \& {Black}}]{schoier05}
{Sch{\"o}ier}, F.~L., {van der Tak}, F.~F.~S., {van Dishoeck}, E.~F., \&
  {Black}, J.~H. 2005, \aap, 432, 369

\bibitem[{{Seidensticker} \& {Schmidt-Kaler}(1989)}]{1989AA...225..192S}
{Seidensticker}, K.~J., \& {Schmidt-Kaler}, T. 1989, \aap, 225, 192

\bibitem[{{Shirley}(2015)}]{2015PASP..127..299S}
{Shirley}, Y.~L. 2015, \pasp, 127, 299

\bibitem[{{Shirley} {et~al.}(2000){Shirley}, {Evans}, {Rawlings}, \&
  {Gregersen}}]{2000ApJS..131..249S}
{Shirley}, Y.~L., {Evans}, II, N.~J., {Rawlings}, J.~M.~C., \& {Gregersen},
  E.~M. 2000, \apjs, 131, 249

\bibitem[{{Spezzi} {et~al.}(2007){Spezzi}, {Alcal{\'a}}, {Frasca}, {Covino}, \&
  {Gandolfi}}]{2007AA...470..281S}
{Spezzi}, L., {Alcal{\'a}}, J.~M., {Frasca}, A., {Covino}, E., \& {Gandolfi},
  D. 2007, \aap, 470, 281

\bibitem[{{Stanke} {et~al.}(2006){Stanke}, {Smith}, {Gredel}, \&
  {Khanzadyan}}]{2006AA...447..609S}
{Stanke}, T., {Smith}, M.~D., {Gredel}, R., \& {Khanzadyan}, T. 2006, \aap,
  447, 609

\bibitem[{{Stetson}(1987)}]{1987PASP...99..191S}
{Stetson}, P.~B. 1987, \pasp, 99, 191

\bibitem[{{Storm} {et~al.}(2014){Storm}, {Mundy}, {Fern{\'a}ndez-L{\'o}pez},
  {Lee}, {Looney}, {Teuben}, {Rosolowsky}, {Arce}, {Ostriker}, {Segura-Cox},
  {Pound}, {Salter}, {Volgenau}, {Shirley}, {Chen}, {Gong}, {Plunkett},
  {Tobin}, {Kwon}, {Isella}, {Kauffmann}, {Tassis}, {Crutcher}, {Gammie}, \&
  {Testi}}]{2014ApJ...794..165S}
{Storm}, S., {Mundy}, L.~G., {Fern{\'a}ndez-L{\'o}pez}, M., {et~al.} 2014,
  \apj, 794, 165

\bibitem[{{Strai\v{z}ys} {et~al.}(1992){Strai\v{z}ys}, {\v{C}ernis},
  {Kazlauskas}, \& {Meistas}}]{1992BaltA...1..149S}
{Strai\v{z}ys}, V., {\v{C}ernis}, K., {Kazlauskas}, A., \& {Meistas}, E. 1992,
  Baltic Astronomy, 1, 149

\bibitem[{{Strai\v{z}ys} {et~al.}(1994){Strai\v{z}ys}, {Claria}, {Piatti}, \&
  {Kaslauskas}}]{1994BaltA...3..199S}
{Strai\v{z}ys}, V., {Claria}, J.~J., {Piatti}, A.~E., \& {Kaslauskas}, A. 1994,
  Baltic Astronomy, 3, 199

\bibitem[{{Stutz} {et~al.}(2008){Stutz}, {Rubin}, {Werner}, {Rieke}, {Bieging},
  {Keene}, {Kang}, {Shirley}, {Su}, {Velusamy}, \&
  {Wilner}}]{2008ApJ...687..389S}
{Stutz}, A.~M., {Rubin}, M., {Werner}, M.~W., {et~al.} 2008, \apj, 687, 389

\bibitem[{{Swinyard} {et~al.}(2014){Swinyard}, {Polehampton}, {Hopwood},
  {Valtchanov}, {Lu}, {Fulton}, {Benielli}, {Imhof}, {Marchili}, {Baluteau},
  {Bendo}, {Ferlet}, {Griffin}, {Lim}, {Makiwa}, {Naylor}, {Orton},
  {Papageorgiou}, {Pearson}, {Schulz}, {Sidher}, {Spencer}, {Wiel}, \&
  {Wu}}]{swinyard14}
{Swinyard}, B.~M., {Polehampton}, E.~T., {Hopwood}, R., {et~al.} 2014, \mnras,
  440, 3658

\bibitem[{{Tafalla} {et~al.}(2015){Tafalla}, {Bachiller}, {Lefloch},
  {Rodr{\'{\i}}guez-Fern{\'a}ndez}, {Codella}, {L{\'o}pez-Sepulcre}, \&
  {Podio}}]{2015AA...573L...2T}
{Tafalla}, M., {Bachiller}, R., {Lefloch}, B., {et~al.} 2015, \aap, 573, L2

\bibitem[{{Terebey} {et~al.}(1990){Terebey}, {Beichman}, {Gautier}, \&
  {Hester}}]{1990ApJ...362L..63T}
{Terebey}, S., {Beichman}, C.~A., {Gautier}, T.~N., \& {Hester}, J.~J. 1990,
  \apjl, 362, L63

\bibitem[{{van der Wiel} {et~al.}(2014){van der Wiel}, {Naylor}, {Kamp},
  {M{\'e}nard}, {Thi}, {Woitke}, {Olofsson}, {Pontoppidan}, {Di Francesco},
  {Glauser}, {Greaves}, \& {Ivison}}]{wiel14}
{van der Wiel}, M.~H.~D., {Naylor}, D.~A., {Kamp}, I., {et~al.} 2014, \mnras,
  444, 3911

\bibitem[{{van Dishoeck} {et~al.}(2011){van Dishoeck}, {Kristensen}, {Benz},
  {Bergin}, {Caselli}, {Cernicharo}, {Herpin}, {Hogerheijde}, {Johnstone},
  {Liseau}, {Nisini}, {Shipman}, {Tafalla}, {van der Tak}, {Wyrowski},
  {Aikawa}, {Bachiller}, {Baudry}, {Benedettini}, {Bjerkeli}, {Blake},
  {Bontemps}, {Braine}, {Brinch}, {Bruderer}, {Chavarr{\'{\i}}a}, {Codella},
  {Daniel}, {de Graauw}, {Deul}, {di Giorgio}, {Dominik}, {Doty}, {Dubernet},
  {Encrenaz}, {Feuchtgruber}, {Fich}, {Frieswijk}, {Fuente}, {Giannini},
  {Goicoechea}, {Helmich}, {Herczeg}, {Jacq}, {J{\o}rgensen}, {Karska},
  {Kaufman}, {Keto}, {Larsson}, {Lefloch}, {Lis}, {Marseille}, {McCoey},
  {Melnick}, {Neufeld}, {Olberg}, {Pagani}, {Pani{\'c}}, {Parise}, {Pearson},
  {Plume}, {Risacher}, {Salter}, {Santiago-Garc{\'{\i}}a}, {Saraceno},
  {St{\"a}uber}, {van Kempen}, {Visser}, {Viti}, {Walmsley}, {Wampfler}, \&
  {Y{\i}ld{\i}z}}]{vandishoeck11}
{van Dishoeck}, E.~F., {Kristensen}, L.~E., {Benz}, A.~O., {et~al.} 2011,
  \pasp, 123, 138

\bibitem[{{van Kempen} {et~al.}(2009{\natexlab{a}}){van Kempen}, {van
  Dishoeck}, {Salter}, {Hogerheijde}, {J{\o}rgensen}, \&
  {Boogert}}]{vankempen09}
{van Kempen}, T.~A., {van Dishoeck}, E.~F., {Salter}, D.~M., {et~al.}
  2009{\natexlab{a}}, \aap, 498, 167

\bibitem[{{van Kempen} {et~al.}(2009{\natexlab{b}}){van Kempen}, {van
  Dishoeck}, {G{\"u}sten}, {Kristensen}, {Schilke}, {Hogerheijde}, {Boland},
  {Menten}, \& {Wyrowski}}]{2009AA...507.1425V}
{van Kempen}, T.~A., {van Dishoeck}, E.~F., {G{\"u}sten}, R., {et~al.}
  2009{\natexlab{b}}, \aap, 507, 1425

\bibitem[{{van Kempen} {et~al.}(2010){van Kempen}, {Green}, {Evans}, {van
  Dishoeck}, {Kristensen}, {Herczeg}, {Mer{\'{\i}}n}, {Lee}, {J{\o}rgensen},
  {Bouwman}, {Acke}, {Adamkovics}, {Augereau}, {Bergin}, {Blake}, {Brown},
  {Carr}, {Chen}, {Cieza}, {Dominik}, {Dullemond}, {Dunham}, {Glassgold},
  {G{\"u}del}, {Harvey}, {Henning}, {Hogerheijde}, {Jaffe}, {Kim}, {Knez},
  {Lacy}, {Maret}, {Meeus}, {Meijerink}, {Mulders}, {Mundy}, {Najita},
  {Olofsson}, {Pontoppidan}, {Salyk}, {Sturm}, {Visser}, {Waters}, {Waelkens},
  \& {Y{\i}ld{\i}z}}]{vankempen10}
{van Kempen}, T.~A., {Green}, J.~D., {Evans}, N.~J., {et~al.} 2010, \aap, 518,
  L128

\bibitem[{{van Kempen et al.}(2010)}]{vankempen10short}
{van Kempen et al.} 2010, \aap, 518, L128

\bibitem[{{Vasta} {et~al.}(2012){Vasta}, {Codella}, {Lorenzani}, {Santangelo},
  {Nisini}, {Giannini}, {Tafalla}, {Liseau}, {van Dishoeck}, \&
  {Kristensen}}]{vasta12}
{Vasta}, M., {Codella}, C., {Lorenzani}, A., {et~al.} 2012, \aap, 537, A98

\bibitem[{{Visser} {et~al.}(2002){Visser}, {Richer}, \&
  {Chandler}}]{2002AJ....124.2756V}
{Visser}, A.~E., {Richer}, J.~S., \& {Chandler}, C.~J. 2002, \aj, 124, 2756

\bibitem[{{Visser} {et~al.}(2012){Visser}, {Kristensen}, {Bruderer}, {van
  Dishoeck}, {Herczeg}, {Brinch}, {Doty}, {Harsono}, \& {Wolfire}}]{visser12}
{Visser}, R., {Kristensen}, L.~E., {Bruderer}, S., {et~al.} 2012, \aap, 537,
  A55

\bibitem[{{Wampfler} {et~al.}(2011){Wampfler}, {Bruderer}, {Kristensen},
  {Chavarr{\'{\i}}a}, {Bergin}, {Benz}, {van Dishoeck}, {Herczeg}, {van der
  Tak}, {Goicoechea}, {Doty}, \& {Herpin}}]{wampfler11}
{Wampfler}, S.~F., {Bruderer}, S., {Kristensen}, L.~E., {et~al.} 2011, \aap,
  531, L16

\bibitem[{{Wampfler} {et~al.}(2013){Wampfler}, {Bruderer}, {Karska}, {Herczeg},
  {van Dishoeck}, {Kristensen}, {Goicoechea}, {Benz}, {Doty}, {McCoey},
  {Baudry}, {Giannini}, \& {Larsson}}]{wampfler13}
{Wampfler}, S.~F., {Bruderer}, S., {Karska}, A., {et~al.} 2013, \aap, 552, A56

\bibitem[{{Ward-Thompson} {et~al.}(2007){Ward-Thompson}, {Andr{\'e}},
  {Crutcher}, {Johnstone}, {Onishi}, \& {Wilson}}]{ward07}
{Ward-Thompson}, D., {Andr{\'e}}, P., {Crutcher}, R., {et~al.} 2007, Protostars
  and Planets V, 33

\bibitem[{Waskom {et~al.}(2014)Waskom, Botvinnik, Hobson, Cole, Halchenko,
  Hoyer, Miles, Augspurger, Yarkoni, Megies, Coelho, Wehner, cynddl, Ziegler,
  diego0020, Zaytsev, Hoppe, Seabold, Cloud, Koskinen, Meyer, Qalieh, \&
  Allan}]{michael_waskom_2014_12710}
Waskom, M., Botvinnik, O., Hobson, P., {et~al.} 2014, seaborn: v0.5.0 (November
  2014), , , doi:10.5281/zenodo.12710

\bibitem[{{Whittet} {et~al.}(1997){Whittet}, {Prusti}, {Franco}, {Gerakines},
  {Kilkenny}, {Larson}, \& {Wesselius}}]{1997AA...327.1194W}
{Whittet}, D.~C.~B., {Prusti}, T., {Franco}, G.~A.~P., {et~al.} 1997, \aap,
  327, 1194

\bibitem[{{Wu} {et~al.}(2007){Wu}, {Dunham}, {Evans}, {Bourke}, \&
  {Young}}]{2007AJ....133.1560W}
{Wu}, J., {Dunham}, M.~M., {Evans}, II, N.~J., {Bourke}, T.~L., \& {Young},
  C.~H. 2007, \aj, 133, 1560

\bibitem[{{Wu} {et~al.}(2009){Wu}, {Takakuwa}, \& {Lim}}]{2009ApJ...698..184W}
{Wu}, P.-F., {Takakuwa}, S., \& {Lim}, J. 2009, \apj, 698, 184

\bibitem[{{Wu} {et~al.}(2013){Wu}, {Polehampton}, {Etxaluze}, {Makiwa},
  {Naylor}, {Salji}, {Swinyard}, {Ferlet}, {van der Wiel}, {Smith}, {Fulton},
  {Griffin}, {Baluteau}, {Benielli}, {Glenn}, {Hopwood}, {Imhof}, {Lim}, {Lu},
  {Panuzzo}, {Pearson}, {Sidher}, \& {Valtchanov}}]{2013AA...556A.116W}
{Wu}, R., {Polehampton}, E.~T., {Etxaluze}, M., {et~al.} 2013, \aap, 556, A116

\bibitem[{{Yamamura} {et~al.}(2010){Yamamura}, {Makiuti}, {Ikeda}, {Fukuda},
  {Oyabu}, {Koga}, \& {White}}]{2010yCat.2298....0Y}
{Yamamura}, I., {Makiuti}, S., {Ikeda}, N., {et~al.} 2010, VizieR Online Data
  Catalog, 2298

\bibitem[{{Yang} {et~al.}(2017){Yang}, {Evans}, {Green}, {Dunham}, \&
  {J{\o}rgensen}}]{yang17}
{Yang}, Y.-L., {Evans}, II, N.~J., {Green}, J.~D., {Dunham}, M.~M., \&
  {J{\o}rgensen}, J.~K. 2017, \apj, 835, 259

\bibitem[{{Y{\i}ld{\i}z} {et~al.}(2012){Y{\i}ld{\i}z}, {Kristensen}, {van
  Dishoeck}, {Belloche}, {van Kempen}, {Hogerheijde}, {G{\"u}sten}, \& {van der
  Marel}}]{yildiz12}
{Y{\i}ld{\i}z}, U.~A., {Kristensen}, L.~E., {van Dishoeck}, E.~F., {et~al.}
  2012, \aap, 542, A86

\bibitem[{{Y{\i}ld{\i}z} {et~al.}(2013){Y{\i}ld{\i}z}, {Kristensen}, {van
  Dishoeck}, {San Jos{\'e}-Garc{\'{\i}}a}, {Karska}, {Harsono}, {Tafalla},
  {Fuente}, {Visser}, {J{\o}rgensen}, \& {Hogerheijde}}]{yildiz13}
---. 2013, \aap, 556, A89

\bibitem[{{Y{\i}ld{\i}z} {et~al.}(2015){Y{\i}ld{\i}z}, {Kristensen}, {van
  Dishoeck}, {Hogerheijde}, {Karska}, {Belloche}, {Endo}, {Frieswijk},
  {G{\"u}sten}, {van Kempen}, {Leurini}, {Nagy}, {P{\'e}rez-Beaupuits},
  {Risacher}, {van der Marel}, {van Weeren}, \&
  {Wyrowski}}]{2015AA...576A.109Y}
---. 2015, \aap, 576, A109

\bibitem[{{Young} \& {Evans}(2005)}]{2005ApJ...627..293Y}
{Young}, C.~H., \& {Evans}, II, N.~J. 2005, \apj, 627, 293

\bibitem[{{Young} {et~al.}(2004){Young}, {J{\o}rgensen}, {Shirley},
  {Kauffmann}, {Huard}, {Lai}, {Lee}, {Crapsi}, {Bourke}, {Dullemond},
  {Brooke}, {Porras}, {Spiesman}, {Allen}, {Blake}, {Evans}, {Harvey},
  {Koerner}, {Mundy}, {Myers}, {Padgett}, {Sargent}, {Stapelfeldt}, {van
  Dishoeck}, {Bertoldi}, {Chapman}, {Cieza}, {DeVries}, {Ridge}, \&
  {Wahhaj}}]{young04}
{Young}, C.~H., {J{\o}rgensen}, J.~K., {Shirley}, Y.~L., {et~al.} 2004, \apjs,
  154, 396

\bibitem[{{Young} {et~al.}(2005){Young}, {Harvey}, {Brooke}, {Chapman},
  {Kauffmann}, {Bertoldi}, {Lai}, {Alcal{\'a}}, {Bourke}, {Spiesman}, {Allen},
  {Blake}, {Evans}, {Koerner}, {Mundy}, {Myers}, {Padgett}, {Salinas},
  {Sargent}, {Stapelfeldt}, {Teuben}, {van Dishoeck}, \&
  {Wahhaj}}]{2005ApJ...628..283Y}
{Young}, K.~E., {Harvey}, P.~M., {Brooke}, T.~Y., {et~al.} 2005, \apj, 628, 283

\bibitem[{{Zacharias} {et~al.}(2004){Zacharias}, {Monet}, {Levine}, {Urban},
  {Gaume}, \& {Wycoff}}]{2004AAS...205.4815Z}
{Zacharias}, N., {Monet}, D.~G., {Levine}, S.~E., {et~al.} 2004, in Bulletin of
  the American Astronomical Society, Vol.~36, American Astronomical Society
  Meeting Abstracts, 1418

\end{thebibliography}

\end{document}